\pdfoutput=1
\documentclass[twocolumn,showkeys,showpacs,preprintnumbers,prd,superscriptaddress,nofootinbib]{revtex4-1}
\usepackage{graphicx}
\usepackage{epsf}
\usepackage{bm}
\usepackage{amsmath}
\usepackage{amsfonts}
\usepackage{amssymb}
\usepackage{epstopdf}
\usepackage{natbib}
\usepackage{hyperref}
\usepackage{color}
\usepackage{verbatim}
\usepackage{multirow}
\usepackage{bm}
\usepackage{hyperref}
\usepackage{url}
\usepackage{orcidlink}
\usepackage{wrapfig}
\usepackage{nicefrac, xfrac}
\usepackage{enumitem}
\usepackage[normalem]{ulem}

\definecolor{navyblue}{rgb}{0.0, 0.0, 0.5}
\definecolor{royalblue}{rgb}{0.25, 0.41, 0.88}
\definecolor{cadmiumgreen}{rgb}{0.0, 0.42, 0.24}
\definecolor{blue-violet}{rgb}{0.54, 0.17, 0.89}
\definecolor{darkviolet}{rgb}{0.58, 0.0, 0.83}
\definecolor{orange(colorwheel)}{rgb}{1.0, 0.5, 0.0}

\usepackage{hyperref}
\hypersetup{
    colorlinks=true, 
    linkcolor=royalblue, 
    citecolor=magenta}

\makeatletter\let\expandableinput\@@input\makeatother

\definecolor{WildStrawberry}{HTML}{EE2967}

\begin{document}

\title{Planck-PR4 anisotropy spectra show (better) consistency with General Relativity}

\author{Enrico Specogna}
\email{especogna1@sheffield.ac.uk}

\author{William Giar\`e}
\email{w.giare@sheffield.ac.uk}

\author{Eleonora Di Valentino}
\email{e.divalentino@sheffield.ac.uk}

\affiliation{School of Mathematical and Physical Sciences, University of Sheffield, Hounsfield Road, Sheffield S3 7RH, United Kingdom}

\begin{abstract}
\noindent We present the results from a series of analyses on two parametric tests of gravity that modify the growth of linear, sub-horizon matter perturbations in the $\Lambda$CDM model. The first test, known as the $(\mu,\Sigma)$ framework, modifies the Poisson and lensing equations from General Relativity (GR). The second test introduces the growth index $\gamma$, which directly affects the time evolution of matter density perturbations. Our study is motivated by results from the analysis of the Planck-PR3 2018 spectra, which indicate a preference for $\Sigma_0 \neq 0$ and $\gamma_0 > 0.55$, both of which deviate from the $\Lambda$CDM predictions at a significance level of $\sim 2.5\sigma$. To clarify the nature of these anomalous results and understand how the lensing anomaly fits into the picture, we analyze the most recent Planck-PR4 spectra extracted from the updated \texttt{NPIPE} maps. Overall, the Planck-PR4 data show better consistency with GR. The updated likelihood \texttt{Camspec} provides constraints on $\Sigma_0$ and $\gamma_0$ that are consistent with GR within $1.5\sigma$ and $2\sigma$, respectively. The updated likelihoods \texttt{HiLLiPoP} and \texttt{LoLLiPoP} show even closer agreement, with all parameter values consistent with a $\Lambda$CDM cosmology within $1\sigma$. This enhanced consistency is closely correlated with the lensing anomaly. Across the different likelihoods, the tendency of $\Sigma_0$ and $\gamma_0$ to drift towards non-standard values matches the observed preference for $A_L > 1$, both of which are significantly reduced or disappear within the Planck-PR4 data.

\end{abstract}



\maketitle


\section{Introduction}
\label{sec:introduction}

At the time of writing, our best understanding of the Universe is summarized by the standard $\Lambda$CDM model of cosmology. This model stands on two pillars of well-established physics: General Relativity (GR), which governs gravitational interactions, and the Standard Model of particle physics (SM), which describes all other fundamental interactions in nature. However, it also relies on three essential components — Inflation, Dark Matter, and Dark Energy — that are crucial for explaining numerous observations ranging from the Cosmic Microwave Background (CMB) radiation to the Large Scale Structure (LSS) of the Universe~\cite{SupernovaSearchTeam:1998fmf,SupernovaCosmologyProject:1998vns,SupernovaSearchTeam:2001qse,SDSS:2003eyi,SDSS:2003lnz,SupernovaSearchTeam:2003cyd,SupernovaCosmologyProject:2003dcn,SDSS:2004kqt,Feng:2004ad,SupernovaSearchTeam:2004lze,SNLS:2005qlf,SDSS:2005xqv,Eisenstein:2006nk,SDSS:2006lmn,Sahni:2006pa,ESSENCE:2007acn,Vikhlinin:2008ym,Stern:2009ep,Sherwin:2011gv,WMAP:2012fli,WMAP:2012nax,BOSS:2012dmf,deJong:2012zb,BOSS:2013rlg,Weinberg:2013agg,BOSS:2013uda,BOSS:2014hwf,SDSS:2014iwm,BOSS:2014hhw,Ross:2014qpa,Moresco:2016mzx,Moresco:2016nqq,Rubin:2016iqe,BOSS:2016wmc,DES:2016jjg,Haridasu:2017lma,DES:2017qwj,Pan-STARRS1:2017jku,Planck:2018nkj,Planck:2018vyg,Gomez-Valent:2018gvm,Yang:2019fjt,ACT:2020frw,ACT:2020gnv,eBOSS:2020yzd,Nadathur:2020kvq,Rose:2020shp,DiValentino:2020evt,eBOSS:2020yzd,KiDS:2020suj,KiDS:2020ghu,SPT-3G:2021eoc,DES:2021wwk,Moresco:2022phi,DES:2022ccp,Brout:2022vxf,ACT:2023kun,Kilo-DegreeSurvey:2023gfr,DESI:2024uvr,DESI:2024kob,DES:2024tys,DES:2024upw,DES:2024hip}. Nevertheless, a comprehensive physical interpretation of these three components remains elusive, both theoretically and experimentally.

Together with these longstanding foundational problems, in the last ten years a few statistically significant tensions and anomalies have challenged the standard cosmological picture. The most puzzling issue is the so-called Hubble tension~\cite{Verde:2019ivm,DiValentino:2020zio,DiValentino:2021izs,Perivolaropoulos:2021jda,Schoneberg:2021qvd,Shah:2021onj,Abdalla:2022yfr,DiValentino:2022fjm,Kamionkowski:2022pkx,Giare:2023xoc,Hu:2023jqc,Verde:2023lmm,DiValentino:2024yew,Perivolaropoulos:2024yxv}: at its core, the value of the Hubble constant ($H_0$) derived from CMB measurements within the $\Lambda$CDM framework~\cite{Planck:2018vyg} differs by more than 5$\sigma$ from direct local measurements reported by the SH0ES team~\cite{Riess:2021jrx, Breuval:2024lsv,Murakami:2023xuy}.\footnote{Over the years, various possibilities have been investigated to explain the Hubble tension, both in terms of observational systematics and new physics beyond the standard cosmological model. Without aiming for a comprehensive review, for discussion on the former we refer to Refs.~\cite{Efstathiou:2020wxn,Mortsell:2021nzg,Mortsell:2021tcx,Riess:2021jrx,Sharon:2023ioz,Murakami:2023xuy,Riess:2023bfx,Bhardwaj:2023mau,Brout:2023wol,Dwomoh:2023bro,Uddin:2023iob,Riess:2024ohe,Freedman:2024eph,Riess:2024vfa} while for examples of models of new physics and their implications, see, e.g., Refs.~\cite{Anchordoqui:2015lqa,Karwal:2016vyq,Zhao:2017urm,Benetti:2017juy,Mortsell:2018mfj,Vagnozzi:2018jhn,Kumar:2018yhh,Yang:2018euj,Banihashemi:2018oxo,Guo:2018ans,Graef:2018fzu,Banihashemi:2018has,Agrawal:2019lmo,Li:2019yem,Yang:2019nhz,Vagnozzi:2019ezj,Visinelli:2019qqu,DiValentino:2019ffd,Escudero:2019gvw,DiValentino:2019jae,Niedermann:2019olb,Sakstein:2019fmf,Ye:2020btb,Hogg:2020rdp,Ballesteros:2020sik,Alestas:2020mvb,Jedamzik:2020krr,Ballardini:2020iws,DiValentino:2020evt,Banerjee:2020xcn,Niedermann:2020dwg,Gonzalez:2020fdy,Braglia:2020auw,RoyChoudhury:2020dmd,Brinckmann:2020bcn,Alestas:2020zol,Gao:2021xnk,Renzi:2021xii,Alestas:2021xes,Karwal:2021vpk,Cyr-Racine:2021oal,Akarsu:2021fol,Niedermann:2021ijp,Saridakis:2021xqy,Sen:2021wld,Herold:2021ksg,Odintsov:2022eqm,Heisenberg:2022lob,Heisenberg:2022gqk,Sharma:2022ifr,Ren:2022aeo,Nunes:2022bhn,Nojiri:2022ski,Schoneberg:2022grr,Joseph:2022jsf,Gomez-Valent:2022bku,Moshafi:2022mva,Odintsov:2022umu,Banerjee:2022ynv,Alvarez:2022wef,Ge:2022qws,Akarsu:2022typ,Gangopadhyay:2022bsh,Schiavone:2022wvq,Gao:2022ahg,Brinckmann:2022ajr,Khodadi:2023ezj,Dahmani:2023bsb,Ben-Dayan:2023rgt,deCruzPerez:2023wzd,Ballardini:2023mzm,Yao:2023ybs,Gangopadhyay:2023nli,Zhai:2023yny,SolaPeracaula:2023swx,Gomez-Valent:2023hov,Ruchika:2023ugh,Vagnozzi:2023nrq,Adil:2023exv,Frion:2023xwq,Akarsu:2023mfb,Escamilla:2023oce,Petronikolou:2023cwu,Sharma:2023kzr,Ben-Dayan:2023htq,Ramadan:2023ivw,Fu:2023tfo,Efstathiou:2023fbn,Montani:2023ywn,Lazkoz:2023oqc,Forconi:2023hsj,Sebastianutti:2023dbt,Benisty:2024lmj,Stahl:2024stz,Shah:2024rme,Giare:2024smz,Giare:2024akf,Giare:2024ytc,Montani:2024pou,Co:2024oek,Akarsu:2024eoo,Yadav:2024duq,Nozari:2024wir,Dwivedi:2024okk,Montani:2024xys,Escamilla:2024xmz,Perivolaropoulos:2024yxv,Yao:2024kex,Giare:2024syw,Toda:2024ncp,Pedrotti:2024kpn,Poulin:2024ken,Simon:2024jmu}.} Additionally, other less significant problems have come to light, such as a well-known 2-3$\sigma$ disagreement between the values of $S_8 = \sigma_8 \sqrt{\Omega_m/0.3}$ derived from CMB measurements by Planck~\cite{Planck:2018vyg} and those inferred from large-scale structure observations~\cite{DiValentino:2020vvd, Nunes:2021ipq, DES:2021bvc, KiDS:2020suj, Li:2023tui, Dalal:2023olq, Kilo-DegreeSurvey:2023gfr, Harnois-Deraps:2024ucb,Armijo:2024ujo,ACT:2024nrz,DES:2024xvm,Lau:2024xrd,Akarsu:2024hsu}.

Foundational problems and longstanding tensions raise the question of whether the $\Lambda$CDM model accurately represents the true paradigm of the Universe or merely serves as a phenomenological, data-driven approximation to a more fundamental scenario that has yet to be fully understood. In other words, could it be that as our data becomes increasingly precise, our parameterization of more than 95\% of the energy density of the Universe — primarily based on simplicity, which, after all, highlights our ignorance regarding these phenomena — becomes inadequate?

Taking a step back and (re)considering the big picture, we cannot help but note that the need to introduce the three unknown components underpinning $\Lambda$CDM cosmology is intricately rooted in the challenge of explaining observations based on the predictions of GR when assuming a SM particle content in the Universe. For instance, the current accelerated expansion of the Universe sharply contrasts with GR’s predictions when only the matter content expected from the SM is considered, leading to the introduction of a cosmological constant term in the Einstein field equations, which is not free from conceptual problems~\cite{Sahni:1999gb,Carroll:2000fy,Peebles:2002gy,Padmanabhan:2002ji,Copeland:2006wr,Caldwell:2009ix,Li:2011sd,Martin:2012bt,Weinberg:1988cp,Krauss:1995yb,Sahni:1999gb,Weinberg:2000yb,Padmanabhan:2002ji,Sahni:2002kh,Yokoyama:2003ii,Nobbenhuis:2004wn,Burgess:2013ara,Joyce:2014kja,Bull:2015stt,Wang:2016lxa,Brustein:1992nk,Witten:2000zk,Kachru:2003aw,Polchinski:2006gy,Danielsson:2018ztv,Zlatev:1998tr,Pavon:2005yx,Martin:2012bt,Velten:2014nra} and has recently been challenged by observations~\cite{DESI:2024mwx,Giare:2024gpk}.\footnote{Readers interested in the latest developments regarding the Dark Energy interpretation of the first-year data release from the DESI collaboration~\cite{DESI:2024mwx} can refer to Refs.~\cite{Wang:2024pui,Efstathiou:2024xcq,Patel:2024odo,Liu:2024gfy,Colgain:2024xqj,Giare:2024ocw} for discussions on potential systematic effects in the data. For physical interpretations of these results, see, e.g., Refs.~\cite{Cortes:2024lgw,Orchard:2024bve,Chudaykin:2024gol,Notari:2024rti,Gialamas:2024lyw,Wang:2024hwd,DESI:2024kob,Wang:2024dka,Giare:2024smz,Carloni:2024zpl,Tada:2024znt,Yin:2024hba,Luongo:2024fww,Park:2024jns,Shlivko:2024llw,Ye:2024ywg,Li:2024qso,Yang:2024kdo,Giare:2024gpk,DESI:2024aqx,Chan-GyungPark:2024brx,Wolf:2024stt,Pourojaghi:2024bxa,Sabogal:2024yha,Jiang:2024xnu,Dinda:2024ktd,Hernandez-Almada:2024ost,Menci:2024hop,Ramadan:2024kmn,Berghaus:2024kra,Qu:2024lpx,Adolf:2024twn, Bhattacharya:2024hep, Bhattacharya:2024kxp}.} Similarly, the CMB and LSS observations contrast with what is expected in GR when only the matter content predicted by the SM is taken into account, providing indirect evidence for cold dark matter. Finally, historically, inflation was introduced to avoid extremely fine-tuned initial conditions in the early Universe~\cite{Guth:1980zm,Linde:1981mu,Albrecht:1982wi,Vilenkin:1983xq} that were necessary to achieve a present-day Universe consistent with observations as it evolved over time — an evolution largely shaped by gravity.

Consequently, when it comes to \textit{hazard} a physical interpretation for these three ingredients, we can follow two main approaches. One option is to work on the right-hand side of the Einstein equations, adding new forms of energy-momentum content to the stress-energy tensor $T_{\mu\nu}$ beyond what the SM predicts (such as introducing scalar fields to drive acceleration in the early or late Universe or a perfect fluid of cold dark matter particles to account for the observed structures). The second option is to work on the left-hand side of the Einstein equations, modifying the Einstein tensor $G_{\mu\nu}$, thereby altering the underlying theory of gravity. In fact, some benchmark scenarios of inflation (such as Starobinsky inflation~\cite{Starobinsky:1980te}), along with many models attempting to interpret dark matter and dark energy, are fundamentally rooted in modified gravity (MG) theories. 

With this premise, it goes without saying that testing whether gravity follows the predictions of GR on cosmological scales is of paramount importance. In the epoch of precision cosmology, CMB observations are a powerful tool for identifying any signatures of MG on such scales. As a matter of fact, the shape of the CMB angular spectra of temperature and polarization anisotropies arises from the dynamics of primordial density fluctuations and is largely determined by the gravitational forces experienced by the coupled photon-baryon fluid before decoupling. Unlike the evolution of late-time structures, such as galaxies or clusters, where nonlinear effects complicate the picture, the CMB primarily involves linear perturbation theory, governed by a system of linearized Einstein-Boltzmann equations. This makes CMB observations largely free from complications introduced by astrophysical processes (aside from modeling foreground contamination). Moreover, the wide range of angular scales covered by the CMB --- from the largest observable scales (such as the quadrupole and dipole) to the smallest scales measured by current and forthcoming ground-based telescopes, such as the Atacama Cosmology Telescope~\cite{ACT:2020gnv,ACT:2023kun,ACT:2023dou} (ACT) and the South Pole Telescope~\cite{SPT-3G:2021eoc,SPT-3G:2022hvq} (SPT) --- allows gravity to be tested across an extensive range of scales through a multitude of different tests.

A particularly interesting test we can perform using CMB data involves assessing whether the gravitational deflection, or lensing, experienced by CMB photons along their paths~\cite{Lewis:2006fu} follows the predictions of GR within a standard cosmological model. At arcminute scales, lensing deflections distort the observed image of the CMB fluctuations, imprinting a distinctive non-Gaussian four-point correlation function (or trispectrum) in both the temperature and polarization anisotropies~\cite{Lewis:2006fu}.\footnote{The CMB lensing signal carries crucial complementary information about late-time processes affecting structure formation, ranging from neutrinos and thermal relics~\cite{Giare:2020vzo,DiValentino:2021imh,DEramo:2022nvb,DiValentino:2022edq,Giare:2023aix} to dark energy and its dynamical properties~\cite{Planck:2018lbu,Ye:2023zel,ACT:2023skz,ACT:2023ipp,ACT:2023kun,ACT:2024okh,Sailer:2024coh,ACT:2024npz}.} Additionally, although gravitational lensing does not alter the overall distribution of primary CMB anisotropies, it leaves distinctive signatures in the spectra of temperature and polarization anisotropies through a convolution between the latter and the CMB lensing potential spectrum~\cite{Lewis:2006fu}. To test whether these effects align with the theoretical predictions, one can introduce a phenomenological parameter $A_L$~\cite{Calabrese:2008rt} which must be $A_L = 1$ as predicted by GR within a $\Lambda$CDM cosmology. This parameter can be used to arbitrarily tune the extent to which $C^{\varphi \varphi}_{\ell}$ (the spectrum of the lensing potential $\varphi$, a redefinition of the Weyl potential that allows for $\vec{\alpha} = \nabla \varphi$, where $\vec{\alpha}$ is the deflection angle of light~\cite{Lewis:2006fu}) affects the spectrum of primary CMB anisotropies~\cite{Calabrese:2008rt}: $ C^{\varphi \varphi}_{\ell} \longrightarrow A_L C^{\varphi \varphi}_{\ell}.$

Interestingly, since the 2013 data release~\cite{Planck:2013pxb, DiValentino:2013mt, Planck:2015fie, Addison:2015wyg, Planck:2016tof, Renzi:2017cbg, DiValentino:2017rcr, Planck:2018vyg, DiValentino:2019qzk, Addison:2023fqc}, the temperature and polarization power spectra of Planck have shown $A_L$ to be greater than unity.\footnote{The presence of the lensing anomaly may hold significant implications when constraining scenarios beyond $\Lambda$CDM, as we argue in this paper in relation to MG scenarios. However, another well-documented example pertains to the curvature density parameter, $\Omega_k$. Since an excess of lensing is naively expected with a higher abundance of cold dark matter, the observed lensing anomaly can be recast into a preference for a closed Universe, as discussed in several recent (and not so recent) works~\cite{Park:2017xbl,Handley:2019tkm,DiValentino:2019qzk,Efstathiou:2020wem,DiValentino:2020hov,Benisty:2020otr,Vagnozzi:2020rcz,Vagnozzi:2020dfn,DiValentino:2020kpf,Yang:2021hxg,Cao:2021ldv,Dhawan:2021mel,Dinda:2021ffa,Gonzalez:2021ojp,Akarsu:2021max,Cao:2022ugh,Glanville:2022xes,Bel:2022iuf,Yang:2022kho,Stevens:2022evv,Favale:2023lnp,Giare:2023ejv}.} To be more quantitative, in the Planck-PR3 2018 data release a $2.8\sigma$ deviation from $A_L = 1$ is found ($A_L = 1.180 \pm 0.065$), see Ref.~\cite{Planck:2018vyg}.\footnote{All the constraints reported in the paper will be at 68\% CL, unless otherwise specified.}

To address the lensing anomaly, a significant range of physical interpretations involving extensions to GR has been explored. In the spirit of hitting multiple targets with the same arrow, many of these approaches are also proposed as promising alternatives for explaining the Universe's late-time accelerated expansion while alleviating other persistent tensions observed across various cosmological and astrophysical datasets~\cite{DiValentino:2021izs, Abdalla:2022yfr, Ishak:2018his, Heisenberg:2018vsk, CANTATA:2021asi, Nojiri:2017ncd}. 

However, when it comes to drawing general conclusions about this plethora of extended theories of gravity, a concrete difficulty arises from the fact that in many MG models, the equations describing the evolution of cosmological perturbations deviate significantly from those of GR, and these deviations can vary among different models. Therefore, parametric tests have been proposed to recast these varying behaviors and model effects in both the dynamics of cosmological perturbations and the growth of cosmic structures. Of particular relevance is the so-called $(\mu,\Sigma)$ framework~\cite{Zhang:2007nk,Amendola:2007rr,Zhao:2008bn}, where $\mu(k,a)$ and $\Sigma(k,a)$ are two arbitrary functions (both null in GR) of the wavenumber $k$ and the scale factor $a$ that alter the predictions for the Poisson and lensing equations. Yet another parametric test of gravity consists in considering the growth index $\gamma$, predicted to be $\gamma = 0.55$ in GR, which captures any potential deviations in the growth of perturbations~\cite{Linder:2005in, Linder:2007hg, Nguyen:2023fip, Specogna:2023nkq}.

Interestingly, Planck-PR3 data measure $\Sigma_0$ (i.e., the current value of $\Sigma$ when it is assumed constant in $k$) to be $\Sigma_0 = 0.27^{+0.15}_{-0.13}$ — over $2\sigma$ away from the null value expected within GR~\cite{Planck:2018vyg}. Regarding the $\gamma$ parameter, the Planck-PR3 data, in combination with Redshift Space Distortions data, yield $\gamma_0 = 0.639^{+0.024}_{-0.025}$~\cite{Nguyen:2023fip}. If taken at face value, these results would provide additional hints for deviations from GR and, more broadly, from a canonical model of structure formation. However, several caveats are in order:

\begin{itemize}

\item Firstly, CMB data show a strong correlation between $A_{\rm L}$ and the $(\mu,\Sigma)$ or $\gamma$ parameters -- see, e.g., Fig.~2 of Ref.~\cite{DiValentino:2015bja}. This correlation recasts the lensing anomaly as an approximate $2\sigma$ preference for $\Sigma_0 \ne 0$ in the Planck data~\cite{Planck:2015bue, DiValentino:2015bja, Planck:2018vyg}, or equivalently (part of) the $4\sigma$ preference for $\gamma \neq 0.55$.

\item Secondly, no indication of a lensing excess has been found in CMB experiments other than Planck, such as ACT and SPT. The value of $A_L$ inferred from these experiments consistently aligns with $A_L = 1$ well within $2$ standard deviations~\cite{ACT:2020gnv,SPT-3G:2022hvq}. Somewhat unsurprisingly, no significant indications of MG dynamics or suppression of structure growth (i.e., $\Sigma_0 \neq 0$ or $\gamma > 0.55$) have been found in the analysis of these ground-based telescope data either, see Refs.~\cite{Specogna:2023nkq,Andrade:2023pws}.

\item Since 2018, the Planck data has undergone extensive reanalyses. The updated Planck-PR4 (\texttt{NPIPE}) CMB maps feature significant advancements, such as enhanced sky coverage at high frequencies, improved processing of time-ordered data, and an approximate 8\% increase in the data used for lensing trispectrum reconstruction~\cite{Planck:2020olo,Carron:2022eyg}. Recent releases of new likelihoods from both \texttt{Camspec}~\cite{Rosenberg:2022sdy} and \texttt{HiLLiPoP}~\cite{Tristram:2023haj} have shown that $A_L$ can be measured to be near unity within the Planck-PR4 (\texttt{NPIPE}) data.\footnote{See, e.g., recent Refs.~\cite{Efstathiou:2023fbn,Chatrchyan:2024xjj,Allali:2024aiv,RoyChoudhury:2024wri} for updated constraints derived from Planck-PR4 based likelihoods on different extended cosmological scenarios.} This lends support to the hypothesis that the lensing anomaly observed in the Planck-PR3 data may simply result from observational systematics, as already argued by ACT and SPT.

\end{itemize}

Fitting all the pieces into place, it seems natural to question whether, and to what extent, the new Planck-PR4 likelihoods can support or relax the indications of MG related to a non-vanishing $\Sigma$ or the suppression of structure growth captured by the parameter $\gamma$. More broadly, this study aims to derive up-to-date constraints on parametric tests of gravity within these two frameworks, using the updated Planck-PR4 likelihoods \footnote{After our work presented here was shared with the community on \texttt{arXiv}, the DESI collaboration also released a series of updated constraints on the $\mu,\Sigma$ parametrization of modified gravity using the \texttt{HiLLiPoP,\,LoLLiPoP} likelihoods \cite{DESI:2024yrg,DESI:2024hhd}, and a different modified gravity software (\texttt{ISiTGR} in \cite{DESI:2024yrg,DESI:2024hhd} instead of \texttt{MGCAMB + Cobaya}).}, and to clarify their strict connection with the lensing anomaly.

The paper is structured as follows. In Sec.~\ref{sec:parametrisations}, we introduce the two parametrizations of gravity under study. In Sec.~\ref{sec:data}, we present the data and the methodology used. In Sec.~\ref{sec:res}, we interpret our results, and in Sec.~\ref{sec:conclusions}, we derive our conclusions.

\section{Parametric Modified Gravity and Perturbation Theory}
\label{sec:parametrisations}

In first-order cosmological perturbation theory, and in the Newtonian gauge, the two scalar potentials $\Psi$ and $\Phi$ perturb the time and spatial components of the zeroth-order flat FLRW metric as follows~\cite{Ma:1995ey}:
\begin{equation}
\label{eq:metric}
    {\rm d}s^2 = a^2[-(1+2\Psi){\rm d}\tau^2+(1-2\Phi){\rm d}{\bf x}^2],
\end{equation}
where ${\rm d}\tau={\rm d}t/a$ is the conformal time. To fully study the perturbations of non-relativistic matter, we need to introduce two further scalars: the density contrast of the matter fluid, $\delta$, and the divergence of its velocity, $v$. In particular, the relations between the metric potentials and $\delta$ are set by the specific theory of gravity considered. In Fourier space, GR gives us the following equalities, representing the Poisson and lensing equations, respectively:
\begin{equation}
    \label{eq:poisson_equation}
    k^2 \Psi = -4\pi G a^2 \rho \Delta,
\end{equation}
\begin{equation}
    \label{eq:weyl_equation}
    k^2(\Psi + \Phi) = -8\pi G a^2 \rho \Delta,
\end{equation}
where $\Delta \equiv \rho \delta + 3\cfrac{\mathcal{H}}{k}(\rho + P)v$ is the comoving density contrast, which depends on the conformal Hubble function $\mathcal{H}$, the fluid's average density $\rho$, and its pressure $P$.

\subsection{The $(\mu,\Sigma)$ Framework}

In MG theories, we often see that Eqs.~(\ref{eq:poisson_equation}, \ref{eq:weyl_equation}) are modified in different ways, usually by the presence of some additional dependence on the scale factor $a$ or the wavenumber $k$~\cite{Hojjati:2011ix}. We can capture such differences in the Poisson and lensing equations within a single, model-independent parametrization, the $(\mu,\Sigma)$ framework~\cite{Caldwell:2007cw, Amendola:2007rr, Hu:2007pj}:
\begin{equation}
    \label{eq:poisson_equation_pert}
    k^2 \Psi = -4\pi G a^2 (\mu+1)\rho \Delta,
\end{equation}
\begin{equation}
    \label{eq:weyl_equation_pert}
    k^2(\Psi + \Phi) = -8\pi G a^2 (\Sigma+1)\rho \Delta,
\end{equation}
where $\mu$ and $\Sigma$ are two arbitrary functions of $a$ and $k$.

As long as we treat linear ($\delta \ll 1$) and sub-horizon ($k \gtrsim \mathcal{H}$) perturbations, Eq.~(\ref{eq:poisson_equation_pert}) represents an adequate description of those models where, for instance, Newton's constant of gravity gains a dependence on time and/or scale that can be introduced through $\mu$: $G \longrightarrow G_\text{eff} \equiv (\mu+1) G$. Similarly, because the Weyl potential $\Psi+\Phi$ directly affects weak lensing~\cite{Zhang:2007nk, DES:2022ccp}, Eq.~(\ref{eq:weyl_equation_pert}) encompasses any deviations in the way gravity deflects light as predicted in GR. However, in models like MOND, which entail extra scalar and tensor perturbations and therefore require more equations than just Eqs.~(\ref{eq:poisson_equation}) and~(\ref{eq:weyl_equation}) and the conservation equations for matter perturbations in GR, this parametrization is not sufficient to describe the theory's growth of structure~\cite{Zhang:2007nk}.\\
More concretely, but not exhaustively, a few examples of what $\mu$ and $\Sigma$ look like in theories of MG can be found in DGP gravity~\cite{Koyama:2005kd}, $f(R)$ theories (where $G_{\rm eff} \sim (1+\frac{df}{dR})^{-1}$~\cite{Tsujikawa:2007gd}), or the Brans-Dicke theory~\cite{Bonvin:2022tii}.

In this work, however, we do not consider any scale factor or wavenumber dependence of the $(\mu,\Sigma)$ functions, restraining ourselves to measuring their current, fixed values: $\mu_0, \Sigma_0$. A detection of either $\mu_0 \neq 0$ or $\Sigma_0 \neq 0$ will indicate a deviation from the $\Lambda$CDM model.

\subsection{The growth index $\gamma$ }

The $(\mu,\Sigma)$ framework is just one example of the MG parametrizations existing in the literature. If we still consider the linear, sub-horizon regime of perturbations, we can also look for deviations away from GR through the so-called \textit{growth factor} $\gamma$~\cite{Linder:2005in, Wang:1998gt}. We write the growth rate $f(a)$ to be:
\begin{equation}
    \label{eq:growth_rate}
    f(a) = \cfrac{d\ln \delta}{d\ln a}.
\end{equation}
Then, $\gamma$ can be defined through the following ansatz, which one can choose to solve Eq.~(\ref{eq:growth_rate}):
\begin{equation}
    \label{eq:growth_factor}
    f(a) = \Omega_m^\gamma,
\end{equation}
where $\Omega_m$ is the density fraction of non-relativistic matter.

According to this definition, since MG models often seem to predict the same expansion history (i.e., they are degenerate in the time evolution of $\Omega_m$), any difference in the growth rate of matter perturbations predicted by alternative theories of gravity has to manifest itself through $\gamma$, which should deviate away from its $\Lambda$CDM model prediction: $\gamma \approx 0.55$ to $0.1\%$ accuracy~\cite{Linder:2005in, Linder:2007hg}. Once again, we can list a few concrete examples for the values taken by $\gamma$ in models other than $\Lambda$CDM. For instance, $\gamma \approx 0.68$ in DGP gravity~\cite{Linder:2005in, Linder:2007hg}, with no more than $2\%$ variation if assumed constant in time. For $f(R)$ theories where $\gamma$ does not suffer excessive dispersion (i.e., dependence on $k$), $\gamma \approx 0.4$~\cite{Tsujikawa:2009ku}, even though it can still show a non-negligible time dependence, up to $\gamma_0' = -0.2$ today~\cite{Gannouji:2008wt}.

As done with the $(\mu,\Sigma)$ parametrization described above, we choose not to assume any time or scale dependence for $\gamma$, opting instead to measure its fixed, current value $\gamma_0$. \footnote{Before going any further, we intend to clarify the nomenclature adopted throughout this article further. We refer to the $(\mu,\Sigma)$ framework or the $\gamma$ model to denote the two generic frameworks introduced in this section. However, as explained, we focus on the case where $\Sigma$ and $\gamma$ do not have any explicit time or scale dependence. Therefore, these functions in our paper are just constants (in time and scale). We refer to them with the subscript 0, which not only denotes their present-day value but also indicates their value at any epoch as they do not change over time.}


\section{Likelihoods and Sampling Methodology}
\label{sec:data}

Since we are primarily interested in re-evaluating the various hints for MG and their connection to the lensing anomaly — all originally identified in the 2018 Planck-PR3 data release~\cite{Planck:2018vyg} — our analysis focuses on Planck CMB measurements of temperature and polarization anisotropies, covering both the PR3 likelihoods and the more recently updated Planck likelihood codes. Specifically, the Planck likelihoods we use are named as follows.
\begin{itemize}[leftmargin=*]
    \item \textbf{Plik;} to reproduce the baseline results documented in the literature (and summarized in the introduction), we begin by considering the \texttt{Plik} likelihood for high-$\ell$ TT, TE, EE PR3 spectra, along with the \texttt{Commander} and \texttt{SimAll} likelihoods for the low-$\ell$ temperature and polarization spectra from the same data release~\cite{Planck:2018vyg}. Specifically, we use:
    \begin{itemize}
    \item the high-$\ell$ \texttt{Plik} likelihood~\cite{Planck:2018vyg} for the TT spectrum in the multipole range $30 \leq \ell \leq 2508$, as well as for the TE and EE spectra at $30 \leq \ell \leq 1996$;
    \item the low-$\ell$ \texttt{Commander} likelihood~\cite{Planck:2018vyg} for the TT spectrum in the multipole range $2 \leq \ell \leq 29$;
    \item the low-$\ell$ \texttt{SimAll} likelihood~\cite{Planck:2018vyg} for the EE spectrum at $2 \leq \ell \leq 29$.
    \end{itemize}
    For simplicity, we refer to this combination as \textbf{Plik}.

    \item \textbf{Camspec;} we replace the Planck-PR3 \texttt{Plik} high-$\ell$ TTTEEE likelihood with the more recent \texttt{Camspec} likelihood~\cite{Rosenberg:2022sdy}, based on the Planck-PR4 \texttt{NPIPE} data release~\cite{Planck:2020olo,Carron:2022eyg}. However, we retain the same low-$\ell$ likelihoods from the \textbf{Plik} combination. Therefore, we use: 
    \begin{itemize} 
    \item the high-$\ell$ \texttt{Camspec} likelihood~\cite{Rosenberg:2022sdy} for the TT, TE, and EE spectra in the multipole range $30 \leq \ell \lesssim 2500$; 
    \item the low-$\ell$ \texttt{Commander} likelihood~\cite{Planck:2018vyg} for the TT spectrum in the multipole range $2 \leq \ell \leq 29$; 
    \item the low-$\ell$ \texttt{SimAll} likelihood~\cite{Planck:2018vyg} for the EE spectrum at $2 \leq \ell \leq 29$. 
    \end{itemize} 
    We refer to this combination as \textbf{Camspec}.

    \item \textbf{Camspec (TT-only);} we use the recent \texttt{Camspec} likelihood~\cite{Rosenberg:2022sdy}, based on the Planck-PR4 \texttt{NPIPE} data release~\cite{Planck:2020olo, Carron:2022eyg}, for the high-$\ell$ temperature anisotropy spectrum, excluding TE and EE information at $\ell > 30$. However, we retain the same low-$\ell$ likelihoods from the \textbf{Plik} combination. Therefore, we use:
    \begin{itemize}
    \item the high-$\ell$ \texttt{Camspec} likelihood~\cite{Rosenberg:2022sdy} for the TT spectrum in the multipole range $30 \leq \ell \lesssim 2500$;
    \item the low-$\ell$ \texttt{Commander} likelihood~\cite{Planck:2018vyg} for the TT spectrum in the multipole range $2 \leq \ell \leq 29$;
    \item the low-$\ell$ \texttt{SimAll} likelihood~\cite{Planck:2018vyg} for the EE spectrum at $2 \leq \ell \leq 29$.
    \end{itemize}
    We refer to this combination as \textbf{Camspec (TT-only)}.
    
    \item \textbf{HiLLiPoP;} we replace the Planck-PR3 \texttt{Plik} high-$\ell$ TTTEEE likelihood with the more recent \texttt{HiLLiPoP} likelihood~\cite{Tristram:2023haj}, based on the Planck-PR4 \texttt{NPIPE} data release~\cite{Planck:2020olo,Carron:2022eyg}. Additionally, we replace the Planck-PR3 \texttt{SimAll} likelihood for E-mode polarization measurements at $\ell<30$ with the Planck-PR4 \texttt{LoLLiPoP} likelihood~\cite{Tristram:2023haj}, while keeping the low-$\ell$ \texttt{Commander} likelihood for the TT spectrum. Therefore, we use:
    \begin{itemize}
    \item the high-$\ell$ \texttt{HiLLiPoP} likelihood~\cite{Tristram:2023haj} for the TT, TE, and EE spectra in the multipole range $\ell \lesssim 2500$;
    \item the low-$\ell$ \texttt{Commander} likelihood~\cite{Planck:2018vyg} for the TT spectrum in the multipole range $2 \leq \ell \leq 29$;
    \item the low-$\ell$ \texttt{LoLLiPoP} likelihood~\cite{Tristram:2023haj} for the EE spectrum at $2 \leq \ell \leq 29$.
    \end{itemize}
    We refer to this combination as \textbf{HiLLiPoP}.

    \item \textbf{HiLLiPoP (TT-only);} we use the recent \texttt{HiLLiPoP} likelihood~\cite{Tristram:2023haj}, based on the Planck-PR4 \texttt{NPIPE} data release~\cite{Planck:2020olo, Carron:2022eyg}, for the temperature anisotropy spectrum, excluding any information on TE and EE at $\ell > 30$. Additionally, we replace the Planck-PR3 \texttt{SimAll} likelihood for E-mode polarization measurements at $\ell < 30$ with the Planck-PR4 \texttt{LoLLiPoP} likelihood~\cite{Tristram:2023haj}, while keeping the low-$\ell$ \texttt{Commander} likelihood for the TT spectrum. Therefore, we use:
    \begin{itemize}
    \item the high-$\ell$ \texttt{HiLLiPoP} likelihood~\cite{Tristram:2023haj} for the TT spectrum in the multipole range $30 \leq \ell \lesssim 2500$;
    \item the low-$\ell$ \texttt{Commander} likelihood~\cite{Planck:2018vyg} for the TT spectrum in the multipole range $2 \leq \ell \leq 29$;
    \item the low-$\ell$ \texttt{LoLLiPoP} likelihood~\cite{Tristram:2023haj} for the EE spectrum at $2 \leq \ell \leq 29$.
    \end{itemize}
    We refer to this combination as \textbf{HiLLiPoP (TT-only)}.
\end{itemize}

\begin{table}[t!]
\begin{tabular}{l  @{\hspace{1.5cm}} c}
\hline
\textbf{Parameter} & \textbf{Prior} \\ 
\hline\hline
$ \Omega_\mathrm{b} h^2  $ & $[0.005, 0.1]$\\ 
$ \Omega_\mathrm{c} h^2  $  & $[0.001, 0.99]$\\ 
$ 100\theta_\mathrm{MC}  $  & $[0.5, 10]$\\ 
$ \tau_\mathrm{reio}  $  & $[0.01, 0.8]$\\ 
$ n_\mathrm{s}  $  & $[0.8, 1.2]$\\ 
$ \log(10^{10} A_\mathrm{s})  $  & $[1.61, 3.91]$ \\ 
\hline
$\mu_0$ & $[-1.5,1.5]$\\
$\Sigma_0$ & $[-1.5,1.5]$\\
\hline
$ \gamma_0  $ & $[0, 1]$\\
\hline
$A_L$ & $[0,10]$\\
\hline \hline
\end{tabular}
\caption{Flat priors imposed on the free parameters considered in this analysis.}
\label{tab:priors}
\end{table}

The six standard model parameters considered in this analysis are: the baryon density ($\Omega_\mathrm{b}h^2$); the density of cold dark matter ($\Omega_\mathrm{c}h^2$); the observed angular size of the sound horizon at recombination ($100\theta_\mathrm{MC}$); the reionization optical depth ($\tau_\mathrm{reio}$); the amplitude of the primordial spectrum of scalar perturbations ($\log(10^{10} A_\mathrm{s})$), and its spectral index ($n_\mathrm{s}$). Along with the parameters of the two MG models introduced in Sec.~\ref{sec:parametrisations}, $\mu_0, \Sigma_0$ for the ($\mu,\Sigma$) framework, and $\gamma_0$ for the $\gamma$ model. In order to clarify the correlation with the lensing anomaly, for each of these models, we consider two cases: one where $A_L$ is fixed to the GR value ($A_L = 1$), and another where $A_L$ is left free to vary. We sampled the parameter space by using the MCMC engine \texttt{Cobaya}~\cite{Torrado:2020dgo}. The Einstein-Boltzmann solver implemented in \texttt{Cobaya} to calculate and evolve perturbations within the ($\mu,\Sigma$) model is \texttt{MGCAMB}~\cite{Hojjati:2011ix, Zhao:2008bn, Zucca:2019xhg, Wang:2023tjj}, while for the $\gamma$ parametrisation of MG, we employed \texttt{CAMB\_GammaPrime\_Growth}~\cite{Nguyen:2023fip}. All the parameters characterising the two MG models considered here have been assumed to have flat priors as listed in Table~\ref{tab:priors}. To assess the convergence of the chains obtained using this approach, we apply the Gelman-Rubin criterion~\cite{Gelman:1992zz}, setting a threshold for chain convergence at $R - 1 < 0.01$. The results are derived from analyzing the MCMC chains using the \texttt{GetDist} code~\cite{Lewis:2019xzd}, after discarding a $30\%$ burn-in fraction of samples from each chain.

\section{Results}
\label{sec:res}

\begin{center}
\begin{table*}[ht!]
\begin{tabular}{l | c | c | c | c | c}
&\textbf{Plik} &  \textbf{Camspec} & \textbf{Camspec (TT-only)} & \textbf{HiLLiPoP} & \textbf{HiLLiPoP (TT-only)}\\
\hline
\boldmath$\Omega_b h^2$ &  $ 0.02258\pm 0.00017 $&$ 0.02231\pm 0.00015 $&$ 0.02253\pm 0.00027 $&$ 0.02232\pm 0.00014 $&$ 0.02233\pm 0.00024$
\\
\boldmath$\Omega_c h^2$ &  $ 0.1182\pm 0.0015 $&$ 0.1187\pm 0.0013 $&$ 0.1162\pm 0.0024 $&$ 0.1183\pm 0.0012 $&$ 0.1179\pm 0.0022$
\\
\boldmath$100\theta_{MC} $ &  $ 1.04112\pm 0.00032 $&$ 1.04086\pm 0.00026 $&$ 1.04133\pm 0.00048 $&$ 1.04091\pm 0.00026 $&$ 1.04110\pm 0.00046$
\\
\boldmath$\tau$ &  $ 0.0507\pm 0.0081 $&$ 0.0493\pm 0.0083 $&$ 0.0512\pm 0.0082 $&$ 0.0570\pm 0.0060 $&$ 0.0573\pm 0.0064$
\\
\boldmath${\rm{ln}}(10^{10} A_s)$ &  $ 3.033\pm 0.017 $&$ 3.028\pm 0.017 $&$ 3.027\pm 0.017 $&$ 3.037\pm 0.014 $&$ 3.036\pm 0.015$
\\
\boldmath$n_s$ &  $ 0.9701\pm 0.0050 $&$ 0.9663\pm 0.0044 $&$ 0.9730\pm 0.0072 $&$ 0.9695\pm 0.0041 $&$ 0.9679\pm 0.0062$
\\
\boldmath$\mu_0$ &  $ 0.08^{+0.45}_{-0.91} $&$ 0.08^{+0.48}_{-0.87} $&$ 0.06^{+0.48}_{-0.84} $&$ 0.02^{+0.38}_{-0.95} $&$ 0.06^{+0.47}_{-0.92}$
\\
\boldmath$\Sigma_0$ &  $ 0.35\pm 0.14 $&$ 0.20^{+0.12}_{-0.14} $&$ 0.38\pm 0.17 $&$ 0.12^{+0.11}_{-0.13} $&$ 0.20^{+0.13}_{-0.18}$
\\
\boldmath $H_0$ [km/s/Mpc] &  $ 68.23\pm 0.70 $&$ 67.74\pm 0.59 $&$ 69.0\pm 1.2 $&$ 67.91\pm 0.57 $&$ 68.13^{+0.94}_{-1.1}$
\\
\boldmath $S_8$ &  $ 0.816^{+0.056}_{-0.083} $&$ 0.823^{+0.055}_{-0.082} $&$ 0.792^{+0.058}_{-0.082} $&$ 0.817^{+0.050}_{-0.087} $&$ 0.814^{+0.059}_{-0.083}$
\\
\hline
\end{tabular}
\caption{1$\sigma$ constraints on the ($\mu,\Sigma$) model when the lensing amplitude $A_L$ is kept fixed to its reference value $A_L=1$. As the likelihood combination employed changes from \textbf{Plik}, through to \textbf{Camspec} and finally to \textbf{HiLLiPoP}, the apparent MG evidence signal in Planck's data disappears, and consistency with $\Sigma_0 = 0$ is re-established. Counting polarization anisotropies out of the analysis (i.e., the TT-only columns) indicates that this signal is predominantly affecting the temperature data.
}
\label{tab_app:ms}
\end{table*}
\end{center}
\begin{center}
\begin{table*}[ht!]
\begin{tabular}{l | c | c | c | c | c}
&\textbf{Plik} &  \textbf{Camspec} & \textbf{Camspec (TT-only)} & \textbf{HiLLiPoP} & \textbf{HiLLiPoP (TT-only)}\\
\hline
\boldmath$\Omega_b h^2$ &  $ 0.02260\pm 0.00017 $&$ 0.02231\pm 0.00016 $&$ 0.02256\pm 0.00029 $&$ 0.02232^{+0.00014}_{-0.00015} $&$ 0.02230\pm 0.00026$
\\
\boldmath$\Omega_c h^2$ &  $ 0.1181\pm 0.0016 $&$ 0.1186\pm 0.0013 $&$ 0.1160\pm 0.0025 $&$ 0.1183\pm 0.0014 $&$ 0.1181\pm 0.0024$
\\
\boldmath$100\theta_{MC} $ &  $ 1.04115\pm 0.00033 $&$ 1.04087\pm 0.00027 $&$ 1.04136\pm 0.00048 $&$ 1.04092\pm 0.00027 $&$ 1.04108\pm 0.00050$
\\
\boldmath$\tau$ &  $ 0.0499\pm 0.0084 $&$ 0.0493\pm 0.0083 $&$ 0.0506^{+0.0088}_{-0.0077} $&$ 0.0577\pm 0.0060 $&$ 0.0579\pm 0.0062$
\\
\boldmath${\rm{ln}}(10^{10} A_s)$ &  $ 3.031\pm 0.018 $&$ 3.028\pm 0.017 $&$ 3.025^{+0.019}_{-0.016} $&$ 3.038\pm 0.015 $&$ 3.038\pm 0.015$
\\
\boldmath$n_s$ &  $ 0.9707\pm 0.0050 $&$ 0.9667\pm 0.0047 $&$ 0.9737\pm 0.0074 $&$ 0.9695\pm 0.0045 $&$ 0.9674\pm 0.0068$
\\
\boldmath$\mu_0$ &  $ 0.04^{+0.42}_{-0.92} $&$ 0.04^{+0.43}_{-0.90} $&$ 0.07^{+0.49}_{-0.86} $&$ 0.01^{+0.42}_{-0.88} $&$ -0.01^{+0.41}_{-0.86}$
\\
\boldmath$\Sigma_0$ &  $ 0.24^{+0.18}_{-0.23} $&$ 0.23^{+0.17}_{-0.22} $&$ 0.26^{+0.19}_{-0.26} $&$ 0.25^{+0.18}_{-0.24} $&$ 0.27^{+0.19}_{-0.23}$
\\
\boldmath $H_0$ [km/s/Mpc] &  $ 68.32\pm 0.72 $&$ 67.78\pm 0.61 $&$ 69.1\pm 1.2 $&$ 67.91^{+0.59}_{-0.65} $&$ 68.0^{+1.0}_{-1.2}$
\\
\boldmath $S_8$ &  $ 0.809^{+0.052}_{-0.083} $&$ 0.817^{+0.052}_{-0.083} $&$ 0.789^{+0.063}_{-0.077} $&$ 0.815^{+0.051}_{-0.083} $&$ 0.810^{+0.054}_{-0.083}$
\\
\boldmath $A_{L}$ &  $ 1.069^{+0.083}_{-0.11} $&$ 0.992^{+0.081}_{-0.097} $&$ 1.076^{+0.093}_{-0.14} $&$ 0.942^{+0.077}_{-0.098} $&$ 0.963^{+0.079}_{-0.11}$
\\
\hline
\end{tabular}
\caption{1$\sigma$ constraints on the ($\mu,\Sigma$) model when the lensing amplitude $A_L$ is left free to vary in the model.}
\label{tab_app:msalens}
\end{table*}
\end{center}
\begin{figure*}[t!]
   \includegraphics[width=1\textwidth]{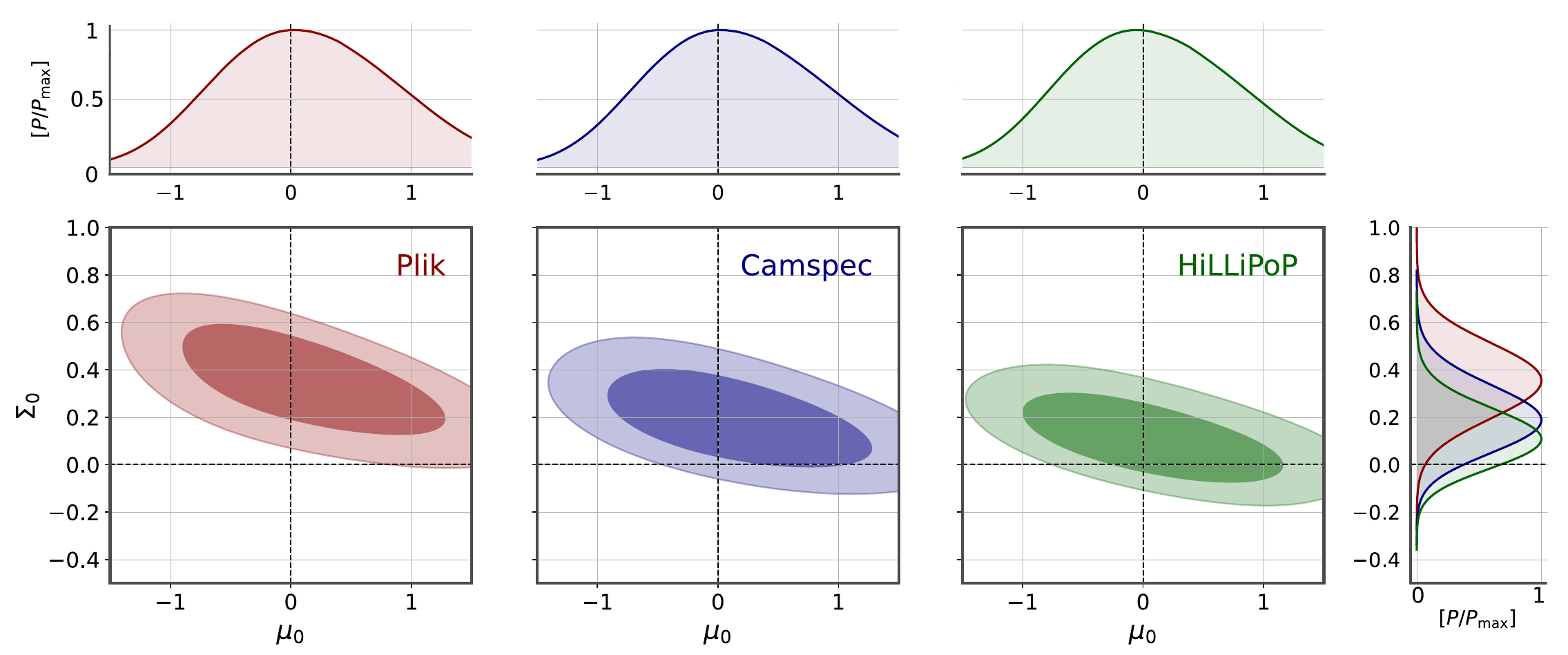}
     \caption{One-dimensional probability distribution functions and two-dimensional marginalized constraints in the ($\mu_0$,$\Sigma_0$) plane for the different datasets listed in Sec.~\ref{sec:data}. The lensing amplitude $A_L$ is kept fixed at its reference value $A_L=1$. The vertical and horizontal dashed lines represent the baseline values $\mu_0=\Sigma_0=0$.}
    \label{fig:1}
\end{figure*}

\begin{figure}[t!]
   \includegraphics[width=0.95\columnwidth]{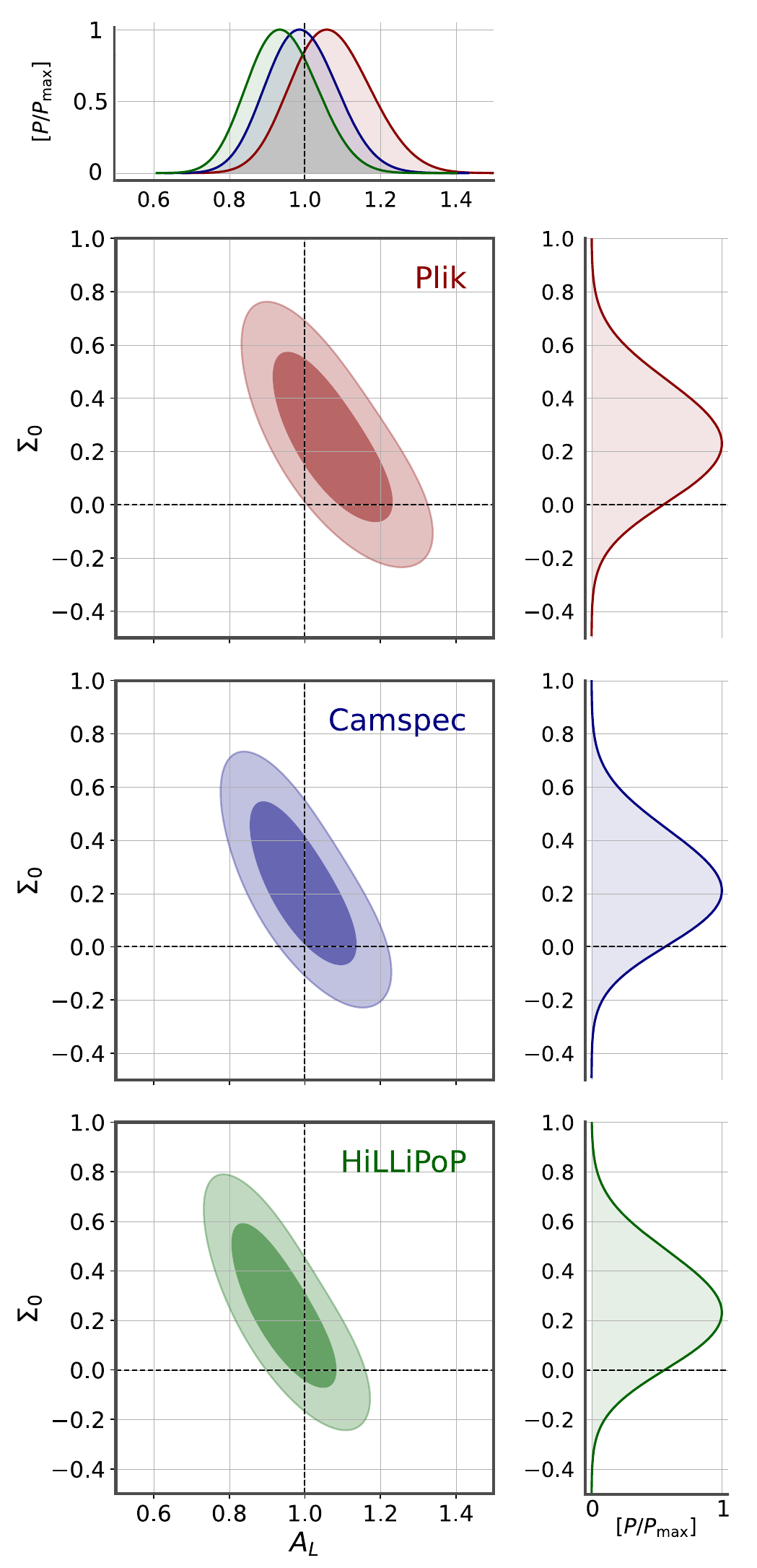}
    \caption{One-dimensional probability distribution functions and two-dimensional marginalized contours in the ($A_L$,$\Sigma_0$) plane for the different datasets listed in Sec.~\ref{sec:data}. The horizontal and vertical dashed lines represent the baseline values $\Sigma_0=0$ and $A_L=1$, respectively. }
    \label{fig:2}
\end{figure}

\begin{figure*}[t!]
   \includegraphics[width=\textwidth]{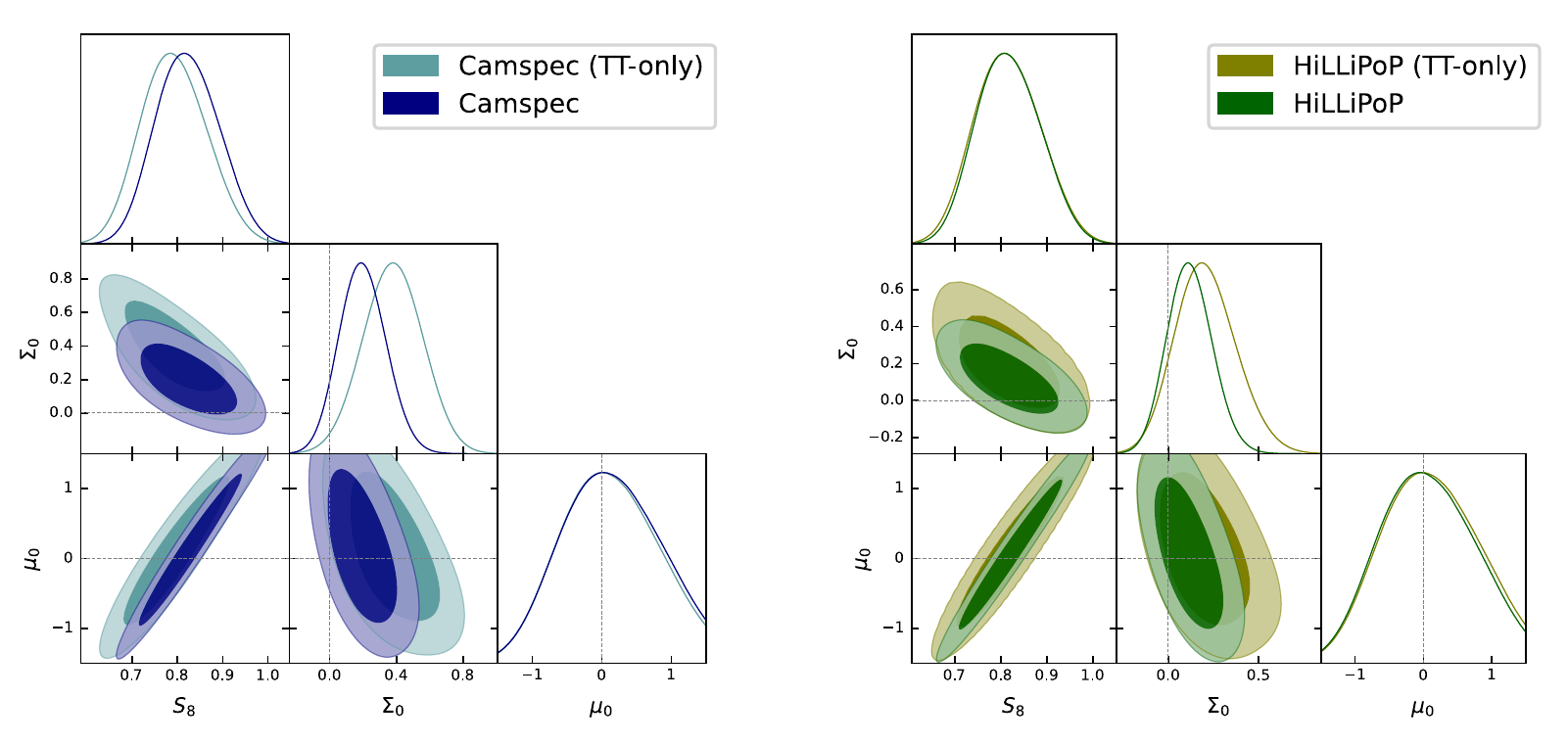}
    \caption{One-dimensional probability distribution functions and two-dimensional marginalized contours for the parameters $\gamma_0$, $\Omega_m$, and $S_8$. The lensing amplitude $A_L$ is kept fixed at its reference value $A_L=1$. In the left side panel, we show the constraints involving the combinations of likelihood \textbf{Camspec} and \textbf{Camspec (TT-only)}, whereas in the right side panel, we show the constraints for the combination of likelihood \textbf{HiLLiPoP} and \textbf{HiLLiPoP (TT-Only)} -- all likelihoods are defined in Sec.~\ref{sec:data}. The vertical dashed lines in both panels correspond to the baseline values $\mu_0=\Sigma_0 = 0$.}
    \label{fig:3}
\end{figure*}


Constraints on the free and derived cosmological parameters characterizing the $(\mu,\Sigma)$ scenario are provided in Tables~\ref{tab_app:ms} and~\ref{tab_app:msalens}, corresponding to the cases where the $A_L$ parameter is fixed and varied, respectively. The most significant findings for this model are discussed in detail in Sec.~\ref{sec:results_mu_sigma}. However, for a quicker overview, the reader may refer directly to Figs.~\ref{fig:1} and~\ref{fig:2}, which visually summarize the main results for this model. Similarly, the extended $\gamma$ model is examined in Sec.~\ref{sec:results_gamma}, with numerical constraints listed in Tables~\ref{tab_app:gamma} and~\ref{tab_app:gammalens} for the fixed and varied $A_L$ cases, respectively. Key aspects of this discussion are visually captured in Figs.~\ref{fig:4} and~\ref{fig:6}.

\subsection{Constraints on the $(\mu,\Sigma)$ framework}
\label{sec:results_mu_sigma}

We start with the constraints obtained from the combinations of likelihoods labeled as \textbf{Plik}. This combination of data is entirely based on the Planck-PR3 anisotropy spectra. Therefore, it is not surprising that we observe (nearly\footnote{This is slightly different from the value measured by the Planck collaboration and reported in the introduction due to a different parametrization and implementation of the model in the Boltzmann code.}) the same preference towards MG documented by the Planck Collaboration~\cite{Planck:2015fie,Planck:2018vyg}. Specifically, when constraining the parameter $\Sigma_0$, we find $\Sigma_0 = 0.35 \pm 0.14$, which is $2.5\sigma$ away from the value expected within GR. Additionally, referring to the leftmost panel of Fig.~\ref{fig:1} -- where we show the 2D marginalized constraints in the ($\mu_0$, $\Sigma_0$) plane along with the one-dimensional probability distribution functions of the two parameters -- we observe that the point (0, 0), which corresponds to a baseline $\Lambda$CDM cosmology, falls well outside the 95\% contours.

As already pointed out in Refs.~\cite{Planck:2015bue, DiValentino:2015bja, Planck:2018vyg, Pogosian:2021mcs, Andrade:2023pws}, this apparent evidence for MG in Planck's PR3 data is closely linked to the lensing anomaly. This can be understood by the effect of $\Sigma_0$ in Eq.~(\ref{eq:weyl_equation_pert}), which alters the strength of CMB lensing. \textbf{Values of $\Sigma_0 > 0$ produce effects similar to those associated with $A_L > 1$, and the two can be recast into one another.} As a result, CMB data are sensitive to both of these changes in much the same way. Unsurprisingly, when $A_L$ is allowed to vary within the ($\mu$,$\Sigma$) framework, a strong negative correlation emerges between $\Sigma_0$ and $A_L$. In this joint analysis, the combined effect of this correlation and the larger uncertainties translates into the $\Sigma_0 \rightarrow 0$ and $A_L \rightarrow 1$ shifts we find with \textbf{Camspec} and \textbf{HiLLiPoP}, diluting both the lensing anomaly and the preference for MG. This is also clearly reflected in the numerical constraints given in Tab.~\ref{tab_app:msalens}. Taken at face value, the results for $A_L$ and $\Sigma_0$ are both consistent with their GR predictions within 2 standard deviations. Nevertheless, referring to the top panel of Fig.~\ref{fig:2} -- where we show the 2D marginalized constraints in the ($A_L$, $\Sigma_0$) plane along with the one-dimensional probability distribution functions of these parameters -- we observe that, due to the strong correlations, the point corresponding to GR ($A_L=1$, $\Sigma_0=0$) still falls outside the 95\% confidence level contours.

When analyzing the new Planck-PR4-based likelihoods, we find that the contours in the ($\mu_0$, $\Sigma_0$) plane shown in Fig.~\ref{fig:1} shift closer to GR as we move from the \textbf{Plik} to \textbf{Camspec}, and finally to the \textbf{HiLLiPoP} likelihood combinations. Overall, moving from the leftmost to the rightmost panel of the figure, we observe that the point ($\mu_0=0$, $\Sigma_0=0$) falls within the $3\sigma$, $2\sigma$, and $1\sigma$ confidence limits for the \textbf{Plik}, \textbf{Camspec}, and \textbf{HiLLiPoP} likelihoods, respectively. At its core, this improved consistency with GR is primarily driven by a shift in the value of $\Sigma_0 \rightarrow 0$, as seen by comparing the one-dimensional probability distribution functions in Fig. \ref{fig:1} for this parameter across the three datasets and confirmed by the numerical results: \textbf{Camspec} yields $\Sigma_0 = 0.20^{+0.12}_{-0.14}$ (1.4$\sigma$ away from GR), while \textbf{HiLLiPoP} gives $\Sigma_0 = 0.12^{+0.11}_{-0.13}$ (consistent with GR within $0.9\sigma$).

We again want to point out to the reader that these results are closely related to the behavior of the lensing anomaly in these new likelihoods. Firstly, the strength of the preference for $A_L > 1$ follows the same decreasing trend as the preference for $\Sigma_0 \neq 0$. Specifically, using \textbf{Camspec} temperature and polarization data reduces the preference for $A_L > 1$ to less than $\sim 1.6\sigma$ (see Tab. 6 of Ref.~\cite{Rosenberg:2022sdy}), matching the extent of the shift in $\Sigma_0$. Using the \textbf{HiLLiPoP} likelihood, the results are consistent with $A_L = 1$ within 1 standard deviation (see Tab. 6 of Ref.~\cite{Tristram:2023haj}) — similar to what happens for $\Sigma_0$. 

Pushing this comparison further, similarly to the $A_L$ anomaly, the MG signal appears to be largely driven by the temperature anisotropy data in both the PR3 and PR4 based likelihoods. As shown by the numerical constraints in Tab.~\ref{tab_app:ms} and the posteriors in Fig.~\ref{fig:3}, for the data combinations \textbf{Camspec (TT-only)} and \textbf{HiLLiPoP (TT-only)} (i.e., those involving only TT measurements at $\ell > 30$), the deviation of $\Sigma_0$ from zero can be more pronounced compared to the full TTTEEE likelihood. Specifically, in \textbf{Camspec}, the preference for $\Sigma_0 \neq 0$ increases from 1.4$\sigma$ to 2.2$\sigma$ when polarization is excluded; see also the light-blue contours in the left-side panel of Fig.~\ref{fig:3}. Conversely, with \textbf{HiLLiPoP (TT-only)}, no significant deviation from GR is found, as illustrated by the olive contours in the right-side panel of Fig.~\ref{fig:3}. This trend closely follows the behavior observed for the lensing anomaly: as reported in Table 6 of Ref.~\cite{Rosenberg:2022sdy}, focusing solely on TT spectrum data in \textbf{Camspec} increases the lensing anomaly up to $2.3\sigma$. However, no significant deviation from $A_L = 1$ is found when considering only the TT spectrum data in \textbf{HiLLiPoP}.

Given the close and expected relationship between $\Sigma_0$ and $A_L$, it is worth examining the impact of varying $A_L$ within the $\mu,\Sigma$ framework using the new PR4-based data combinations. The results are summarized in Tab.~\ref{tab_app:msalens}. Across all Planck-PR4 anisotropy spectrum datasets, the numerical results for both $A_L$ and $\Sigma_0$ remain consistent with the GR values within one standard deviation. Additionally, referring to the middle and bottom panels of Fig.~\ref{fig:2}, we see that GR comfortably falls within the 95\% CL contours. This contrasts with the results from \textbf{Plik}, shown in the top panel of Fig.~\ref{fig:2}, reinforcing the interpretation that the earlier signals observed in Planck-PR3 spectra were likely due to systematic effects that have since been mitigated (if not erased) in the new data release.

We conclude this section with a final important remark: for \textbf{Plik}, \textbf{Camspec}, and \textbf{HiLLiPoP}, we find $\mu_0$ to be well within the GR limit (see both Fig.~\ref{fig:1} and Fig.~\ref{fig:3}), albeit with much wider error bars compared to $\Sigma_0$. This can be attributed to the limited impact of this parameter on the CMB spectra, reflecting the physical interpretation of the $\mu$ function as $G_{\rm eff}$. Fundamentally, the $\mu$ parameter influences the growth of structure; for instance, a higher $\mu$ corresponds to larger values of $\sigma_8$. This strong correlation with the growth parameter $S_8$ is illustrated by the $\mu_0 - S_8$ contour in Fig.~\ref{fig:3}. However, since the CMB constrains modified gravity (MG) primarily through the integrated Sachs-Wolfe effect and CMB lensing, it is unable to tightly constrain $\mu_0$.

\subsection{Constraints on the growth index $\gamma$}
\label{sec:results_gamma}

We anticipate that modifying the growth of perturbations using the $\gamma$ parametrization will not significantly alter the conclusions drawn in the previous subsection with the ($\mu,\Sigma$) framework.

As usual, to establish a reference case, we first derive results from the Planck-PR3 based spectra using the data combination \textbf{Plik}. In this instance, we obtain $\gamma = 0.840^{+0.11}_{-0.074}$ -- approximately 3.9$\sigma$ away from what one might naively expect within GR. This finding aligns with the recently debated claims of deviations from a baseline structure growth history, as reviewed in the introduction.

\begin{figure}[ht!]
   \includegraphics[width=0.95\columnwidth]{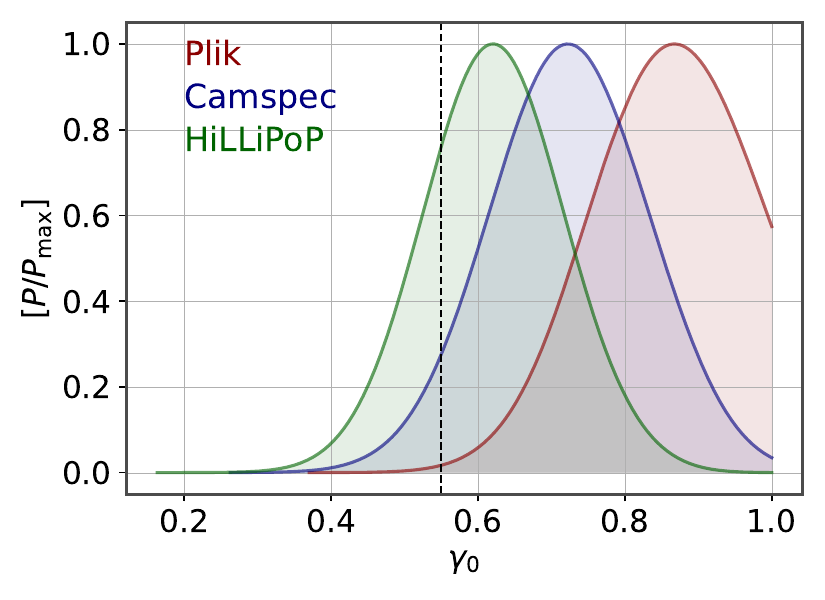}
     \caption{One-dimensional marginalized posterior distributions for $\gamma_0$ are presented for the three different datasets listed in Sec.~\ref{sec:data}. The lensing amplitude $A_L$ is kept fixed at its reference value $A_L=1$. The vertical dashed line represents the baseline value $\gamma_0 = 0.55$.}
    \label{fig:4}
\end{figure}

\begin{figure*}[ht!]
   \includegraphics[width=\textwidth]{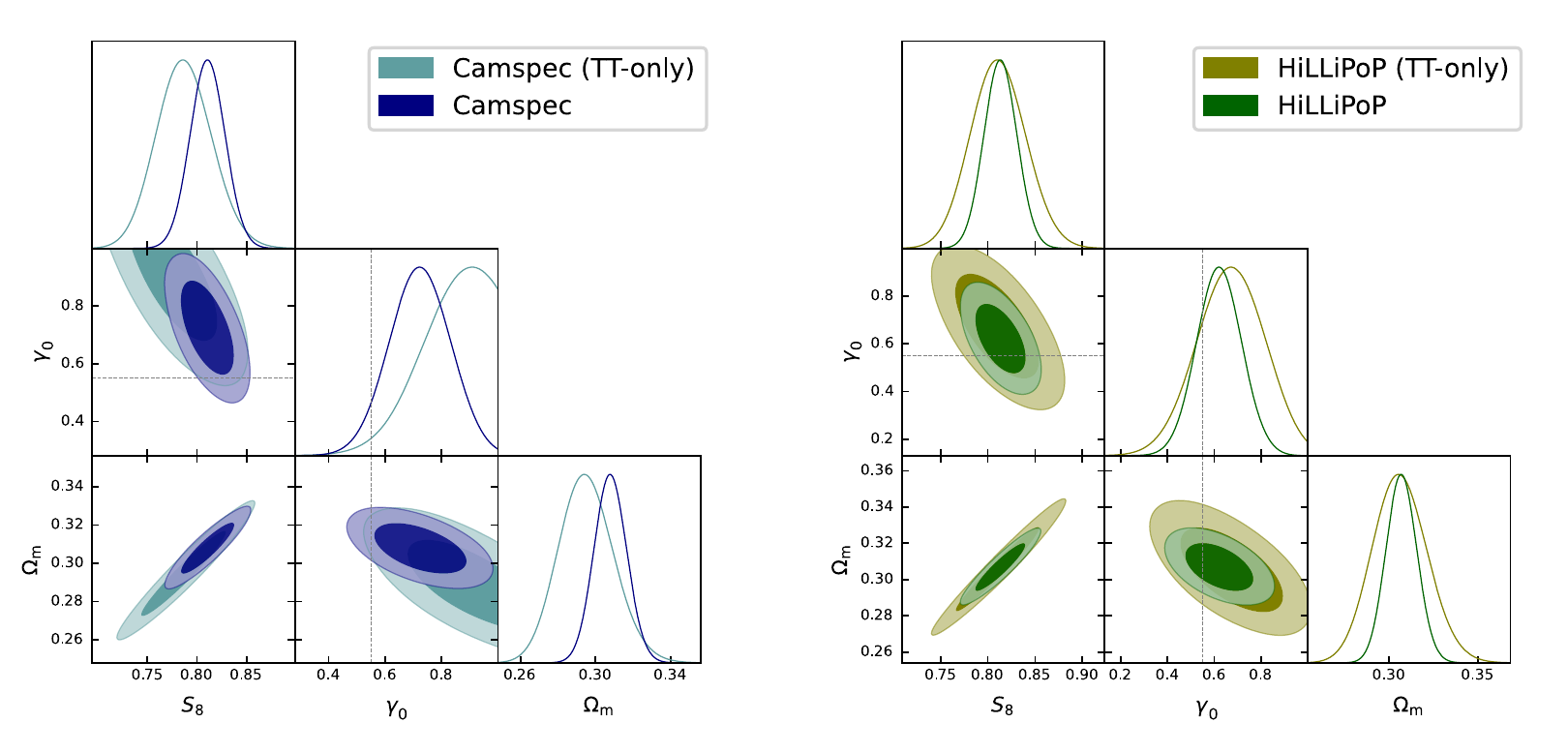}
     \caption{One-dimensional probability distribution functions and two-dimensional marginalized contours for the parameters $\gamma_0$, $\Omega_m$, and $S_8$. The lensing amplitude $A_L$ is kept fixed at its reference value $A_L=1$. In the left-side panel, we show the constraints involving the combinations of likelihood \textbf{Camspec} and \textbf{Camspec (TT-only)}, whereas in the right-side panel, we show the constraints for the combination of likelihood \textbf{HiLLiPoP} and \textbf{HiLLiPoP (TT-only)} -- all likelihoods are defined in Sec.~\ref{sec:data}. The vertical dashed lines in both panels correspond to the baseline value $\gamma_0 = 0.55$.}
    \label{fig:5}
\end{figure*}

However, as seen in Tab.~\ref{tab_app:gamma} and Fig.~\ref{fig:4}, where we display the one-dimensional probability distribution function for the parameter $\gamma_0$ as obtained from the three different datasets\footnote{An interesting difference between the two theoretical setups concerns the correlation with other matter clustering quantities. For instance, as shown in Fig.~\ref{fig:5}, we find a weaker degeneracy between the MG parameter $\gamma_0$ and quantities such as $S_8$, compared to what we observed for $\mu_0$, which also impacts the growth of cosmic structures.}, the indication of beyond-GR phenomenology is found to be reduced to the $2\sigma$ level in the Planck-PR4-based combination of likelihoods \textbf{Camspec} (which gives $\gamma_0 = 0.72 \pm 0.10$) and disappears entirely with \textbf{HiLLiPoP}, which shows consistency with GR well within one standard deviation, yielding $\gamma_0 = 0.621 \pm 0.090$.

\begin{center}
\begin{table*}[ht!]
   \begin{tabular}{l | c | c | c | c | c}
&\textbf{Plik} &  \textbf{Camspec} & \textbf{Camspec (TT-only)} & \textbf{HiLLiPoP} & \textbf{HiLLiPoP (TT-only)}\\
\hline
\boldmath$\Omega_b h^2$ &  $ 0.02258\pm 0.00017 $&$ 0.02232\pm 0.00016 $&$ 0.02249^{+0.00028}_{-0.00025} $&$ 0.02230\pm 0.00015 $&$ 0.02228\pm 0.00026$
\\
\boldmath$\Omega_c h^2$ &  $ 0.1182\pm 0.0015 $&$ 0.1186\pm 0.0013 $&$ 0.1165\pm 0.0023 $&$ 0.1184\pm 0.0013 $&$ 0.1182\pm 0.0023$
\\
\boldmath$100\theta_{MC} $ &  $ 1.04112\pm 0.00032 $&$ 1.04088\pm 0.00027 $&$ 1.04128\pm 0.00046 $&$ 1.04091\pm 0.00027 $&$ 1.04104\pm 0.00048$
\\
\boldmath$\tau$ &  $ 0.0499\pm 0.0082 $&$ 0.0487\pm 0.0083 $&$ 0.0506\pm 0.0085 $&$ 0.0572\pm 0.0063 $&$ 0.0572\pm 0.0064$
\\
\boldmath${\rm{ln}}(10^{10} A_s)$ &  $ 3.031\pm 0.017 $&$ 3.027^{+0.018}_{-0.016} $&$ 3.026\pm 0.017 $&$ 3.036\pm 0.015 $&$ 3.036\pm 0.015$
\\
\boldmath$n_s$ &  $ 0.9704\pm 0.0048 $&$ 0.9672\pm 0.0045 $&$ 0.9722^{+0.0071}_{-0.0063} $&$ 0.9691\pm 0.0044 $&$ 0.9670\pm 0.0068$
\\
\boldmath$\gamma_0$ &  $ 0.840^{+0.11}_{-0.074} $&$ 0.72\pm 0.10 $&$ 0.827^{+0.16}_{-0.051} $&$ 0.621\pm 0.090 $ & $0.66\pm 0.13$
\\
\boldmath $H_0$ [km/s/Mpc] &  $ 68.24\pm 0.69 $&$ 67.81\pm 0.61 $&$ 68.8\pm 1.1 $&$ 67.87\pm 0.61 $&$ 68.0\pm 1.1$
\\
\boldmath$S_8$ &  $ 0.806\pm 0.018 $&$ 0.811\pm 0.017 $&$ 0.787^{+0.024}_{-0.028} $&$ 0.813\pm 0.016 $&$ 0.811\pm 0.027$
\\
\hline
\end{tabular}
\caption{$1\sigma$ constraints on the $\gamma$ model when the lensing amplitude $A_L$ is kept fixed to its reference value $A_L=1$. As the likelihood combination employed changes from \textbf{Plik}, through to \textbf{Camspec} and finally to \textbf{HiLLiPoP}, the apparent MG evidence signal in Planck-PR3 data disappears, and consistency with $\gamma_0 = 0.55$ is re-established. Counting polarization anisotropies out of the analysis (i.e., the TT-only columns) indicates that this signal is predominantly affecting the temperature data.}
\label{tab_app:gamma}
\end{table*}
\end{center}
\begin{center}
\begin{table*}[ht!]
   \begin{tabular}{l | c | c | c | c | c}
  &\textbf{Plik} &  \textbf{Camspec} & \textbf{Camspec (TT-only)} & \textbf{HiLLiPoP} & \textbf{HiLLiPoP (TT-only)}\\
\hline
\boldmath$\Omega_b h^2$ &  $ 0.02260\pm 0.00017 $&$ 0.02257\pm 0.00028 $&$ 0.02232\pm 0.00016 $&$ 0.02229\pm 0.00025 $&$ 0.02230\pm 0.00015$
\\
\boldmath$\Omega_c h^2$ &  $ 0.1180\pm 0.0015 $&$ 0.1159\pm 0.0023 $&$ 0.1186\pm 0.0014 $&$ 0.1181\pm 0.0024 $&$ 0.1183\pm 0.0013$
\\
\boldmath$100\theta_{MC} $ &  $ 1.04115\pm 0.00032 $&$ 1.04138\pm 0.00047 $&$ 1.04088\pm 0.00027 $&$ 1.04107\pm 0.00047 $&$ 1.04091\pm 0.00027$
\\
\boldmath$\tau$ &  $ 0.0494^{+0.0087}_{-0.0075} $&$ 0.0505\pm 0.0085 $&$ 0.0487^{+0.0086}_{-0.0072} $&$ 0.0572\pm 0.0063 $&$ 0.0571\pm 0.0062$
\\
\boldmath${\rm{ln}}(10^{10} A_s)$ &  $ 3.029^{+0.018}_{-0.016} $&$ 3.025^{+0.018}_{-0.016} $&$ 3.027^{+0.018}_{-0.015} $&$ 3.036\pm 0.015 $&$ 3.036\pm 0.015$
\\
\boldmath$n_s$ &  $ 0.9708\pm 0.0048 $&$ 0.9739\pm 0.0070 $&$ 0.9670\pm 0.0046 $&$ 0.9674\pm 0.0068 $&$ 0.9692\pm 0.0044$
\\
\boldmath$\gamma_0$ &  $ < 0.597 $&$ < 0.610 $&$ < 0.621 $&$ < 0.603 $&$ < 0.605$
\\
\boldmath $H_0$ [km/s/Mpc] &  $ 68.34\pm 0.71 $&$ 69.2\pm 1.1 $&$ 67.81\pm 0.62 $&$ 68.0\pm 1.1 $&$ 67.89\pm 0.62$
\\
\boldmath$S_8$ &  $ 0.803\pm 0.019 $&$ 0.780\pm 0.028 $&$ 0.811\pm 0.017 $&$ 0.809^{+0.026}_{-0.029} $&$ 0.813\pm 0.017$
\\
\boldmath$A_{L}$ &  $ 1.31^{+0.19}_{-0.33} $&$ 1.32^{+0.20}_{-0.35} $&$ 1.20^{+0.17}_{-0.30} $&$ 1.18^{+0.17}_{-0.30} $&$ 1.14^{+0.16}_{-0.28}$
\\
\hline
\end{tabular}
\caption{$1\sigma$ constraints on the $\gamma$ model  when the lensing amplitude $A_L$ is left free to vary.}
\label{tab_app:gammalens}
\end{table*}
\end{center}

\begin{figure}[ht!]
   \includegraphics[width=0.7 \columnwidth]{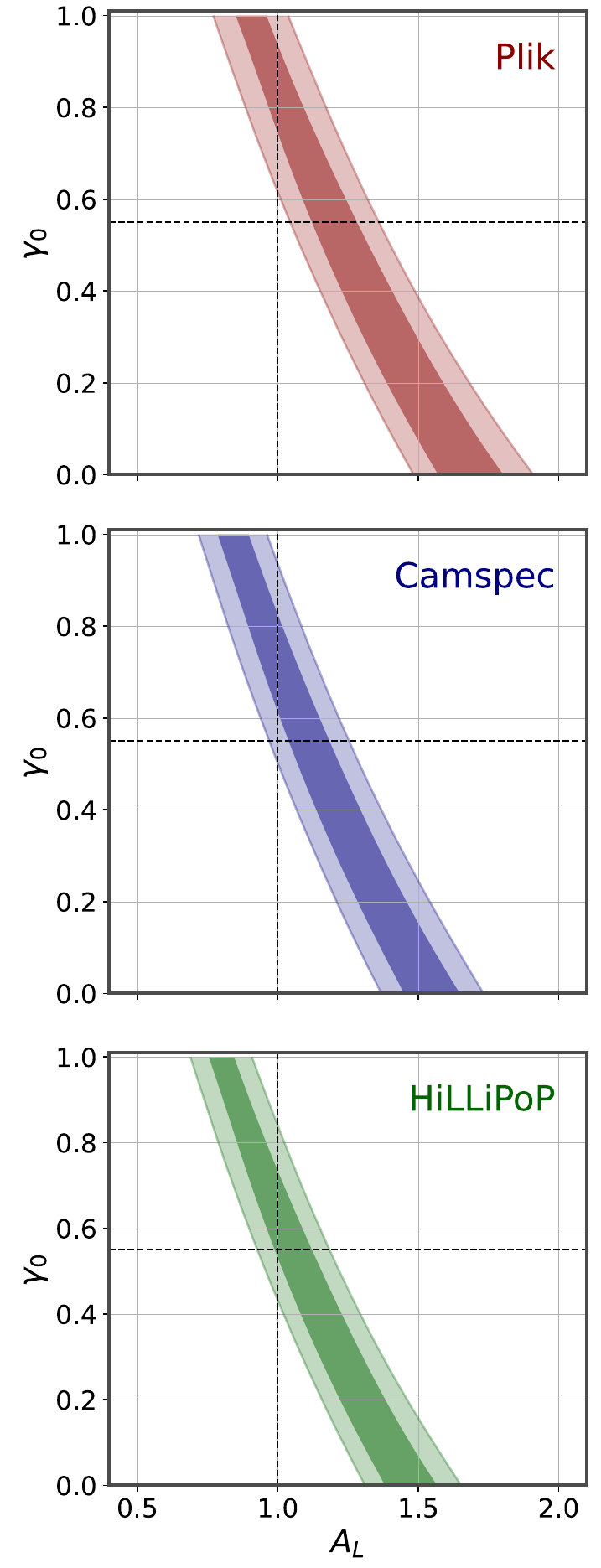}
     \caption{Two-dimensional marginalized constraints in the ($A_L$, $\gamma_0$) plane for the different datasets listed in Sec.~\ref{sec:data}. The horizontal and vertical dashed lines represent the baseline values $\gamma_0 = 0.55$ and $A_L = 1$, respectively. }
    \label{fig:6}
\end{figure}

As shown in Table~\ref{tab_app:gamma} and Fig.~\ref{fig:5}, when using a TT-only likelihood instead of the full TTTEEE high-$\ell$ combination, we observe a shift in \textbf{Camspec (TT-only)} towards values of $\gamma_0 > 0.55$. Specifically, we find $\gamma_0 = 0.827^{+0.16}_{-0.051}$, which shifts further away from the GR expectation, as can be seen by comparing the dark blue contours  with the light blue contours in the left-side panel of Fig.~\ref{fig:5}. In contrast, the value of $\gamma_0$ inferred from the \textbf{HiLLiPoP (TT-only)} data combination remains consistent with $\Lambda$CDM within $1\sigma$. Aside from a significant increase in the uncertainty of parameter value inference, no notable discrepancies are observed between the constraints inferred from the full TT, TE, EE combinations of spectra (dark green contours in the right panel of Fig.~\ref{fig:5}) and those obtained by retaining only TT data (olive green contours in the right panel of Fig.~\ref{fig:5}). 

As discussed in the previous subsection, this trend mirrors the behavior of the lensing anomaly across the two different combinations of TT-only data, suggesting that the results for $\gamma_0$ can be significantly influenced by the excess lensing smoothing observed in the acoustic peaks of temperature anisotropy spectra. Therefore, following the same approach as in the ($\mu$-$\Sigma$) framework, it is worthwhile to investigate the relationship between $\gamma_0$ and $A_L$ when the latter is allowed to vary freely in the model. The main conclusions from previous sections largely hold: $\gamma_0$ exhibits a strong correlation with $A_L$, which is illustrated for the three datasets in Fig.~\ref{fig:6}.

Interestingly, allowing $A_L$ to vary precludes achieving a two-tail constraint on $\gamma_0$. From \textbf{Plik}, we obtain joint constraints of $A_L = 1.31^{+0.19}_{-0.33}$ and $\gamma_0 < 0.597$. Taken at face value, both parameters remain consistent with GR within two standard deviations. However, as shown in the top panel of Fig.~\ref{fig:6}, the point $A_L = 1$ and $\gamma_0 = 0.55$ still lies outside the 2D marginalized probability contours in the ($A_L$, $\gamma_0$) plane. Turning to the Planck-PR4-based likelihoods, \textbf{Camspec} yields $A_L = 1.32^{+0.20}_{-0.35}$ and $\gamma_0 < 0.610$, while \textbf{HiLLiPoP} gives $\gamma_0 < 0.603$ and $A_L = 1.18^{+0.17}_{-0.30}$. When interpreted at face value, these results indicate that both $A_L$ and $\gamma_0$ are consistent with their $\Lambda$CDM values. As seen in Fig.~\ref{fig:6}, the key difference between the results inferred from the Planck-PR4 spectra and those based on \textbf{Plik} is that both \textbf{Camspec} and \textbf{HiLLiPoP} produce contours such that the standard cosmological model now falls within the 95\% confidence region in the ($A_L$,$\gamma_0$) plane. In other words, we can recover the standard $\Lambda$CDM model without reducing one anomaly at the expense of exacerbating another.

\section{Discussion And Conclusion}
\label{sec:conclusions}

Precise CMB observations are a powerful tool for testing whether gravity follows GR on cosmic scales. On the brighter side, CMB physics primarily relies on linear perturbation theory, governed by a system of linearized Einstein-Boltzmann equations, which makes it largely immune to complications from astrophysical processes (beyond foreground contamination) or nonlinear dynamics that can complicate the analysis of late-time structures. Consequently, over the past two decades, numerous extended theories of gravity have been tested against CMB observations.

However, on the downside, in extended theories of gravity, the equations describing the evolution of cosmological perturbations (and the growth of cosmic structures) can deviate significantly from GR in ways that differ from model to model. To face this complexity, various parametric tests have been developed to recast these behaviors and establish broader frameworks for testing GR predictions. Within these frameworks, analyses of the Planck-PR3 2018 spectra have revealed multiple hints of potential deviations from the baseline predictions of a $\Lambda$CDM cosmology.

The first longstanding issue concerns the amplitude of the lensing spectrum as inferred from the smoothing of acoustic peaks in the temperature and polarization spectra. Since the 2013 data release, the temperature and polarization power spectra from Planck have shown an excess of lensing, leading to a value of the phenomenological parameter $A_L$ greater than unity. In the Planck-PR3 2018 data release, $A_L$ is measured to be $A_L = 1.180 \pm 0.065$ -- i.e., $2.8\sigma$ away from the $\Lambda$CDM predictions.
 
Another interesting hint of deviation from GR appears when altering the predictions for the Poisson and lensing equations within the so-called ($\mu,\Sigma$) framework, where $\mu(k,a)$ and $\Sigma(k,a)$ are two arbitrary functions of the wavenumber $k$ and the scale factor $a$, both null in GR. Assuming them to be constant both in $a$ and $k$, Planck-PR3 data measure $\Sigma_0 = 0.35 \pm 0.14$, which is $2.5\sigma$ away from the expected value.

Finally, yet another test that produces a non-null result involves measuring the growth index $\gamma$ that captures potential deviations in the growth of perturbations. This parameter is predicted to be $\gamma = 0.55$ in GR and measured to be $\gamma = 0.840^{+0.011}_{-0.074}$ with Planck-PR3 data.

Although these results are certainly interesting, possibly suggesting departures from a $\Lambda$CDM model of structure formation, they come with a good number of caveats and warnings that we already pointed out in the introduction and further stress here: firstly, no indication of a lensing excess, $\Sigma_0 > 0$, or $\gamma \ne 0.55$ has been found in CMB experiments other than Planck, such as ACT and SPT. Secondly, within the Planck-PR3 spectra, a strong correlation between $A_{\rm L}$ and the $\Sigma_0$ or $\gamma_0$ parameters is observed, suggesting that these three problems may have a common root -- most probably the excess smoothing in the TT Planck-PR3 spectrum, which is the main driver of the lensing anomaly. Finally, the latest versions of \texttt{CamSpec}~\cite{Rosenberg:2022sdy} and \texttt{HiLLiPoP}~\cite{Tristram:2023haj} (both based on Planck-PR4 \texttt{NPIPE} maps) show a much better agreement with $A_{\rm L} = 1$, pointing at an observational systematic issue in the Planck-PR3 spectra and immediately calling into question all the other hints for modified gravity discussed thus far.

In an attempt to clarify once and for all whether these anomalous results should be considered genuine indicators of new physics beyond GR or whether they simply arise from observational systematics (while understanding how \textit{exactly} the lensing anomaly fits into the picture), we present in this paper a re-evaluation of the modified gravity constraints inferred from the most recent Planck-PR4 spectra, derived from the new \texttt{NPIPE} data release. We stated the main conclusion of our analysis right in the title of this paper: \textit{Planck-PR4 spectra show (better) agreement with GR}. Specifically:
\begin{itemize}
\item For the combination of likelihoods labeled as \textbf{Camspec} (which is primarily based on the \texttt{Camspec} likelihoods for the high-$\ell$ TT, TE, and EE spectra), we find that the constraint $\Sigma_0 = 0.20^{+0.12}_{-0.14}$ lies within 1.5$\sigma$ of GR, see also Fig.~\ref{fig:1}. Similarly, the indication of beyond-GR phenomenology in terms of non-canonical values of the growth index is reduced to approximately $2\sigma$, with the numerical constraint reading $\gamma_0 = 0.72 \pm 0.10$, see also Fig.~\ref{fig:4}.
\item For the combination of likelihoods labeled as \textbf{HiLLiPoP} (which relies mainly on the \texttt{HiLLiPoP} likelihoods for the high-$\ell$ TT, TE, and EE spectra, along with the \texttt{LoLLiPoP} low-$\ell$ likelihood for E-mode polarization measurements), we find \textit{close} agreement with GR. Within the ($\mu,\Sigma$) framework, we obtain $\Sigma_0 = 0.12^{+0.11}_{-0.13}$, while for the growth index, we find $\gamma_0 = 0.621 \pm 0.090$ -- both consistent with the baseline values within one standard deviation (see Fig.~\ref{fig:1} and Fig.~\ref{fig:4}, again).
\end{itemize}

The better consistency with the predictions of GR found is closely correlated to the lensing anomaly. Indeed, across the different likelihoods, the shift towards values of $\Sigma_0 \neq 0$ and $\gamma_0 > 0.55$ follows the same trend as the reduced preference for $A_L > 1$. Specifically:
\begin{itemize}
    \item Using \textbf{Camspec} for temperature and polarization data shifts the lensing anomaly to less than $1.6\sigma$ (see Tab. 6 of Ref.~\cite{Rosenberg:2022sdy}), which roughly corresponds to the shifts observed both in $\Sigma_0$ and $\gamma_0$ -- see Tab.~\ref{tab_app:ms} and Tab.~\ref{tab_app:gamma}, respectively.
    \item Using the \textbf{HiLLiPoP} combination, the results regarding the lensing amplitude are consistent with $A_L = 1$ within one standard deviation (see, e.g., Tab. 6 of Ref.~\cite{Tristram:2023haj}). Similarly, $\Sigma_0$ and $\gamma_0$ align with their respective baseline values within one standard deviation.
\end{itemize}

Pushing this comparison further, we find that, similar to the lensing anomaly, the shifts in the values of $\Sigma_0$ and $\gamma_0$ observed in \textbf{Plik} and, to some extent, in \textbf{Camspec} are largely driven by the temperature anisotropy data. Specifically:
\begin{itemize}
    \item Focusing on \textbf{Camspec (TT-only)} (i.e., a combination of data involving only the TT spectrum at $\ell > 30$ from the \texttt{Camspec} likelihood), the deviation of $\Sigma_0$ from zero is more pronounced compared to the full \texttt{Camspec} TTTEEE likelihood, increasing from 1.4$\sigma$ to 2.2$\sigma$, see also Tab.~\ref{tab_app:ms} and Fig.~\ref{fig:3}. Notably, in this case, a significant shift in $\gamma_0 = 0.827^{+0.16}_{-0.051}$ is also observed, pulling this parameter away from the GR expectations, see also Tab.~\ref{tab_app:gamma} and Fig.~\ref{fig:5}.
    
    \item Focusing on \textbf{HiLLiPoP (TT-only)} (i.e., a combination of data involving only the TT spectrum at $\ell > 30$ from the \texttt{HiLLiPoP} likelihood), no significant deviation from GR is observed in either $\Sigma_0$ or $\gamma_0$ -- see also Fig.~\ref{fig:3} and Fig.~\ref{fig:5}. This is easily explained by noting that, for the same dataset, $A_L$ remains consistent with unity.
\end{itemize}

Digging deeper into the correlation with the lensing anomaly, we vary the parameter $A_L$ within both the ($\mu,\Sigma$) framework and the extended model involving the growth index $\gamma$. In these extended parameter spaces, the correlations among parameters are such that the numerical constraints on \underline{all} modified gravity indicators (i.e., $A_L$, $\Sigma_0$, and $\gamma_0$) remain consistent within two standard deviations of GR. This holds true even for the Planck-PR3-based data combination \textbf{Plik}. However, the joint marginalized constraints in the ($A_L$, $\Sigma_0$) plane shown in Fig.~\ref{fig:2} and in the ($A_L$, $\gamma_0$) plane presented in Fig.~\ref{fig:6} reveal that, for \textbf{Plik}, the point corresponding to $\Lambda$CDM consistently falls outside the 95\% CL contours in both cases. In contrast, $\Lambda$CDM comfortably falls within the 95\% CL contours for both \textbf{Camspec} and \textbf{HiLLiPoP}, underscoring that in these latter cases, Planck-PR4 data agree with the standard $\Lambda$CDM model without needing one anomaly to be resolved at the expense of another.

Wrapping everything up, there is a strong basis to conclude that a significant portion of the hints for modified theories of gravity identified and debated over the past few years might stem from observational systematics in the Planck-PR3 data, likely rooted in the well-documented excess smoothing in the high-$\ell$ peaks of the TT spectrum. To further validate this interpretation, it would be interesting to adapt the \texttt{Plik} high-$\ell$ likelihood to the Planck-PR4 data and reassess Planck's 2018 conclusions in their entirety.

\begin{acknowledgments}
\noindent W.G. is supported by the Lancaster–Sheffield Consortium for Fundamental Physics under STFC grant: ST/X000621/1. E.D.V. is supported by a Royal Society Dorothy Hodgkin Research Fellowship. This article is based upon work from the COST Action CA21136 - ``Addressing observational tensions in cosmology with systematics and fundamental physics (CosmoVerse)'', supported by COST - ``European Cooperation in Science and Technology''.
We acknowledge the IT Services at The University of Sheffield for the provision of services for High Performance Computing.
\end{acknowledgments}

\bibliography{refs}

\begin{thebibliography}{368}%
\makeatletter
\providecommand \@ifxundefined [1]{%
 \@ifx{#1\undefined}
}%
\providecommand \@ifnum [1]{%
 \ifnum #1\expandafter \@firstoftwo
 \else \expandafter \@secondoftwo
 \fi
}%
\providecommand \@ifx [1]{%
 \ifx #1\expandafter \@firstoftwo
 \else \expandafter \@secondoftwo
 \fi
}%
\providecommand \natexlab [1]{#1}%
\providecommand \enquote  [1]{``#1''}%
\providecommand \bibnamefont  [1]{#1}%
\providecommand \bibfnamefont [1]{#1}%
\providecommand \citenamefont [1]{#1}%
\providecommand \href@noop [0]{\@secondoftwo}%
\providecommand \href [0]{\begingroup \@sanitize@url \@href}%
\providecommand \@href[1]{\@@startlink{#1}\@@href}%
\providecommand \@@href[1]{\endgroup#1\@@endlink}%
\providecommand \@sanitize@url [0]{\catcode `\\12\catcode `\$12\catcode `\&12\catcode `\#12\catcode `\^12\catcode `\_12\catcode `\%12\relax}%
\providecommand \@@startlink[1]{}%
\providecommand \@@endlink[0]{}%
\providecommand \url  [0]{\begingroup\@sanitize@url \@url }%
\providecommand \@url [1]{\endgroup\@href {#1}{\urlprefix }}%
\providecommand \urlprefix  [0]{URL }%
\providecommand \Eprint [0]{\href }%
\providecommand \doibase [0]{http://dx.doi.org/}%
\providecommand \selectlanguage [0]{\@gobble}%
\providecommand \bibinfo  [0]{\@secondoftwo}%
\providecommand \bibfield  [0]{\@secondoftwo}%
\providecommand \translation [1]{[#1]}%
\providecommand \BibitemOpen [0]{}%
\providecommand \bibitemStop [0]{}%
\providecommand \bibitemNoStop [0]{.\EOS\space}%
\providecommand \EOS [0]{\spacefactor3000\relax}%
\providecommand \BibitemShut  [1]{\csname bibitem#1\endcsname}%
\let\auto@bib@innerbib\@empty
\bibitem [{\citenamefont {Riess}\ \emph {et~al.}(1998)\citenamefont {Riess} \emph {et~al.}}]{SupernovaSearchTeam:1998fmf}%
  \BibitemOpen
  \bibfield  {author} {\bibinfo {author} {\bibfnamefont {A.~G.}\ \bibnamefont {Riess}} \emph {et~al.} (\bibinfo {collaboration} {Supernova Search Team}),\ }\href {\doibase 10.1086/300499} {\bibfield  {journal} {\bibinfo  {journal} {Astron. J.}\ }\textbf {\bibinfo {volume} {116}},\ \bibinfo {pages} {1009} (\bibinfo {year} {1998})},\ \Eprint {http://arxiv.org/abs/astro-ph/9805201} {arXiv:astro-ph/9805201} \BibitemShut {NoStop}%
\bibitem [{\citenamefont {Perlmutter}\ \emph {et~al.}(1999)\citenamefont {Perlmutter} \emph {et~al.}}]{SupernovaCosmologyProject:1998vns}%
  \BibitemOpen
  \bibfield  {author} {\bibinfo {author} {\bibfnamefont {S.}~\bibnamefont {Perlmutter}} \emph {et~al.} (\bibinfo {collaboration} {Supernova Cosmology Project}),\ }\href {\doibase 10.1086/307221} {\bibfield  {journal} {\bibinfo  {journal} {Astrophys. J.}\ }\textbf {\bibinfo {volume} {517}},\ \bibinfo {pages} {565} (\bibinfo {year} {1999})},\ \Eprint {http://arxiv.org/abs/astro-ph/9812133} {arXiv:astro-ph/9812133} \BibitemShut {NoStop}%
\bibitem [{\citenamefont {Riess}\ \emph {et~al.}(2001)\citenamefont {Riess} \emph {et~al.}}]{SupernovaSearchTeam:2001qse}%
  \BibitemOpen
  \bibfield  {author} {\bibinfo {author} {\bibfnamefont {A.~G.}\ \bibnamefont {Riess}} \emph {et~al.} (\bibinfo {collaboration} {Supernova Search Team}),\ }\href {\doibase 10.1086/322348} {\bibfield  {journal} {\bibinfo  {journal} {Astrophys. J.}\ }\textbf {\bibinfo {volume} {560}},\ \bibinfo {pages} {49} (\bibinfo {year} {2001})},\ \Eprint {http://arxiv.org/abs/astro-ph/0104455} {arXiv:astro-ph/0104455} \BibitemShut {NoStop}%
\bibitem [{\citenamefont {Tegmark}\ \emph {et~al.}(2004)\citenamefont {Tegmark} \emph {et~al.}}]{SDSS:2003eyi}%
  \BibitemOpen
  \bibfield  {author} {\bibinfo {author} {\bibfnamefont {M.}~\bibnamefont {Tegmark}} \emph {et~al.} (\bibinfo {collaboration} {SDSS}),\ }\href {\doibase 10.1103/PhysRevD.69.103501} {\bibfield  {journal} {\bibinfo  {journal} {Phys. Rev. D}\ }\textbf {\bibinfo {volume} {69}},\ \bibinfo {pages} {103501} (\bibinfo {year} {2004})},\ \Eprint {http://arxiv.org/abs/astro-ph/0310723} {arXiv:astro-ph/0310723} \BibitemShut {NoStop}%
\bibitem [{\citenamefont {Scranton}\ \emph {et~al.}(2003)\citenamefont {Scranton} \emph {et~al.}}]{SDSS:2003lnz}%
  \BibitemOpen
  \bibfield  {author} {\bibinfo {author} {\bibfnamefont {R.}~\bibnamefont {Scranton}} \emph {et~al.} (\bibinfo {collaboration} {SDSS}),\ }\href@noop {} {\  (\bibinfo {year} {2003})},\ \Eprint {http://arxiv.org/abs/astro-ph/0307335} {arXiv:astro-ph/0307335} \BibitemShut {NoStop}%
\bibitem [{\citenamefont {Tonry}\ \emph {et~al.}(2003)\citenamefont {Tonry} \emph {et~al.}}]{SupernovaSearchTeam:2003cyd}%
  \BibitemOpen
  \bibfield  {author} {\bibinfo {author} {\bibfnamefont {J.~L.}\ \bibnamefont {Tonry}} \emph {et~al.} (\bibinfo {collaboration} {Supernova Search Team}),\ }\href {\doibase 10.1086/376865} {\bibfield  {journal} {\bibinfo  {journal} {Astrophys. J.}\ }\textbf {\bibinfo {volume} {594}},\ \bibinfo {pages} {1} (\bibinfo {year} {2003})},\ \Eprint {http://arxiv.org/abs/astro-ph/0305008} {arXiv:astro-ph/0305008} \BibitemShut {NoStop}%
\bibitem [{\citenamefont {Knop}\ \emph {et~al.}(2003)\citenamefont {Knop} \emph {et~al.}}]{SupernovaCosmologyProject:2003dcn}%
  \BibitemOpen
  \bibfield  {author} {\bibinfo {author} {\bibfnamefont {R.~A.}\ \bibnamefont {Knop}} \emph {et~al.} (\bibinfo {collaboration} {Supernova Cosmology Project}),\ }\href {\doibase 10.1086/378560} {\bibfield  {journal} {\bibinfo  {journal} {Astrophys. J.}\ }\textbf {\bibinfo {volume} {598}},\ \bibinfo {pages} {102} (\bibinfo {year} {2003})},\ \Eprint {http://arxiv.org/abs/astro-ph/0309368} {arXiv:astro-ph/0309368} \BibitemShut {NoStop}%
\bibitem [{\citenamefont {Seljak}\ \emph {et~al.}(2005)\citenamefont {Seljak} \emph {et~al.}}]{SDSS:2004kqt}%
  \BibitemOpen
  \bibfield  {author} {\bibinfo {author} {\bibfnamefont {U.}~\bibnamefont {Seljak}} \emph {et~al.} (\bibinfo {collaboration} {SDSS}),\ }\href {\doibase 10.1103/PhysRevD.71.103515} {\bibfield  {journal} {\bibinfo  {journal} {Phys. Rev. D}\ }\textbf {\bibinfo {volume} {71}},\ \bibinfo {pages} {103515} (\bibinfo {year} {2005})},\ \Eprint {http://arxiv.org/abs/astro-ph/0407372} {arXiv:astro-ph/0407372} \BibitemShut {NoStop}%
\bibitem [{\citenamefont {Feng}\ \emph {et~al.}(2005)\citenamefont {Feng}, \citenamefont {Wang},\ and\ \citenamefont {Zhang}}]{Feng:2004ad}%
  \BibitemOpen
  \bibfield  {author} {\bibinfo {author} {\bibfnamefont {B.}~\bibnamefont {Feng}}, \bibinfo {author} {\bibfnamefont {X.-L.}\ \bibnamefont {Wang}}, \ and\ \bibinfo {author} {\bibfnamefont {X.-M.}\ \bibnamefont {Zhang}},\ }\href {\doibase 10.1016/j.physletb.2004.12.071} {\bibfield  {journal} {\bibinfo  {journal} {Phys. Lett. B}\ }\textbf {\bibinfo {volume} {607}},\ \bibinfo {pages} {35} (\bibinfo {year} {2005})},\ \Eprint {http://arxiv.org/abs/astro-ph/0404224} {arXiv:astro-ph/0404224} \BibitemShut {NoStop}%
\bibitem [{\citenamefont {Riess}\ \emph {et~al.}(2004)\citenamefont {Riess} \emph {et~al.}}]{SupernovaSearchTeam:2004lze}%
  \BibitemOpen
  \bibfield  {author} {\bibinfo {author} {\bibfnamefont {A.~G.}\ \bibnamefont {Riess}} \emph {et~al.} (\bibinfo {collaboration} {Supernova Search Team}),\ }\href {\doibase 10.1086/383612} {\bibfield  {journal} {\bibinfo  {journal} {Astrophys. J.}\ }\textbf {\bibinfo {volume} {607}},\ \bibinfo {pages} {665} (\bibinfo {year} {2004})},\ \Eprint {http://arxiv.org/abs/astro-ph/0402512} {arXiv:astro-ph/0402512} \BibitemShut {NoStop}%
\bibitem [{\citenamefont {Astier}\ \emph {et~al.}(2006)\citenamefont {Astier} \emph {et~al.}}]{SNLS:2005qlf}%
  \BibitemOpen
  \bibfield  {author} {\bibinfo {author} {\bibfnamefont {P.}~\bibnamefont {Astier}} \emph {et~al.} (\bibinfo {collaboration} {SNLS}),\ }\href {\doibase 10.1051/0004-6361:20054185} {\bibfield  {journal} {\bibinfo  {journal} {Astron. Astrophys.}\ }\textbf {\bibinfo {volume} {447}},\ \bibinfo {pages} {31} (\bibinfo {year} {2006})},\ \Eprint {http://arxiv.org/abs/astro-ph/0510447} {arXiv:astro-ph/0510447} \BibitemShut {NoStop}%
\bibitem [{\citenamefont {Eisenstein}\ \emph {et~al.}(2005)\citenamefont {Eisenstein} \emph {et~al.}}]{SDSS:2005xqv}%
  \BibitemOpen
  \bibfield  {author} {\bibinfo {author} {\bibfnamefont {D.~J.}\ \bibnamefont {Eisenstein}} \emph {et~al.} (\bibinfo {collaboration} {SDSS}),\ }\href {\doibase 10.1086/466512} {\bibfield  {journal} {\bibinfo  {journal} {Astrophys. J.}\ }\textbf {\bibinfo {volume} {633}},\ \bibinfo {pages} {560} (\bibinfo {year} {2005})},\ \Eprint {http://arxiv.org/abs/astro-ph/0501171} {arXiv:astro-ph/0501171} \BibitemShut {NoStop}%
\bibitem [{\citenamefont {Eisenstein}\ \emph {et~al.}(2007)\citenamefont {Eisenstein}, \citenamefont {Seo}, \citenamefont {Sirko},\ and\ \citenamefont {Spergel}}]{Eisenstein:2006nk}%
  \BibitemOpen
  \bibfield  {author} {\bibinfo {author} {\bibfnamefont {D.~J.}\ \bibnamefont {Eisenstein}}, \bibinfo {author} {\bibfnamefont {H.-j.}\ \bibnamefont {Seo}}, \bibinfo {author} {\bibfnamefont {E.}~\bibnamefont {Sirko}}, \ and\ \bibinfo {author} {\bibfnamefont {D.}~\bibnamefont {Spergel}},\ }\href {\doibase 10.1086/518712} {\bibfield  {journal} {\bibinfo  {journal} {Astrophys. J.}\ }\textbf {\bibinfo {volume} {664}},\ \bibinfo {pages} {675} (\bibinfo {year} {2007})},\ \Eprint {http://arxiv.org/abs/astro-ph/0604362} {arXiv:astro-ph/0604362} \BibitemShut {NoStop}%
\bibitem [{\citenamefont {Tegmark}\ \emph {et~al.}(2006)\citenamefont {Tegmark} \emph {et~al.}}]{SDSS:2006lmn}%
  \BibitemOpen
  \bibfield  {author} {\bibinfo {author} {\bibfnamefont {M.}~\bibnamefont {Tegmark}} \emph {et~al.} (\bibinfo {collaboration} {SDSS}),\ }\href {\doibase 10.1103/PhysRevD.74.123507} {\bibfield  {journal} {\bibinfo  {journal} {Phys. Rev. D}\ }\textbf {\bibinfo {volume} {74}},\ \bibinfo {pages} {123507} (\bibinfo {year} {2006})},\ \Eprint {http://arxiv.org/abs/astro-ph/0608632} {arXiv:astro-ph/0608632} \BibitemShut {NoStop}%
\bibitem [{\citenamefont {Sahni}\ and\ \citenamefont {Starobinsky}(2006)}]{Sahni:2006pa}%
  \BibitemOpen
  \bibfield  {author} {\bibinfo {author} {\bibfnamefont {V.}~\bibnamefont {Sahni}}\ and\ \bibinfo {author} {\bibfnamefont {A.}~\bibnamefont {Starobinsky}},\ }\href {\doibase 10.1142/S0218271806009704} {\bibfield  {journal} {\bibinfo  {journal} {Int. J. Mod. Phys. D}\ }\textbf {\bibinfo {volume} {15}},\ \bibinfo {pages} {2105} (\bibinfo {year} {2006})},\ \Eprint {http://arxiv.org/abs/astro-ph/0610026} {arXiv:astro-ph/0610026} \BibitemShut {NoStop}%
\bibitem [{\citenamefont {Wood-Vasey}\ \emph {et~al.}(2007)\citenamefont {Wood-Vasey} \emph {et~al.}}]{ESSENCE:2007acn}%
  \BibitemOpen
  \bibfield  {author} {\bibinfo {author} {\bibfnamefont {W.~M.}\ \bibnamefont {Wood-Vasey}} \emph {et~al.} (\bibinfo {collaboration} {ESSENCE}),\ }\href {\doibase 10.1086/518642} {\bibfield  {journal} {\bibinfo  {journal} {Astrophys. J.}\ }\textbf {\bibinfo {volume} {666}},\ \bibinfo {pages} {694} (\bibinfo {year} {2007})},\ \Eprint {http://arxiv.org/abs/astro-ph/0701041} {arXiv:astro-ph/0701041} \BibitemShut {NoStop}%
\bibitem [{\citenamefont {Vikhlinin}\ \emph {et~al.}(2009)\citenamefont {Vikhlinin} \emph {et~al.}}]{Vikhlinin:2008ym}%
  \BibitemOpen
  \bibfield  {author} {\bibinfo {author} {\bibfnamefont {A.}~\bibnamefont {Vikhlinin}} \emph {et~al.},\ }\href {\doibase 10.1088/0004-637X/692/2/1060} {\bibfield  {journal} {\bibinfo  {journal} {Astrophys. J.}\ }\textbf {\bibinfo {volume} {692}},\ \bibinfo {pages} {1060} (\bibinfo {year} {2009})},\ \Eprint {http://arxiv.org/abs/0812.2720} {arXiv:0812.2720 [astro-ph]} \BibitemShut {NoStop}%
\bibitem [{\citenamefont {Stern}\ \emph {et~al.}(2010)\citenamefont {Stern}, \citenamefont {Jimenez}, \citenamefont {Verde}, \citenamefont {Kamionkowski},\ and\ \citenamefont {Stanford}}]{Stern:2009ep}%
  \BibitemOpen
  \bibfield  {author} {\bibinfo {author} {\bibfnamefont {D.}~\bibnamefont {Stern}}, \bibinfo {author} {\bibfnamefont {R.}~\bibnamefont {Jimenez}}, \bibinfo {author} {\bibfnamefont {L.}~\bibnamefont {Verde}}, \bibinfo {author} {\bibfnamefont {M.}~\bibnamefont {Kamionkowski}}, \ and\ \bibinfo {author} {\bibfnamefont {S.~A.}\ \bibnamefont {Stanford}},\ }\href {\doibase 10.1088/1475-7516/2010/02/008} {\bibfield  {journal} {\bibinfo  {journal} {JCAP}\ }\textbf {\bibinfo {volume} {02}},\ \bibinfo {pages} {008} (\bibinfo {year} {2010})},\ \Eprint {http://arxiv.org/abs/0907.3149} {arXiv:0907.3149 [astro-ph.CO]} \BibitemShut {NoStop}%
\bibitem [{\citenamefont {Sherwin}\ \emph {et~al.}(2011)\citenamefont {Sherwin} \emph {et~al.}}]{Sherwin:2011gv}%
  \BibitemOpen
  \bibfield  {author} {\bibinfo {author} {\bibfnamefont {B.~D.}\ \bibnamefont {Sherwin}} \emph {et~al.},\ }\href {\doibase 10.1103/PhysRevLett.107.021302} {\bibfield  {journal} {\bibinfo  {journal} {Phys. Rev. Lett.}\ }\textbf {\bibinfo {volume} {107}},\ \bibinfo {pages} {021302} (\bibinfo {year} {2011})},\ \Eprint {http://arxiv.org/abs/1105.0419} {arXiv:1105.0419 [astro-ph.CO]} \BibitemShut {NoStop}%
\bibitem [{\citenamefont {Bennett}\ \emph {et~al.}(2013)\citenamefont {Bennett} \emph {et~al.}}]{WMAP:2012fli}%
  \BibitemOpen
  \bibfield  {author} {\bibinfo {author} {\bibfnamefont {C.~L.}\ \bibnamefont {Bennett}} \emph {et~al.} (\bibinfo {collaboration} {WMAP}),\ }\href {\doibase 10.1088/0067-0049/208/2/20} {\bibfield  {journal} {\bibinfo  {journal} {Astrophys. J. Suppl.}\ }\textbf {\bibinfo {volume} {208}},\ \bibinfo {pages} {20} (\bibinfo {year} {2013})},\ \Eprint {http://arxiv.org/abs/1212.5225} {arXiv:1212.5225 [astro-ph.CO]} \BibitemShut {NoStop}%
\bibitem [{\citenamefont {Hinshaw}\ \emph {et~al.}(2013)\citenamefont {Hinshaw} \emph {et~al.}}]{WMAP:2012nax}%
  \BibitemOpen
  \bibfield  {author} {\bibinfo {author} {\bibfnamefont {G.}~\bibnamefont {Hinshaw}} \emph {et~al.} (\bibinfo {collaboration} {WMAP}),\ }\href {\doibase 10.1088/0067-0049/208/2/19} {\bibfield  {journal} {\bibinfo  {journal} {Astrophys. J. Suppl.}\ }\textbf {\bibinfo {volume} {208}},\ \bibinfo {pages} {19} (\bibinfo {year} {2013})},\ \Eprint {http://arxiv.org/abs/1212.5226} {arXiv:1212.5226 [astro-ph.CO]} \BibitemShut {NoStop}%
\bibitem [{\citenamefont {Dawson}\ \emph {et~al.}(2013)\citenamefont {Dawson} \emph {et~al.}}]{BOSS:2012dmf}%
  \BibitemOpen
  \bibfield  {author} {\bibinfo {author} {\bibfnamefont {K.~S.}\ \bibnamefont {Dawson}} \emph {et~al.} (\bibinfo {collaboration} {BOSS}),\ }\href {\doibase 10.1088/0004-6256/145/1/10} {\bibfield  {journal} {\bibinfo  {journal} {Astron. J.}\ }\textbf {\bibinfo {volume} {145}},\ \bibinfo {pages} {10} (\bibinfo {year} {2013})},\ \Eprint {http://arxiv.org/abs/1208.0022} {arXiv:1208.0022 [astro-ph.CO]} \BibitemShut {NoStop}%
\bibitem [{\citenamefont {de~Jong}\ \emph {et~al.}(2013)\citenamefont {de~Jong}, \citenamefont {Verdoes~Kleijn}, \citenamefont {Kuijken},\ and\ \citenamefont {Valentijn}}]{deJong:2012zb}%
  \BibitemOpen
  \bibfield  {author} {\bibinfo {author} {\bibfnamefont {J.~T.~A.}\ \bibnamefont {de~Jong}}, \bibinfo {author} {\bibfnamefont {G.~A.}\ \bibnamefont {Verdoes~Kleijn}}, \bibinfo {author} {\bibfnamefont {K.~H.}\ \bibnamefont {Kuijken}}, \ and\ \bibinfo {author} {\bibfnamefont {E.~A.}\ \bibnamefont {Valentijn}} (\bibinfo {collaboration} {Astro-WISE, KiDS}),\ }\href {\doibase 10.1007/s10686-012-9306-1} {\bibfield  {journal} {\bibinfo  {journal} {Exper. Astron.}\ }\textbf {\bibinfo {volume} {35}},\ \bibinfo {pages} {25} (\bibinfo {year} {2013})},\ \Eprint {http://arxiv.org/abs/1206.1254} {arXiv:1206.1254 [astro-ph.CO]} \BibitemShut {NoStop}%
\bibitem [{\citenamefont {Anderson}\ \emph {et~al.}(2014)\citenamefont {Anderson} \emph {et~al.}}]{BOSS:2013rlg}%
  \BibitemOpen
  \bibfield  {author} {\bibinfo {author} {\bibfnamefont {L.}~\bibnamefont {Anderson}} \emph {et~al.} (\bibinfo {collaboration} {BOSS}),\ }\href {\doibase 10.1093/mnras/stu523} {\bibfield  {journal} {\bibinfo  {journal} {Mon. Not. Roy. Astron. Soc.}\ }\textbf {\bibinfo {volume} {441}},\ \bibinfo {pages} {24} (\bibinfo {year} {2014})},\ \Eprint {http://arxiv.org/abs/1312.4877} {arXiv:1312.4877 [astro-ph.CO]} \BibitemShut {NoStop}%
\bibitem [{\citenamefont {Weinberg}\ \emph {et~al.}(2013)\citenamefont {Weinberg}, \citenamefont {Mortonson}, \citenamefont {Eisenstein}, \citenamefont {Hirata}, \citenamefont {Riess},\ and\ \citenamefont {Rozo}}]{Weinberg:2013agg}%
  \BibitemOpen
  \bibfield  {author} {\bibinfo {author} {\bibfnamefont {D.~H.}\ \bibnamefont {Weinberg}}, \bibinfo {author} {\bibfnamefont {M.~J.}\ \bibnamefont {Mortonson}}, \bibinfo {author} {\bibfnamefont {D.~J.}\ \bibnamefont {Eisenstein}}, \bibinfo {author} {\bibfnamefont {C.}~\bibnamefont {Hirata}}, \bibinfo {author} {\bibfnamefont {A.~G.}\ \bibnamefont {Riess}}, \ and\ \bibinfo {author} {\bibfnamefont {E.}~\bibnamefont {Rozo}},\ }\href {\doibase 10.1016/j.physrep.2013.05.001} {\bibfield  {journal} {\bibinfo  {journal} {Phys. Rept.}\ }\textbf {\bibinfo {volume} {530}},\ \bibinfo {pages} {87} (\bibinfo {year} {2013})},\ \Eprint {http://arxiv.org/abs/1201.2434} {arXiv:1201.2434 [astro-ph.CO]} \BibitemShut {NoStop}%
\bibitem [{\citenamefont {Beutler}\ \emph {et~al.}(2014)\citenamefont {Beutler} \emph {et~al.}}]{BOSS:2013uda}%
  \BibitemOpen
  \bibfield  {author} {\bibinfo {author} {\bibfnamefont {F.}~\bibnamefont {Beutler}} \emph {et~al.} (\bibinfo {collaboration} {BOSS}),\ }\href {\doibase 10.1093/mnras/stu1051} {\bibfield  {journal} {\bibinfo  {journal} {Mon. Not. Roy. Astron. Soc.}\ }\textbf {\bibinfo {volume} {443}},\ \bibinfo {pages} {1065} (\bibinfo {year} {2014})},\ \Eprint {http://arxiv.org/abs/1312.4611} {arXiv:1312.4611 [astro-ph.CO]} \BibitemShut {NoStop}%
\bibitem [{\citenamefont {Delubac}\ \emph {et~al.}(2015)\citenamefont {Delubac} \emph {et~al.}}]{BOSS:2014hwf}%
  \BibitemOpen
  \bibfield  {author} {\bibinfo {author} {\bibfnamefont {T.}~\bibnamefont {Delubac}} \emph {et~al.} (\bibinfo {collaboration} {BOSS}),\ }\href {\doibase 10.1051/0004-6361/201423969} {\bibfield  {journal} {\bibinfo  {journal} {Astron. Astrophys.}\ }\textbf {\bibinfo {volume} {574}},\ \bibinfo {pages} {A59} (\bibinfo {year} {2015})},\ \Eprint {http://arxiv.org/abs/1404.1801} {arXiv:1404.1801 [astro-ph.CO]} \BibitemShut {NoStop}%
\bibitem [{\citenamefont {Betoule}\ \emph {et~al.}(2014)\citenamefont {Betoule} \emph {et~al.}}]{SDSS:2014iwm}%
  \BibitemOpen
  \bibfield  {author} {\bibinfo {author} {\bibfnamefont {M.}~\bibnamefont {Betoule}} \emph {et~al.} (\bibinfo {collaboration} {SDSS}),\ }\href {\doibase 10.1051/0004-6361/201423413} {\bibfield  {journal} {\bibinfo  {journal} {Astron. Astrophys.}\ }\textbf {\bibinfo {volume} {568}},\ \bibinfo {pages} {A22} (\bibinfo {year} {2014})},\ \Eprint {http://arxiv.org/abs/1401.4064} {arXiv:1401.4064 [astro-ph.CO]} \BibitemShut {NoStop}%
\bibitem [{\citenamefont {Aubourg}\ \emph {et~al.}(2015)\citenamefont {Aubourg} \emph {et~al.}}]{BOSS:2014hhw}%
  \BibitemOpen
  \bibfield  {author} {\bibinfo {author} {\bibfnamefont {E.}~\bibnamefont {Aubourg}} \emph {et~al.} (\bibinfo {collaboration} {BOSS}),\ }\href {\doibase 10.1103/PhysRevD.92.123516} {\bibfield  {journal} {\bibinfo  {journal} {Phys. Rev. D}\ }\textbf {\bibinfo {volume} {92}},\ \bibinfo {pages} {123516} (\bibinfo {year} {2015})},\ \Eprint {http://arxiv.org/abs/1411.1074} {arXiv:1411.1074 [astro-ph.CO]} \BibitemShut {NoStop}%
\bibitem [{\citenamefont {Ross}\ \emph {et~al.}(2015)\citenamefont {Ross}, \citenamefont {Samushia}, \citenamefont {Howlett}, \citenamefont {Percival}, \citenamefont {Burden},\ and\ \citenamefont {Manera}}]{Ross:2014qpa}%
  \BibitemOpen
  \bibfield  {author} {\bibinfo {author} {\bibfnamefont {A.~J.}\ \bibnamefont {Ross}}, \bibinfo {author} {\bibfnamefont {L.}~\bibnamefont {Samushia}}, \bibinfo {author} {\bibfnamefont {C.}~\bibnamefont {Howlett}}, \bibinfo {author} {\bibfnamefont {W.~J.}\ \bibnamefont {Percival}}, \bibinfo {author} {\bibfnamefont {A.}~\bibnamefont {Burden}}, \ and\ \bibinfo {author} {\bibfnamefont {M.}~\bibnamefont {Manera}},\ }\href {\doibase 10.1093/mnras/stv154} {\bibfield  {journal} {\bibinfo  {journal} {Mon. Not. Roy. Astron. Soc.}\ }\textbf {\bibinfo {volume} {449}},\ \bibinfo {pages} {835} (\bibinfo {year} {2015})},\ \Eprint {http://arxiv.org/abs/1409.3242} {arXiv:1409.3242 [astro-ph.CO]} \BibitemShut {NoStop}%
\bibitem [{\citenamefont {Moresco}\ \emph {et~al.}(2016{\natexlab{a}})\citenamefont {Moresco}, \citenamefont {Pozzetti}, \citenamefont {Cimatti}, \citenamefont {Jimenez}, \citenamefont {Maraston}, \citenamefont {Verde}, \citenamefont {Thomas}, \citenamefont {Citro}, \citenamefont {Tojeiro},\ and\ \citenamefont {Wilkinson}}]{Moresco:2016mzx}%
  \BibitemOpen
  \bibfield  {author} {\bibinfo {author} {\bibfnamefont {M.}~\bibnamefont {Moresco}}, \bibinfo {author} {\bibfnamefont {L.}~\bibnamefont {Pozzetti}}, \bibinfo {author} {\bibfnamefont {A.}~\bibnamefont {Cimatti}}, \bibinfo {author} {\bibfnamefont {R.}~\bibnamefont {Jimenez}}, \bibinfo {author} {\bibfnamefont {C.}~\bibnamefont {Maraston}}, \bibinfo {author} {\bibfnamefont {L.}~\bibnamefont {Verde}}, \bibinfo {author} {\bibfnamefont {D.}~\bibnamefont {Thomas}}, \bibinfo {author} {\bibfnamefont {A.}~\bibnamefont {Citro}}, \bibinfo {author} {\bibfnamefont {R.}~\bibnamefont {Tojeiro}}, \ and\ \bibinfo {author} {\bibfnamefont {D.}~\bibnamefont {Wilkinson}},\ }\href {\doibase 10.1088/1475-7516/2016/05/014} {\bibfield  {journal} {\bibinfo  {journal} {JCAP}\ }\textbf {\bibinfo {volume} {05}},\ \bibinfo {pages} {014} (\bibinfo {year} {2016}{\natexlab{a}})},\ \Eprint {http://arxiv.org/abs/1601.01701} {arXiv:1601.01701 [astro-ph.CO]} \BibitemShut {NoStop}%
\bibitem [{\citenamefont {Moresco}\ \emph {et~al.}(2016{\natexlab{b}})\citenamefont {Moresco}, \citenamefont {Jimenez}, \citenamefont {Verde}, \citenamefont {Cimatti}, \citenamefont {Pozzetti}, \citenamefont {Maraston},\ and\ \citenamefont {Thomas}}]{Moresco:2016nqq}%
  \BibitemOpen
  \bibfield  {author} {\bibinfo {author} {\bibfnamefont {M.}~\bibnamefont {Moresco}}, \bibinfo {author} {\bibfnamefont {R.}~\bibnamefont {Jimenez}}, \bibinfo {author} {\bibfnamefont {L.}~\bibnamefont {Verde}}, \bibinfo {author} {\bibfnamefont {A.}~\bibnamefont {Cimatti}}, \bibinfo {author} {\bibfnamefont {L.}~\bibnamefont {Pozzetti}}, \bibinfo {author} {\bibfnamefont {C.}~\bibnamefont {Maraston}}, \ and\ \bibinfo {author} {\bibfnamefont {D.}~\bibnamefont {Thomas}},\ }\href {\doibase 10.1088/1475-7516/2016/12/039} {\bibfield  {journal} {\bibinfo  {journal} {JCAP}\ }\textbf {\bibinfo {volume} {12}},\ \bibinfo {pages} {039} (\bibinfo {year} {2016}{\natexlab{b}})},\ \Eprint {http://arxiv.org/abs/1604.00183} {arXiv:1604.00183 [astro-ph.CO]} \BibitemShut {NoStop}%
\bibitem [{\citenamefont {Rubin}\ and\ \citenamefont {Hayden}(2016)}]{Rubin:2016iqe}%
  \BibitemOpen
  \bibfield  {author} {\bibinfo {author} {\bibfnamefont {D.}~\bibnamefont {Rubin}}\ and\ \bibinfo {author} {\bibfnamefont {B.}~\bibnamefont {Hayden}},\ }\href {\doibase 10.3847/2041-8213/833/2/L30} {\bibfield  {journal} {\bibinfo  {journal} {Astrophys. J. Lett.}\ }\textbf {\bibinfo {volume} {833}},\ \bibinfo {pages} {L30} (\bibinfo {year} {2016})},\ \Eprint {http://arxiv.org/abs/1610.08972} {arXiv:1610.08972 [astro-ph.CO]} \BibitemShut {NoStop}%
\bibitem [{\citenamefont {Alam}\ \emph {et~al.}(2017)\citenamefont {Alam} \emph {et~al.}}]{BOSS:2016wmc}%
  \BibitemOpen
  \bibfield  {author} {\bibinfo {author} {\bibfnamefont {S.}~\bibnamefont {Alam}} \emph {et~al.} (\bibinfo {collaboration} {BOSS}),\ }\href {\doibase 10.1093/mnras/stx721} {\bibfield  {journal} {\bibinfo  {journal} {Mon. Not. Roy. Astron. Soc.}\ }\textbf {\bibinfo {volume} {470}},\ \bibinfo {pages} {2617} (\bibinfo {year} {2017})},\ \Eprint {http://arxiv.org/abs/1607.03155} {arXiv:1607.03155 [astro-ph.CO]} \BibitemShut {NoStop}%
\bibitem [{\citenamefont {Abbott}\ \emph {et~al.}(2016)\citenamefont {Abbott} \emph {et~al.}}]{DES:2016jjg}%
  \BibitemOpen
  \bibfield  {author} {\bibinfo {author} {\bibfnamefont {T.}~\bibnamefont {Abbott}} \emph {et~al.} (\bibinfo {collaboration} {DES}),\ }\href {\doibase 10.1093/mnras/stw641} {\bibfield  {journal} {\bibinfo  {journal} {Mon. Not. Roy. Astron. Soc.}\ }\textbf {\bibinfo {volume} {460}},\ \bibinfo {pages} {1270} (\bibinfo {year} {2016})},\ \Eprint {http://arxiv.org/abs/1601.00329} {arXiv:1601.00329 [astro-ph.CO]} \BibitemShut {NoStop}%
\bibitem [{\citenamefont {Haridasu}\ \emph {et~al.}(2017)\citenamefont {Haridasu}, \citenamefont {Lukovi\'c}, \citenamefont {D'Agostino},\ and\ \citenamefont {Vittorio}}]{Haridasu:2017lma}%
  \BibitemOpen
  \bibfield  {author} {\bibinfo {author} {\bibfnamefont {B.~S.}\ \bibnamefont {Haridasu}}, \bibinfo {author} {\bibfnamefont {V.~V.}\ \bibnamefont {Lukovi\'c}}, \bibinfo {author} {\bibfnamefont {R.}~\bibnamefont {D'Agostino}}, \ and\ \bibinfo {author} {\bibfnamefont {N.}~\bibnamefont {Vittorio}},\ }\href {\doibase 10.1051/0004-6361/201730469} {\bibfield  {journal} {\bibinfo  {journal} {Astron. Astrophys.}\ }\textbf {\bibinfo {volume} {600}},\ \bibinfo {pages} {L1} (\bibinfo {year} {2017})},\ \Eprint {http://arxiv.org/abs/1702.08244} {arXiv:1702.08244 [astro-ph.CO]} \BibitemShut {NoStop}%
\bibitem [{\citenamefont {Troxel}\ \emph {et~al.}(2018)\citenamefont {Troxel} \emph {et~al.}}]{DES:2017qwj}%
  \BibitemOpen
  \bibfield  {author} {\bibinfo {author} {\bibfnamefont {M.~A.}\ \bibnamefont {Troxel}} \emph {et~al.} (\bibinfo {collaboration} {DES}),\ }\href {\doibase 10.1103/PhysRevD.98.043528} {\bibfield  {journal} {\bibinfo  {journal} {Phys. Rev. D}\ }\textbf {\bibinfo {volume} {98}},\ \bibinfo {pages} {043528} (\bibinfo {year} {2018})},\ \Eprint {http://arxiv.org/abs/1708.01538} {arXiv:1708.01538 [astro-ph.CO]} \BibitemShut {NoStop}%
\bibitem [{\citenamefont {Scolnic}\ \emph {et~al.}(2018)\citenamefont {Scolnic} \emph {et~al.}}]{Pan-STARRS1:2017jku}%
  \BibitemOpen
  \bibfield  {author} {\bibinfo {author} {\bibfnamefont {D.~M.}\ \bibnamefont {Scolnic}} \emph {et~al.} (\bibinfo {collaboration} {Pan-STARRS1}),\ }\href {\doibase 10.3847/1538-4357/aab9bb} {\bibfield  {journal} {\bibinfo  {journal} {Astrophys. J.}\ }\textbf {\bibinfo {volume} {859}},\ \bibinfo {pages} {101} (\bibinfo {year} {2018})},\ \Eprint {http://arxiv.org/abs/1710.00845} {arXiv:1710.00845 [astro-ph.CO]} \BibitemShut {NoStop}%
\bibitem [{\citenamefont {Aghanim}\ \emph {et~al.}(2020{\natexlab{a}})\citenamefont {Aghanim} \emph {et~al.}}]{Planck:2018nkj}%
  \BibitemOpen
  \bibfield  {author} {\bibinfo {author} {\bibfnamefont {N.}~\bibnamefont {Aghanim}} \emph {et~al.} (\bibinfo {collaboration} {Planck}),\ }\href {\doibase 10.1051/0004-6361/201833880} {\bibfield  {journal} {\bibinfo  {journal} {Astron. Astrophys.}\ }\textbf {\bibinfo {volume} {641}},\ \bibinfo {pages} {A1} (\bibinfo {year} {2020}{\natexlab{a}})},\ \Eprint {http://arxiv.org/abs/1807.06205} {arXiv:1807.06205 [astro-ph.CO]} \BibitemShut {NoStop}%
\bibitem [{\citenamefont {Aghanim}\ \emph {et~al.}(2020{\natexlab{b}})\citenamefont {Aghanim} \emph {et~al.}}]{Planck:2018vyg}%
  \BibitemOpen
  \bibfield  {author} {\bibinfo {author} {\bibfnamefont {N.}~\bibnamefont {Aghanim}} \emph {et~al.} (\bibinfo {collaboration} {Planck}),\ }\href {\doibase 10.1051/0004-6361/201833910} {\bibfield  {journal} {\bibinfo  {journal} {Astron. Astrophys.}\ }\textbf {\bibinfo {volume} {641}},\ \bibinfo {pages} {A6} (\bibinfo {year} {2020}{\natexlab{b}})},\ \bibinfo {note} {[Erratum: Astron.Astrophys. 652, C4 (2021)]},\ \Eprint {http://arxiv.org/abs/1807.06209} {arXiv:1807.06209 [astro-ph.CO]} \BibitemShut {NoStop}%
\bibitem [{\citenamefont {G\'omez-Valent}(2019)}]{Gomez-Valent:2018gvm}%
  \BibitemOpen
  \bibfield  {author} {\bibinfo {author} {\bibfnamefont {A.}~\bibnamefont {G\'omez-Valent}},\ }\href {\doibase 10.1088/1475-7516/2019/05/026} {\bibfield  {journal} {\bibinfo  {journal} {JCAP}\ }\textbf {\bibinfo {volume} {05}},\ \bibinfo {pages} {026} (\bibinfo {year} {2019})},\ \Eprint {http://arxiv.org/abs/1810.02278} {arXiv:1810.02278 [astro-ph.CO]} \BibitemShut {NoStop}%
\bibitem [{\citenamefont {Yang}\ and\ \citenamefont {Gong}(2020)}]{Yang:2019fjt}%
  \BibitemOpen
  \bibfield  {author} {\bibinfo {author} {\bibfnamefont {Y.}~\bibnamefont {Yang}}\ and\ \bibinfo {author} {\bibfnamefont {Y.}~\bibnamefont {Gong}},\ }\href {\doibase 10.1088/1475-7516/2020/06/059} {\bibfield  {journal} {\bibinfo  {journal} {JCAP}\ }\textbf {\bibinfo {volume} {06}},\ \bibinfo {pages} {059} (\bibinfo {year} {2020})},\ \Eprint {http://arxiv.org/abs/1912.07375} {arXiv:1912.07375 [astro-ph.CO]} \BibitemShut {NoStop}%
\bibitem [{\citenamefont {Choi}\ \emph {et~al.}(2020)\citenamefont {Choi} \emph {et~al.}}]{ACT:2020frw}%
  \BibitemOpen
  \bibfield  {author} {\bibinfo {author} {\bibfnamefont {S.~K.}\ \bibnamefont {Choi}} \emph {et~al.} (\bibinfo {collaboration} {ACT}),\ }\href {\doibase 10.1088/1475-7516/2020/12/045} {\bibfield  {journal} {\bibinfo  {journal} {JCAP}\ }\textbf {\bibinfo {volume} {12}},\ \bibinfo {pages} {045} (\bibinfo {year} {2020})},\ \Eprint {http://arxiv.org/abs/2007.07289} {arXiv:2007.07289 [astro-ph.CO]} \BibitemShut {NoStop}%
\bibitem [{\citenamefont {Aiola}\ \emph {et~al.}(2020)\citenamefont {Aiola} \emph {et~al.}}]{ACT:2020gnv}%
  \BibitemOpen
  \bibfield  {author} {\bibinfo {author} {\bibfnamefont {S.}~\bibnamefont {Aiola}} \emph {et~al.} (\bibinfo {collaboration} {ACT}),\ }\href {\doibase 10.1088/1475-7516/2020/12/047} {\bibfield  {journal} {\bibinfo  {journal} {JCAP}\ }\textbf {\bibinfo {volume} {12}},\ \bibinfo {pages} {047} (\bibinfo {year} {2020})},\ \Eprint {http://arxiv.org/abs/2007.07288} {arXiv:2007.07288 [astro-ph.CO]} \BibitemShut {NoStop}%
\bibitem [{\citenamefont {Alam}\ \emph {et~al.}(2021)\citenamefont {Alam} \emph {et~al.}}]{eBOSS:2020yzd}%
  \BibitemOpen
  \bibfield  {author} {\bibinfo {author} {\bibfnamefont {S.}~\bibnamefont {Alam}} \emph {et~al.} (\bibinfo {collaboration} {eBOSS}),\ }\href {\doibase 10.1103/PhysRevD.103.083533} {\bibfield  {journal} {\bibinfo  {journal} {Phys. Rev. D}\ }\textbf {\bibinfo {volume} {103}},\ \bibinfo {pages} {083533} (\bibinfo {year} {2021})},\ \Eprint {http://arxiv.org/abs/2007.08991} {arXiv:2007.08991 [astro-ph.CO]} \BibitemShut {NoStop}%
\bibitem [{\citenamefont {Nadathur}\ \emph {et~al.}(2020)\citenamefont {Nadathur}, \citenamefont {Percival}, \citenamefont {Beutler},\ and\ \citenamefont {Winther}}]{Nadathur:2020kvq}%
  \BibitemOpen
  \bibfield  {author} {\bibinfo {author} {\bibfnamefont {S.}~\bibnamefont {Nadathur}}, \bibinfo {author} {\bibfnamefont {W.~J.}\ \bibnamefont {Percival}}, \bibinfo {author} {\bibfnamefont {F.}~\bibnamefont {Beutler}}, \ and\ \bibinfo {author} {\bibfnamefont {H.}~\bibnamefont {Winther}},\ }\href {\doibase 10.1103/PhysRevLett.124.221301} {\bibfield  {journal} {\bibinfo  {journal} {Phys. Rev. Lett.}\ }\textbf {\bibinfo {volume} {124}},\ \bibinfo {pages} {221301} (\bibinfo {year} {2020})},\ \Eprint {http://arxiv.org/abs/2001.11044} {arXiv:2001.11044 [astro-ph.CO]} \BibitemShut {NoStop}%
\bibitem [{\citenamefont {Rose}\ \emph {et~al.}(2020)\citenamefont {Rose}, \citenamefont {Rubin}, \citenamefont {Cikota}, \citenamefont {Deustua}, \citenamefont {Dixon}, \citenamefont {Fruchter}, \citenamefont {Jones}, \citenamefont {Riess},\ and\ \citenamefont {Scolnic}}]{Rose:2020shp}%
  \BibitemOpen
  \bibfield  {author} {\bibinfo {author} {\bibfnamefont {B.~M.}\ \bibnamefont {Rose}}, \bibinfo {author} {\bibfnamefont {D.}~\bibnamefont {Rubin}}, \bibinfo {author} {\bibfnamefont {A.}~\bibnamefont {Cikota}}, \bibinfo {author} {\bibfnamefont {S.~E.}\ \bibnamefont {Deustua}}, \bibinfo {author} {\bibfnamefont {S.}~\bibnamefont {Dixon}}, \bibinfo {author} {\bibfnamefont {A.}~\bibnamefont {Fruchter}}, \bibinfo {author} {\bibfnamefont {D.~O.}\ \bibnamefont {Jones}}, \bibinfo {author} {\bibfnamefont {A.~G.}\ \bibnamefont {Riess}}, \ and\ \bibinfo {author} {\bibfnamefont {D.~M.}\ \bibnamefont {Scolnic}},\ }\href {\doibase 10.3847/2041-8213/ab94ad} {\bibfield  {journal} {\bibinfo  {journal} {Astrophys. J. Lett.}\ }\textbf {\bibinfo {volume} {896}},\ \bibinfo {pages} {L4} (\bibinfo {year} {2020})},\ \Eprint {http://arxiv.org/abs/2002.12382} {arXiv:2002.12382 [astro-ph.CO]} \BibitemShut {NoStop}%
\bibitem [{\citenamefont {Di~Valentino}\ \emph {et~al.}(2020{\natexlab{a}})\citenamefont {Di~Valentino}, \citenamefont {Gariazzo}, \citenamefont {Mena},\ and\ \citenamefont {Vagnozzi}}]{DiValentino:2020evt}%
  \BibitemOpen
  \bibfield  {author} {\bibinfo {author} {\bibfnamefont {E.}~\bibnamefont {Di~Valentino}}, \bibinfo {author} {\bibfnamefont {S.}~\bibnamefont {Gariazzo}}, \bibinfo {author} {\bibfnamefont {O.}~\bibnamefont {Mena}}, \ and\ \bibinfo {author} {\bibfnamefont {S.}~\bibnamefont {Vagnozzi}},\ }\href {\doibase 10.1088/1475-7516/2020/07/045} {\bibfield  {journal} {\bibinfo  {journal} {JCAP}\ }\textbf {\bibinfo {volume} {07}},\ \bibinfo {pages} {045} (\bibinfo {year} {2020}{\natexlab{a}})},\ \Eprint {http://arxiv.org/abs/2005.02062} {arXiv:2005.02062 [astro-ph.CO]} \BibitemShut {NoStop}%
\bibitem [{\citenamefont {Asgari}\ \emph {et~al.}(2021)\citenamefont {Asgari} \emph {et~al.}}]{KiDS:2020suj}%
  \BibitemOpen
  \bibfield  {author} {\bibinfo {author} {\bibfnamefont {M.}~\bibnamefont {Asgari}} \emph {et~al.} (\bibinfo {collaboration} {KiDS}),\ }\href {\doibase 10.1051/0004-6361/202039070} {\bibfield  {journal} {\bibinfo  {journal} {Astron. Astrophys.}\ }\textbf {\bibinfo {volume} {645}},\ \bibinfo {pages} {A104} (\bibinfo {year} {2021})},\ \Eprint {http://arxiv.org/abs/2007.15633} {arXiv:2007.15633 [astro-ph.CO]} \BibitemShut {NoStop}%
\bibitem [{\citenamefont {Tr\"oster}\ \emph {et~al.}(2021)\citenamefont {Tr\"oster} \emph {et~al.}}]{KiDS:2020ghu}%
  \BibitemOpen
  \bibfield  {author} {\bibinfo {author} {\bibfnamefont {T.}~\bibnamefont {Tr\"oster}} \emph {et~al.} (\bibinfo {collaboration} {KiDS}),\ }\href {\doibase 10.1051/0004-6361/202039805} {\bibfield  {journal} {\bibinfo  {journal} {Astron. Astrophys.}\ }\textbf {\bibinfo {volume} {649}},\ \bibinfo {pages} {A88} (\bibinfo {year} {2021})},\ \Eprint {http://arxiv.org/abs/2010.16416} {arXiv:2010.16416 [astro-ph.CO]} \BibitemShut {NoStop}%
\bibitem [{\citenamefont {Dutcher}\ \emph {et~al.}(2021)\citenamefont {Dutcher} \emph {et~al.}}]{SPT-3G:2021eoc}%
  \BibitemOpen
  \bibfield  {author} {\bibinfo {author} {\bibfnamefont {D.}~\bibnamefont {Dutcher}} \emph {et~al.} (\bibinfo {collaboration} {SPT-3G}),\ }\href {\doibase 10.1103/PhysRevD.104.022003} {\bibfield  {journal} {\bibinfo  {journal} {Phys. Rev. D}\ }\textbf {\bibinfo {volume} {104}},\ \bibinfo {pages} {022003} (\bibinfo {year} {2021})},\ \Eprint {http://arxiv.org/abs/2101.01684} {arXiv:2101.01684 [astro-ph.CO]} \BibitemShut {NoStop}%
\bibitem [{\citenamefont {Abbott}\ \emph {et~al.}(2022)\citenamefont {Abbott} \emph {et~al.}}]{DES:2021wwk}%
  \BibitemOpen
  \bibfield  {author} {\bibinfo {author} {\bibfnamefont {T.~M.~C.}\ \bibnamefont {Abbott}} \emph {et~al.} (\bibinfo {collaboration} {DES}),\ }\href {\doibase 10.1103/PhysRevD.105.023520} {\bibfield  {journal} {\bibinfo  {journal} {Phys. Rev. D}\ }\textbf {\bibinfo {volume} {105}},\ \bibinfo {pages} {023520} (\bibinfo {year} {2022})},\ \Eprint {http://arxiv.org/abs/2105.13549} {arXiv:2105.13549 [astro-ph.CO]} \BibitemShut {NoStop}%
\bibitem [{\citenamefont {Moresco}\ \emph {et~al.}(2022)\citenamefont {Moresco} \emph {et~al.}}]{Moresco:2022phi}%
  \BibitemOpen
  \bibfield  {author} {\bibinfo {author} {\bibfnamefont {M.}~\bibnamefont {Moresco}} \emph {et~al.},\ }\href {\doibase 10.1007/s41114-022-00040-z} {\bibfield  {journal} {\bibinfo  {journal} {Living Rev. Rel.}\ }\textbf {\bibinfo {volume} {25}},\ \bibinfo {pages} {6} (\bibinfo {year} {2022})},\ \Eprint {http://arxiv.org/abs/2201.07241} {arXiv:2201.07241 [astro-ph.CO]} \BibitemShut {NoStop}%
\bibitem [{\citenamefont {Abbott}\ \emph {et~al.}(2023{\natexlab{a}})\citenamefont {Abbott} \emph {et~al.}}]{DES:2022ccp}%
  \BibitemOpen
  \bibfield  {author} {\bibinfo {author} {\bibfnamefont {T.~M.~C.}\ \bibnamefont {Abbott}} \emph {et~al.} (\bibinfo {collaboration} {DES}),\ }\href {\doibase 10.1103/PhysRevD.107.083504} {\bibfield  {journal} {\bibinfo  {journal} {Phys. Rev. D}\ }\textbf {\bibinfo {volume} {107}},\ \bibinfo {pages} {083504} (\bibinfo {year} {2023}{\natexlab{a}})},\ \Eprint {http://arxiv.org/abs/2207.05766} {arXiv:2207.05766 [astro-ph.CO]} \BibitemShut {NoStop}%
\bibitem [{\citenamefont {Brout}\ \emph {et~al.}(2022)\citenamefont {Brout} \emph {et~al.}}]{Brout:2022vxf}%
  \BibitemOpen
  \bibfield  {author} {\bibinfo {author} {\bibfnamefont {D.}~\bibnamefont {Brout}} \emph {et~al.},\ }\href {\doibase 10.3847/1538-4357/ac8e04} {\bibfield  {journal} {\bibinfo  {journal} {Astrophys. J.}\ }\textbf {\bibinfo {volume} {938}},\ \bibinfo {pages} {110} (\bibinfo {year} {2022})},\ \Eprint {http://arxiv.org/abs/2202.04077} {arXiv:2202.04077 [astro-ph.CO]} \BibitemShut {NoStop}%
\bibitem [{\citenamefont {Madhavacheril}\ \emph {et~al.}(2024)\citenamefont {Madhavacheril} \emph {et~al.}}]{ACT:2023kun}%
  \BibitemOpen
  \bibfield  {author} {\bibinfo {author} {\bibfnamefont {M.~S.}\ \bibnamefont {Madhavacheril}} \emph {et~al.} (\bibinfo {collaboration} {ACT}),\ }\href {\doibase 10.3847/1538-4357/acff5f} {\bibfield  {journal} {\bibinfo  {journal} {Astrophys. J.}\ }\textbf {\bibinfo {volume} {962}},\ \bibinfo {pages} {113} (\bibinfo {year} {2024})},\ \Eprint {http://arxiv.org/abs/2304.05203} {arXiv:2304.05203 [astro-ph.CO]} \BibitemShut {NoStop}%
\bibitem [{\citenamefont {Abbott}\ \emph {et~al.}(2023{\natexlab{b}})\citenamefont {Abbott} \emph {et~al.}}]{Kilo-DegreeSurvey:2023gfr}%
  \BibitemOpen
  \bibfield  {author} {\bibinfo {author} {\bibfnamefont {T.~M.~C.}\ \bibnamefont {Abbott}} \emph {et~al.} (\bibinfo {collaboration} {Kilo-Degree Survey, DES}),\ }\href {\doibase 10.21105/astro.2305.17173} {\bibfield  {journal} {\bibinfo  {journal} {Open J. Astrophys.}\ }\textbf {\bibinfo {volume} {6}},\ \bibinfo {pages} {2305.17173} (\bibinfo {year} {2023}{\natexlab{b}})},\ \Eprint {http://arxiv.org/abs/2305.17173} {arXiv:2305.17173 [astro-ph.CO]} \BibitemShut {NoStop}%
\bibitem [{\citenamefont {Adame}\ \emph {et~al.}(2024{\natexlab{a}})\citenamefont {Adame} \emph {et~al.}}]{DESI:2024uvr}%
  \BibitemOpen
  \bibfield  {author} {\bibinfo {author} {\bibfnamefont {A.~G.}\ \bibnamefont {Adame}} \emph {et~al.} (\bibinfo {collaboration} {DESI}),\ }\href@noop {} {\  (\bibinfo {year} {2024}{\natexlab{a}})},\ \Eprint {http://arxiv.org/abs/2404.03000} {arXiv:2404.03000 [astro-ph.CO]} \BibitemShut {NoStop}%
\bibitem [{\citenamefont {Lodha}\ \emph {et~al.}(2024)\citenamefont {Lodha} \emph {et~al.}}]{DESI:2024kob}%
  \BibitemOpen
  \bibfield  {author} {\bibinfo {author} {\bibfnamefont {K.}~\bibnamefont {Lodha}} \emph {et~al.} (\bibinfo {collaboration} {DESI}),\ }\href@noop {} {\  (\bibinfo {year} {2024})},\ \Eprint {http://arxiv.org/abs/2405.13588} {arXiv:2405.13588 [astro-ph.CO]} \BibitemShut {NoStop}%
\bibitem [{\citenamefont {Abbott}\ \emph {et~al.}(2024)\citenamefont {Abbott} \emph {et~al.}}]{DES:2024tys}%
  \BibitemOpen
  \bibfield  {author} {\bibinfo {author} {\bibfnamefont {T.~M.~C.}\ \bibnamefont {Abbott}} \emph {et~al.} (\bibinfo {collaboration} {Abbott, DES: T. M. C, DES}),\ }\href {\doibase 10.3847/2041-8213/ad6f9f} {\bibfield  {journal} {\bibinfo  {journal} {Astrophys. J. Lett.}\ }\textbf {\bibinfo {volume} {973}},\ \bibinfo {pages} {L14} (\bibinfo {year} {2024})},\ \Eprint {http://arxiv.org/abs/2401.02929} {arXiv:2401.02929 [astro-ph.CO]} \BibitemShut {NoStop}%
\bibitem [{\citenamefont {S\'anchez}\ \emph {et~al.}(2024)\citenamefont {S\'anchez} \emph {et~al.}}]{DES:2024upw}%
  \BibitemOpen
  \bibfield  {author} {\bibinfo {author} {\bibfnamefont {B.~O.}\ \bibnamefont {S\'anchez}} \emph {et~al.} (\bibinfo {collaboration} {DES}),\ }\href {\doibase 10.3847/1538-4357/ad739a} {\bibfield  {journal} {\bibinfo  {journal} {Astrophys. J.}\ }\textbf {\bibinfo {volume} {975}},\ \bibinfo {pages} {5} (\bibinfo {year} {2024})},\ \Eprint {http://arxiv.org/abs/2406.05046} {arXiv:2406.05046 [astro-ph.CO]} \BibitemShut {NoStop}%
\bibitem [{\citenamefont {Vincenzi}\ \emph {et~al.}(2024)\citenamefont {Vincenzi} \emph {et~al.}}]{DES:2024hip}%
  \BibitemOpen
  \bibfield  {author} {\bibinfo {author} {\bibfnamefont {M.}~\bibnamefont {Vincenzi}} \emph {et~al.} (\bibinfo {collaboration} {DES}),\ }\href {\doibase 10.3847/1538-4357/ad5e6c} {\bibfield  {journal} {\bibinfo  {journal} {Astrophys. J.}\ }\textbf {\bibinfo {volume} {975}},\ \bibinfo {pages} {86} (\bibinfo {year} {2024})},\ \Eprint {http://arxiv.org/abs/2401.02945} {arXiv:2401.02945 [astro-ph.CO]} \BibitemShut {NoStop}%
\bibitem [{\citenamefont {Verde}\ \emph {et~al.}(2019)\citenamefont {Verde}, \citenamefont {Treu},\ and\ \citenamefont {Riess}}]{Verde:2019ivm}%
  \BibitemOpen
  \bibfield  {author} {\bibinfo {author} {\bibfnamefont {L.}~\bibnamefont {Verde}}, \bibinfo {author} {\bibfnamefont {T.}~\bibnamefont {Treu}}, \ and\ \bibinfo {author} {\bibfnamefont {A.~G.}\ \bibnamefont {Riess}},\ }\href {\doibase 10.1038/s41550-019-0902-0} {\bibfield  {journal} {\bibinfo  {journal} {Nature Astron.}\ }\textbf {\bibinfo {volume} {3}},\ \bibinfo {pages} {891} (\bibinfo {year} {2019})},\ \Eprint {http://arxiv.org/abs/1907.10625} {arXiv:1907.10625 [astro-ph.CO]} \BibitemShut {NoStop}%
\bibitem [{\citenamefont {Di~Valentino}\ \emph {et~al.}(2021{\natexlab{a}})\citenamefont {Di~Valentino} \emph {et~al.}}]{DiValentino:2020zio}%
  \BibitemOpen
  \bibfield  {author} {\bibinfo {author} {\bibfnamefont {E.}~\bibnamefont {Di~Valentino}} \emph {et~al.},\ }\href {\doibase 10.1016/j.astropartphys.2021.102605} {\bibfield  {journal} {\bibinfo  {journal} {Astropart. Phys.}\ }\textbf {\bibinfo {volume} {131}},\ \bibinfo {pages} {102605} (\bibinfo {year} {2021}{\natexlab{a}})},\ \Eprint {http://arxiv.org/abs/2008.11284} {arXiv:2008.11284 [astro-ph.CO]} \BibitemShut {NoStop}%
\bibitem [{\citenamefont {Di~Valentino}\ \emph {et~al.}(2021{\natexlab{b}})\citenamefont {Di~Valentino}, \citenamefont {Mena}, \citenamefont {Pan}, \citenamefont {Visinelli}, \citenamefont {Yang}, \citenamefont {Melchiorri}, \citenamefont {Mota}, \citenamefont {Riess},\ and\ \citenamefont {Silk}}]{DiValentino:2021izs}%
  \BibitemOpen
  \bibfield  {author} {\bibinfo {author} {\bibfnamefont {E.}~\bibnamefont {Di~Valentino}}, \bibinfo {author} {\bibfnamefont {O.}~\bibnamefont {Mena}}, \bibinfo {author} {\bibfnamefont {S.}~\bibnamefont {Pan}}, \bibinfo {author} {\bibfnamefont {L.}~\bibnamefont {Visinelli}}, \bibinfo {author} {\bibfnamefont {W.}~\bibnamefont {Yang}}, \bibinfo {author} {\bibfnamefont {A.}~\bibnamefont {Melchiorri}}, \bibinfo {author} {\bibfnamefont {D.~F.}\ \bibnamefont {Mota}}, \bibinfo {author} {\bibfnamefont {A.~G.}\ \bibnamefont {Riess}}, \ and\ \bibinfo {author} {\bibfnamefont {J.}~\bibnamefont {Silk}},\ }\href {\doibase 10.1088/1361-6382/ac086d} {\bibfield  {journal} {\bibinfo  {journal} {Class. Quant. Grav.}\ }\textbf {\bibinfo {volume} {38}},\ \bibinfo {pages} {153001} (\bibinfo {year} {2021}{\natexlab{b}})},\ \Eprint {http://arxiv.org/abs/2103.01183} {arXiv:2103.01183 [astro-ph.CO]} \BibitemShut {NoStop}%
\bibitem [{\citenamefont {Perivolaropoulos}\ and\ \citenamefont {Skara}(2022)}]{Perivolaropoulos:2021jda}%
  \BibitemOpen
  \bibfield  {author} {\bibinfo {author} {\bibfnamefont {L.}~\bibnamefont {Perivolaropoulos}}\ and\ \bibinfo {author} {\bibfnamefont {F.}~\bibnamefont {Skara}},\ }\href {\doibase 10.1016/j.newar.2022.101659} {\bibfield  {journal} {\bibinfo  {journal} {New Astron. Rev.}\ }\textbf {\bibinfo {volume} {95}},\ \bibinfo {pages} {101659} (\bibinfo {year} {2022})},\ \Eprint {http://arxiv.org/abs/2105.05208} {arXiv:2105.05208 [astro-ph.CO]} \BibitemShut {NoStop}%
\bibitem [{\citenamefont {Sch\"oneberg}\ \emph {et~al.}(2022)\citenamefont {Sch\"oneberg}, \citenamefont {Franco~Abell\'an}, \citenamefont {P\'erez~S\'anchez}, \citenamefont {Witte}, \citenamefont {Poulin},\ and\ \citenamefont {Lesgourgues}}]{Schoneberg:2021qvd}%
  \BibitemOpen
  \bibfield  {author} {\bibinfo {author} {\bibfnamefont {N.}~\bibnamefont {Sch\"oneberg}}, \bibinfo {author} {\bibfnamefont {G.}~\bibnamefont {Franco~Abell\'an}}, \bibinfo {author} {\bibfnamefont {A.}~\bibnamefont {P\'erez~S\'anchez}}, \bibinfo {author} {\bibfnamefont {S.~J.}\ \bibnamefont {Witte}}, \bibinfo {author} {\bibfnamefont {V.}~\bibnamefont {Poulin}}, \ and\ \bibinfo {author} {\bibfnamefont {J.}~\bibnamefont {Lesgourgues}},\ }\href {\doibase 10.1016/j.physrep.2022.07.001} {\bibfield  {journal} {\bibinfo  {journal} {Phys. Rept.}\ }\textbf {\bibinfo {volume} {984}},\ \bibinfo {pages} {1} (\bibinfo {year} {2022})},\ \Eprint {http://arxiv.org/abs/2107.10291} {arXiv:2107.10291 [astro-ph.CO]} \BibitemShut {NoStop}%
\bibitem [{\citenamefont {Shah}\ \emph {et~al.}(2021)\citenamefont {Shah}, \citenamefont {Lemos},\ and\ \citenamefont {Lahav}}]{Shah:2021onj}%
  \BibitemOpen
  \bibfield  {author} {\bibinfo {author} {\bibfnamefont {P.}~\bibnamefont {Shah}}, \bibinfo {author} {\bibfnamefont {P.}~\bibnamefont {Lemos}}, \ and\ \bibinfo {author} {\bibfnamefont {O.}~\bibnamefont {Lahav}},\ }\href {\doibase 10.1007/s00159-021-00137-4} {\bibfield  {journal} {\bibinfo  {journal} {Astron. Astrophys. Rev.}\ }\textbf {\bibinfo {volume} {29}},\ \bibinfo {pages} {9} (\bibinfo {year} {2021})},\ \Eprint {http://arxiv.org/abs/2109.01161} {arXiv:2109.01161 [astro-ph.CO]} \BibitemShut {NoStop}%
\bibitem [{\citenamefont {Abdalla}\ \emph {et~al.}(2022)\citenamefont {Abdalla} \emph {et~al.}}]{Abdalla:2022yfr}%
  \BibitemOpen
  \bibfield  {author} {\bibinfo {author} {\bibfnamefont {E.}~\bibnamefont {Abdalla}} \emph {et~al.},\ }\href {\doibase 10.1016/j.jheap.2022.04.002} {\bibfield  {journal} {\bibinfo  {journal} {JHEAp}\ }\textbf {\bibinfo {volume} {34}},\ \bibinfo {pages} {49} (\bibinfo {year} {2022})},\ \Eprint {http://arxiv.org/abs/2203.06142} {arXiv:2203.06142 [astro-ph.CO]} \BibitemShut {NoStop}%
\bibitem [{\citenamefont {Di~Valentino}(2022)}]{DiValentino:2022fjm}%
  \BibitemOpen
  \bibfield  {author} {\bibinfo {author} {\bibfnamefont {E.}~\bibnamefont {Di~Valentino}},\ }\href {\doibase 10.3390/universe8080399} {\bibfield  {journal} {\bibinfo  {journal} {Universe}\ }\textbf {\bibinfo {volume} {8}},\ \bibinfo {pages} {399} (\bibinfo {year} {2022})}\BibitemShut {NoStop}%
\bibitem [{\citenamefont {Kamionkowski}\ and\ \citenamefont {Riess}(2023)}]{Kamionkowski:2022pkx}%
  \BibitemOpen
  \bibfield  {author} {\bibinfo {author} {\bibfnamefont {M.}~\bibnamefont {Kamionkowski}}\ and\ \bibinfo {author} {\bibfnamefont {A.~G.}\ \bibnamefont {Riess}},\ }\href {\doibase 10.1146/annurev-nucl-111422-024107} {\bibfield  {journal} {\bibinfo  {journal} {Ann. Rev. Nucl. Part. Sci.}\ }\textbf {\bibinfo {volume} {73}},\ \bibinfo {pages} {153} (\bibinfo {year} {2023})},\ \Eprint {http://arxiv.org/abs/2211.04492} {arXiv:2211.04492 [astro-ph.CO]} \BibitemShut {NoStop}%
\bibitem [{\citenamefont {Giar\`e}(2023)}]{Giare:2023xoc}%
  \BibitemOpen
  \bibfield  {author} {\bibinfo {author} {\bibfnamefont {W.}~\bibnamefont {Giar\`e}},\ }\href {\doibase 10.1007/978-981-99-0177-7$\_$36} {\  (\bibinfo {year} {2023}),\ 10.1007/978-981-99-0177-7$\_$36},\ \Eprint {http://arxiv.org/abs/2305.16919} {arXiv:2305.16919 [astro-ph.CO]} \BibitemShut {NoStop}%
\bibitem [{\citenamefont {Hu}\ and\ \citenamefont {Wang}(2023)}]{Hu:2023jqc}%
  \BibitemOpen
  \bibfield  {author} {\bibinfo {author} {\bibfnamefont {J.-P.}\ \bibnamefont {Hu}}\ and\ \bibinfo {author} {\bibfnamefont {F.-Y.}\ \bibnamefont {Wang}},\ }\href {\doibase 10.3390/universe9020094} {\bibfield  {journal} {\bibinfo  {journal} {Universe}\ }\textbf {\bibinfo {volume} {9}},\ \bibinfo {pages} {94} (\bibinfo {year} {2023})},\ \Eprint {http://arxiv.org/abs/2302.05709} {arXiv:2302.05709 [astro-ph.CO]} \BibitemShut {NoStop}%
\bibitem [{\citenamefont {Verde}\ \emph {et~al.}(2023)\citenamefont {Verde}, \citenamefont {Sch\"oneberg},\ and\ \citenamefont {Gil-Mar\'\i{}n}}]{Verde:2023lmm}%
  \BibitemOpen
  \bibfield  {author} {\bibinfo {author} {\bibfnamefont {L.}~\bibnamefont {Verde}}, \bibinfo {author} {\bibfnamefont {N.}~\bibnamefont {Sch\"oneberg}}, \ and\ \bibinfo {author} {\bibfnamefont {H.}~\bibnamefont {Gil-Mar\'\i{}n}},\ }\href@noop {} {\  (\bibinfo {year} {2023})},\ \Eprint {http://arxiv.org/abs/2311.13305} {arXiv:2311.13305 [astro-ph.CO]} \BibitemShut {NoStop}%
\bibitem [{\citenamefont {Di~Valentino}\ and\ \citenamefont {Brout}(2024)}]{DiValentino:2024yew}%
  \BibitemOpen
  \bibinfo {editor} {\bibfnamefont {E.}~\bibnamefont {Di~Valentino}}\ and\ \bibinfo {editor} {\bibfnamefont {D.}~\bibnamefont {Brout}},\ eds.,\ \href {\doibase 10.1007/978-981-99-0177-7} {\emph {\bibinfo {title} {{The Hubble Constant Tension}}}},\ Springer Series in Astrophysics and Cosmology\ (\bibinfo  {publisher} {Springer},\ \bibinfo {year} {2024})\BibitemShut {NoStop}%
\bibitem [{\citenamefont {Perivolaropoulos}(2024)}]{Perivolaropoulos:2024yxv}%
  \BibitemOpen
  \bibfield  {author} {\bibinfo {author} {\bibfnamefont {L.}~\bibnamefont {Perivolaropoulos}},\ }\href@noop {} {\  (\bibinfo {year} {2024})},\ \Eprint {http://arxiv.org/abs/2408.11031} {arXiv:2408.11031 [astro-ph.CO]} \BibitemShut {NoStop}%
\bibitem [{\citenamefont {Riess}\ \emph {et~al.}(2022)\citenamefont {Riess} \emph {et~al.}}]{Riess:2021jrx}%
  \BibitemOpen
  \bibfield  {author} {\bibinfo {author} {\bibfnamefont {A.~G.}\ \bibnamefont {Riess}} \emph {et~al.},\ }\href {\doibase 10.3847/2041-8213/ac5c5b} {\bibfield  {journal} {\bibinfo  {journal} {Astrophys. J. Lett.}\ }\textbf {\bibinfo {volume} {934}},\ \bibinfo {pages} {L7} (\bibinfo {year} {2022})},\ \Eprint {http://arxiv.org/abs/2112.04510} {arXiv:2112.04510 [astro-ph.CO]} \BibitemShut {NoStop}%
\bibitem [{\citenamefont {Breuval}\ \emph {et~al.}(2024)\citenamefont {Breuval}, \citenamefont {Riess}, \citenamefont {Casertano}, \citenamefont {Yuan}, \citenamefont {Macri}, \citenamefont {Romaniello}, \citenamefont {Murakami}, \citenamefont {Scolnic}, \citenamefont {Anand},\ and\ \citenamefont {Soszy\'nski}}]{Breuval:2024lsv}%
  \BibitemOpen
  \bibfield  {author} {\bibinfo {author} {\bibfnamefont {L.}~\bibnamefont {Breuval}}, \bibinfo {author} {\bibfnamefont {A.~G.}\ \bibnamefont {Riess}}, \bibinfo {author} {\bibfnamefont {S.}~\bibnamefont {Casertano}}, \bibinfo {author} {\bibfnamefont {W.}~\bibnamefont {Yuan}}, \bibinfo {author} {\bibfnamefont {L.~M.}\ \bibnamefont {Macri}}, \bibinfo {author} {\bibfnamefont {M.}~\bibnamefont {Romaniello}}, \bibinfo {author} {\bibfnamefont {Y.~S.}\ \bibnamefont {Murakami}}, \bibinfo {author} {\bibfnamefont {D.}~\bibnamefont {Scolnic}}, \bibinfo {author} {\bibfnamefont {G.~S.}\ \bibnamefont {Anand}}, \ and\ \bibinfo {author} {\bibfnamefont {I.}~\bibnamefont {Soszy\'nski}},\ }\href {\doibase 10.3847/1538-4357/ad630e} {\bibfield  {journal} {\bibinfo  {journal} {Astrophys. J.}\ }\textbf {\bibinfo {volume} {973}},\ \bibinfo {pages} {30} (\bibinfo {year} {2024})},\ \Eprint {http://arxiv.org/abs/2404.08038} {arXiv:2404.08038 [astro-ph.CO]} \BibitemShut {NoStop}%
\bibitem [{\citenamefont {Murakami}\ \emph {et~al.}(2023)\citenamefont {Murakami}, \citenamefont {Riess}, \citenamefont {Stahl}, \citenamefont {Kenworthy}, \citenamefont {Pluck}, \citenamefont {Macoretta}, \citenamefont {Brout}, \citenamefont {Jones}, \citenamefont {Scolnic},\ and\ \citenamefont {Filippenko}}]{Murakami:2023xuy}%
  \BibitemOpen
  \bibfield  {author} {\bibinfo {author} {\bibfnamefont {Y.~S.}\ \bibnamefont {Murakami}}, \bibinfo {author} {\bibfnamefont {A.~G.}\ \bibnamefont {Riess}}, \bibinfo {author} {\bibfnamefont {B.~E.}\ \bibnamefont {Stahl}}, \bibinfo {author} {\bibfnamefont {W.~D.}\ \bibnamefont {Kenworthy}}, \bibinfo {author} {\bibfnamefont {D.-M.~A.}\ \bibnamefont {Pluck}}, \bibinfo {author} {\bibfnamefont {A.}~\bibnamefont {Macoretta}}, \bibinfo {author} {\bibfnamefont {D.}~\bibnamefont {Brout}}, \bibinfo {author} {\bibfnamefont {D.~O.}\ \bibnamefont {Jones}}, \bibinfo {author} {\bibfnamefont {D.~M.}\ \bibnamefont {Scolnic}}, \ and\ \bibinfo {author} {\bibfnamefont {A.~V.}\ \bibnamefont {Filippenko}},\ }\href {\doibase 10.1088/1475-7516/2023/11/046} {\bibfield  {journal} {\bibinfo  {journal} {JCAP}\ }\textbf {\bibinfo {volume} {11}},\ \bibinfo {pages} {046} (\bibinfo {year} {2023})},\ \Eprint {http://arxiv.org/abs/2306.00070} {arXiv:2306.00070 [astro-ph.CO]} \BibitemShut {NoStop}%
\bibitem [{\citenamefont {Efstathiou}(2020)}]{Efstathiou:2020wxn}%
  \BibitemOpen
  \bibfield  {author} {\bibinfo {author} {\bibfnamefont {G.}~\bibnamefont {Efstathiou}},\ }\href@noop {} {\  (\bibinfo {year} {2020})},\ \Eprint {http://arxiv.org/abs/2007.10716} {arXiv:2007.10716 [astro-ph.CO]} \BibitemShut {NoStop}%
\bibitem [{\citenamefont {Mortsell}\ \emph {et~al.}(2022{\natexlab{a}})\citenamefont {Mortsell}, \citenamefont {Goobar}, \citenamefont {Johansson},\ and\ \citenamefont {Dhawan}}]{Mortsell:2021nzg}%
  \BibitemOpen
  \bibfield  {author} {\bibinfo {author} {\bibfnamefont {E.}~\bibnamefont {Mortsell}}, \bibinfo {author} {\bibfnamefont {A.}~\bibnamefont {Goobar}}, \bibinfo {author} {\bibfnamefont {J.}~\bibnamefont {Johansson}}, \ and\ \bibinfo {author} {\bibfnamefont {S.}~\bibnamefont {Dhawan}},\ }\href {\doibase 10.3847/1538-4357/ac756e} {\bibfield  {journal} {\bibinfo  {journal} {Astrophys. J.}\ }\textbf {\bibinfo {volume} {933}},\ \bibinfo {pages} {212} (\bibinfo {year} {2022}{\natexlab{a}})},\ \Eprint {http://arxiv.org/abs/2105.11461} {arXiv:2105.11461 [astro-ph.CO]} \BibitemShut {NoStop}%
\bibitem [{\citenamefont {Mortsell}\ \emph {et~al.}(2022{\natexlab{b}})\citenamefont {Mortsell}, \citenamefont {Goobar}, \citenamefont {Johansson},\ and\ \citenamefont {Dhawan}}]{Mortsell:2021tcx}%
  \BibitemOpen
  \bibfield  {author} {\bibinfo {author} {\bibfnamefont {E.}~\bibnamefont {Mortsell}}, \bibinfo {author} {\bibfnamefont {A.}~\bibnamefont {Goobar}}, \bibinfo {author} {\bibfnamefont {J.}~\bibnamefont {Johansson}}, \ and\ \bibinfo {author} {\bibfnamefont {S.}~\bibnamefont {Dhawan}},\ }\href {\doibase 10.3847/1538-4357/ac7c19} {\bibfield  {journal} {\bibinfo  {journal} {Astrophys. J.}\ }\textbf {\bibinfo {volume} {935}},\ \bibinfo {pages} {58} (\bibinfo {year} {2022}{\natexlab{b}})},\ \Eprint {http://arxiv.org/abs/2106.09400} {arXiv:2106.09400 [astro-ph.CO]} \BibitemShut {NoStop}%
\bibitem [{\citenamefont {Sharon}\ \emph {et~al.}(2024)\citenamefont {Sharon}, \citenamefont {Kushnir}, \citenamefont {Yuan}, \citenamefont {Macri},\ and\ \citenamefont {Riess}}]{Sharon:2023ioz}%
  \BibitemOpen
  \bibfield  {author} {\bibinfo {author} {\bibfnamefont {A.}~\bibnamefont {Sharon}}, \bibinfo {author} {\bibfnamefont {D.}~\bibnamefont {Kushnir}}, \bibinfo {author} {\bibfnamefont {W.}~\bibnamefont {Yuan}}, \bibinfo {author} {\bibfnamefont {L.}~\bibnamefont {Macri}}, \ and\ \bibinfo {author} {\bibfnamefont {A.}~\bibnamefont {Riess}},\ }\href {\doibase 10.1093/mnras/stae451} {\bibfield  {journal} {\bibinfo  {journal} {Mon. Not. Roy. Astron. Soc.}\ }\textbf {\bibinfo {volume} {528}},\ \bibinfo {pages} {6861} (\bibinfo {year} {2024})},\ \Eprint {http://arxiv.org/abs/2305.14435} {arXiv:2305.14435 [astro-ph.CO]} \BibitemShut {NoStop}%
\bibitem [{\citenamefont {Riess}\ \emph {et~al.}(2023)\citenamefont {Riess}, \citenamefont {Anand}, \citenamefont {Yuan}, \citenamefont {Casertano}, \citenamefont {Dolphin}, \citenamefont {Macri}, \citenamefont {Breuval}, \citenamefont {Scolnic}, \citenamefont {Perrin},\ and\ \citenamefont {Anderson}}]{Riess:2023bfx}%
  \BibitemOpen
  \bibfield  {author} {\bibinfo {author} {\bibfnamefont {A.~G.}\ \bibnamefont {Riess}}, \bibinfo {author} {\bibfnamefont {G.~S.}\ \bibnamefont {Anand}}, \bibinfo {author} {\bibfnamefont {W.}~\bibnamefont {Yuan}}, \bibinfo {author} {\bibfnamefont {S.}~\bibnamefont {Casertano}}, \bibinfo {author} {\bibfnamefont {A.}~\bibnamefont {Dolphin}}, \bibinfo {author} {\bibfnamefont {L.~M.}\ \bibnamefont {Macri}}, \bibinfo {author} {\bibfnamefont {L.}~\bibnamefont {Breuval}}, \bibinfo {author} {\bibfnamefont {D.}~\bibnamefont {Scolnic}}, \bibinfo {author} {\bibfnamefont {M.}~\bibnamefont {Perrin}}, \ and\ \bibinfo {author} {\bibfnamefont {R.~I.}\ \bibnamefont {Anderson}},\ }\href {\doibase 10.3847/2041-8213/acf769} {\bibfield  {journal} {\bibinfo  {journal} {Astrophys. J. Lett.}\ }\textbf {\bibinfo {volume} {956}},\ \bibinfo {pages} {L18} (\bibinfo {year} {2023})},\ \Eprint {http://arxiv.org/abs/2307.15806} {arXiv:2307.15806 [astro-ph.CO]} \BibitemShut {NoStop}%
\bibitem [{\citenamefont {Bhardwaj}\ \emph {et~al.}(2023)\citenamefont {Bhardwaj} \emph {et~al.}}]{Bhardwaj:2023mau}%
  \BibitemOpen
  \bibfield  {author} {\bibinfo {author} {\bibfnamefont {A.}~\bibnamefont {Bhardwaj}} \emph {et~al.},\ }\href {\doibase 10.3847/2041-8213/acf710} {\bibfield  {journal} {\bibinfo  {journal} {Astrophys. J. Lett.}\ }\textbf {\bibinfo {volume} {955}},\ \bibinfo {pages} {L13} (\bibinfo {year} {2023})},\ \Eprint {http://arxiv.org/abs/2309.03263} {arXiv:2309.03263 [astro-ph.SR]} \BibitemShut {NoStop}%
\bibitem [{\citenamefont {Brout}\ and\ \citenamefont {Riess}(2023)}]{Brout:2023wol}%
  \BibitemOpen
  \bibfield  {author} {\bibinfo {author} {\bibfnamefont {D.}~\bibnamefont {Brout}}\ and\ \bibinfo {author} {\bibfnamefont {A.}~\bibnamefont {Riess}},\ }\href@noop {} {\  (\bibinfo {year} {2023})},\ \Eprint {http://arxiv.org/abs/2311.08253} {arXiv:2311.08253 [astro-ph.CO]} \BibitemShut {NoStop}%
\bibitem [{\citenamefont {Dwomoh}\ \emph {et~al.}(2024)\citenamefont {Dwomoh}, \citenamefont {Peterson}, \citenamefont {Scolnic}, \citenamefont {Ashall}, \citenamefont {DerKacy}, \citenamefont {Do}, \citenamefont {Johansson}, \citenamefont {Jones}, \citenamefont {Riess},\ and\ \citenamefont {Shappee}}]{Dwomoh:2023bro}%
  \BibitemOpen
  \bibfield  {author} {\bibinfo {author} {\bibfnamefont {A.~M.}\ \bibnamefont {Dwomoh}}, \bibinfo {author} {\bibfnamefont {E.~R.}\ \bibnamefont {Peterson}}, \bibinfo {author} {\bibfnamefont {D.}~\bibnamefont {Scolnic}}, \bibinfo {author} {\bibfnamefont {C.}~\bibnamefont {Ashall}}, \bibinfo {author} {\bibfnamefont {J.~M.}\ \bibnamefont {DerKacy}}, \bibinfo {author} {\bibfnamefont {A.}~\bibnamefont {Do}}, \bibinfo {author} {\bibfnamefont {J.}~\bibnamefont {Johansson}}, \bibinfo {author} {\bibfnamefont {D.~O.}\ \bibnamefont {Jones}}, \bibinfo {author} {\bibfnamefont {A.~G.}\ \bibnamefont {Riess}}, \ and\ \bibinfo {author} {\bibfnamefont {B.~J.}\ \bibnamefont {Shappee}},\ }\href {\doibase 10.3847/1538-4357/ad1ff5} {\bibfield  {journal} {\bibinfo  {journal} {Astrophys. J.}\ }\textbf {\bibinfo {volume} {965}},\ \bibinfo {pages} {90} (\bibinfo {year} {2024})},\ \Eprint {http://arxiv.org/abs/2311.06178} {arXiv:2311.06178 [astro-ph.CO]} \BibitemShut {NoStop}%
\bibitem [{\citenamefont {Uddin}\ \emph {et~al.}(2024)\citenamefont {Uddin} \emph {et~al.}}]{Uddin:2023iob}%
  \BibitemOpen
  \bibfield  {author} {\bibinfo {author} {\bibfnamefont {S.~A.}\ \bibnamefont {Uddin}} \emph {et~al.},\ }\href {\doibase 10.3847/1538-4357/ad3e63} {\bibfield  {journal} {\bibinfo  {journal} {Astrophys. J.}\ }\textbf {\bibinfo {volume} {970}},\ \bibinfo {pages} {72} (\bibinfo {year} {2024})},\ \Eprint {http://arxiv.org/abs/2308.01875} {arXiv:2308.01875 [astro-ph.CO]} \BibitemShut {NoStop}%
\bibitem [{\citenamefont {Riess}\ \emph {et~al.}(2024{\natexlab{a}})\citenamefont {Riess}, \citenamefont {Anand}, \citenamefont {Yuan}, \citenamefont {Casertano}, \citenamefont {Dolphin}, \citenamefont {Macri}, \citenamefont {Breuval}, \citenamefont {Scolnic}, \citenamefont {Perrin},\ and\ \citenamefont {Anderson}}]{Riess:2024ohe}%
  \BibitemOpen
  \bibfield  {author} {\bibinfo {author} {\bibfnamefont {A.~G.}\ \bibnamefont {Riess}}, \bibinfo {author} {\bibfnamefont {G.~S.}\ \bibnamefont {Anand}}, \bibinfo {author} {\bibfnamefont {W.}~\bibnamefont {Yuan}}, \bibinfo {author} {\bibfnamefont {S.}~\bibnamefont {Casertano}}, \bibinfo {author} {\bibfnamefont {A.}~\bibnamefont {Dolphin}}, \bibinfo {author} {\bibfnamefont {L.~M.}\ \bibnamefont {Macri}}, \bibinfo {author} {\bibfnamefont {L.}~\bibnamefont {Breuval}}, \bibinfo {author} {\bibfnamefont {D.}~\bibnamefont {Scolnic}}, \bibinfo {author} {\bibfnamefont {M.}~\bibnamefont {Perrin}}, \ and\ \bibinfo {author} {\bibfnamefont {I.~R.}\ \bibnamefont {Anderson}},\ }\href {\doibase 10.3847/2041-8213/ad1ddd} {\bibfield  {journal} {\bibinfo  {journal} {Astrophys. J. Lett.}\ }\textbf {\bibinfo {volume} {962}},\ \bibinfo {pages} {L17} (\bibinfo {year} {2024}{\natexlab{a}})},\ \Eprint {http://arxiv.org/abs/2401.04773} {arXiv:2401.04773 [astro-ph.CO]} \BibitemShut {NoStop}%
\bibitem [{\citenamefont {Freedman}\ \emph {et~al.}(2024)\citenamefont {Freedman}, \citenamefont {Madore}, \citenamefont {Jang}, \citenamefont {Hoyt}, \citenamefont {Lee},\ and\ \citenamefont {Owens}}]{Freedman:2024eph}%
  \BibitemOpen
  \bibfield  {author} {\bibinfo {author} {\bibfnamefont {W.~L.}\ \bibnamefont {Freedman}}, \bibinfo {author} {\bibfnamefont {B.~F.}\ \bibnamefont {Madore}}, \bibinfo {author} {\bibfnamefont {I.~S.}\ \bibnamefont {Jang}}, \bibinfo {author} {\bibfnamefont {T.~J.}\ \bibnamefont {Hoyt}}, \bibinfo {author} {\bibfnamefont {A.~J.}\ \bibnamefont {Lee}}, \ and\ \bibinfo {author} {\bibfnamefont {K.~A.}\ \bibnamefont {Owens}},\ }\href@noop {} {\  (\bibinfo {year} {2024})},\ \Eprint {http://arxiv.org/abs/2408.06153} {arXiv:2408.06153 [astro-ph.CO]} \BibitemShut {NoStop}%
\bibitem [{\citenamefont {Riess}\ \emph {et~al.}(2024{\natexlab{b}})\citenamefont {Riess} \emph {et~al.}}]{Riess:2024vfa}%
  \BibitemOpen
  \bibfield  {author} {\bibinfo {author} {\bibfnamefont {A.~G.}\ \bibnamefont {Riess}} \emph {et~al.},\ }\href@noop {} {\  (\bibinfo {year} {2024}{\natexlab{b}})},\ \Eprint {http://arxiv.org/abs/2408.11770} {arXiv:2408.11770 [astro-ph.CO]} \BibitemShut {NoStop}%
\bibitem [{\citenamefont {Anchordoqui}\ \emph {et~al.}(2015)\citenamefont {Anchordoqui}, \citenamefont {Barger}, \citenamefont {Goldberg}, \citenamefont {Huang}, \citenamefont {Marfatia}, \citenamefont {da~Silva},\ and\ \citenamefont {Weiler}}]{Anchordoqui:2015lqa}%
  \BibitemOpen
  \bibfield  {author} {\bibinfo {author} {\bibfnamefont {L.~A.}\ \bibnamefont {Anchordoqui}}, \bibinfo {author} {\bibfnamefont {V.}~\bibnamefont {Barger}}, \bibinfo {author} {\bibfnamefont {H.}~\bibnamefont {Goldberg}}, \bibinfo {author} {\bibfnamefont {X.}~\bibnamefont {Huang}}, \bibinfo {author} {\bibfnamefont {D.}~\bibnamefont {Marfatia}}, \bibinfo {author} {\bibfnamefont {L.~H.~M.}\ \bibnamefont {da~Silva}}, \ and\ \bibinfo {author} {\bibfnamefont {T.~J.}\ \bibnamefont {Weiler}},\ }\href {\doibase 10.1103/PhysRevD.94.069901} {\bibfield  {journal} {\bibinfo  {journal} {Phys. Rev. D}\ }\textbf {\bibinfo {volume} {92}},\ \bibinfo {pages} {061301} (\bibinfo {year} {2015})},\ \bibinfo {note} {[Erratum: Phys.Rev.D 94, 069901 (2016)]},\ \Eprint {http://arxiv.org/abs/1506.08788} {arXiv:1506.08788 [hep-ph]} \BibitemShut {NoStop}%
\bibitem [{\citenamefont {Karwal}\ and\ \citenamefont {Kamionkowski}(2016)}]{Karwal:2016vyq}%
  \BibitemOpen
  \bibfield  {author} {\bibinfo {author} {\bibfnamefont {T.}~\bibnamefont {Karwal}}\ and\ \bibinfo {author} {\bibfnamefont {M.}~\bibnamefont {Kamionkowski}},\ }\href {\doibase 10.1103/PhysRevD.94.103523} {\bibfield  {journal} {\bibinfo  {journal} {Phys. Rev. D}\ }\textbf {\bibinfo {volume} {94}},\ \bibinfo {pages} {103523} (\bibinfo {year} {2016})},\ \Eprint {http://arxiv.org/abs/1608.01309} {arXiv:1608.01309 [astro-ph.CO]} \BibitemShut {NoStop}%
\bibitem [{\citenamefont {Zhao}\ \emph {et~al.}(2017)\citenamefont {Zhao}, \citenamefont {He}, \citenamefont {Zhang},\ and\ \citenamefont {Zhang}}]{Zhao:2017urm}%
  \BibitemOpen
  \bibfield  {author} {\bibinfo {author} {\bibfnamefont {M.-M.}\ \bibnamefont {Zhao}}, \bibinfo {author} {\bibfnamefont {D.-Z.}\ \bibnamefont {He}}, \bibinfo {author} {\bibfnamefont {J.-F.}\ \bibnamefont {Zhang}}, \ and\ \bibinfo {author} {\bibfnamefont {X.}~\bibnamefont {Zhang}},\ }\href {\doibase 10.1103/PhysRevD.96.043520} {\bibfield  {journal} {\bibinfo  {journal} {Phys. Rev. D}\ }\textbf {\bibinfo {volume} {96}},\ \bibinfo {pages} {043520} (\bibinfo {year} {2017})},\ \Eprint {http://arxiv.org/abs/1703.08456} {arXiv:1703.08456 [astro-ph.CO]} \BibitemShut {NoStop}%
\bibitem [{\citenamefont {Benetti}\ \emph {et~al.}(2018)\citenamefont {Benetti}, \citenamefont {Graef},\ and\ \citenamefont {Alcaniz}}]{Benetti:2017juy}%
  \BibitemOpen
  \bibfield  {author} {\bibinfo {author} {\bibfnamefont {M.}~\bibnamefont {Benetti}}, \bibinfo {author} {\bibfnamefont {L.~L.}\ \bibnamefont {Graef}}, \ and\ \bibinfo {author} {\bibfnamefont {J.~S.}\ \bibnamefont {Alcaniz}},\ }\href {\doibase 10.1088/1475-7516/2018/07/066} {\bibfield  {journal} {\bibinfo  {journal} {JCAP}\ }\textbf {\bibinfo {volume} {07}},\ \bibinfo {pages} {066} (\bibinfo {year} {2018})},\ \Eprint {http://arxiv.org/abs/1712.00677} {arXiv:1712.00677 [astro-ph.CO]} \BibitemShut {NoStop}%
\bibitem [{\citenamefont {M\"ortsell}\ and\ \citenamefont {Dhawan}(2018)}]{Mortsell:2018mfj}%
  \BibitemOpen
  \bibfield  {author} {\bibinfo {author} {\bibfnamefont {E.}~\bibnamefont {M\"ortsell}}\ and\ \bibinfo {author} {\bibfnamefont {S.}~\bibnamefont {Dhawan}},\ }\href {\doibase 10.1088/1475-7516/2018/09/025} {\bibfield  {journal} {\bibinfo  {journal} {JCAP}\ }\textbf {\bibinfo {volume} {09}},\ \bibinfo {pages} {025} (\bibinfo {year} {2018})},\ \Eprint {http://arxiv.org/abs/1801.07260} {arXiv:1801.07260 [astro-ph.CO]} \BibitemShut {NoStop}%
\bibitem [{\citenamefont {Vagnozzi}\ \emph {et~al.}(2018)\citenamefont {Vagnozzi}, \citenamefont {Dhawan}, \citenamefont {Gerbino}, \citenamefont {Freese}, \citenamefont {Goobar},\ and\ \citenamefont {Mena}}]{Vagnozzi:2018jhn}%
  \BibitemOpen
  \bibfield  {author} {\bibinfo {author} {\bibfnamefont {S.}~\bibnamefont {Vagnozzi}}, \bibinfo {author} {\bibfnamefont {S.}~\bibnamefont {Dhawan}}, \bibinfo {author} {\bibfnamefont {M.}~\bibnamefont {Gerbino}}, \bibinfo {author} {\bibfnamefont {K.}~\bibnamefont {Freese}}, \bibinfo {author} {\bibfnamefont {A.}~\bibnamefont {Goobar}}, \ and\ \bibinfo {author} {\bibfnamefont {O.}~\bibnamefont {Mena}},\ }\href {\doibase 10.1103/PhysRevD.98.083501} {\bibfield  {journal} {\bibinfo  {journal} {Phys. Rev. D}\ }\textbf {\bibinfo {volume} {98}},\ \bibinfo {pages} {083501} (\bibinfo {year} {2018})},\ \Eprint {http://arxiv.org/abs/1801.08553} {arXiv:1801.08553 [astro-ph.CO]} \BibitemShut {NoStop}%
\bibitem [{\citenamefont {Kumar}\ \emph {et~al.}(2018)\citenamefont {Kumar}, \citenamefont {Nunes},\ and\ \citenamefont {Yadav}}]{Kumar:2018yhh}%
  \BibitemOpen
  \bibfield  {author} {\bibinfo {author} {\bibfnamefont {S.}~\bibnamefont {Kumar}}, \bibinfo {author} {\bibfnamefont {R.~C.}\ \bibnamefont {Nunes}}, \ and\ \bibinfo {author} {\bibfnamefont {S.~K.}\ \bibnamefont {Yadav}},\ }\href {\doibase 10.1103/PhysRevD.98.043521} {\bibfield  {journal} {\bibinfo  {journal} {Phys. Rev. D}\ }\textbf {\bibinfo {volume} {98}},\ \bibinfo {pages} {043521} (\bibinfo {year} {2018})},\ \Eprint {http://arxiv.org/abs/1803.10229} {arXiv:1803.10229 [astro-ph.CO]} \BibitemShut {NoStop}%
\bibitem [{\citenamefont {Yang}\ \emph {et~al.}(2018)\citenamefont {Yang}, \citenamefont {Pan}, \citenamefont {Di~Valentino}, \citenamefont {Nunes}, \citenamefont {Vagnozzi},\ and\ \citenamefont {Mota}}]{Yang:2018euj}%
  \BibitemOpen
  \bibfield  {author} {\bibinfo {author} {\bibfnamefont {W.}~\bibnamefont {Yang}}, \bibinfo {author} {\bibfnamefont {S.}~\bibnamefont {Pan}}, \bibinfo {author} {\bibfnamefont {E.}~\bibnamefont {Di~Valentino}}, \bibinfo {author} {\bibfnamefont {R.~C.}\ \bibnamefont {Nunes}}, \bibinfo {author} {\bibfnamefont {S.}~\bibnamefont {Vagnozzi}}, \ and\ \bibinfo {author} {\bibfnamefont {D.~F.}\ \bibnamefont {Mota}},\ }\href {\doibase 10.1088/1475-7516/2018/09/019} {\bibfield  {journal} {\bibinfo  {journal} {JCAP}\ }\textbf {\bibinfo {volume} {09}},\ \bibinfo {pages} {019} (\bibinfo {year} {2018})},\ \Eprint {http://arxiv.org/abs/1805.08252} {arXiv:1805.08252 [astro-ph.CO]} \BibitemShut {NoStop}%
\bibitem [{\citenamefont {Banihashemi}\ \emph {et~al.}(2020)\citenamefont {Banihashemi}, \citenamefont {Khosravi},\ and\ \citenamefont {Shirazi}}]{Banihashemi:2018oxo}%
  \BibitemOpen
  \bibfield  {author} {\bibinfo {author} {\bibfnamefont {A.}~\bibnamefont {Banihashemi}}, \bibinfo {author} {\bibfnamefont {N.}~\bibnamefont {Khosravi}}, \ and\ \bibinfo {author} {\bibfnamefont {A.~H.}\ \bibnamefont {Shirazi}},\ }\href {\doibase 10.1103/PhysRevD.101.123521} {\bibfield  {journal} {\bibinfo  {journal} {Phys. Rev. D}\ }\textbf {\bibinfo {volume} {101}},\ \bibinfo {pages} {123521} (\bibinfo {year} {2020})},\ \Eprint {http://arxiv.org/abs/1808.02472} {arXiv:1808.02472 [astro-ph.CO]} \BibitemShut {NoStop}%
\bibitem [{\citenamefont {Guo}\ \emph {et~al.}(2019)\citenamefont {Guo}, \citenamefont {Zhang},\ and\ \citenamefont {Zhang}}]{Guo:2018ans}%
  \BibitemOpen
  \bibfield  {author} {\bibinfo {author} {\bibfnamefont {R.-Y.}\ \bibnamefont {Guo}}, \bibinfo {author} {\bibfnamefont {J.-F.}\ \bibnamefont {Zhang}}, \ and\ \bibinfo {author} {\bibfnamefont {X.}~\bibnamefont {Zhang}},\ }\href {\doibase 10.1088/1475-7516/2019/02/054} {\bibfield  {journal} {\bibinfo  {journal} {JCAP}\ }\textbf {\bibinfo {volume} {02}},\ \bibinfo {pages} {054} (\bibinfo {year} {2019})},\ \Eprint {http://arxiv.org/abs/1809.02340} {arXiv:1809.02340 [astro-ph.CO]} \BibitemShut {NoStop}%
\bibitem [{\citenamefont {Graef}\ \emph {et~al.}(2019)\citenamefont {Graef}, \citenamefont {Benetti},\ and\ \citenamefont {Alcaniz}}]{Graef:2018fzu}%
  \BibitemOpen
  \bibfield  {author} {\bibinfo {author} {\bibfnamefont {L.~L.}\ \bibnamefont {Graef}}, \bibinfo {author} {\bibfnamefont {M.}~\bibnamefont {Benetti}}, \ and\ \bibinfo {author} {\bibfnamefont {J.~S.}\ \bibnamefont {Alcaniz}},\ }\href {\doibase 10.1103/PhysRevD.99.043519} {\bibfield  {journal} {\bibinfo  {journal} {Phys. Rev. D}\ }\textbf {\bibinfo {volume} {99}},\ \bibinfo {pages} {043519} (\bibinfo {year} {2019})},\ \Eprint {http://arxiv.org/abs/1809.04501} {arXiv:1809.04501 [astro-ph.CO]} \BibitemShut {NoStop}%
\bibitem [{\citenamefont {Banihashemi}\ \emph {et~al.}(2019)\citenamefont {Banihashemi}, \citenamefont {Khosravi},\ and\ \citenamefont {Shirazi}}]{Banihashemi:2018has}%
  \BibitemOpen
  \bibfield  {author} {\bibinfo {author} {\bibfnamefont {A.}~\bibnamefont {Banihashemi}}, \bibinfo {author} {\bibfnamefont {N.}~\bibnamefont {Khosravi}}, \ and\ \bibinfo {author} {\bibfnamefont {A.~H.}\ \bibnamefont {Shirazi}},\ }\href {\doibase 10.1103/PhysRevD.99.083509} {\bibfield  {journal} {\bibinfo  {journal} {Phys. Rev. D}\ }\textbf {\bibinfo {volume} {99}},\ \bibinfo {pages} {083509} (\bibinfo {year} {2019})},\ \Eprint {http://arxiv.org/abs/1810.11007} {arXiv:1810.11007 [astro-ph.CO]} \BibitemShut {NoStop}%
\bibitem [{\citenamefont {Agrawal}\ \emph {et~al.}(2023)\citenamefont {Agrawal}, \citenamefont {Cyr-Racine}, \citenamefont {Pinner},\ and\ \citenamefont {Randall}}]{Agrawal:2019lmo}%
  \BibitemOpen
  \bibfield  {author} {\bibinfo {author} {\bibfnamefont {P.}~\bibnamefont {Agrawal}}, \bibinfo {author} {\bibfnamefont {F.-Y.}\ \bibnamefont {Cyr-Racine}}, \bibinfo {author} {\bibfnamefont {D.}~\bibnamefont {Pinner}}, \ and\ \bibinfo {author} {\bibfnamefont {L.}~\bibnamefont {Randall}},\ }\href {\doibase 10.1016/j.dark.2023.101347} {\bibfield  {journal} {\bibinfo  {journal} {Phys. Dark Univ.}\ }\textbf {\bibinfo {volume} {42}},\ \bibinfo {pages} {101347} (\bibinfo {year} {2023})},\ \Eprint {http://arxiv.org/abs/1904.01016} {arXiv:1904.01016 [astro-ph.CO]} \BibitemShut {NoStop}%
\bibitem [{\citenamefont {Li}\ and\ \citenamefont {Shafieloo}(2019)}]{Li:2019yem}%
  \BibitemOpen
  \bibfield  {author} {\bibinfo {author} {\bibfnamefont {X.}~\bibnamefont {Li}}\ and\ \bibinfo {author} {\bibfnamefont {A.}~\bibnamefont {Shafieloo}},\ }\href {\doibase 10.3847/2041-8213/ab3e09} {\bibfield  {journal} {\bibinfo  {journal} {Astrophys. J. Lett.}\ }\textbf {\bibinfo {volume} {883}},\ \bibinfo {pages} {L3} (\bibinfo {year} {2019})},\ \Eprint {http://arxiv.org/abs/1906.08275} {arXiv:1906.08275 [astro-ph.CO]} \BibitemShut {NoStop}%
\bibitem [{\citenamefont {Yang}\ \emph {et~al.}(2019)\citenamefont {Yang}, \citenamefont {Pan}, \citenamefont {Vagnozzi}, \citenamefont {Di~Valentino}, \citenamefont {Mota},\ and\ \citenamefont {Capozziello}}]{Yang:2019nhz}%
  \BibitemOpen
  \bibfield  {author} {\bibinfo {author} {\bibfnamefont {W.}~\bibnamefont {Yang}}, \bibinfo {author} {\bibfnamefont {S.}~\bibnamefont {Pan}}, \bibinfo {author} {\bibfnamefont {S.}~\bibnamefont {Vagnozzi}}, \bibinfo {author} {\bibfnamefont {E.}~\bibnamefont {Di~Valentino}}, \bibinfo {author} {\bibfnamefont {D.~F.}\ \bibnamefont {Mota}}, \ and\ \bibinfo {author} {\bibfnamefont {S.}~\bibnamefont {Capozziello}},\ }\href {\doibase 10.1088/1475-7516/2019/11/044} {\bibfield  {journal} {\bibinfo  {journal} {JCAP}\ }\textbf {\bibinfo {volume} {11}},\ \bibinfo {pages} {044} (\bibinfo {year} {2019})},\ \Eprint {http://arxiv.org/abs/1907.05344} {arXiv:1907.05344 [astro-ph.CO]} \BibitemShut {NoStop}%
\bibitem [{\citenamefont {Vagnozzi}(2020)}]{Vagnozzi:2019ezj}%
  \BibitemOpen
  \bibfield  {author} {\bibinfo {author} {\bibfnamefont {S.}~\bibnamefont {Vagnozzi}},\ }\href {\doibase 10.1103/PhysRevD.102.023518} {\bibfield  {journal} {\bibinfo  {journal} {Phys. Rev. D}\ }\textbf {\bibinfo {volume} {102}},\ \bibinfo {pages} {023518} (\bibinfo {year} {2020})},\ \Eprint {http://arxiv.org/abs/1907.07569} {arXiv:1907.07569 [astro-ph.CO]} \BibitemShut {NoStop}%
\bibitem [{\citenamefont {Visinelli}\ \emph {et~al.}(2019)\citenamefont {Visinelli}, \citenamefont {Vagnozzi},\ and\ \citenamefont {Danielsson}}]{Visinelli:2019qqu}%
  \BibitemOpen
  \bibfield  {author} {\bibinfo {author} {\bibfnamefont {L.}~\bibnamefont {Visinelli}}, \bibinfo {author} {\bibfnamefont {S.}~\bibnamefont {Vagnozzi}}, \ and\ \bibinfo {author} {\bibfnamefont {U.}~\bibnamefont {Danielsson}},\ }\href {\doibase 10.3390/sym11081035} {\bibfield  {journal} {\bibinfo  {journal} {Symmetry}\ }\textbf {\bibinfo {volume} {11}},\ \bibinfo {pages} {1035} (\bibinfo {year} {2019})},\ \Eprint {http://arxiv.org/abs/1907.07953} {arXiv:1907.07953 [astro-ph.CO]} \BibitemShut {NoStop}%
\bibitem [{\citenamefont {Di~Valentino}\ \emph {et~al.}(2020{\natexlab{b}})\citenamefont {Di~Valentino}, \citenamefont {Melchiorri}, \citenamefont {Mena},\ and\ \citenamefont {Vagnozzi}}]{DiValentino:2019ffd}%
  \BibitemOpen
  \bibfield  {author} {\bibinfo {author} {\bibfnamefont {E.}~\bibnamefont {Di~Valentino}}, \bibinfo {author} {\bibfnamefont {A.}~\bibnamefont {Melchiorri}}, \bibinfo {author} {\bibfnamefont {O.}~\bibnamefont {Mena}}, \ and\ \bibinfo {author} {\bibfnamefont {S.}~\bibnamefont {Vagnozzi}},\ }\href {\doibase 10.1016/j.dark.2020.100666} {\bibfield  {journal} {\bibinfo  {journal} {Phys. Dark Univ.}\ }\textbf {\bibinfo {volume} {30}},\ \bibinfo {pages} {100666} (\bibinfo {year} {2020}{\natexlab{b}})},\ \Eprint {http://arxiv.org/abs/1908.04281} {arXiv:1908.04281 [astro-ph.CO]} \BibitemShut {NoStop}%
\bibitem [{\citenamefont {Escudero}\ and\ \citenamefont {Witte}(2020)}]{Escudero:2019gvw}%
  \BibitemOpen
  \bibfield  {author} {\bibinfo {author} {\bibfnamefont {M.}~\bibnamefont {Escudero}}\ and\ \bibinfo {author} {\bibfnamefont {S.~J.}\ \bibnamefont {Witte}},\ }\href {\doibase 10.1140/epjc/s10052-020-7854-5} {\bibfield  {journal} {\bibinfo  {journal} {Eur. Phys. J. C}\ }\textbf {\bibinfo {volume} {80}},\ \bibinfo {pages} {294} (\bibinfo {year} {2020})},\ \Eprint {http://arxiv.org/abs/1909.04044} {arXiv:1909.04044 [astro-ph.CO]} \BibitemShut {NoStop}%
\bibitem [{\citenamefont {Di~Valentino}\ \emph {et~al.}(2020{\natexlab{c}})\citenamefont {Di~Valentino}, \citenamefont {Melchiorri}, \citenamefont {Mena},\ and\ \citenamefont {Vagnozzi}}]{DiValentino:2019jae}%
  \BibitemOpen
  \bibfield  {author} {\bibinfo {author} {\bibfnamefont {E.}~\bibnamefont {Di~Valentino}}, \bibinfo {author} {\bibfnamefont {A.}~\bibnamefont {Melchiorri}}, \bibinfo {author} {\bibfnamefont {O.}~\bibnamefont {Mena}}, \ and\ \bibinfo {author} {\bibfnamefont {S.}~\bibnamefont {Vagnozzi}},\ }\href {\doibase 10.1103/PhysRevD.101.063502} {\bibfield  {journal} {\bibinfo  {journal} {Phys. Rev. D}\ }\textbf {\bibinfo {volume} {101}},\ \bibinfo {pages} {063502} (\bibinfo {year} {2020}{\natexlab{c}})},\ \Eprint {http://arxiv.org/abs/1910.09853} {arXiv:1910.09853 [astro-ph.CO]} \BibitemShut {NoStop}%
\bibitem [{\citenamefont {Niedermann}\ and\ \citenamefont {Sloth}(2021)}]{Niedermann:2019olb}%
  \BibitemOpen
  \bibfield  {author} {\bibinfo {author} {\bibfnamefont {F.}~\bibnamefont {Niedermann}}\ and\ \bibinfo {author} {\bibfnamefont {M.~S.}\ \bibnamefont {Sloth}},\ }\href {\doibase 10.1103/PhysRevD.103.L041303} {\bibfield  {journal} {\bibinfo  {journal} {Phys. Rev. D}\ }\textbf {\bibinfo {volume} {103}},\ \bibinfo {pages} {L041303} (\bibinfo {year} {2021})},\ \Eprint {http://arxiv.org/abs/1910.10739} {arXiv:1910.10739 [astro-ph.CO]} \BibitemShut {NoStop}%
\bibitem [{\citenamefont {Sakstein}\ and\ \citenamefont {Trodden}(2020)}]{Sakstein:2019fmf}%
  \BibitemOpen
  \bibfield  {author} {\bibinfo {author} {\bibfnamefont {J.}~\bibnamefont {Sakstein}}\ and\ \bibinfo {author} {\bibfnamefont {M.}~\bibnamefont {Trodden}},\ }\href {\doibase 10.1103/PhysRevLett.124.161301} {\bibfield  {journal} {\bibinfo  {journal} {Phys. Rev. Lett.}\ }\textbf {\bibinfo {volume} {124}},\ \bibinfo {pages} {161301} (\bibinfo {year} {2020})},\ \Eprint {http://arxiv.org/abs/1911.11760} {arXiv:1911.11760 [astro-ph.CO]} \BibitemShut {NoStop}%
\bibitem [{\citenamefont {Ye}\ and\ \citenamefont {Piao}(2020)}]{Ye:2020btb}%
  \BibitemOpen
  \bibfield  {author} {\bibinfo {author} {\bibfnamefont {G.}~\bibnamefont {Ye}}\ and\ \bibinfo {author} {\bibfnamefont {Y.-S.}\ \bibnamefont {Piao}},\ }\href {\doibase 10.1103/PhysRevD.101.083507} {\bibfield  {journal} {\bibinfo  {journal} {Phys. Rev. D}\ }\textbf {\bibinfo {volume} {101}},\ \bibinfo {pages} {083507} (\bibinfo {year} {2020})},\ \Eprint {http://arxiv.org/abs/2001.02451} {arXiv:2001.02451 [astro-ph.CO]} \BibitemShut {NoStop}%
\bibitem [{\citenamefont {Hogg}\ \emph {et~al.}(2020)\citenamefont {Hogg}, \citenamefont {Bruni}, \citenamefont {Crittenden}, \citenamefont {Martinelli},\ and\ \citenamefont {Peirone}}]{Hogg:2020rdp}%
  \BibitemOpen
  \bibfield  {author} {\bibinfo {author} {\bibfnamefont {N.~B.}\ \bibnamefont {Hogg}}, \bibinfo {author} {\bibfnamefont {M.}~\bibnamefont {Bruni}}, \bibinfo {author} {\bibfnamefont {R.}~\bibnamefont {Crittenden}}, \bibinfo {author} {\bibfnamefont {M.}~\bibnamefont {Martinelli}}, \ and\ \bibinfo {author} {\bibfnamefont {S.}~\bibnamefont {Peirone}},\ }\href {\doibase 10.1016/j.dark.2020.100583} {\bibfield  {journal} {\bibinfo  {journal} {Phys. Dark Univ.}\ }\textbf {\bibinfo {volume} {29}},\ \bibinfo {pages} {100583} (\bibinfo {year} {2020})},\ \Eprint {http://arxiv.org/abs/2002.10449} {arXiv:2002.10449 [astro-ph.CO]} \BibitemShut {NoStop}%
\bibitem [{\citenamefont {Ballesteros}\ \emph {et~al.}(2020)\citenamefont {Ballesteros}, \citenamefont {Notari},\ and\ \citenamefont {Rompineve}}]{Ballesteros:2020sik}%
  \BibitemOpen
  \bibfield  {author} {\bibinfo {author} {\bibfnamefont {G.}~\bibnamefont {Ballesteros}}, \bibinfo {author} {\bibfnamefont {A.}~\bibnamefont {Notari}}, \ and\ \bibinfo {author} {\bibfnamefont {F.}~\bibnamefont {Rompineve}},\ }\href {\doibase 10.1088/1475-7516/2020/11/024} {\bibfield  {journal} {\bibinfo  {journal} {JCAP}\ }\textbf {\bibinfo {volume} {11}},\ \bibinfo {pages} {024} (\bibinfo {year} {2020})},\ \Eprint {http://arxiv.org/abs/2004.05049} {arXiv:2004.05049 [astro-ph.CO]} \BibitemShut {NoStop}%
\bibitem [{\citenamefont {Alestas}\ \emph {et~al.}(2020)\citenamefont {Alestas}, \citenamefont {Kazantzidis},\ and\ \citenamefont {Perivolaropoulos}}]{Alestas:2020mvb}%
  \BibitemOpen
  \bibfield  {author} {\bibinfo {author} {\bibfnamefont {G.}~\bibnamefont {Alestas}}, \bibinfo {author} {\bibfnamefont {L.}~\bibnamefont {Kazantzidis}}, \ and\ \bibinfo {author} {\bibfnamefont {L.}~\bibnamefont {Perivolaropoulos}},\ }\href {\doibase 10.1103/PhysRevD.101.123516} {\bibfield  {journal} {\bibinfo  {journal} {Phys. Rev. D}\ }\textbf {\bibinfo {volume} {101}},\ \bibinfo {pages} {123516} (\bibinfo {year} {2020})},\ \Eprint {http://arxiv.org/abs/2004.08363} {arXiv:2004.08363 [astro-ph.CO]} \BibitemShut {NoStop}%
\bibitem [{\citenamefont {Jedamzik}\ and\ \citenamefont {Pogosian}(2020)}]{Jedamzik:2020krr}%
  \BibitemOpen
  \bibfield  {author} {\bibinfo {author} {\bibfnamefont {K.}~\bibnamefont {Jedamzik}}\ and\ \bibinfo {author} {\bibfnamefont {L.}~\bibnamefont {Pogosian}},\ }\href {\doibase 10.1103/PhysRevLett.125.181302} {\bibfield  {journal} {\bibinfo  {journal} {Phys. Rev. Lett.}\ }\textbf {\bibinfo {volume} {125}},\ \bibinfo {pages} {181302} (\bibinfo {year} {2020})},\ \Eprint {http://arxiv.org/abs/2004.09487} {arXiv:2004.09487 [astro-ph.CO]} \BibitemShut {NoStop}%
\bibitem [{\citenamefont {Ballardini}\ \emph {et~al.}(2020)\citenamefont {Ballardini}, \citenamefont {Braglia}, \citenamefont {Finelli}, \citenamefont {Paoletti}, \citenamefont {Starobinsky},\ and\ \citenamefont {Umilt\`a}}]{Ballardini:2020iws}%
  \BibitemOpen
  \bibfield  {author} {\bibinfo {author} {\bibfnamefont {M.}~\bibnamefont {Ballardini}}, \bibinfo {author} {\bibfnamefont {M.}~\bibnamefont {Braglia}}, \bibinfo {author} {\bibfnamefont {F.}~\bibnamefont {Finelli}}, \bibinfo {author} {\bibfnamefont {D.}~\bibnamefont {Paoletti}}, \bibinfo {author} {\bibfnamefont {A.~A.}\ \bibnamefont {Starobinsky}}, \ and\ \bibinfo {author} {\bibfnamefont {C.}~\bibnamefont {Umilt\`a}},\ }\href {\doibase 10.1088/1475-7516/2020/10/044} {\bibfield  {journal} {\bibinfo  {journal} {JCAP}\ }\textbf {\bibinfo {volume} {10}},\ \bibinfo {pages} {044} (\bibinfo {year} {2020})},\ \Eprint {http://arxiv.org/abs/2004.14349} {arXiv:2004.14349 [astro-ph.CO]} \BibitemShut {NoStop}%
\bibitem [{\citenamefont {Banerjee}\ \emph {et~al.}(2021)\citenamefont {Banerjee}, \citenamefont {Cai}, \citenamefont {Heisenberg}, \citenamefont {Colg\'ain}, \citenamefont {Sheikh-Jabbari},\ and\ \citenamefont {Yang}}]{Banerjee:2020xcn}%
  \BibitemOpen
  \bibfield  {author} {\bibinfo {author} {\bibfnamefont {A.}~\bibnamefont {Banerjee}}, \bibinfo {author} {\bibfnamefont {H.}~\bibnamefont {Cai}}, \bibinfo {author} {\bibfnamefont {L.}~\bibnamefont {Heisenberg}}, \bibinfo {author} {\bibfnamefont {E.~O.}\ \bibnamefont {Colg\'ain}}, \bibinfo {author} {\bibfnamefont {M.~M.}\ \bibnamefont {Sheikh-Jabbari}}, \ and\ \bibinfo {author} {\bibfnamefont {T.}~\bibnamefont {Yang}},\ }\href {\doibase 10.1103/PhysRevD.103.L081305} {\bibfield  {journal} {\bibinfo  {journal} {Phys. Rev. D}\ }\textbf {\bibinfo {volume} {103}},\ \bibinfo {pages} {L081305} (\bibinfo {year} {2021})},\ \Eprint {http://arxiv.org/abs/2006.00244} {arXiv:2006.00244 [astro-ph.CO]} \BibitemShut {NoStop}%
\bibitem [{\citenamefont {Niedermann}\ and\ \citenamefont {Sloth}(2020)}]{Niedermann:2020dwg}%
  \BibitemOpen
  \bibfield  {author} {\bibinfo {author} {\bibfnamefont {F.}~\bibnamefont {Niedermann}}\ and\ \bibinfo {author} {\bibfnamefont {M.~S.}\ \bibnamefont {Sloth}},\ }\href {\doibase 10.1103/PhysRevD.102.063527} {\bibfield  {journal} {\bibinfo  {journal} {Phys. Rev. D}\ }\textbf {\bibinfo {volume} {102}},\ \bibinfo {pages} {063527} (\bibinfo {year} {2020})},\ \Eprint {http://arxiv.org/abs/2006.06686} {arXiv:2006.06686 [astro-ph.CO]} \BibitemShut {NoStop}%
\bibitem [{\citenamefont {Gonzalez}\ \emph {et~al.}(2020)\citenamefont {Gonzalez}, \citenamefont {Hertzberg},\ and\ \citenamefont {Rompineve}}]{Gonzalez:2020fdy}%
  \BibitemOpen
  \bibfield  {author} {\bibinfo {author} {\bibfnamefont {M.}~\bibnamefont {Gonzalez}}, \bibinfo {author} {\bibfnamefont {M.~P.}\ \bibnamefont {Hertzberg}}, \ and\ \bibinfo {author} {\bibfnamefont {F.}~\bibnamefont {Rompineve}},\ }\href {\doibase 10.1088/1475-7516/2020/10/028} {\bibfield  {journal} {\bibinfo  {journal} {JCAP}\ }\textbf {\bibinfo {volume} {10}},\ \bibinfo {pages} {028} (\bibinfo {year} {2020})},\ \Eprint {http://arxiv.org/abs/2006.13959} {arXiv:2006.13959 [astro-ph.CO]} \BibitemShut {NoStop}%
\bibitem [{\citenamefont {Braglia}\ \emph {et~al.}(2021)\citenamefont {Braglia}, \citenamefont {Ballardini}, \citenamefont {Finelli},\ and\ \citenamefont {Koyama}}]{Braglia:2020auw}%
  \BibitemOpen
  \bibfield  {author} {\bibinfo {author} {\bibfnamefont {M.}~\bibnamefont {Braglia}}, \bibinfo {author} {\bibfnamefont {M.}~\bibnamefont {Ballardini}}, \bibinfo {author} {\bibfnamefont {F.}~\bibnamefont {Finelli}}, \ and\ \bibinfo {author} {\bibfnamefont {K.}~\bibnamefont {Koyama}},\ }\href {\doibase 10.1103/PhysRevD.103.043528} {\bibfield  {journal} {\bibinfo  {journal} {Phys. Rev. D}\ }\textbf {\bibinfo {volume} {103}},\ \bibinfo {pages} {043528} (\bibinfo {year} {2021})},\ \Eprint {http://arxiv.org/abs/2011.12934} {arXiv:2011.12934 [astro-ph.CO]} \BibitemShut {NoStop}%
\bibitem [{\citenamefont {Roy~Choudhury}\ \emph {et~al.}(2021)\citenamefont {Roy~Choudhury}, \citenamefont {Hannestad},\ and\ \citenamefont {Tram}}]{RoyChoudhury:2020dmd}%
  \BibitemOpen
  \bibfield  {author} {\bibinfo {author} {\bibfnamefont {S.}~\bibnamefont {Roy~Choudhury}}, \bibinfo {author} {\bibfnamefont {S.}~\bibnamefont {Hannestad}}, \ and\ \bibinfo {author} {\bibfnamefont {T.}~\bibnamefont {Tram}},\ }\href {\doibase 10.1088/1475-7516/2021/03/084} {\bibfield  {journal} {\bibinfo  {journal} {JCAP}\ }\textbf {\bibinfo {volume} {03}},\ \bibinfo {pages} {084} (\bibinfo {year} {2021})},\ \Eprint {http://arxiv.org/abs/2012.07519} {arXiv:2012.07519 [astro-ph.CO]} \BibitemShut {NoStop}%
\bibitem [{\citenamefont {Brinckmann}\ \emph {et~al.}(2021)\citenamefont {Brinckmann}, \citenamefont {Chang},\ and\ \citenamefont {LoVerde}}]{Brinckmann:2020bcn}%
  \BibitemOpen
  \bibfield  {author} {\bibinfo {author} {\bibfnamefont {T.}~\bibnamefont {Brinckmann}}, \bibinfo {author} {\bibfnamefont {J.~H.}\ \bibnamefont {Chang}}, \ and\ \bibinfo {author} {\bibfnamefont {M.}~\bibnamefont {LoVerde}},\ }\href {\doibase 10.1103/PhysRevD.104.063523} {\bibfield  {journal} {\bibinfo  {journal} {Phys. Rev. D}\ }\textbf {\bibinfo {volume} {104}},\ \bibinfo {pages} {063523} (\bibinfo {year} {2021})},\ \Eprint {http://arxiv.org/abs/2012.11830} {arXiv:2012.11830 [astro-ph.CO]} \BibitemShut {NoStop}%
\bibitem [{\citenamefont {Alestas}\ \emph {et~al.}(2021)\citenamefont {Alestas}, \citenamefont {Kazantzidis},\ and\ \citenamefont {Perivolaropoulos}}]{Alestas:2020zol}%
  \BibitemOpen
  \bibfield  {author} {\bibinfo {author} {\bibfnamefont {G.}~\bibnamefont {Alestas}}, \bibinfo {author} {\bibfnamefont {L.}~\bibnamefont {Kazantzidis}}, \ and\ \bibinfo {author} {\bibfnamefont {L.}~\bibnamefont {Perivolaropoulos}},\ }\href {\doibase 10.1103/PhysRevD.103.083517} {\bibfield  {journal} {\bibinfo  {journal} {Phys. Rev. D}\ }\textbf {\bibinfo {volume} {103}},\ \bibinfo {pages} {083517} (\bibinfo {year} {2021})},\ \Eprint {http://arxiv.org/abs/2012.13932} {arXiv:2012.13932 [astro-ph.CO]} \BibitemShut {NoStop}%
\bibitem [{\citenamefont {Gao}\ \emph {et~al.}(2021)\citenamefont {Gao}, \citenamefont {Zhao}, \citenamefont {Xue},\ and\ \citenamefont {Zhang}}]{Gao:2021xnk}%
  \BibitemOpen
  \bibfield  {author} {\bibinfo {author} {\bibfnamefont {L.-Y.}\ \bibnamefont {Gao}}, \bibinfo {author} {\bibfnamefont {Z.-W.}\ \bibnamefont {Zhao}}, \bibinfo {author} {\bibfnamefont {S.-S.}\ \bibnamefont {Xue}}, \ and\ \bibinfo {author} {\bibfnamefont {X.}~\bibnamefont {Zhang}},\ }\href {\doibase 10.1088/1475-7516/2021/07/005} {\bibfield  {journal} {\bibinfo  {journal} {JCAP}\ }\textbf {\bibinfo {volume} {07}},\ \bibinfo {pages} {005} (\bibinfo {year} {2021})},\ \Eprint {http://arxiv.org/abs/2101.10714} {arXiv:2101.10714 [astro-ph.CO]} \BibitemShut {NoStop}%
\bibitem [{\citenamefont {Renzi}\ \emph {et~al.}(2022)\citenamefont {Renzi}, \citenamefont {Hogg},\ and\ \citenamefont {Giar\`e}}]{Renzi:2021xii}%
  \BibitemOpen
  \bibfield  {author} {\bibinfo {author} {\bibfnamefont {F.}~\bibnamefont {Renzi}}, \bibinfo {author} {\bibfnamefont {N.~B.}\ \bibnamefont {Hogg}}, \ and\ \bibinfo {author} {\bibfnamefont {W.}~\bibnamefont {Giar\`e}},\ }\href {\doibase 10.1093/mnras/stac1030} {\bibfield  {journal} {\bibinfo  {journal} {Mon. Not. Roy. Astron. Soc.}\ }\textbf {\bibinfo {volume} {513}},\ \bibinfo {pages} {4004} (\bibinfo {year} {2022})},\ \Eprint {http://arxiv.org/abs/2112.05701} {arXiv:2112.05701 [astro-ph.CO]} \BibitemShut {NoStop}%
\bibitem [{\citenamefont {Alestas}\ and\ \citenamefont {Perivolaropoulos}(2021)}]{Alestas:2021xes}%
  \BibitemOpen
  \bibfield  {author} {\bibinfo {author} {\bibfnamefont {G.}~\bibnamefont {Alestas}}\ and\ \bibinfo {author} {\bibfnamefont {L.}~\bibnamefont {Perivolaropoulos}},\ }\href {\doibase 10.1093/mnras/stab1070} {\bibfield  {journal} {\bibinfo  {journal} {Mon. Not. Roy. Astron. Soc.}\ }\textbf {\bibinfo {volume} {504}},\ \bibinfo {pages} {3956} (\bibinfo {year} {2021})},\ \Eprint {http://arxiv.org/abs/2103.04045} {arXiv:2103.04045 [astro-ph.CO]} \BibitemShut {NoStop}%
\bibitem [{\citenamefont {Karwal}\ \emph {et~al.}(2022)\citenamefont {Karwal}, \citenamefont {Raveri}, \citenamefont {Jain}, \citenamefont {Khoury},\ and\ \citenamefont {Trodden}}]{Karwal:2021vpk}%
  \BibitemOpen
  \bibfield  {author} {\bibinfo {author} {\bibfnamefont {T.}~\bibnamefont {Karwal}}, \bibinfo {author} {\bibfnamefont {M.}~\bibnamefont {Raveri}}, \bibinfo {author} {\bibfnamefont {B.}~\bibnamefont {Jain}}, \bibinfo {author} {\bibfnamefont {J.}~\bibnamefont {Khoury}}, \ and\ \bibinfo {author} {\bibfnamefont {M.}~\bibnamefont {Trodden}},\ }\href {\doibase 10.1103/PhysRevD.105.063535} {\bibfield  {journal} {\bibinfo  {journal} {Phys. Rev. D}\ }\textbf {\bibinfo {volume} {105}},\ \bibinfo {pages} {063535} (\bibinfo {year} {2022})},\ \Eprint {http://arxiv.org/abs/2106.13290} {arXiv:2106.13290 [astro-ph.CO]} \BibitemShut {NoStop}%
\bibitem [{\citenamefont {Cyr-Racine}\ \emph {et~al.}(2022)\citenamefont {Cyr-Racine}, \citenamefont {Ge},\ and\ \citenamefont {Knox}}]{Cyr-Racine:2021oal}%
  \BibitemOpen
  \bibfield  {author} {\bibinfo {author} {\bibfnamefont {F.-Y.}\ \bibnamefont {Cyr-Racine}}, \bibinfo {author} {\bibfnamefont {F.}~\bibnamefont {Ge}}, \ and\ \bibinfo {author} {\bibfnamefont {L.}~\bibnamefont {Knox}},\ }\href {\doibase 10.1103/PhysRevLett.128.201301} {\bibfield  {journal} {\bibinfo  {journal} {Phys. Rev. Lett.}\ }\textbf {\bibinfo {volume} {128}},\ \bibinfo {pages} {201301} (\bibinfo {year} {2022})},\ \Eprint {http://arxiv.org/abs/2107.13000} {arXiv:2107.13000 [astro-ph.CO]} \BibitemShut {NoStop}%
\bibitem [{\citenamefont {Akarsu}\ \emph {et~al.}(2021)\citenamefont {Akarsu}, \citenamefont {Kumar}, \citenamefont {\"Oz\"ulker},\ and\ \citenamefont {Vazquez}}]{Akarsu:2021fol}%
  \BibitemOpen
  \bibfield  {author} {\bibinfo {author} {\bibfnamefont {O.}~\bibnamefont {Akarsu}}, \bibinfo {author} {\bibfnamefont {S.}~\bibnamefont {Kumar}}, \bibinfo {author} {\bibfnamefont {E.}~\bibnamefont {\"Oz\"ulker}}, \ and\ \bibinfo {author} {\bibfnamefont {J.~A.}\ \bibnamefont {Vazquez}},\ }\href {\doibase 10.1103/PhysRevD.104.123512} {\bibfield  {journal} {\bibinfo  {journal} {Phys. Rev. D}\ }\textbf {\bibinfo {volume} {104}},\ \bibinfo {pages} {123512} (\bibinfo {year} {2021})},\ \Eprint {http://arxiv.org/abs/2108.09239} {arXiv:2108.09239 [astro-ph.CO]} \BibitemShut {NoStop}%
\bibitem [{\citenamefont {Niedermann}\ and\ \citenamefont {Sloth}(2022)}]{Niedermann:2021ijp}%
  \BibitemOpen
  \bibfield  {author} {\bibinfo {author} {\bibfnamefont {F.}~\bibnamefont {Niedermann}}\ and\ \bibinfo {author} {\bibfnamefont {M.~S.}\ \bibnamefont {Sloth}},\ }\href {\doibase 10.1016/j.physletb.2022.137555} {\bibfield  {journal} {\bibinfo  {journal} {Phys. Lett. B}\ }\textbf {\bibinfo {volume} {835}},\ \bibinfo {pages} {137555} (\bibinfo {year} {2022})},\ \Eprint {http://arxiv.org/abs/2112.00759} {arXiv:2112.00759 [hep-ph]} \BibitemShut {NoStop}%
\bibitem [{\citenamefont {Saridakis}\ \emph {et~al.}(2023)\citenamefont {Saridakis}, \citenamefont {Yang}, \citenamefont {Pan}, \citenamefont {Anagnostopoulos},\ and\ \citenamefont {Basilakos}}]{Saridakis:2021xqy}%
  \BibitemOpen
  \bibfield  {author} {\bibinfo {author} {\bibfnamefont {E.~N.}\ \bibnamefont {Saridakis}}, \bibinfo {author} {\bibfnamefont {W.}~\bibnamefont {Yang}}, \bibinfo {author} {\bibfnamefont {S.}~\bibnamefont {Pan}}, \bibinfo {author} {\bibfnamefont {F.~K.}\ \bibnamefont {Anagnostopoulos}}, \ and\ \bibinfo {author} {\bibfnamefont {S.}~\bibnamefont {Basilakos}},\ }\href {\doibase 10.1016/j.nuclphysb.2022.116042} {\bibfield  {journal} {\bibinfo  {journal} {Nucl. Phys. B}\ }\textbf {\bibinfo {volume} {986}},\ \bibinfo {pages} {116042} (\bibinfo {year} {2023})},\ \Eprint {http://arxiv.org/abs/2112.08330} {arXiv:2112.08330 [astro-ph.CO]} \BibitemShut {NoStop}%
\bibitem [{\citenamefont {Sen}\ \emph {et~al.}(2022)\citenamefont {Sen}, \citenamefont {Adil},\ and\ \citenamefont {Sen}}]{Sen:2021wld}%
  \BibitemOpen
  \bibfield  {author} {\bibinfo {author} {\bibfnamefont {A.~A.}\ \bibnamefont {Sen}}, \bibinfo {author} {\bibfnamefont {S.~A.}\ \bibnamefont {Adil}}, \ and\ \bibinfo {author} {\bibfnamefont {S.}~\bibnamefont {Sen}},\ }\href {\doibase 10.1093/mnras/stac2796} {\bibfield  {journal} {\bibinfo  {journal} {Mon. Not. Roy. Astron. Soc.}\ }\textbf {\bibinfo {volume} {518}},\ \bibinfo {pages} {1098} (\bibinfo {year} {2022})},\ \Eprint {http://arxiv.org/abs/2112.10641} {arXiv:2112.10641 [astro-ph.CO]} \BibitemShut {NoStop}%
\bibitem [{\citenamefont {Herold}\ \emph {et~al.}(2022)\citenamefont {Herold}, \citenamefont {Ferreira},\ and\ \citenamefont {Komatsu}}]{Herold:2021ksg}%
  \BibitemOpen
  \bibfield  {author} {\bibinfo {author} {\bibfnamefont {L.}~\bibnamefont {Herold}}, \bibinfo {author} {\bibfnamefont {E.~G.~M.}\ \bibnamefont {Ferreira}}, \ and\ \bibinfo {author} {\bibfnamefont {E.}~\bibnamefont {Komatsu}},\ }\href {\doibase 10.3847/2041-8213/ac63a3} {\bibfield  {journal} {\bibinfo  {journal} {Astrophys. J. Lett.}\ }\textbf {\bibinfo {volume} {929}},\ \bibinfo {pages} {L16} (\bibinfo {year} {2022})},\ \Eprint {http://arxiv.org/abs/2112.12140} {arXiv:2112.12140 [astro-ph.CO]} \BibitemShut {NoStop}%
\bibitem [{\citenamefont {Odintsov}\ and\ \citenamefont {Oikonomou}(2022{\natexlab{a}})}]{Odintsov:2022eqm}%
  \BibitemOpen
  \bibfield  {author} {\bibinfo {author} {\bibfnamefont {S.~D.}\ \bibnamefont {Odintsov}}\ and\ \bibinfo {author} {\bibfnamefont {V.~K.}\ \bibnamefont {Oikonomou}},\ }\href {\doibase 10.1209/0295-5075/ac52dc} {\bibfield  {journal} {\bibinfo  {journal} {EPL}\ }\textbf {\bibinfo {volume} {137}},\ \bibinfo {pages} {39001} (\bibinfo {year} {2022}{\natexlab{a}})},\ \Eprint {http://arxiv.org/abs/2201.07647} {arXiv:2201.07647 [gr-qc]} \BibitemShut {NoStop}%
\bibitem [{\citenamefont {Heisenberg}\ \emph {et~al.}(2023)\citenamefont {Heisenberg}, \citenamefont {Villarrubia-Rojo},\ and\ \citenamefont {Zosso}}]{Heisenberg:2022lob}%
  \BibitemOpen
  \bibfield  {author} {\bibinfo {author} {\bibfnamefont {L.}~\bibnamefont {Heisenberg}}, \bibinfo {author} {\bibfnamefont {H.}~\bibnamefont {Villarrubia-Rojo}}, \ and\ \bibinfo {author} {\bibfnamefont {J.}~\bibnamefont {Zosso}},\ }\href {\doibase 10.1016/j.dark.2022.101163} {\bibfield  {journal} {\bibinfo  {journal} {Phys. Dark Univ.}\ }\textbf {\bibinfo {volume} {39}},\ \bibinfo {pages} {101163} (\bibinfo {year} {2023})},\ \Eprint {http://arxiv.org/abs/2201.11623} {arXiv:2201.11623 [astro-ph.CO]} \BibitemShut {NoStop}%
\bibitem [{\citenamefont {Heisenberg}\ \emph {et~al.}(2022)\citenamefont {Heisenberg}, \citenamefont {Villarrubia-Rojo},\ and\ \citenamefont {Zosso}}]{Heisenberg:2022gqk}%
  \BibitemOpen
  \bibfield  {author} {\bibinfo {author} {\bibfnamefont {L.}~\bibnamefont {Heisenberg}}, \bibinfo {author} {\bibfnamefont {H.}~\bibnamefont {Villarrubia-Rojo}}, \ and\ \bibinfo {author} {\bibfnamefont {J.}~\bibnamefont {Zosso}},\ }\href {\doibase 10.1103/PhysRevD.106.043503} {\bibfield  {journal} {\bibinfo  {journal} {Phys. Rev. D}\ }\textbf {\bibinfo {volume} {106}},\ \bibinfo {pages} {043503} (\bibinfo {year} {2022})},\ \Eprint {http://arxiv.org/abs/2202.01202} {arXiv:2202.01202 [astro-ph.CO]} \BibitemShut {NoStop}%
\bibitem [{\citenamefont {Sharma}\ \emph {et~al.}(2022)\citenamefont {Sharma}, \citenamefont {Pandey},\ and\ \citenamefont {Das}}]{Sharma:2022ifr}%
  \BibitemOpen
  \bibfield  {author} {\bibinfo {author} {\bibfnamefont {R.~K.}\ \bibnamefont {Sharma}}, \bibinfo {author} {\bibfnamefont {K.~L.}\ \bibnamefont {Pandey}}, \ and\ \bibinfo {author} {\bibfnamefont {S.}~\bibnamefont {Das}},\ }\href {\doibase 10.3847/1538-4357/ac7a33} {\bibfield  {journal} {\bibinfo  {journal} {Astrophys. J.}\ }\textbf {\bibinfo {volume} {934}},\ \bibinfo {pages} {113} (\bibinfo {year} {2022})},\ \Eprint {http://arxiv.org/abs/2202.01749} {arXiv:2202.01749 [astro-ph.CO]} \BibitemShut {NoStop}%
\bibitem [{\citenamefont {Ren}\ \emph {et~al.}(2022)\citenamefont {Ren}, \citenamefont {Yan}, \citenamefont {Zhao}, \citenamefont {Cai},\ and\ \citenamefont {Saridakis}}]{Ren:2022aeo}%
  \BibitemOpen
  \bibfield  {author} {\bibinfo {author} {\bibfnamefont {X.}~\bibnamefont {Ren}}, \bibinfo {author} {\bibfnamefont {S.-F.}\ \bibnamefont {Yan}}, \bibinfo {author} {\bibfnamefont {Y.}~\bibnamefont {Zhao}}, \bibinfo {author} {\bibfnamefont {Y.-F.}\ \bibnamefont {Cai}}, \ and\ \bibinfo {author} {\bibfnamefont {E.~N.}\ \bibnamefont {Saridakis}},\ }\href {\doibase 10.3847/1538-4357/ac6ba5} {\bibfield  {journal} {\bibinfo  {journal} {Astrophys. J.}\ }\textbf {\bibinfo {volume} {932}},\ \bibinfo {pages} {131} (\bibinfo {year} {2022})},\ \Eprint {http://arxiv.org/abs/2203.01926} {arXiv:2203.01926 [astro-ph.CO]} \BibitemShut {NoStop}%
\bibitem [{\citenamefont {Nunes}\ \emph {et~al.}(2022)\citenamefont {Nunes}, \citenamefont {Vagnozzi}, \citenamefont {Kumar}, \citenamefont {Di~Valentino},\ and\ \citenamefont {Mena}}]{Nunes:2022bhn}%
  \BibitemOpen
  \bibfield  {author} {\bibinfo {author} {\bibfnamefont {R.~C.}\ \bibnamefont {Nunes}}, \bibinfo {author} {\bibfnamefont {S.}~\bibnamefont {Vagnozzi}}, \bibinfo {author} {\bibfnamefont {S.}~\bibnamefont {Kumar}}, \bibinfo {author} {\bibfnamefont {E.}~\bibnamefont {Di~Valentino}}, \ and\ \bibinfo {author} {\bibfnamefont {O.}~\bibnamefont {Mena}},\ }\href {\doibase 10.1103/PhysRevD.105.123506} {\bibfield  {journal} {\bibinfo  {journal} {Phys. Rev. D}\ }\textbf {\bibinfo {volume} {105}},\ \bibinfo {pages} {123506} (\bibinfo {year} {2022})},\ \Eprint {http://arxiv.org/abs/2203.08093} {arXiv:2203.08093 [astro-ph.CO]} \BibitemShut {NoStop}%
\bibitem [{\citenamefont {Nojiri}\ \emph {et~al.}(2022)\citenamefont {Nojiri}, \citenamefont {Odintsov},\ and\ \citenamefont {Oikonomou}}]{Nojiri:2022ski}%
  \BibitemOpen
  \bibfield  {author} {\bibinfo {author} {\bibfnamefont {S.}~\bibnamefont {Nojiri}}, \bibinfo {author} {\bibfnamefont {S.~D.}\ \bibnamefont {Odintsov}}, \ and\ \bibinfo {author} {\bibfnamefont {V.~K.}\ \bibnamefont {Oikonomou}},\ }\href {\doibase 10.1016/j.nuclphysb.2022.115850} {\bibfield  {journal} {\bibinfo  {journal} {Nucl. Phys. B}\ }\textbf {\bibinfo {volume} {980}},\ \bibinfo {pages} {115850} (\bibinfo {year} {2022})},\ \Eprint {http://arxiv.org/abs/2205.11681} {arXiv:2205.11681 [gr-qc]} \BibitemShut {NoStop}%
\bibitem [{\citenamefont {Sch\"oneberg}\ and\ \citenamefont {Franco~Abell\'an}(2022)}]{Schoneberg:2022grr}%
  \BibitemOpen
  \bibfield  {author} {\bibinfo {author} {\bibfnamefont {N.}~\bibnamefont {Sch\"oneberg}}\ and\ \bibinfo {author} {\bibfnamefont {G.}~\bibnamefont {Franco~Abell\'an}},\ }\href {\doibase 10.1088/1475-7516/2022/12/001} {\bibfield  {journal} {\bibinfo  {journal} {JCAP}\ }\textbf {\bibinfo {volume} {12}},\ \bibinfo {pages} {001} (\bibinfo {year} {2022})},\ \Eprint {http://arxiv.org/abs/2206.11276} {arXiv:2206.11276 [astro-ph.CO]} \BibitemShut {NoStop}%
\bibitem [{\citenamefont {Joseph}\ \emph {et~al.}(2023)\citenamefont {Joseph}, \citenamefont {Aloni}, \citenamefont {Schmaltz}, \citenamefont {Sivarajan},\ and\ \citenamefont {Weiner}}]{Joseph:2022jsf}%
  \BibitemOpen
  \bibfield  {author} {\bibinfo {author} {\bibfnamefont {M.}~\bibnamefont {Joseph}}, \bibinfo {author} {\bibfnamefont {D.}~\bibnamefont {Aloni}}, \bibinfo {author} {\bibfnamefont {M.}~\bibnamefont {Schmaltz}}, \bibinfo {author} {\bibfnamefont {E.~N.}\ \bibnamefont {Sivarajan}}, \ and\ \bibinfo {author} {\bibfnamefont {N.}~\bibnamefont {Weiner}},\ }\href {\doibase 10.1103/PhysRevD.108.023520} {\bibfield  {journal} {\bibinfo  {journal} {Phys. Rev. D}\ }\textbf {\bibinfo {volume} {108}},\ \bibinfo {pages} {023520} (\bibinfo {year} {2023})},\ \Eprint {http://arxiv.org/abs/2207.03500} {arXiv:2207.03500 [astro-ph.CO]} \BibitemShut {NoStop}%
\bibitem [{\citenamefont {G\'omez-Valent}\ \emph {et~al.}(2022)\citenamefont {G\'omez-Valent}, \citenamefont {Zheng}, \citenamefont {Amendola}, \citenamefont {Wetterich},\ and\ \citenamefont {Pettorino}}]{Gomez-Valent:2022bku}%
  \BibitemOpen
  \bibfield  {author} {\bibinfo {author} {\bibfnamefont {A.}~\bibnamefont {G\'omez-Valent}}, \bibinfo {author} {\bibfnamefont {Z.}~\bibnamefont {Zheng}}, \bibinfo {author} {\bibfnamefont {L.}~\bibnamefont {Amendola}}, \bibinfo {author} {\bibfnamefont {C.}~\bibnamefont {Wetterich}}, \ and\ \bibinfo {author} {\bibfnamefont {V.}~\bibnamefont {Pettorino}},\ }\href {\doibase 10.1103/PhysRevD.106.103522} {\bibfield  {journal} {\bibinfo  {journal} {Phys. Rev. D}\ }\textbf {\bibinfo {volume} {106}},\ \bibinfo {pages} {103522} (\bibinfo {year} {2022})},\ \Eprint {http://arxiv.org/abs/2207.14487} {arXiv:2207.14487 [astro-ph.CO]} \BibitemShut {NoStop}%
\bibitem [{\citenamefont {Moshafi}\ \emph {et~al.}(2022)\citenamefont {Moshafi}, \citenamefont {Firouzjahi},\ and\ \citenamefont {Talebian}}]{Moshafi:2022mva}%
  \BibitemOpen
  \bibfield  {author} {\bibinfo {author} {\bibfnamefont {H.}~\bibnamefont {Moshafi}}, \bibinfo {author} {\bibfnamefont {H.}~\bibnamefont {Firouzjahi}}, \ and\ \bibinfo {author} {\bibfnamefont {A.}~\bibnamefont {Talebian}},\ }\href {\doibase 10.3847/1538-4357/ac9c58} {\bibfield  {journal} {\bibinfo  {journal} {Astrophys. J.}\ }\textbf {\bibinfo {volume} {940}},\ \bibinfo {pages} {121} (\bibinfo {year} {2022})},\ \Eprint {http://arxiv.org/abs/2208.05583} {arXiv:2208.05583 [astro-ph.CO]} \BibitemShut {NoStop}%
\bibitem [{\citenamefont {Odintsov}\ and\ \citenamefont {Oikonomou}(2022{\natexlab{b}})}]{Odintsov:2022umu}%
  \BibitemOpen
  \bibfield  {author} {\bibinfo {author} {\bibfnamefont {S.~D.}\ \bibnamefont {Odintsov}}\ and\ \bibinfo {author} {\bibfnamefont {V.~K.}\ \bibnamefont {Oikonomou}},\ }\href {\doibase 10.1209/0295-5075/ac8a13} {\bibfield  {journal} {\bibinfo  {journal} {EPL}\ }\textbf {\bibinfo {volume} {139}},\ \bibinfo {pages} {59003} (\bibinfo {year} {2022}{\natexlab{b}})},\ \Eprint {http://arxiv.org/abs/2208.07972} {arXiv:2208.07972 [gr-qc]} \BibitemShut {NoStop}%
\bibitem [{\citenamefont {Banerjee}\ \emph {et~al.}(2023)\citenamefont {Banerjee}, \citenamefont {Petronikolou},\ and\ \citenamefont {Saridakis}}]{Banerjee:2022ynv}%
  \BibitemOpen
  \bibfield  {author} {\bibinfo {author} {\bibfnamefont {S.}~\bibnamefont {Banerjee}}, \bibinfo {author} {\bibfnamefont {M.}~\bibnamefont {Petronikolou}}, \ and\ \bibinfo {author} {\bibfnamefont {E.~N.}\ \bibnamefont {Saridakis}},\ }\href {\doibase 10.1103/PhysRevD.108.024012} {\bibfield  {journal} {\bibinfo  {journal} {Phys. Rev. D}\ }\textbf {\bibinfo {volume} {108}},\ \bibinfo {pages} {024012} (\bibinfo {year} {2023})},\ \Eprint {http://arxiv.org/abs/2209.02426} {arXiv:2209.02426 [gr-qc]} \BibitemShut {NoStop}%
\bibitem [{\citenamefont {Alvarez}\ \emph {et~al.}(2024)\citenamefont {Alvarez}, \citenamefont {Koch}, \citenamefont {Laporte},\ and\ \citenamefont {Rincon}}]{Alvarez:2022wef}%
  \BibitemOpen
  \bibfield  {author} {\bibinfo {author} {\bibfnamefont {P.~D.}\ \bibnamefont {Alvarez}}, \bibinfo {author} {\bibfnamefont {B.}~\bibnamefont {Koch}}, \bibinfo {author} {\bibfnamefont {C.}~\bibnamefont {Laporte}}, \ and\ \bibinfo {author} {\bibfnamefont {A.}~\bibnamefont {Rincon}},\ }\href {\doibase 10.1016/j.dark.2024.101531} {\bibfield  {journal} {\bibinfo  {journal} {Phys. Dark Univ.}\ }\textbf {\bibinfo {volume} {45}},\ \bibinfo {pages} {101531} (\bibinfo {year} {2024})},\ \Eprint {http://arxiv.org/abs/2210.11853} {arXiv:2210.11853 [gr-qc]} \BibitemShut {NoStop}%
\bibitem [{\citenamefont {Ge}\ \emph {et~al.}(2023)\citenamefont {Ge}, \citenamefont {Cyr-Racine},\ and\ \citenamefont {Knox}}]{Ge:2022qws}%
  \BibitemOpen
  \bibfield  {author} {\bibinfo {author} {\bibfnamefont {F.}~\bibnamefont {Ge}}, \bibinfo {author} {\bibfnamefont {F.-Y.}\ \bibnamefont {Cyr-Racine}}, \ and\ \bibinfo {author} {\bibfnamefont {L.}~\bibnamefont {Knox}},\ }\href {\doibase 10.1103/PhysRevD.107.023517} {\bibfield  {journal} {\bibinfo  {journal} {Phys. Rev. D}\ }\textbf {\bibinfo {volume} {107}},\ \bibinfo {pages} {023517} (\bibinfo {year} {2023})},\ \Eprint {http://arxiv.org/abs/2210.16335} {arXiv:2210.16335 [astro-ph.CO]} \BibitemShut {NoStop}%
\bibitem [{\citenamefont {Akarsu}\ \emph {et~al.}(2023{\natexlab{a}})\citenamefont {Akarsu}, \citenamefont {Kumar}, \citenamefont {\"Oz\"ulker}, \citenamefont {Vazquez},\ and\ \citenamefont {Yadav}}]{Akarsu:2022typ}%
  \BibitemOpen
  \bibfield  {author} {\bibinfo {author} {\bibfnamefont {O.}~\bibnamefont {Akarsu}}, \bibinfo {author} {\bibfnamefont {S.}~\bibnamefont {Kumar}}, \bibinfo {author} {\bibfnamefont {E.}~\bibnamefont {\"Oz\"ulker}}, \bibinfo {author} {\bibfnamefont {J.~A.}\ \bibnamefont {Vazquez}}, \ and\ \bibinfo {author} {\bibfnamefont {A.}~\bibnamefont {Yadav}},\ }\href {\doibase 10.1103/PhysRevD.108.023513} {\bibfield  {journal} {\bibinfo  {journal} {Phys. Rev. D}\ }\textbf {\bibinfo {volume} {108}},\ \bibinfo {pages} {023513} (\bibinfo {year} {2023}{\natexlab{a}})},\ \Eprint {http://arxiv.org/abs/2211.05742} {arXiv:2211.05742 [astro-ph.CO]} \BibitemShut {NoStop}%
\bibitem [{\citenamefont {Gangopadhyay}\ \emph {et~al.}(2023{\natexlab{a}})\citenamefont {Gangopadhyay}, \citenamefont {Pacif}, \citenamefont {Sami},\ and\ \citenamefont {Sharma}}]{Gangopadhyay:2022bsh}%
  \BibitemOpen
  \bibfield  {author} {\bibinfo {author} {\bibfnamefont {M.~R.}\ \bibnamefont {Gangopadhyay}}, \bibinfo {author} {\bibfnamefont {S.~K.~J.}\ \bibnamefont {Pacif}}, \bibinfo {author} {\bibfnamefont {M.}~\bibnamefont {Sami}}, \ and\ \bibinfo {author} {\bibfnamefont {M.~K.}\ \bibnamefont {Sharma}},\ }\href {\doibase 10.3390/universe9020083} {\bibfield  {journal} {\bibinfo  {journal} {Universe}\ }\textbf {\bibinfo {volume} {9}},\ \bibinfo {pages} {83} (\bibinfo {year} {2023}{\natexlab{a}})},\ \Eprint {http://arxiv.org/abs/2211.12041} {arXiv:2211.12041 [gr-qc]} \BibitemShut {NoStop}%
\bibitem [{\citenamefont {Schiavone}\ \emph {et~al.}(2023)\citenamefont {Schiavone}, \citenamefont {Montani},\ and\ \citenamefont {Bombacigno}}]{Schiavone:2022wvq}%
  \BibitemOpen
  \bibfield  {author} {\bibinfo {author} {\bibfnamefont {T.}~\bibnamefont {Schiavone}}, \bibinfo {author} {\bibfnamefont {G.}~\bibnamefont {Montani}}, \ and\ \bibinfo {author} {\bibfnamefont {F.}~\bibnamefont {Bombacigno}},\ }\href {\doibase 10.1093/mnrasl/slad041} {\bibfield  {journal} {\bibinfo  {journal} {Mon. Not. Roy. Astron. Soc.}\ }\textbf {\bibinfo {volume} {522}},\ \bibinfo {pages} {L72} (\bibinfo {year} {2023})},\ \Eprint {http://arxiv.org/abs/2211.16737} {arXiv:2211.16737 [gr-qc]} \BibitemShut {NoStop}%
\bibitem [{\citenamefont {Gao}\ \emph {et~al.}(2024)\citenamefont {Gao}, \citenamefont {Xue},\ and\ \citenamefont {Zhang}}]{Gao:2022ahg}%
  \BibitemOpen
  \bibfield  {author} {\bibinfo {author} {\bibfnamefont {L.-Y.}\ \bibnamefont {Gao}}, \bibinfo {author} {\bibfnamefont {S.-S.}\ \bibnamefont {Xue}}, \ and\ \bibinfo {author} {\bibfnamefont {X.}~\bibnamefont {Zhang}},\ }\href {\doibase 10.1088/1674-1137/ad2b52} {\bibfield  {journal} {\bibinfo  {journal} {Chin. Phys. C}\ }\textbf {\bibinfo {volume} {48}},\ \bibinfo {pages} {051001} (\bibinfo {year} {2024})},\ \Eprint {http://arxiv.org/abs/2212.13146} {arXiv:2212.13146 [astro-ph.CO]} \BibitemShut {NoStop}%
\bibitem [{\citenamefont {Brinckmann}\ \emph {et~al.}(2023)\citenamefont {Brinckmann}, \citenamefont {Chang}, \citenamefont {Du},\ and\ \citenamefont {LoVerde}}]{Brinckmann:2022ajr}%
  \BibitemOpen
  \bibfield  {author} {\bibinfo {author} {\bibfnamefont {T.}~\bibnamefont {Brinckmann}}, \bibinfo {author} {\bibfnamefont {J.~H.}\ \bibnamefont {Chang}}, \bibinfo {author} {\bibfnamefont {P.}~\bibnamefont {Du}}, \ and\ \bibinfo {author} {\bibfnamefont {M.}~\bibnamefont {LoVerde}},\ }\href {\doibase 10.1103/PhysRevD.107.123517} {\bibfield  {journal} {\bibinfo  {journal} {Phys. Rev. D}\ }\textbf {\bibinfo {volume} {107}},\ \bibinfo {pages} {123517} (\bibinfo {year} {2023})},\ \Eprint {http://arxiv.org/abs/2212.13264} {arXiv:2212.13264 [astro-ph.CO]} \BibitemShut {NoStop}%
\bibitem [{\citenamefont {Khodadi}\ and\ \citenamefont {Schreck}(2023)}]{Khodadi:2023ezj}%
  \BibitemOpen
  \bibfield  {author} {\bibinfo {author} {\bibfnamefont {M.}~\bibnamefont {Khodadi}}\ and\ \bibinfo {author} {\bibfnamefont {M.}~\bibnamefont {Schreck}},\ }\href {\doibase 10.1016/j.dark.2023.101170} {\bibfield  {journal} {\bibinfo  {journal} {Phys. Dark Univ.}\ }\textbf {\bibinfo {volume} {39}},\ \bibinfo {pages} {101170} (\bibinfo {year} {2023})},\ \Eprint {http://arxiv.org/abs/2301.03883} {arXiv:2301.03883 [gr-qc]} \BibitemShut {NoStop}%
\bibitem [{\citenamefont {Dahmani}\ \emph {et~al.}(2023)\citenamefont {Dahmani}, \citenamefont {Bouali}, \citenamefont {El~Bojaddaini}, \citenamefont {Errahmani},\ and\ \citenamefont {Ouali}}]{Dahmani:2023bsb}%
  \BibitemOpen
  \bibfield  {author} {\bibinfo {author} {\bibfnamefont {S.}~\bibnamefont {Dahmani}}, \bibinfo {author} {\bibfnamefont {A.}~\bibnamefont {Bouali}}, \bibinfo {author} {\bibfnamefont {I.}~\bibnamefont {El~Bojaddaini}}, \bibinfo {author} {\bibfnamefont {A.}~\bibnamefont {Errahmani}}, \ and\ \bibinfo {author} {\bibfnamefont {T.}~\bibnamefont {Ouali}},\ }\href {\doibase 10.1016/j.dark.2023.101266} {\bibfield  {journal} {\bibinfo  {journal} {Phys. Dark Univ.}\ }\textbf {\bibinfo {volume} {42}},\ \bibinfo {pages} {101266} (\bibinfo {year} {2023})},\ \Eprint {http://arxiv.org/abs/2301.04200} {arXiv:2301.04200 [astro-ph.CO]} \BibitemShut {NoStop}%
\bibitem [{\citenamefont {Ben-Dayan}\ and\ \citenamefont {Kumar}(2023)}]{Ben-Dayan:2023rgt}%
  \BibitemOpen
  \bibfield  {author} {\bibinfo {author} {\bibfnamefont {I.}~\bibnamefont {Ben-Dayan}}\ and\ \bibinfo {author} {\bibfnamefont {U.}~\bibnamefont {Kumar}},\ }\href {\doibase 10.1088/1475-7516/2023/12/047} {\bibfield  {journal} {\bibinfo  {journal} {JCAP}\ }\textbf {\bibinfo {volume} {12}},\ \bibinfo {pages} {047} (\bibinfo {year} {2023})},\ \Eprint {http://arxiv.org/abs/2302.00067} {arXiv:2302.00067 [astro-ph.CO]} \BibitemShut {NoStop}%
\bibitem [{\citenamefont {de~Cruz~Perez}\ and\ \citenamefont {Sola~Peracaula}(2024)}]{deCruzPerez:2023wzd}%
  \BibitemOpen
  \bibfield  {author} {\bibinfo {author} {\bibfnamefont {J.}~\bibnamefont {de~Cruz~Perez}}\ and\ \bibinfo {author} {\bibfnamefont {J.}~\bibnamefont {Sola~Peracaula}},\ }\href {\doibase 10.1016/j.dark.2023.101406} {\bibfield  {journal} {\bibinfo  {journal} {Phys. Dark Univ.}\ }\textbf {\bibinfo {volume} {43}},\ \bibinfo {pages} {101406} (\bibinfo {year} {2024})},\ \Eprint {http://arxiv.org/abs/2302.04807} {arXiv:2302.04807 [astro-ph.CO]} \BibitemShut {NoStop}%
\bibitem [{\citenamefont {Ballardini}\ \emph {et~al.}(2023)\citenamefont {Ballardini}, \citenamefont {Ferrari},\ and\ \citenamefont {Finelli}}]{Ballardini:2023mzm}%
  \BibitemOpen
  \bibfield  {author} {\bibinfo {author} {\bibfnamefont {M.}~\bibnamefont {Ballardini}}, \bibinfo {author} {\bibfnamefont {A.~G.}\ \bibnamefont {Ferrari}}, \ and\ \bibinfo {author} {\bibfnamefont {F.}~\bibnamefont {Finelli}},\ }\href {\doibase 10.1088/1475-7516/2023/04/029} {\bibfield  {journal} {\bibinfo  {journal} {JCAP}\ }\textbf {\bibinfo {volume} {04}},\ \bibinfo {pages} {029} (\bibinfo {year} {2023})},\ \Eprint {http://arxiv.org/abs/2302.05291} {arXiv:2302.05291 [astro-ph.CO]} \BibitemShut {NoStop}%
\bibitem [{\citenamefont {Yao}\ \emph {et~al.}(2024{\natexlab{a}})\citenamefont {Yao}, \citenamefont {Wang},\ and\ \citenamefont {Meng}}]{Yao:2023ybs}%
  \BibitemOpen
  \bibfield  {author} {\bibinfo {author} {\bibfnamefont {Y.-H.}\ \bibnamefont {Yao}}, \bibinfo {author} {\bibfnamefont {J.-C.}\ \bibnamefont {Wang}}, \ and\ \bibinfo {author} {\bibfnamefont {X.-H.}\ \bibnamefont {Meng}},\ }\href {\doibase 10.1103/PhysRevD.109.063502} {\bibfield  {journal} {\bibinfo  {journal} {Phys. Rev. D}\ }\textbf {\bibinfo {volume} {109}},\ \bibinfo {pages} {063502} (\bibinfo {year} {2024}{\natexlab{a}})},\ \Eprint {http://arxiv.org/abs/2303.00961} {arXiv:2303.00961 [astro-ph.CO]} \BibitemShut {NoStop}%
\bibitem [{\citenamefont {Gangopadhyay}\ \emph {et~al.}(2023{\natexlab{b}})\citenamefont {Gangopadhyay}, \citenamefont {Sami},\ and\ \citenamefont {Sharma}}]{Gangopadhyay:2023nli}%
  \BibitemOpen
  \bibfield  {author} {\bibinfo {author} {\bibfnamefont {M.~R.}\ \bibnamefont {Gangopadhyay}}, \bibinfo {author} {\bibfnamefont {M.}~\bibnamefont {Sami}}, \ and\ \bibinfo {author} {\bibfnamefont {M.~K.}\ \bibnamefont {Sharma}},\ }\href {\doibase 10.1103/PhysRevD.108.103526} {\bibfield  {journal} {\bibinfo  {journal} {Phys. Rev. D}\ }\textbf {\bibinfo {volume} {108}},\ \bibinfo {pages} {103526} (\bibinfo {year} {2023}{\natexlab{b}})},\ \Eprint {http://arxiv.org/abs/2303.07301} {arXiv:2303.07301 [astro-ph.CO]} \BibitemShut {NoStop}%
\bibitem [{\citenamefont {Zhai}\ \emph {et~al.}(2023)\citenamefont {Zhai}, \citenamefont {Giar\`e}, \citenamefont {van~de Bruck}, \citenamefont {Di~Valentino}, \citenamefont {Mena},\ and\ \citenamefont {Nunes}}]{Zhai:2023yny}%
  \BibitemOpen
  \bibfield  {author} {\bibinfo {author} {\bibfnamefont {Y.}~\bibnamefont {Zhai}}, \bibinfo {author} {\bibfnamefont {W.}~\bibnamefont {Giar\`e}}, \bibinfo {author} {\bibfnamefont {C.}~\bibnamefont {van~de Bruck}}, \bibinfo {author} {\bibfnamefont {E.}~\bibnamefont {Di~Valentino}}, \bibinfo {author} {\bibfnamefont {O.}~\bibnamefont {Mena}}, \ and\ \bibinfo {author} {\bibfnamefont {R.~C.}\ \bibnamefont {Nunes}},\ }\href {\doibase 10.1088/1475-7516/2023/07/032} {\bibfield  {journal} {\bibinfo  {journal} {JCAP}\ }\textbf {\bibinfo {volume} {07}},\ \bibinfo {pages} {032} (\bibinfo {year} {2023})},\ \Eprint {http://arxiv.org/abs/2303.08201} {arXiv:2303.08201 [astro-ph.CO]} \BibitemShut {NoStop}%
\bibitem [{\citenamefont {Sola~Peracaula}\ \emph {et~al.}(2023)\citenamefont {Sola~Peracaula}, \citenamefont {Gomez-Valent}, \citenamefont {de~Cruz~Perez},\ and\ \citenamefont {Moreno-Pulido}}]{SolaPeracaula:2023swx}%
  \BibitemOpen
  \bibfield  {author} {\bibinfo {author} {\bibfnamefont {J.}~\bibnamefont {Sola~Peracaula}}, \bibinfo {author} {\bibfnamefont {A.}~\bibnamefont {Gomez-Valent}}, \bibinfo {author} {\bibfnamefont {J.}~\bibnamefont {de~Cruz~Perez}}, \ and\ \bibinfo {author} {\bibfnamefont {C.}~\bibnamefont {Moreno-Pulido}},\ }\href {\doibase 10.3390/universe9060262} {\bibfield  {journal} {\bibinfo  {journal} {Universe}\ }\textbf {\bibinfo {volume} {9}},\ \bibinfo {pages} {262} (\bibinfo {year} {2023})},\ \Eprint {http://arxiv.org/abs/2304.11157} {arXiv:2304.11157 [astro-ph.CO]} \BibitemShut {NoStop}%
\bibitem [{\citenamefont {G\'omez-Valent}\ \emph {et~al.}(2024)\citenamefont {G\'omez-Valent}, \citenamefont {Mavromatos},\ and\ \citenamefont {Sol\`a~Peracaula}}]{Gomez-Valent:2023hov}%
  \BibitemOpen
  \bibfield  {author} {\bibinfo {author} {\bibfnamefont {A.}~\bibnamefont {G\'omez-Valent}}, \bibinfo {author} {\bibfnamefont {N.~E.}\ \bibnamefont {Mavromatos}}, \ and\ \bibinfo {author} {\bibfnamefont {J.}~\bibnamefont {Sol\`a~Peracaula}},\ }\href {\doibase 10.1088/1361-6382/ad0fb8} {\bibfield  {journal} {\bibinfo  {journal} {Class. Quant. Grav.}\ }\textbf {\bibinfo {volume} {41}},\ \bibinfo {pages} {015026} (\bibinfo {year} {2024})},\ \Eprint {http://arxiv.org/abs/2305.15774} {arXiv:2305.15774 [gr-qc]} \BibitemShut {NoStop}%
\bibitem [{\citenamefont {Ruchika}\ \emph {et~al.}(2024)\citenamefont {Ruchika}, \citenamefont {Rathore}, \citenamefont {Roy~Choudhury},\ and\ \citenamefont {Rentala}}]{Ruchika:2023ugh}%
  \BibitemOpen
  \bibfield  {author} {\bibinfo {author} {\bibnamefont {Ruchika}}, \bibinfo {author} {\bibfnamefont {H.}~\bibnamefont {Rathore}}, \bibinfo {author} {\bibfnamefont {S.}~\bibnamefont {Roy~Choudhury}}, \ and\ \bibinfo {author} {\bibfnamefont {V.}~\bibnamefont {Rentala}},\ }\href {\doibase 10.1088/1475-7516/2024/06/056} {\bibfield  {journal} {\bibinfo  {journal} {JCAP}\ }\textbf {\bibinfo {volume} {06}},\ \bibinfo {pages} {056} (\bibinfo {year} {2024})},\ \Eprint {http://arxiv.org/abs/2306.05450} {arXiv:2306.05450 [astro-ph.CO]} \BibitemShut {NoStop}%
\bibitem [{\citenamefont {Vagnozzi}(2023)}]{Vagnozzi:2023nrq}%
  \BibitemOpen
  \bibfield  {author} {\bibinfo {author} {\bibfnamefont {S.}~\bibnamefont {Vagnozzi}},\ }\href {\doibase 10.3390/universe9090393} {\bibfield  {journal} {\bibinfo  {journal} {Universe}\ }\textbf {\bibinfo {volume} {9}},\ \bibinfo {pages} {393} (\bibinfo {year} {2023})},\ \Eprint {http://arxiv.org/abs/2308.16628} {arXiv:2308.16628 [astro-ph.CO]} \BibitemShut {NoStop}%
\bibitem [{\citenamefont {Adil}\ \emph {et~al.}(2024)\citenamefont {Adil}, \citenamefont {Akarsu}, \citenamefont {Di~Valentino}, \citenamefont {Nunes}, \citenamefont {\"Oz\"ulker}, \citenamefont {Sen},\ and\ \citenamefont {Specogna}}]{Adil:2023exv}%
  \BibitemOpen
  \bibfield  {author} {\bibinfo {author} {\bibfnamefont {S.~A.}\ \bibnamefont {Adil}}, \bibinfo {author} {\bibfnamefont {O.}~\bibnamefont {Akarsu}}, \bibinfo {author} {\bibfnamefont {E.}~\bibnamefont {Di~Valentino}}, \bibinfo {author} {\bibfnamefont {R.~C.}\ \bibnamefont {Nunes}}, \bibinfo {author} {\bibfnamefont {E.}~\bibnamefont {\"Oz\"ulker}}, \bibinfo {author} {\bibfnamefont {A.~A.}\ \bibnamefont {Sen}}, \ and\ \bibinfo {author} {\bibfnamefont {E.}~\bibnamefont {Specogna}},\ }\href {\doibase 10.1103/PhysRevD.109.023527} {\bibfield  {journal} {\bibinfo  {journal} {Phys. Rev. D}\ }\textbf {\bibinfo {volume} {109}},\ \bibinfo {pages} {023527} (\bibinfo {year} {2024})},\ \Eprint {http://arxiv.org/abs/2306.08046} {arXiv:2306.08046 [astro-ph.CO]} \BibitemShut {NoStop}%
\bibitem [{\citenamefont {Frion}\ \emph {et~al.}(2023)\citenamefont {Frion}, \citenamefont {Camarena}, \citenamefont {Giani}, \citenamefont {Miranda}, \citenamefont {Bertacca}, \citenamefont {Marra},\ and\ \citenamefont {Piattella}}]{Frion:2023xwq}%
  \BibitemOpen
  \bibfield  {author} {\bibinfo {author} {\bibfnamefont {E.}~\bibnamefont {Frion}}, \bibinfo {author} {\bibfnamefont {D.}~\bibnamefont {Camarena}}, \bibinfo {author} {\bibfnamefont {L.}~\bibnamefont {Giani}}, \bibinfo {author} {\bibfnamefont {T.}~\bibnamefont {Miranda}}, \bibinfo {author} {\bibfnamefont {D.}~\bibnamefont {Bertacca}}, \bibinfo {author} {\bibfnamefont {V.}~\bibnamefont {Marra}}, \ and\ \bibinfo {author} {\bibfnamefont {O.~F.}\ \bibnamefont {Piattella}},\ }\href {\doibase 10.21105/astro.2307.06320} {\  (\bibinfo {year} {2023}),\ 10.21105/astro.2307.06320},\ \Eprint {http://arxiv.org/abs/2307.06320} {arXiv:2307.06320 [astro-ph.CO]} \BibitemShut {NoStop}%
\bibitem [{\citenamefont {Akarsu}\ \emph {et~al.}(2023{\natexlab{b}})\citenamefont {Akarsu}, \citenamefont {Di~Valentino}, \citenamefont {Kumar}, \citenamefont {Nunes}, \citenamefont {Vazquez},\ and\ \citenamefont {Yadav}}]{Akarsu:2023mfb}%
  \BibitemOpen
  \bibfield  {author} {\bibinfo {author} {\bibfnamefont {O.}~\bibnamefont {Akarsu}}, \bibinfo {author} {\bibfnamefont {E.}~\bibnamefont {Di~Valentino}}, \bibinfo {author} {\bibfnamefont {S.}~\bibnamefont {Kumar}}, \bibinfo {author} {\bibfnamefont {R.~C.}\ \bibnamefont {Nunes}}, \bibinfo {author} {\bibfnamefont {J.~A.}\ \bibnamefont {Vazquez}}, \ and\ \bibinfo {author} {\bibfnamefont {A.}~\bibnamefont {Yadav}},\ }\href@noop {} {\  (\bibinfo {year} {2023}{\natexlab{b}})},\ \Eprint {http://arxiv.org/abs/2307.10899} {arXiv:2307.10899 [astro-ph.CO]} \BibitemShut {NoStop}%
\bibitem [{\citenamefont {Escamilla}\ \emph {et~al.}(2024{\natexlab{a}})\citenamefont {Escamilla}, \citenamefont {Giar\`e}, \citenamefont {Di~Valentino}, \citenamefont {Nunes},\ and\ \citenamefont {Vagnozzi}}]{Escamilla:2023oce}%
  \BibitemOpen
  \bibfield  {author} {\bibinfo {author} {\bibfnamefont {L.~A.}\ \bibnamefont {Escamilla}}, \bibinfo {author} {\bibfnamefont {W.}~\bibnamefont {Giar\`e}}, \bibinfo {author} {\bibfnamefont {E.}~\bibnamefont {Di~Valentino}}, \bibinfo {author} {\bibfnamefont {R.~C.}\ \bibnamefont {Nunes}}, \ and\ \bibinfo {author} {\bibfnamefont {S.}~\bibnamefont {Vagnozzi}},\ }\href {\doibase 10.1088/1475-7516/2024/05/091} {\bibfield  {journal} {\bibinfo  {journal} {JCAP}\ }\textbf {\bibinfo {volume} {05}},\ \bibinfo {pages} {091} (\bibinfo {year} {2024}{\natexlab{a}})},\ \Eprint {http://arxiv.org/abs/2307.14802} {arXiv:2307.14802 [astro-ph.CO]} \BibitemShut {NoStop}%
\bibitem [{\citenamefont {Petronikolou}\ and\ \citenamefont {Saridakis}(2023)}]{Petronikolou:2023cwu}%
  \BibitemOpen
  \bibfield  {author} {\bibinfo {author} {\bibfnamefont {M.}~\bibnamefont {Petronikolou}}\ and\ \bibinfo {author} {\bibfnamefont {E.~N.}\ \bibnamefont {Saridakis}},\ }\href {\doibase 10.3390/universe9090397} {\bibfield  {journal} {\bibinfo  {journal} {Universe}\ }\textbf {\bibinfo {volume} {9}},\ \bibinfo {pages} {397} (\bibinfo {year} {2023})},\ \Eprint {http://arxiv.org/abs/2308.16044} {arXiv:2308.16044 [gr-qc]} \BibitemShut {NoStop}%
\bibitem [{\citenamefont {Sharma}\ \emph {et~al.}(2024)\citenamefont {Sharma}, \citenamefont {Das},\ and\ \citenamefont {Poulin}}]{Sharma:2023kzr}%
  \BibitemOpen
  \bibfield  {author} {\bibinfo {author} {\bibfnamefont {R.~K.}\ \bibnamefont {Sharma}}, \bibinfo {author} {\bibfnamefont {S.}~\bibnamefont {Das}}, \ and\ \bibinfo {author} {\bibfnamefont {V.}~\bibnamefont {Poulin}},\ }\href {\doibase 10.1103/PhysRevD.109.043530} {\bibfield  {journal} {\bibinfo  {journal} {Phys. Rev. D}\ }\textbf {\bibinfo {volume} {109}},\ \bibinfo {pages} {043530} (\bibinfo {year} {2024})},\ \Eprint {http://arxiv.org/abs/2309.00401} {arXiv:2309.00401 [astro-ph.CO]} \BibitemShut {NoStop}%
\bibitem [{\citenamefont {Ben-Dayan}\ and\ \citenamefont {Kumar}(2024)}]{Ben-Dayan:2023htq}%
  \BibitemOpen
  \bibfield  {author} {\bibinfo {author} {\bibfnamefont {I.}~\bibnamefont {Ben-Dayan}}\ and\ \bibinfo {author} {\bibfnamefont {U.}~\bibnamefont {Kumar}},\ }\href {\doibase 10.1140/epjc/s10052-024-12488-0} {\bibfield  {journal} {\bibinfo  {journal} {Eur. Phys. J. C}\ }\textbf {\bibinfo {volume} {84}},\ \bibinfo {pages} {167} (\bibinfo {year} {2024})},\ \Eprint {http://arxiv.org/abs/2310.03092} {arXiv:2310.03092 [astro-ph.CO]} \BibitemShut {NoStop}%
\bibitem [{\citenamefont {Ramadan}\ \emph {et~al.}(2024{\natexlab{a}})\citenamefont {Ramadan}, \citenamefont {Karwal},\ and\ \citenamefont {Sakstein}}]{Ramadan:2023ivw}%
  \BibitemOpen
  \bibfield  {author} {\bibinfo {author} {\bibfnamefont {O.~F.}\ \bibnamefont {Ramadan}}, \bibinfo {author} {\bibfnamefont {T.}~\bibnamefont {Karwal}}, \ and\ \bibinfo {author} {\bibfnamefont {J.}~\bibnamefont {Sakstein}},\ }\href {\doibase 10.1103/PhysRevD.109.063525} {\bibfield  {journal} {\bibinfo  {journal} {Phys. Rev. D}\ }\textbf {\bibinfo {volume} {109}},\ \bibinfo {pages} {063525} (\bibinfo {year} {2024}{\natexlab{a}})},\ \Eprint {http://arxiv.org/abs/2309.08082} {arXiv:2309.08082 [astro-ph.CO]} \BibitemShut {NoStop}%
\bibitem [{\citenamefont {Fu}\ and\ \citenamefont {Wang}(2024)}]{Fu:2023tfo}%
  \BibitemOpen
  \bibfield  {author} {\bibinfo {author} {\bibfnamefont {C.}~\bibnamefont {Fu}}\ and\ \bibinfo {author} {\bibfnamefont {S.-J.}\ \bibnamefont {Wang}},\ }\href {\doibase 10.1103/PhysRevD.109.L041304} {\bibfield  {journal} {\bibinfo  {journal} {Phys. Rev. D}\ }\textbf {\bibinfo {volume} {109}},\ \bibinfo {pages} {L041304} (\bibinfo {year} {2024})},\ \Eprint {http://arxiv.org/abs/2310.12932} {arXiv:2310.12932 [astro-ph.CO]} \BibitemShut {NoStop}%
\bibitem [{\citenamefont {Efstathiou}\ \emph {et~al.}(2024)\citenamefont {Efstathiou}, \citenamefont {Rosenberg},\ and\ \citenamefont {Poulin}}]{Efstathiou:2023fbn}%
  \BibitemOpen
  \bibfield  {author} {\bibinfo {author} {\bibfnamefont {G.}~\bibnamefont {Efstathiou}}, \bibinfo {author} {\bibfnamefont {E.}~\bibnamefont {Rosenberg}}, \ and\ \bibinfo {author} {\bibfnamefont {V.}~\bibnamefont {Poulin}},\ }\href {\doibase 10.1103/PhysRevLett.132.221002} {\bibfield  {journal} {\bibinfo  {journal} {Phys. Rev. Lett.}\ }\textbf {\bibinfo {volume} {132}},\ \bibinfo {pages} {221002} (\bibinfo {year} {2024})},\ \Eprint {http://arxiv.org/abs/2311.00524} {arXiv:2311.00524 [astro-ph.CO]} \BibitemShut {NoStop}%
\bibitem [{\citenamefont {Montani}\ \emph {et~al.}(2024{\natexlab{a}})\citenamefont {Montani}, \citenamefont {Carlevaro},\ and\ \citenamefont {Dainotti}}]{Montani:2023ywn}%
  \BibitemOpen
  \bibfield  {author} {\bibinfo {author} {\bibfnamefont {G.}~\bibnamefont {Montani}}, \bibinfo {author} {\bibfnamefont {N.}~\bibnamefont {Carlevaro}}, \ and\ \bibinfo {author} {\bibfnamefont {M.~G.}\ \bibnamefont {Dainotti}},\ }\href {\doibase 10.1016/j.dark.2024.101486} {\bibfield  {journal} {\bibinfo  {journal} {Phys. Dark Univ.}\ }\textbf {\bibinfo {volume} {44}},\ \bibinfo {pages} {101486} (\bibinfo {year} {2024}{\natexlab{a}})},\ \Eprint {http://arxiv.org/abs/2311.04822} {arXiv:2311.04822 [gr-qc]} \BibitemShut {NoStop}%
\bibitem [{\citenamefont {Lazkoz}\ \emph {et~al.}(2024)\citenamefont {Lazkoz}, \citenamefont {Salzano}, \citenamefont {Fernandez-Jambrina},\ and\ \citenamefont {Bouhmadi-L\'opez}}]{Lazkoz:2023oqc}%
  \BibitemOpen
  \bibfield  {author} {\bibinfo {author} {\bibfnamefont {R.}~\bibnamefont {Lazkoz}}, \bibinfo {author} {\bibfnamefont {V.}~\bibnamefont {Salzano}}, \bibinfo {author} {\bibfnamefont {L.}~\bibnamefont {Fernandez-Jambrina}}, \ and\ \bibinfo {author} {\bibfnamefont {M.}~\bibnamefont {Bouhmadi-L\'opez}},\ }\href {\doibase 10.1016/j.dark.2024.101511} {\bibfield  {journal} {\bibinfo  {journal} {Phys. Dark Univ.}\ }\textbf {\bibinfo {volume} {45}},\ \bibinfo {pages} {101511} (\bibinfo {year} {2024})},\ \Eprint {http://arxiv.org/abs/2311.10526} {arXiv:2311.10526 [astro-ph.CO]} \BibitemShut {NoStop}%
\bibitem [{\citenamefont {Forconi}\ \emph {et~al.}(2024)\citenamefont {Forconi}, \citenamefont {Giar\`e}, \citenamefont {Mena}, \citenamefont {Ruchika}, \citenamefont {Di~Valentino}, \citenamefont {Melchiorri},\ and\ \citenamefont {Nunes}}]{Forconi:2023hsj}%
  \BibitemOpen
  \bibfield  {author} {\bibinfo {author} {\bibfnamefont {M.}~\bibnamefont {Forconi}}, \bibinfo {author} {\bibfnamefont {W.}~\bibnamefont {Giar\`e}}, \bibinfo {author} {\bibfnamefont {O.}~\bibnamefont {Mena}}, \bibinfo {author} {\bibnamefont {Ruchika}}, \bibinfo {author} {\bibfnamefont {E.}~\bibnamefont {Di~Valentino}}, \bibinfo {author} {\bibfnamefont {A.}~\bibnamefont {Melchiorri}}, \ and\ \bibinfo {author} {\bibfnamefont {R.~C.}\ \bibnamefont {Nunes}},\ }\href {\doibase 10.1088/1475-7516/2024/05/097} {\bibfield  {journal} {\bibinfo  {journal} {JCAP}\ }\textbf {\bibinfo {volume} {05}},\ \bibinfo {pages} {097} (\bibinfo {year} {2024})},\ \Eprint {http://arxiv.org/abs/2312.11074} {arXiv:2312.11074 [astro-ph.CO]} \BibitemShut {NoStop}%
\bibitem [{\citenamefont {Sebastianutti}\ \emph {et~al.}(2024)\citenamefont {Sebastianutti}, \citenamefont {Hogg},\ and\ \citenamefont {Bruni}}]{Sebastianutti:2023dbt}%
  \BibitemOpen
  \bibfield  {author} {\bibinfo {author} {\bibfnamefont {M.}~\bibnamefont {Sebastianutti}}, \bibinfo {author} {\bibfnamefont {N.~B.}\ \bibnamefont {Hogg}}, \ and\ \bibinfo {author} {\bibfnamefont {M.}~\bibnamefont {Bruni}},\ }\href {\doibase 10.1016/j.dark.2024.101546} {\bibfield  {journal} {\bibinfo  {journal} {Phys. Dark Univ.}\ }\textbf {\bibinfo {volume} {46}},\ \bibinfo {pages} {101546} (\bibinfo {year} {2024})},\ \Eprint {http://arxiv.org/abs/2312.14123} {arXiv:2312.14123 [astro-ph.CO]} \BibitemShut {NoStop}%
\bibitem [{\citenamefont {Benisty}\ \emph {et~al.}(2024)\citenamefont {Benisty}, \citenamefont {Pan}, \citenamefont {Staicova}, \citenamefont {Di~Valentino},\ and\ \citenamefont {Nunes}}]{Benisty:2024lmj}%
  \BibitemOpen
  \bibfield  {author} {\bibinfo {author} {\bibfnamefont {D.}~\bibnamefont {Benisty}}, \bibinfo {author} {\bibfnamefont {S.}~\bibnamefont {Pan}}, \bibinfo {author} {\bibfnamefont {D.}~\bibnamefont {Staicova}}, \bibinfo {author} {\bibfnamefont {E.}~\bibnamefont {Di~Valentino}}, \ and\ \bibinfo {author} {\bibfnamefont {R.~C.}\ \bibnamefont {Nunes}},\ }\href {\doibase 10.1051/0004-6361/202449883} {\bibfield  {journal} {\bibinfo  {journal} {Astron. Astrophys.}\ }\textbf {\bibinfo {volume} {688}},\ \bibinfo {pages} {A156} (\bibinfo {year} {2024})},\ \Eprint {http://arxiv.org/abs/2403.00056} {arXiv:2403.00056 [astro-ph.CO]} \BibitemShut {NoStop}%
\bibitem [{\citenamefont {Stahl}\ \emph {et~al.}(2024)\citenamefont {Stahl}, \citenamefont {Famaey}, \citenamefont {Ibata}, \citenamefont {Hahn}, \citenamefont {Martinet},\ and\ \citenamefont {Montandon}}]{Stahl:2024stz}%
  \BibitemOpen
  \bibfield  {author} {\bibinfo {author} {\bibfnamefont {C.}~\bibnamefont {Stahl}}, \bibinfo {author} {\bibfnamefont {B.}~\bibnamefont {Famaey}}, \bibinfo {author} {\bibfnamefont {R.}~\bibnamefont {Ibata}}, \bibinfo {author} {\bibfnamefont {O.}~\bibnamefont {Hahn}}, \bibinfo {author} {\bibfnamefont {N.}~\bibnamefont {Martinet}}, \ and\ \bibinfo {author} {\bibfnamefont {T.}~\bibnamefont {Montandon}},\ }\href {\doibase 10.1103/PhysRevD.110.063501} {\bibfield  {journal} {\bibinfo  {journal} {Phys. Rev. D}\ }\textbf {\bibinfo {volume} {110}},\ \bibinfo {pages} {063501} (\bibinfo {year} {2024})},\ \Eprint {http://arxiv.org/abs/2404.03244} {arXiv:2404.03244 [astro-ph.CO]} \BibitemShut {NoStop}%
\bibitem [{\citenamefont {Shah}\ \emph {et~al.}(2024)\citenamefont {Shah}, \citenamefont {Mukherjee},\ and\ \citenamefont {Pal}}]{Shah:2024rme}%
  \BibitemOpen
  \bibfield  {author} {\bibinfo {author} {\bibfnamefont {R.}~\bibnamefont {Shah}}, \bibinfo {author} {\bibfnamefont {P.}~\bibnamefont {Mukherjee}}, \ and\ \bibinfo {author} {\bibfnamefont {S.}~\bibnamefont {Pal}},\ }\href@noop {} {\  (\bibinfo {year} {2024})},\ \Eprint {http://arxiv.org/abs/2404.06396} {arXiv:2404.06396 [astro-ph.CO]} \BibitemShut {NoStop}%
\bibitem [{\citenamefont {Giar\`e}\ \emph {et~al.}(2024{\natexlab{a}})\citenamefont {Giar\`e}, \citenamefont {Sabogal}, \citenamefont {Nunes},\ and\ \citenamefont {Di~Valentino}}]{Giare:2024smz}%
  \BibitemOpen
  \bibfield  {author} {\bibinfo {author} {\bibfnamefont {W.}~\bibnamefont {Giar\`e}}, \bibinfo {author} {\bibfnamefont {M.~A.}\ \bibnamefont {Sabogal}}, \bibinfo {author} {\bibfnamefont {R.~C.}\ \bibnamefont {Nunes}}, \ and\ \bibinfo {author} {\bibfnamefont {E.}~\bibnamefont {Di~Valentino}},\ }\href@noop {} {\  (\bibinfo {year} {2024}{\natexlab{a}})},\ \Eprint {http://arxiv.org/abs/2404.15232} {arXiv:2404.15232 [astro-ph.CO]} \BibitemShut {NoStop}%
\bibitem [{\citenamefont {Giar\`e}(2024{\natexlab{a}})}]{Giare:2024akf}%
  \BibitemOpen
  \bibfield  {author} {\bibinfo {author} {\bibfnamefont {W.}~\bibnamefont {Giar\`e}},\ }\href {\doibase 10.1103/PhysRevD.109.123545} {\bibfield  {journal} {\bibinfo  {journal} {Phys. Rev. D}\ }\textbf {\bibinfo {volume} {109}},\ \bibinfo {pages} {123545} (\bibinfo {year} {2024}{\natexlab{a}})},\ \Eprint {http://arxiv.org/abs/2404.12779} {arXiv:2404.12779 [astro-ph.CO]} \BibitemShut {NoStop}%
\bibitem [{\citenamefont {Giar\`e}\ \emph {et~al.}(2024{\natexlab{b}})\citenamefont {Giar\`e}, \citenamefont {Zhai}, \citenamefont {Pan}, \citenamefont {Di~Valentino}, \citenamefont {Nunes},\ and\ \citenamefont {van~de Bruck}}]{Giare:2024ytc}%
  \BibitemOpen
  \bibfield  {author} {\bibinfo {author} {\bibfnamefont {W.}~\bibnamefont {Giar\`e}}, \bibinfo {author} {\bibfnamefont {Y.}~\bibnamefont {Zhai}}, \bibinfo {author} {\bibfnamefont {S.}~\bibnamefont {Pan}}, \bibinfo {author} {\bibfnamefont {E.}~\bibnamefont {Di~Valentino}}, \bibinfo {author} {\bibfnamefont {R.~C.}\ \bibnamefont {Nunes}}, \ and\ \bibinfo {author} {\bibfnamefont {C.}~\bibnamefont {van~de Bruck}},\ }\href {\doibase 10.1103/PhysRevD.110.063527} {\bibfield  {journal} {\bibinfo  {journal} {Phys. Rev. D}\ }\textbf {\bibinfo {volume} {110}},\ \bibinfo {pages} {063527} (\bibinfo {year} {2024}{\natexlab{b}})},\ \Eprint {http://arxiv.org/abs/2404.02110} {arXiv:2404.02110 [astro-ph.CO]} \BibitemShut {NoStop}%
\bibitem [{\citenamefont {Montani}\ \emph {et~al.}(2024{\natexlab{b}})\citenamefont {Montani}, \citenamefont {Carlevaro}, \citenamefont {Escamilla},\ and\ \citenamefont {Di~Valentino}}]{Montani:2024pou}%
  \BibitemOpen
  \bibfield  {author} {\bibinfo {author} {\bibfnamefont {G.}~\bibnamefont {Montani}}, \bibinfo {author} {\bibfnamefont {N.}~\bibnamefont {Carlevaro}}, \bibinfo {author} {\bibfnamefont {L.~A.}\ \bibnamefont {Escamilla}}, \ and\ \bibinfo {author} {\bibfnamefont {E.}~\bibnamefont {Di~Valentino}},\ }\href@noop {} {\  (\bibinfo {year} {2024}{\natexlab{b}})},\ \Eprint {http://arxiv.org/abs/2404.15977} {arXiv:2404.15977 [gr-qc]} \BibitemShut {NoStop}%
\bibitem [{\citenamefont {Co}\ \emph {et~al.}(2024)\citenamefont {Co}, \citenamefont {Fernandez}, \citenamefont {Ghalsasi}, \citenamefont {Harigaya},\ and\ \citenamefont {Shelton}}]{Co:2024oek}%
  \BibitemOpen
  \bibfield  {author} {\bibinfo {author} {\bibfnamefont {R.~T.}\ \bibnamefont {Co}}, \bibinfo {author} {\bibfnamefont {N.}~\bibnamefont {Fernandez}}, \bibinfo {author} {\bibfnamefont {A.}~\bibnamefont {Ghalsasi}}, \bibinfo {author} {\bibfnamefont {K.}~\bibnamefont {Harigaya}}, \ and\ \bibinfo {author} {\bibfnamefont {J.}~\bibnamefont {Shelton}},\ }\href {\doibase 10.1103/PhysRevD.110.083534} {\bibfield  {journal} {\bibinfo  {journal} {Phys. Rev. D}\ }\textbf {\bibinfo {volume} {110}},\ \bibinfo {pages} {083534} (\bibinfo {year} {2024})},\ \Eprint {http://arxiv.org/abs/2405.12268} {arXiv:2405.12268 [hep-ph]} \BibitemShut {NoStop}%
\bibitem [{\citenamefont {Akarsu}\ \emph {et~al.}(2024{\natexlab{a}})\citenamefont {Akarsu}, \citenamefont {De~Felice}, \citenamefont {Di~Valentino}, \citenamefont {Kumar}, \citenamefont {Nunes}, \citenamefont {Ozulker}, \citenamefont {Vazquez},\ and\ \citenamefont {Yadav}}]{Akarsu:2024eoo}%
  \BibitemOpen
  \bibfield  {author} {\bibinfo {author} {\bibfnamefont {O.}~\bibnamefont {Akarsu}}, \bibinfo {author} {\bibfnamefont {A.}~\bibnamefont {De~Felice}}, \bibinfo {author} {\bibfnamefont {E.}~\bibnamefont {Di~Valentino}}, \bibinfo {author} {\bibfnamefont {S.}~\bibnamefont {Kumar}}, \bibinfo {author} {\bibfnamefont {R.~C.}\ \bibnamefont {Nunes}}, \bibinfo {author} {\bibfnamefont {E.}~\bibnamefont {Ozulker}}, \bibinfo {author} {\bibfnamefont {J.~A.}\ \bibnamefont {Vazquez}}, \ and\ \bibinfo {author} {\bibfnamefont {A.}~\bibnamefont {Yadav}},\ }\href@noop {} {\  (\bibinfo {year} {2024}{\natexlab{a}})},\ \Eprint {http://arxiv.org/abs/2406.07526} {arXiv:2406.07526 [astro-ph.CO]} \BibitemShut {NoStop}%
\bibitem [{\citenamefont {Yadav}\ \emph {et~al.}(2024)\citenamefont {Yadav}, \citenamefont {Kumar}, \citenamefont {Kibris},\ and\ \citenamefont {Akarsu}}]{Yadav:2024duq}%
  \BibitemOpen
  \bibfield  {author} {\bibinfo {author} {\bibfnamefont {A.}~\bibnamefont {Yadav}}, \bibinfo {author} {\bibfnamefont {S.}~\bibnamefont {Kumar}}, \bibinfo {author} {\bibfnamefont {C.}~\bibnamefont {Kibris}}, \ and\ \bibinfo {author} {\bibfnamefont {O.}~\bibnamefont {Akarsu}},\ }\href@noop {} {\  (\bibinfo {year} {2024})},\ \Eprint {http://arxiv.org/abs/2406.18496} {arXiv:2406.18496 [astro-ph.CO]} \BibitemShut {NoStop}%
\bibitem [{\citenamefont {Nozari}\ \emph {et~al.}(2024)\citenamefont {Nozari}, \citenamefont {Saghafi},\ and\ \citenamefont {Hajebrahimi}}]{Nozari:2024wir}%
  \BibitemOpen
  \bibfield  {author} {\bibinfo {author} {\bibfnamefont {K.}~\bibnamefont {Nozari}}, \bibinfo {author} {\bibfnamefont {S.}~\bibnamefont {Saghafi}}, \ and\ \bibinfo {author} {\bibfnamefont {M.}~\bibnamefont {Hajebrahimi}},\ }\href {\doibase 10.1016/j.dark.2024.101571} {\bibfield  {journal} {\bibinfo  {journal} {Phys. Dark Univ.}\ }\textbf {\bibinfo {volume} {46}},\ \bibinfo {pages} {101571} (\bibinfo {year} {2024})},\ \Eprint {http://arxiv.org/abs/2407.01961} {arXiv:2407.01961 [gr-qc]} \BibitemShut {NoStop}%
\bibitem [{\citenamefont {Dwivedi}\ and\ \citenamefont {H\"og\r{a}s}(2024)}]{Dwivedi:2024okk}%
  \BibitemOpen
  \bibfield  {author} {\bibinfo {author} {\bibfnamefont {S.}~\bibnamefont {Dwivedi}}\ and\ \bibinfo {author} {\bibfnamefont {M.}~\bibnamefont {H\"og\r{a}s}},\ }\href@noop {} {\  (\bibinfo {year} {2024})},\ \Eprint {http://arxiv.org/abs/2407.04322} {arXiv:2407.04322 [astro-ph.CO]} \BibitemShut {NoStop}%
\bibitem [{\citenamefont {Montani}\ \emph {et~al.}(2024{\natexlab{c}})\citenamefont {Montani}, \citenamefont {Carlevaro},\ and\ \citenamefont {De~Angelis}}]{Montani:2024xys}%
  \BibitemOpen
  \bibfield  {author} {\bibinfo {author} {\bibfnamefont {G.}~\bibnamefont {Montani}}, \bibinfo {author} {\bibfnamefont {N.}~\bibnamefont {Carlevaro}}, \ and\ \bibinfo {author} {\bibfnamefont {M.}~\bibnamefont {De~Angelis}},\ }\href {\doibase 10.3390/e26080662} {\bibfield  {journal} {\bibinfo  {journal} {Entropy}\ }\textbf {\bibinfo {volume} {26}},\ \bibinfo {pages} {662} (\bibinfo {year} {2024}{\natexlab{c}})},\ \Eprint {http://arxiv.org/abs/2407.12409} {arXiv:2407.12409 [gr-qc]} \BibitemShut {NoStop}%
\bibitem [{\citenamefont {Escamilla}\ \emph {et~al.}(2024{\natexlab{b}})\citenamefont {Escamilla}, \citenamefont {Fiorucci}, \citenamefont {Montani},\ and\ \citenamefont {Di~Valentino}}]{Escamilla:2024xmz}%
  \BibitemOpen
  \bibfield  {author} {\bibinfo {author} {\bibfnamefont {L.~A.}\ \bibnamefont {Escamilla}}, \bibinfo {author} {\bibfnamefont {D.}~\bibnamefont {Fiorucci}}, \bibinfo {author} {\bibfnamefont {G.}~\bibnamefont {Montani}}, \ and\ \bibinfo {author} {\bibfnamefont {E.}~\bibnamefont {Di~Valentino}},\ }\href {\doibase 10.1016/j.dark.2024.101652} {\bibfield  {journal} {\bibinfo  {journal} {Phys. Dark Univ.}\ }\textbf {\bibinfo {volume} {46}},\ \bibinfo {pages} {101652} (\bibinfo {year} {2024}{\natexlab{b}})},\ \Eprint {http://arxiv.org/abs/2408.04354} {arXiv:2408.04354 [astro-ph.CO]} \BibitemShut {NoStop}%
\bibitem [{\citenamefont {Yao}\ \emph {et~al.}(2024{\natexlab{b}})\citenamefont {Yao}, \citenamefont {Liu}, \citenamefont {Huang}, \citenamefont {Wang},\ and\ \citenamefont {Su}}]{Yao:2024kex}%
  \BibitemOpen
  \bibfield  {author} {\bibinfo {author} {\bibfnamefont {Y.-H.}\ \bibnamefont {Yao}}, \bibinfo {author} {\bibfnamefont {J.-Q.}\ \bibnamefont {Liu}}, \bibinfo {author} {\bibfnamefont {Z.-Q.}\ \bibnamefont {Huang}}, \bibinfo {author} {\bibfnamefont {J.-C.}\ \bibnamefont {Wang}}, \ and\ \bibinfo {author} {\bibfnamefont {Y.}~\bibnamefont {Su}},\ }\href@noop {} {\  (\bibinfo {year} {2024}{\natexlab{b}})},\ \Eprint {http://arxiv.org/abs/2409.04678} {arXiv:2409.04678 [astro-ph.CO]} \BibitemShut {NoStop}%
\bibitem [{\citenamefont {Giar\`e}\ \emph {et~al.}(2024{\natexlab{c}})\citenamefont {Giar\`e}, \citenamefont {Betts}, \citenamefont {van~de Bruck},\ and\ \citenamefont {Di~Valentino}}]{Giare:2024syw}%
  \BibitemOpen
  \bibfield  {author} {\bibinfo {author} {\bibfnamefont {W.}~\bibnamefont {Giar\`e}}, \bibinfo {author} {\bibfnamefont {J.}~\bibnamefont {Betts}}, \bibinfo {author} {\bibfnamefont {C.}~\bibnamefont {van~de Bruck}}, \ and\ \bibinfo {author} {\bibfnamefont {E.}~\bibnamefont {Di~Valentino}},\ }\href@noop {} {\  (\bibinfo {year} {2024}{\natexlab{c}})},\ \Eprint {http://arxiv.org/abs/2406.07493} {arXiv:2406.07493 [astro-ph.CO]} \BibitemShut {NoStop}%
\bibitem [{\citenamefont {Toda}\ \emph {et~al.}(2024)\citenamefont {Toda}, \citenamefont {Giar\`e}, \citenamefont {\"Oz\"ulker}, \citenamefont {Di~Valentino},\ and\ \citenamefont {Vagnozzi}}]{Toda:2024ncp}%
  \BibitemOpen
  \bibfield  {author} {\bibinfo {author} {\bibfnamefont {Y.}~\bibnamefont {Toda}}, \bibinfo {author} {\bibfnamefont {W.}~\bibnamefont {Giar\`e}}, \bibinfo {author} {\bibfnamefont {E.}~\bibnamefont {\"Oz\"ulker}}, \bibinfo {author} {\bibfnamefont {E.}~\bibnamefont {Di~Valentino}}, \ and\ \bibinfo {author} {\bibfnamefont {S.}~\bibnamefont {Vagnozzi}},\ }\href {\doibase 10.1016/j.dark.2024.101676} {\bibfield  {journal} {\bibinfo  {journal} {Phys. Dark Univ.}\ }\textbf {\bibinfo {volume} {46}},\ \bibinfo {pages} {101676} (\bibinfo {year} {2024})},\ \Eprint {http://arxiv.org/abs/2407.01173} {arXiv:2407.01173 [astro-ph.CO]} \BibitemShut {NoStop}%
\bibitem [{\citenamefont {Pedrotti}\ \emph {et~al.}(2024)\citenamefont {Pedrotti}, \citenamefont {Jiang}, \citenamefont {Escamilla}, \citenamefont {da~Costa},\ and\ \citenamefont {Vagnozzi}}]{Pedrotti:2024kpn}%
  \BibitemOpen
  \bibfield  {author} {\bibinfo {author} {\bibfnamefont {D.}~\bibnamefont {Pedrotti}}, \bibinfo {author} {\bibfnamefont {J.-Q.}\ \bibnamefont {Jiang}}, \bibinfo {author} {\bibfnamefont {L.~A.}\ \bibnamefont {Escamilla}}, \bibinfo {author} {\bibfnamefont {S.~S.}\ \bibnamefont {da~Costa}}, \ and\ \bibinfo {author} {\bibfnamefont {S.}~\bibnamefont {Vagnozzi}},\ }\href@noop {} {\  (\bibinfo {year} {2024})},\ \Eprint {http://arxiv.org/abs/2408.04530} {arXiv:2408.04530 [astro-ph.CO]} \BibitemShut {NoStop}%
\bibitem [{\citenamefont {Poulin}\ \emph {et~al.}(2024)\citenamefont {Poulin}, \citenamefont {Smith}, \citenamefont {Calder\'on},\ and\ \citenamefont {Simon}}]{Poulin:2024ken}%
  \BibitemOpen
  \bibfield  {author} {\bibinfo {author} {\bibfnamefont {V.}~\bibnamefont {Poulin}}, \bibinfo {author} {\bibfnamefont {T.~L.}\ \bibnamefont {Smith}}, \bibinfo {author} {\bibfnamefont {R.}~\bibnamefont {Calder\'on}}, \ and\ \bibinfo {author} {\bibfnamefont {T.}~\bibnamefont {Simon}},\ }\href@noop {} {\  (\bibinfo {year} {2024})},\ \Eprint {http://arxiv.org/abs/2407.18292} {arXiv:2407.18292 [astro-ph.CO]} \BibitemShut {NoStop}%
\bibitem [{\citenamefont {Simon}\ \emph {et~al.}(2024)\citenamefont {Simon}, \citenamefont {Adi}, \citenamefont {Bernal}, \citenamefont {Kovetz}, \citenamefont {Poulin},\ and\ \citenamefont {Smith}}]{Simon:2024jmu}%
  \BibitemOpen
  \bibfield  {author} {\bibinfo {author} {\bibfnamefont {T.}~\bibnamefont {Simon}}, \bibinfo {author} {\bibfnamefont {T.}~\bibnamefont {Adi}}, \bibinfo {author} {\bibfnamefont {J.~L.}\ \bibnamefont {Bernal}}, \bibinfo {author} {\bibfnamefont {E.~D.}\ \bibnamefont {Kovetz}}, \bibinfo {author} {\bibfnamefont {V.}~\bibnamefont {Poulin}}, \ and\ \bibinfo {author} {\bibfnamefont {T.~L.}\ \bibnamefont {Smith}},\ }\href@noop {} {\  (\bibinfo {year} {2024})},\ \Eprint {http://arxiv.org/abs/2410.21459} {arXiv:2410.21459 [astro-ph.CO]} \BibitemShut {NoStop}%
\bibitem [{\citenamefont {Di~Valentino}\ \emph {et~al.}(2021{\natexlab{c}})\citenamefont {Di~Valentino} \emph {et~al.}}]{DiValentino:2020vvd}%
  \BibitemOpen
  \bibfield  {author} {\bibinfo {author} {\bibfnamefont {E.}~\bibnamefont {Di~Valentino}} \emph {et~al.},\ }\href {\doibase 10.1016/j.astropartphys.2021.102604} {\bibfield  {journal} {\bibinfo  {journal} {Astropart. Phys.}\ }\textbf {\bibinfo {volume} {131}},\ \bibinfo {pages} {102604} (\bibinfo {year} {2021}{\natexlab{c}})},\ \Eprint {http://arxiv.org/abs/2008.11285} {arXiv:2008.11285 [astro-ph.CO]} \BibitemShut {NoStop}%
\bibitem [{\citenamefont {Nunes}\ and\ \citenamefont {Vagnozzi}(2021)}]{Nunes:2021ipq}%
  \BibitemOpen
  \bibfield  {author} {\bibinfo {author} {\bibfnamefont {R.~C.}\ \bibnamefont {Nunes}}\ and\ \bibinfo {author} {\bibfnamefont {S.}~\bibnamefont {Vagnozzi}},\ }\href {\doibase 10.1093/mnras/stab1613} {\bibfield  {journal} {\bibinfo  {journal} {Mon. Not. Roy. Astron. Soc.}\ }\textbf {\bibinfo {volume} {505}},\ \bibinfo {pages} {5427} (\bibinfo {year} {2021})},\ \Eprint {http://arxiv.org/abs/2106.01208} {arXiv:2106.01208 [astro-ph.CO]} \BibitemShut {NoStop}%
\bibitem [{\citenamefont {Amon}\ \emph {et~al.}(2022)\citenamefont {Amon} \emph {et~al.}}]{DES:2021bvc}%
  \BibitemOpen
  \bibfield  {author} {\bibinfo {author} {\bibfnamefont {A.}~\bibnamefont {Amon}} \emph {et~al.} (\bibinfo {collaboration} {DES}),\ }\href {\doibase 10.1103/PhysRevD.105.023514} {\bibfield  {journal} {\bibinfo  {journal} {Phys. Rev. D}\ }\textbf {\bibinfo {volume} {105}},\ \bibinfo {pages} {023514} (\bibinfo {year} {2022})},\ \Eprint {http://arxiv.org/abs/2105.13543} {arXiv:2105.13543 [astro-ph.CO]} \BibitemShut {NoStop}%
\bibitem [{\citenamefont {Li}\ \emph {et~al.}(2023)\citenamefont {Li} \emph {et~al.}}]{Li:2023tui}%
  \BibitemOpen
  \bibfield  {author} {\bibinfo {author} {\bibfnamefont {X.}~\bibnamefont {Li}} \emph {et~al.},\ }\href {\doibase 10.1103/PhysRevD.108.123518} {\bibfield  {journal} {\bibinfo  {journal} {Phys. Rev. D}\ }\textbf {\bibinfo {volume} {108}},\ \bibinfo {pages} {123518} (\bibinfo {year} {2023})},\ \Eprint {http://arxiv.org/abs/2304.00702} {arXiv:2304.00702 [astro-ph.CO]} \BibitemShut {NoStop}%
\bibitem [{\citenamefont {Dalal}\ \emph {et~al.}(2023)\citenamefont {Dalal} \emph {et~al.}}]{Dalal:2023olq}%
  \BibitemOpen
  \bibfield  {author} {\bibinfo {author} {\bibfnamefont {R.}~\bibnamefont {Dalal}} \emph {et~al.},\ }\href {\doibase 10.1103/PhysRevD.108.123519} {\bibfield  {journal} {\bibinfo  {journal} {Phys. Rev. D}\ }\textbf {\bibinfo {volume} {108}},\ \bibinfo {pages} {123519} (\bibinfo {year} {2023})},\ \Eprint {http://arxiv.org/abs/2304.00701} {arXiv:2304.00701 [astro-ph.CO]} \BibitemShut {NoStop}%
\bibitem [{\citenamefont {Harnois-Deraps}\ \emph {et~al.}(2024)\citenamefont {Harnois-Deraps} \emph {et~al.}}]{Harnois-Deraps:2024ucb}%
  \BibitemOpen
  \bibfield  {author} {\bibinfo {author} {\bibfnamefont {J.}~\bibnamefont {Harnois-Deraps}} \emph {et~al.},\ }\href@noop {} {\  (\bibinfo {year} {2024})},\ \Eprint {http://arxiv.org/abs/2405.10312} {arXiv:2405.10312 [astro-ph.CO]} \BibitemShut {NoStop}%
\bibitem [{\citenamefont {Armijo}\ \emph {et~al.}(2024)\citenamefont {Armijo}, \citenamefont {Marques}, \citenamefont {Novaes}, \citenamefont {Thiele}, \citenamefont {Cowell}, \citenamefont {Grand\'on}, \citenamefont {Shirasaki},\ and\ \citenamefont {Liu}}]{Armijo:2024ujo}%
  \BibitemOpen
  \bibfield  {author} {\bibinfo {author} {\bibfnamefont {J.}~\bibnamefont {Armijo}}, \bibinfo {author} {\bibfnamefont {G.~A.}\ \bibnamefont {Marques}}, \bibinfo {author} {\bibfnamefont {C.~P.}\ \bibnamefont {Novaes}}, \bibinfo {author} {\bibfnamefont {L.}~\bibnamefont {Thiele}}, \bibinfo {author} {\bibfnamefont {J.~A.}\ \bibnamefont {Cowell}}, \bibinfo {author} {\bibfnamefont {D.}~\bibnamefont {Grand\'on}}, \bibinfo {author} {\bibfnamefont {M.}~\bibnamefont {Shirasaki}}, \ and\ \bibinfo {author} {\bibfnamefont {J.}~\bibnamefont {Liu}},\ }\href@noop {} {\  (\bibinfo {year} {2024})},\ \Eprint {http://arxiv.org/abs/2410.00401} {arXiv:2410.00401 [astro-ph.CO]} \BibitemShut {NoStop}%
\bibitem [{\citenamefont {Qu}\ \emph {et~al.}(2024{\natexlab{a}})\citenamefont {Qu} \emph {et~al.}}]{ACT:2024nrz}%
  \BibitemOpen
  \bibfield  {author} {\bibinfo {author} {\bibfnamefont {F.~J.}\ \bibnamefont {Qu}} \emph {et~al.} (\bibinfo {collaboration} {ACT, DESI}),\ }\href@noop {} {\  (\bibinfo {year} {2024}{\natexlab{a}})},\ \Eprint {http://arxiv.org/abs/2410.10808} {arXiv:2410.10808 [astro-ph.CO]} \BibitemShut {NoStop}%
\bibitem [{\citenamefont {McCullough}\ \emph {et~al.}(2024)\citenamefont {McCullough} \emph {et~al.}}]{DES:2024xvm}%
  \BibitemOpen
  \bibfield  {author} {\bibinfo {author} {\bibfnamefont {J.}~\bibnamefont {McCullough}} \emph {et~al.} (\bibinfo {collaboration} {DES}),\ }\href@noop {} {\  (\bibinfo {year} {2024})},\ \Eprint {http://arxiv.org/abs/2410.22272} {arXiv:2410.22272 [astro-ph.CO]} \BibitemShut {NoStop}%
\bibitem [{\citenamefont {Lau}\ \emph {et~al.}(2024)\citenamefont {Lau}, \citenamefont {Bogd\'an}, \citenamefont {Nagai}, \citenamefont {Cappelluti},\ and\ \citenamefont {Shirasaki}}]{Lau:2024xrd}%
  \BibitemOpen
  \bibfield  {author} {\bibinfo {author} {\bibfnamefont {E.~T.}\ \bibnamefont {Lau}}, \bibinfo {author} {\bibfnamefont {A.}~\bibnamefont {Bogd\'an}}, \bibinfo {author} {\bibfnamefont {D.}~\bibnamefont {Nagai}}, \bibinfo {author} {\bibfnamefont {N.}~\bibnamefont {Cappelluti}}, \ and\ \bibinfo {author} {\bibfnamefont {M.}~\bibnamefont {Shirasaki}},\ }\href@noop {} {\  (\bibinfo {year} {2024})},\ \Eprint {http://arxiv.org/abs/2410.22397} {arXiv:2410.22397 [astro-ph.CO]} \BibitemShut {NoStop}%
\bibitem [{\citenamefont {Akarsu}\ \emph {et~al.}(2024{\natexlab{b}})\citenamefont {Akarsu}, \citenamefont {Colg\'ain}, \citenamefont {Sen},\ and\ \citenamefont {Sheikh-Jabbari}}]{Akarsu:2024hsu}%
  \BibitemOpen
  \bibfield  {author} {\bibinfo {author} {\bibfnamefont {O.}~\bibnamefont {Akarsu}}, \bibinfo {author} {\bibfnamefont {E.~O.}\ \bibnamefont {Colg\'ain}}, \bibinfo {author} {\bibfnamefont {A.~A.}\ \bibnamefont {Sen}}, \ and\ \bibinfo {author} {\bibfnamefont {M.~M.}\ \bibnamefont {Sheikh-Jabbari}},\ }\href@noop {} {\  (\bibinfo {year} {2024}{\natexlab{b}})},\ \Eprint {http://arxiv.org/abs/2410.23134} {arXiv:2410.23134 [astro-ph.CO]} \BibitemShut {NoStop}%
\bibitem [{\citenamefont {Sahni}\ and\ \citenamefont {Starobinsky}(2000)}]{Sahni:1999gb}%
  \BibitemOpen
  \bibfield  {author} {\bibinfo {author} {\bibfnamefont {V.}~\bibnamefont {Sahni}}\ and\ \bibinfo {author} {\bibfnamefont {A.~A.}\ \bibnamefont {Starobinsky}},\ }\href {\doibase 10.1142/S0218271800000542} {\bibfield  {journal} {\bibinfo  {journal} {Int. J. Mod. Phys. D}\ }\textbf {\bibinfo {volume} {9}},\ \bibinfo {pages} {373} (\bibinfo {year} {2000})},\ \Eprint {http://arxiv.org/abs/astro-ph/9904398} {arXiv:astro-ph/9904398} \BibitemShut {NoStop}%
\bibitem [{\citenamefont {Carroll}(2001)}]{Carroll:2000fy}%
  \BibitemOpen
  \bibfield  {author} {\bibinfo {author} {\bibfnamefont {S.~M.}\ \bibnamefont {Carroll}},\ }\href {\doibase 10.12942/lrr-2001-1} {\bibfield  {journal} {\bibinfo  {journal} {Living Rev. Rel.}\ }\textbf {\bibinfo {volume} {4}},\ \bibinfo {pages} {1} (\bibinfo {year} {2001})},\ \Eprint {http://arxiv.org/abs/astro-ph/0004075} {arXiv:astro-ph/0004075} \BibitemShut {NoStop}%
\bibitem [{\citenamefont {Peebles}\ and\ \citenamefont {Ratra}(2003)}]{Peebles:2002gy}%
  \BibitemOpen
  \bibfield  {author} {\bibinfo {author} {\bibfnamefont {P.~J.~E.}\ \bibnamefont {Peebles}}\ and\ \bibinfo {author} {\bibfnamefont {B.}~\bibnamefont {Ratra}},\ }\href {\doibase 10.1103/RevModPhys.75.559} {\bibfield  {journal} {\bibinfo  {journal} {Rev. Mod. Phys.}\ }\textbf {\bibinfo {volume} {75}},\ \bibinfo {pages} {559} (\bibinfo {year} {2003})},\ \Eprint {http://arxiv.org/abs/astro-ph/0207347} {arXiv:astro-ph/0207347} \BibitemShut {NoStop}%
\bibitem [{\citenamefont {Padmanabhan}(2003)}]{Padmanabhan:2002ji}%
  \BibitemOpen
  \bibfield  {author} {\bibinfo {author} {\bibfnamefont {T.}~\bibnamefont {Padmanabhan}},\ }\href {\doibase 10.1016/S0370-1573(03)00120-0} {\bibfield  {journal} {\bibinfo  {journal} {Phys. Rept.}\ }\textbf {\bibinfo {volume} {380}},\ \bibinfo {pages} {235} (\bibinfo {year} {2003})},\ \Eprint {http://arxiv.org/abs/hep-th/0212290} {arXiv:hep-th/0212290} \BibitemShut {NoStop}%
\bibitem [{\citenamefont {Copeland}\ \emph {et~al.}(2006)\citenamefont {Copeland}, \citenamefont {Sami},\ and\ \citenamefont {Tsujikawa}}]{Copeland:2006wr}%
  \BibitemOpen
  \bibfield  {author} {\bibinfo {author} {\bibfnamefont {E.~J.}\ \bibnamefont {Copeland}}, \bibinfo {author} {\bibfnamefont {M.}~\bibnamefont {Sami}}, \ and\ \bibinfo {author} {\bibfnamefont {S.}~\bibnamefont {Tsujikawa}},\ }\href {\doibase 10.1142/S021827180600942X} {\bibfield  {journal} {\bibinfo  {journal} {Int. J. Mod. Phys. D}\ }\textbf {\bibinfo {volume} {15}},\ \bibinfo {pages} {1753} (\bibinfo {year} {2006})},\ \Eprint {http://arxiv.org/abs/hep-th/0603057} {arXiv:hep-th/0603057} \BibitemShut {NoStop}%
\bibitem [{\citenamefont {Caldwell}\ and\ \citenamefont {Kamionkowski}(2009)}]{Caldwell:2009ix}%
  \BibitemOpen
  \bibfield  {author} {\bibinfo {author} {\bibfnamefont {R.~R.}\ \bibnamefont {Caldwell}}\ and\ \bibinfo {author} {\bibfnamefont {M.}~\bibnamefont {Kamionkowski}},\ }\href {\doibase 10.1146/annurev-nucl-010709-151330} {\bibfield  {journal} {\bibinfo  {journal} {Ann. Rev. Nucl. Part. Sci.}\ }\textbf {\bibinfo {volume} {59}},\ \bibinfo {pages} {397} (\bibinfo {year} {2009})},\ \Eprint {http://arxiv.org/abs/0903.0866} {arXiv:0903.0866 [astro-ph.CO]} \BibitemShut {NoStop}%
\bibitem [{\citenamefont {Li}\ \emph {et~al.}(2011)\citenamefont {Li}, \citenamefont {Li}, \citenamefont {Wang},\ and\ \citenamefont {Wang}}]{Li:2011sd}%
  \BibitemOpen
  \bibfield  {author} {\bibinfo {author} {\bibfnamefont {M.}~\bibnamefont {Li}}, \bibinfo {author} {\bibfnamefont {X.-D.}\ \bibnamefont {Li}}, \bibinfo {author} {\bibfnamefont {S.}~\bibnamefont {Wang}}, \ and\ \bibinfo {author} {\bibfnamefont {Y.}~\bibnamefont {Wang}},\ }\href {\doibase 10.1088/0253-6102/56/3/24} {\bibfield  {journal} {\bibinfo  {journal} {Commun. Theor. Phys.}\ }\textbf {\bibinfo {volume} {56}},\ \bibinfo {pages} {525} (\bibinfo {year} {2011})},\ \Eprint {http://arxiv.org/abs/1103.5870} {arXiv:1103.5870 [astro-ph.CO]} \BibitemShut {NoStop}%
\bibitem [{\citenamefont {Martin}(2012)}]{Martin:2012bt}%
  \BibitemOpen
  \bibfield  {author} {\bibinfo {author} {\bibfnamefont {J.}~\bibnamefont {Martin}},\ }\href {\doibase 10.1016/j.crhy.2012.04.008} {\bibfield  {journal} {\bibinfo  {journal} {Comptes Rendus Physique}\ }\textbf {\bibinfo {volume} {13}},\ \bibinfo {pages} {566} (\bibinfo {year} {2012})},\ \Eprint {http://arxiv.org/abs/1205.3365} {arXiv:1205.3365 [astro-ph.CO]} \BibitemShut {NoStop}%
\bibitem [{\citenamefont {Weinberg}(1989)}]{Weinberg:1988cp}%
  \BibitemOpen
  \bibfield  {author} {\bibinfo {author} {\bibfnamefont {S.}~\bibnamefont {Weinberg}},\ }\href {\doibase 10.1103/RevModPhys.61.1} {\bibfield  {journal} {\bibinfo  {journal} {Rev. Mod. Phys.}\ }\textbf {\bibinfo {volume} {61}},\ \bibinfo {pages} {1} (\bibinfo {year} {1989})}\BibitemShut {NoStop}%
\bibitem [{\citenamefont {Krauss}\ and\ \citenamefont {Turner}(1995)}]{Krauss:1995yb}%
  \BibitemOpen
  \bibfield  {author} {\bibinfo {author} {\bibfnamefont {L.~M.}\ \bibnamefont {Krauss}}\ and\ \bibinfo {author} {\bibfnamefont {M.~S.}\ \bibnamefont {Turner}},\ }\href {\doibase 10.1007/BF02108229} {\bibfield  {journal} {\bibinfo  {journal} {Gen. Rel. Grav.}\ }\textbf {\bibinfo {volume} {27}},\ \bibinfo {pages} {1137} (\bibinfo {year} {1995})},\ \Eprint {http://arxiv.org/abs/astro-ph/9504003} {arXiv:astro-ph/9504003} \BibitemShut {NoStop}%
\bibitem [{\citenamefont {Weinberg}(2000)}]{Weinberg:2000yb}%
  \BibitemOpen
  \bibfield  {author} {\bibinfo {author} {\bibfnamefont {S.}~\bibnamefont {Weinberg}},\ }in\ \href@noop {} {\emph {\bibinfo {booktitle} {{4th International Symposium on Sources and Detection of Dark Matter in the Universe (DM 2000)}}}}\ (\bibinfo {year} {2000})\ pp.\ \bibinfo {pages} {18--26},\ \Eprint {http://arxiv.org/abs/astro-ph/0005265} {arXiv:astro-ph/0005265} \BibitemShut {NoStop}%
\bibitem [{\citenamefont {Sahni}(2002)}]{Sahni:2002kh}%
  \BibitemOpen
  \bibfield  {author} {\bibinfo {author} {\bibfnamefont {V.}~\bibnamefont {Sahni}},\ }\href {\doibase 10.1088/0264-9381/19/13/304} {\bibfield  {journal} {\bibinfo  {journal} {Class. Quant. Grav.}\ }\textbf {\bibinfo {volume} {19}},\ \bibinfo {pages} {3435} (\bibinfo {year} {2002})},\ \Eprint {http://arxiv.org/abs/astro-ph/0202076} {arXiv:astro-ph/0202076} \BibitemShut {NoStop}%
\bibitem [{\citenamefont {Yokoyama}(2003)}]{Yokoyama:2003ii}%
  \BibitemOpen
  \bibfield  {author} {\bibinfo {author} {\bibfnamefont {J.}~\bibnamefont {Yokoyama}},\ }in\ \href@noop {} {\emph {\bibinfo {booktitle} {{12th Workshop on General Relativity and Gravitation}}}}\ (\bibinfo {year} {2003})\ \Eprint {http://arxiv.org/abs/gr-qc/0305068} {arXiv:gr-qc/0305068} \BibitemShut {NoStop}%
\bibitem [{\citenamefont {Nobbenhuis}(2006)}]{Nobbenhuis:2004wn}%
  \BibitemOpen
  \bibfield  {author} {\bibinfo {author} {\bibfnamefont {S.}~\bibnamefont {Nobbenhuis}},\ }\href {\doibase 10.1007/s10701-005-9042-8} {\bibfield  {journal} {\bibinfo  {journal} {Found. Phys.}\ }\textbf {\bibinfo {volume} {36}},\ \bibinfo {pages} {613} (\bibinfo {year} {2006})},\ \Eprint {http://arxiv.org/abs/gr-qc/0411093} {arXiv:gr-qc/0411093} \BibitemShut {NoStop}%
\bibitem [{\citenamefont {Burgess}(2015)}]{Burgess:2013ara}%
  \BibitemOpen
  \bibfield  {author} {\bibinfo {author} {\bibfnamefont {C.~P.}\ \bibnamefont {Burgess}},\ }in\ \href {\doibase 10.1093/acprof:oso/9780198728856.003.0004} {\emph {\bibinfo {booktitle} {{100e Ecole d'Ete de Physique: Post-Planck Cosmology}}}}\ (\bibinfo {year} {2015})\ pp.\ \bibinfo {pages} {149--197},\ \Eprint {http://arxiv.org/abs/1309.4133} {arXiv:1309.4133 [hep-th]} \BibitemShut {NoStop}%
\bibitem [{\citenamefont {Joyce}\ \emph {et~al.}(2015)\citenamefont {Joyce}, \citenamefont {Jain}, \citenamefont {Khoury},\ and\ \citenamefont {Trodden}}]{Joyce:2014kja}%
  \BibitemOpen
  \bibfield  {author} {\bibinfo {author} {\bibfnamefont {A.}~\bibnamefont {Joyce}}, \bibinfo {author} {\bibfnamefont {B.}~\bibnamefont {Jain}}, \bibinfo {author} {\bibfnamefont {J.}~\bibnamefont {Khoury}}, \ and\ \bibinfo {author} {\bibfnamefont {M.}~\bibnamefont {Trodden}},\ }\href {\doibase 10.1016/j.physrep.2014.12.002} {\bibfield  {journal} {\bibinfo  {journal} {Phys. Rept.}\ }\textbf {\bibinfo {volume} {568}},\ \bibinfo {pages} {1} (\bibinfo {year} {2015})},\ \Eprint {http://arxiv.org/abs/1407.0059} {arXiv:1407.0059 [astro-ph.CO]} \BibitemShut {NoStop}%
\bibitem [{\citenamefont {Bull}\ \emph {et~al.}(2016)\citenamefont {Bull} \emph {et~al.}}]{Bull:2015stt}%
  \BibitemOpen
  \bibfield  {author} {\bibinfo {author} {\bibfnamefont {P.}~\bibnamefont {Bull}} \emph {et~al.},\ }\href {\doibase 10.1016/j.dark.2016.02.001} {\bibfield  {journal} {\bibinfo  {journal} {Phys. Dark Univ.}\ }\textbf {\bibinfo {volume} {12}},\ \bibinfo {pages} {56} (\bibinfo {year} {2016})},\ \Eprint {http://arxiv.org/abs/1512.05356} {arXiv:1512.05356 [astro-ph.CO]} \BibitemShut {NoStop}%
\bibitem [{\citenamefont {Wang}\ \emph {et~al.}(2016)\citenamefont {Wang}, \citenamefont {Abdalla}, \citenamefont {Atrio-Barandela},\ and\ \citenamefont {Pavon}}]{Wang:2016lxa}%
  \BibitemOpen
  \bibfield  {author} {\bibinfo {author} {\bibfnamefont {B.}~\bibnamefont {Wang}}, \bibinfo {author} {\bibfnamefont {E.}~\bibnamefont {Abdalla}}, \bibinfo {author} {\bibfnamefont {F.}~\bibnamefont {Atrio-Barandela}}, \ and\ \bibinfo {author} {\bibfnamefont {D.}~\bibnamefont {Pavon}},\ }\href {\doibase 10.1088/0034-4885/79/9/096901} {\bibfield  {journal} {\bibinfo  {journal} {Rept. Prog. Phys.}\ }\textbf {\bibinfo {volume} {79}},\ \bibinfo {pages} {096901} (\bibinfo {year} {2016})},\ \Eprint {http://arxiv.org/abs/1603.08299} {arXiv:1603.08299 [astro-ph.CO]} \BibitemShut {NoStop}%
\bibitem [{\citenamefont {Brustein}\ and\ \citenamefont {Steinhardt}(1993)}]{Brustein:1992nk}%
  \BibitemOpen
  \bibfield  {author} {\bibinfo {author} {\bibfnamefont {R.}~\bibnamefont {Brustein}}\ and\ \bibinfo {author} {\bibfnamefont {P.~J.}\ \bibnamefont {Steinhardt}},\ }\href {\doibase 10.1016/0370-2693(93)90384-T} {\bibfield  {journal} {\bibinfo  {journal} {Phys. Lett. B}\ }\textbf {\bibinfo {volume} {302}},\ \bibinfo {pages} {196} (\bibinfo {year} {1993})},\ \Eprint {http://arxiv.org/abs/hep-th/9212049} {arXiv:hep-th/9212049} \BibitemShut {NoStop}%
\bibitem [{\citenamefont {Witten}(2000)}]{Witten:2000zk}%
  \BibitemOpen
  \bibfield  {author} {\bibinfo {author} {\bibfnamefont {E.}~\bibnamefont {Witten}},\ }in\ \href@noop {} {\emph {\bibinfo {booktitle} {{4th International Symposium on Sources and Detection of Dark Matter in the Universe (DM 2000)}}}}\ (\bibinfo {year} {2000})\ pp.\ \bibinfo {pages} {27--36},\ \Eprint {http://arxiv.org/abs/hep-ph/0002297} {arXiv:hep-ph/0002297} \BibitemShut {NoStop}%
\bibitem [{\citenamefont {Kachru}\ \emph {et~al.}(2003)\citenamefont {Kachru}, \citenamefont {Kallosh}, \citenamefont {Linde},\ and\ \citenamefont {Trivedi}}]{Kachru:2003aw}%
  \BibitemOpen
  \bibfield  {author} {\bibinfo {author} {\bibfnamefont {S.}~\bibnamefont {Kachru}}, \bibinfo {author} {\bibfnamefont {R.}~\bibnamefont {Kallosh}}, \bibinfo {author} {\bibfnamefont {A.~D.}\ \bibnamefont {Linde}}, \ and\ \bibinfo {author} {\bibfnamefont {S.~P.}\ \bibnamefont {Trivedi}},\ }\href {\doibase 10.1103/PhysRevD.68.046005} {\bibfield  {journal} {\bibinfo  {journal} {Phys. Rev. D}\ }\textbf {\bibinfo {volume} {68}},\ \bibinfo {pages} {046005} (\bibinfo {year} {2003})},\ \Eprint {http://arxiv.org/abs/hep-th/0301240} {arXiv:hep-th/0301240} \BibitemShut {NoStop}%
\bibitem [{\citenamefont {Polchinski}(2006)}]{Polchinski:2006gy}%
  \BibitemOpen
  \bibfield  {author} {\bibinfo {author} {\bibfnamefont {J.}~\bibnamefont {Polchinski}},\ }in\ \href@noop {} {\emph {\bibinfo {booktitle} {{23rd Solvay Conference in Physics: The Quantum Structure of Space and Time}}}}\ (\bibinfo {year} {2006})\ pp.\ \bibinfo {pages} {216--236},\ \Eprint {http://arxiv.org/abs/hep-th/0603249} {arXiv:hep-th/0603249} \BibitemShut {NoStop}%
\bibitem [{\citenamefont {Danielsson}\ and\ \citenamefont {Van~Riet}(2018)}]{Danielsson:2018ztv}%
  \BibitemOpen
  \bibfield  {author} {\bibinfo {author} {\bibfnamefont {U.~H.}\ \bibnamefont {Danielsson}}\ and\ \bibinfo {author} {\bibfnamefont {T.}~\bibnamefont {Van~Riet}},\ }\href {\doibase 10.1142/S0218271818300070} {\bibfield  {journal} {\bibinfo  {journal} {Int. J. Mod. Phys. D}\ }\textbf {\bibinfo {volume} {27}},\ \bibinfo {pages} {1830007} (\bibinfo {year} {2018})},\ \Eprint {http://arxiv.org/abs/1804.01120} {arXiv:1804.01120 [hep-th]} \BibitemShut {NoStop}%
\bibitem [{\citenamefont {Zlatev}\ \emph {et~al.}(1999)\citenamefont {Zlatev}, \citenamefont {Wang},\ and\ \citenamefont {Steinhardt}}]{Zlatev:1998tr}%
  \BibitemOpen
  \bibfield  {author} {\bibinfo {author} {\bibfnamefont {I.}~\bibnamefont {Zlatev}}, \bibinfo {author} {\bibfnamefont {L.-M.}\ \bibnamefont {Wang}}, \ and\ \bibinfo {author} {\bibfnamefont {P.~J.}\ \bibnamefont {Steinhardt}},\ }\href {\doibase 10.1103/PhysRevLett.82.896} {\bibfield  {journal} {\bibinfo  {journal} {Phys. Rev. Lett.}\ }\textbf {\bibinfo {volume} {82}},\ \bibinfo {pages} {896} (\bibinfo {year} {1999})},\ \Eprint {http://arxiv.org/abs/astro-ph/9807002} {arXiv:astro-ph/9807002} \BibitemShut {NoStop}%
\bibitem [{\citenamefont {Pavon}\ and\ \citenamefont {Zimdahl}(2005)}]{Pavon:2005yx}%
  \BibitemOpen
  \bibfield  {author} {\bibinfo {author} {\bibfnamefont {D.}~\bibnamefont {Pavon}}\ and\ \bibinfo {author} {\bibfnamefont {W.}~\bibnamefont {Zimdahl}},\ }\href {\doibase 10.1016/j.physletb.2005.08.134} {\bibfield  {journal} {\bibinfo  {journal} {Phys. Lett. B}\ }\textbf {\bibinfo {volume} {628}},\ \bibinfo {pages} {206} (\bibinfo {year} {2005})},\ \Eprint {http://arxiv.org/abs/gr-qc/0505020} {arXiv:gr-qc/0505020} \BibitemShut {NoStop}%
\bibitem [{\citenamefont {Velten}\ \emph {et~al.}(2014)\citenamefont {Velten}, \citenamefont {vom Marttens},\ and\ \citenamefont {Zimdahl}}]{Velten:2014nra}%
  \BibitemOpen
  \bibfield  {author} {\bibinfo {author} {\bibfnamefont {H.~E.~S.}\ \bibnamefont {Velten}}, \bibinfo {author} {\bibfnamefont {R.~F.}\ \bibnamefont {vom Marttens}}, \ and\ \bibinfo {author} {\bibfnamefont {W.}~\bibnamefont {Zimdahl}},\ }\href {\doibase 10.1140/epjc/s10052-014-3160-4} {\bibfield  {journal} {\bibinfo  {journal} {Eur. Phys. J. C}\ }\textbf {\bibinfo {volume} {74}},\ \bibinfo {pages} {3160} (\bibinfo {year} {2014})},\ \Eprint {http://arxiv.org/abs/1410.2509} {arXiv:1410.2509 [astro-ph.CO]} \BibitemShut {NoStop}%
\bibitem [{\citenamefont {Adame}\ \emph {et~al.}(2024{\natexlab{b}})\citenamefont {Adame} \emph {et~al.}}]{DESI:2024mwx}%
  \BibitemOpen
  \bibfield  {author} {\bibinfo {author} {\bibfnamefont {A.~G.}\ \bibnamefont {Adame}} \emph {et~al.} (\bibinfo {collaboration} {DESI}),\ }\href@noop {} {\  (\bibinfo {year} {2024}{\natexlab{b}})},\ \Eprint {http://arxiv.org/abs/2404.03002} {arXiv:2404.03002 [astro-ph.CO]} \BibitemShut {NoStop}%
\bibitem [{\citenamefont {Giar\`e}\ \emph {et~al.}(2024{\natexlab{d}})\citenamefont {Giar\`e}, \citenamefont {Najafi}, \citenamefont {Pan}, \citenamefont {Di~Valentino},\ and\ \citenamefont {Firouzjaee}}]{Giare:2024gpk}%
  \BibitemOpen
  \bibfield  {author} {\bibinfo {author} {\bibfnamefont {W.}~\bibnamefont {Giar\`e}}, \bibinfo {author} {\bibfnamefont {M.}~\bibnamefont {Najafi}}, \bibinfo {author} {\bibfnamefont {S.}~\bibnamefont {Pan}}, \bibinfo {author} {\bibfnamefont {E.}~\bibnamefont {Di~Valentino}}, \ and\ \bibinfo {author} {\bibfnamefont {J.~T.}\ \bibnamefont {Firouzjaee}},\ }\href {\doibase 10.1088/1475-7516/2024/10/035} {\bibfield  {journal} {\bibinfo  {journal} {JCAP}\ }\textbf {\bibinfo {volume} {10}},\ \bibinfo {pages} {035} (\bibinfo {year} {2024}{\natexlab{d}})},\ \Eprint {http://arxiv.org/abs/2407.16689} {arXiv:2407.16689 [astro-ph.CO]} \BibitemShut {NoStop}%
\bibitem [{\citenamefont {Wang}\ \emph {et~al.}(2024{\natexlab{a}})\citenamefont {Wang}, \citenamefont {Lin}, \citenamefont {Ding},\ and\ \citenamefont {Hu}}]{Wang:2024pui}%
  \BibitemOpen
  \bibfield  {author} {\bibinfo {author} {\bibfnamefont {Z.}~\bibnamefont {Wang}}, \bibinfo {author} {\bibfnamefont {S.}~\bibnamefont {Lin}}, \bibinfo {author} {\bibfnamefont {Z.}~\bibnamefont {Ding}}, \ and\ \bibinfo {author} {\bibfnamefont {B.}~\bibnamefont {Hu}},\ }\href@noop {} {\  (\bibinfo {year} {2024}{\natexlab{a}})},\ \Eprint {http://arxiv.org/abs/2405.02168} {arXiv:2405.02168 [astro-ph.CO]} \BibitemShut {NoStop}%
\bibitem [{\citenamefont {Efstathiou}(2024)}]{Efstathiou:2024xcq}%
  \BibitemOpen
  \bibfield  {author} {\bibinfo {author} {\bibfnamefont {G.}~\bibnamefont {Efstathiou}},\ }\href@noop {} {\  (\bibinfo {year} {2024})},\ \Eprint {http://arxiv.org/abs/2408.07175} {arXiv:2408.07175 [astro-ph.CO]} \BibitemShut {NoStop}%
\bibitem [{\citenamefont {Patel}\ \emph {et~al.}(2024)\citenamefont {Patel}, \citenamefont {Chakraborty},\ and\ \citenamefont {Amendola}}]{Patel:2024odo}%
  \BibitemOpen
  \bibfield  {author} {\bibinfo {author} {\bibfnamefont {V.}~\bibnamefont {Patel}}, \bibinfo {author} {\bibfnamefont {A.}~\bibnamefont {Chakraborty}}, \ and\ \bibinfo {author} {\bibfnamefont {L.}~\bibnamefont {Amendola}},\ }\href@noop {} {\  (\bibinfo {year} {2024})},\ \Eprint {http://arxiv.org/abs/2407.06586} {arXiv:2407.06586 [astro-ph.CO]} \BibitemShut {NoStop}%
\bibitem [{\citenamefont {Liu}\ \emph {et~al.}(2024)\citenamefont {Liu}, \citenamefont {Wang},\ and\ \citenamefont {Zhao}}]{Liu:2024gfy}%
  \BibitemOpen
  \bibfield  {author} {\bibinfo {author} {\bibfnamefont {G.}~\bibnamefont {Liu}}, \bibinfo {author} {\bibfnamefont {Y.}~\bibnamefont {Wang}}, \ and\ \bibinfo {author} {\bibfnamefont {W.}~\bibnamefont {Zhao}},\ }\href@noop {} {\  (\bibinfo {year} {2024})},\ \Eprint {http://arxiv.org/abs/2407.04385} {arXiv:2407.04385 [astro-ph.CO]} \BibitemShut {NoStop}%
\bibitem [{\citenamefont {Colg\'ain}\ \emph {et~al.}(2024)\citenamefont {Colg\'ain}, \citenamefont {Dainotti}, \citenamefont {Capozziello}, \citenamefont {Pourojaghi}, \citenamefont {Sheikh-Jabbari},\ and\ \citenamefont {Stojkovic}}]{Colgain:2024xqj}%
  \BibitemOpen
  \bibfield  {author} {\bibinfo {author} {\bibfnamefont {E.~O.}\ \bibnamefont {Colg\'ain}}, \bibinfo {author} {\bibfnamefont {M.~G.}\ \bibnamefont {Dainotti}}, \bibinfo {author} {\bibfnamefont {S.}~\bibnamefont {Capozziello}}, \bibinfo {author} {\bibfnamefont {S.}~\bibnamefont {Pourojaghi}}, \bibinfo {author} {\bibfnamefont {M.~M.}\ \bibnamefont {Sheikh-Jabbari}}, \ and\ \bibinfo {author} {\bibfnamefont {D.}~\bibnamefont {Stojkovic}},\ }\href@noop {} {\  (\bibinfo {year} {2024})},\ \Eprint {http://arxiv.org/abs/2404.08633} {arXiv:2404.08633 [astro-ph.CO]} \BibitemShut {NoStop}%
\bibitem [{\citenamefont {Giar\`e}(2024{\natexlab{b}})}]{Giare:2024ocw}%
  \BibitemOpen
  \bibfield  {author} {\bibinfo {author} {\bibfnamefont {W.}~\bibnamefont {Giar\`e}},\ }\href@noop {} {\  (\bibinfo {year} {2024}{\natexlab{b}})},\ \Eprint {http://arxiv.org/abs/2409.17074} {arXiv:2409.17074 [astro-ph.CO]} \BibitemShut {NoStop}%
\bibitem [{\citenamefont {Cort\^es}\ and\ \citenamefont {Liddle}(2024)}]{Cortes:2024lgw}%
  \BibitemOpen
  \bibfield  {author} {\bibinfo {author} {\bibfnamefont {M.}~\bibnamefont {Cort\^es}}\ and\ \bibinfo {author} {\bibfnamefont {A.~R.}\ \bibnamefont {Liddle}},\ }\href@noop {} {\  (\bibinfo {year} {2024})},\ \Eprint {http://arxiv.org/abs/2404.08056} {arXiv:2404.08056 [astro-ph.CO]} \BibitemShut {NoStop}%
\bibitem [{\citenamefont {Orchard}\ and\ \citenamefont {C\'ardenas}(2024)}]{Orchard:2024bve}%
  \BibitemOpen
  \bibfield  {author} {\bibinfo {author} {\bibfnamefont {L.}~\bibnamefont {Orchard}}\ and\ \bibinfo {author} {\bibfnamefont {V.~H.}\ \bibnamefont {C\'ardenas}},\ }\href {\doibase 10.1016/j.dark.2024.101678} {\bibfield  {journal} {\bibinfo  {journal} {Phys. Dark Univ.}\ }\textbf {\bibinfo {volume} {46}},\ \bibinfo {pages} {101678} (\bibinfo {year} {2024})},\ \Eprint {http://arxiv.org/abs/2407.05579} {arXiv:2407.05579 [astro-ph.CO]} \BibitemShut {NoStop}%
\bibitem [{\citenamefont {Chudaykin}\ and\ \citenamefont {Kunz}(2024)}]{Chudaykin:2024gol}%
  \BibitemOpen
  \bibfield  {author} {\bibinfo {author} {\bibfnamefont {A.}~\bibnamefont {Chudaykin}}\ and\ \bibinfo {author} {\bibfnamefont {M.}~\bibnamefont {Kunz}},\ }\href@noop {} {\  (\bibinfo {year} {2024})},\ \Eprint {http://arxiv.org/abs/2407.02558} {arXiv:2407.02558 [astro-ph.CO]} \BibitemShut {NoStop}%
\bibitem [{\citenamefont {Notari}\ \emph {et~al.}(2024)\citenamefont {Notari}, \citenamefont {Redi},\ and\ \citenamefont {Tesi}}]{Notari:2024rti}%
  \BibitemOpen
  \bibfield  {author} {\bibinfo {author} {\bibfnamefont {A.}~\bibnamefont {Notari}}, \bibinfo {author} {\bibfnamefont {M.}~\bibnamefont {Redi}}, \ and\ \bibinfo {author} {\bibfnamefont {A.}~\bibnamefont {Tesi}},\ }\href@noop {} {\  (\bibinfo {year} {2024})},\ \Eprint {http://arxiv.org/abs/2406.08459} {arXiv:2406.08459 [astro-ph.CO]} \BibitemShut {NoStop}%
\bibitem [{\citenamefont {Gialamas}\ \emph {et~al.}(2024)\citenamefont {Gialamas}, \citenamefont {H\"utsi}, \citenamefont {Kannike}, \citenamefont {Racioppi}, \citenamefont {Raidal}, \citenamefont {Vasar},\ and\ \citenamefont {Veerm\"ae}}]{Gialamas:2024lyw}%
  \BibitemOpen
  \bibfield  {author} {\bibinfo {author} {\bibfnamefont {I.~D.}\ \bibnamefont {Gialamas}}, \bibinfo {author} {\bibfnamefont {G.}~\bibnamefont {H\"utsi}}, \bibinfo {author} {\bibfnamefont {K.}~\bibnamefont {Kannike}}, \bibinfo {author} {\bibfnamefont {A.}~\bibnamefont {Racioppi}}, \bibinfo {author} {\bibfnamefont {M.}~\bibnamefont {Raidal}}, \bibinfo {author} {\bibfnamefont {M.}~\bibnamefont {Vasar}}, \ and\ \bibinfo {author} {\bibfnamefont {H.}~\bibnamefont {Veerm\"ae}},\ }\href@noop {} {\  (\bibinfo {year} {2024})},\ \Eprint {http://arxiv.org/abs/2406.07533} {arXiv:2406.07533 [astro-ph.CO]} \BibitemShut {NoStop}%
\bibitem [{\citenamefont {Wang}\ \emph {et~al.}(2024{\natexlab{b}})\citenamefont {Wang}, \citenamefont {Peng},\ and\ \citenamefont {Piao}}]{Wang:2024hwd}%
  \BibitemOpen
  \bibfield  {author} {\bibinfo {author} {\bibfnamefont {H.}~\bibnamefont {Wang}}, \bibinfo {author} {\bibfnamefont {Z.-Y.}\ \bibnamefont {Peng}}, \ and\ \bibinfo {author} {\bibfnamefont {Y.-S.}\ \bibnamefont {Piao}},\ }\href@noop {} {\  (\bibinfo {year} {2024}{\natexlab{b}})},\ \Eprint {http://arxiv.org/abs/2406.03395} {arXiv:2406.03395 [astro-ph.CO]} \BibitemShut {NoStop}%
\bibitem [{\citenamefont {Wang}\ and\ \citenamefont {Piao}(2024)}]{Wang:2024dka}%
  \BibitemOpen
  \bibfield  {author} {\bibinfo {author} {\bibfnamefont {H.}~\bibnamefont {Wang}}\ and\ \bibinfo {author} {\bibfnamefont {Y.-S.}\ \bibnamefont {Piao}},\ }\href@noop {} {\  (\bibinfo {year} {2024})},\ \Eprint {http://arxiv.org/abs/2404.18579} {arXiv:2404.18579 [astro-ph.CO]} \BibitemShut {NoStop}%
\bibitem [{\citenamefont {Carloni}\ \emph {et~al.}(2024)\citenamefont {Carloni}, \citenamefont {Luongo},\ and\ \citenamefont {Muccino}}]{Carloni:2024zpl}%
  \BibitemOpen
  \bibfield  {author} {\bibinfo {author} {\bibfnamefont {Y.}~\bibnamefont {Carloni}}, \bibinfo {author} {\bibfnamefont {O.}~\bibnamefont {Luongo}}, \ and\ \bibinfo {author} {\bibfnamefont {M.}~\bibnamefont {Muccino}},\ }\href@noop {} {\  (\bibinfo {year} {2024})},\ \Eprint {http://arxiv.org/abs/2404.12068} {arXiv:2404.12068 [astro-ph.CO]} \BibitemShut {NoStop}%
\bibitem [{\citenamefont {Tada}\ and\ \citenamefont {Terada}(2024)}]{Tada:2024znt}%
  \BibitemOpen
  \bibfield  {author} {\bibinfo {author} {\bibfnamefont {Y.}~\bibnamefont {Tada}}\ and\ \bibinfo {author} {\bibfnamefont {T.}~\bibnamefont {Terada}},\ }\href {\doibase 10.1103/PhysRevD.109.L121305} {\bibfield  {journal} {\bibinfo  {journal} {Phys. Rev. D}\ }\textbf {\bibinfo {volume} {109}},\ \bibinfo {pages} {L121305} (\bibinfo {year} {2024})},\ \Eprint {http://arxiv.org/abs/2404.05722} {arXiv:2404.05722 [astro-ph.CO]} \BibitemShut {NoStop}%
\bibitem [{\citenamefont {Yin}(2024)}]{Yin:2024hba}%
  \BibitemOpen
  \bibfield  {author} {\bibinfo {author} {\bibfnamefont {W.}~\bibnamefont {Yin}},\ }\href {\doibase 10.1007/JHEP05(2024)327} {\bibfield  {journal} {\bibinfo  {journal} {JHEP}\ }\textbf {\bibinfo {volume} {05}},\ \bibinfo {pages} {327} (\bibinfo {year} {2024})},\ \Eprint {http://arxiv.org/abs/2404.06444} {arXiv:2404.06444 [hep-ph]} \BibitemShut {NoStop}%
\bibitem [{\citenamefont {Luongo}\ and\ \citenamefont {Muccino}(2024)}]{Luongo:2024fww}%
  \BibitemOpen
  \bibfield  {author} {\bibinfo {author} {\bibfnamefont {O.}~\bibnamefont {Luongo}}\ and\ \bibinfo {author} {\bibfnamefont {M.}~\bibnamefont {Muccino}},\ }\href {\doibase 10.1051/0004-6361/202450512} {\  (\bibinfo {year} {2024}),\ 10.1051/0004-6361/202450512},\ \Eprint {http://arxiv.org/abs/2404.07070} {arXiv:2404.07070 [astro-ph.CO]} \BibitemShut {NoStop}%
\bibitem [{\citenamefont {Park}\ \emph {et~al.}(2024{\natexlab{a}})\citenamefont {Park}, \citenamefont {de~Cruz~P\'erez},\ and\ \citenamefont {Ratra}}]{Park:2024jns}%
  \BibitemOpen
  \bibfield  {author} {\bibinfo {author} {\bibfnamefont {C.-G.}\ \bibnamefont {Park}}, \bibinfo {author} {\bibfnamefont {J.}~\bibnamefont {de~Cruz~P\'erez}}, \ and\ \bibinfo {author} {\bibfnamefont {B.}~\bibnamefont {Ratra}},\ }\href@noop {} {\  (\bibinfo {year} {2024}{\natexlab{a}})},\ \Eprint {http://arxiv.org/abs/2405.00502} {arXiv:2405.00502 [astro-ph.CO]} \BibitemShut {NoStop}%
\bibitem [{\citenamefont {Shlivko}\ and\ \citenamefont {Steinhardt}(2024)}]{Shlivko:2024llw}%
  \BibitemOpen
  \bibfield  {author} {\bibinfo {author} {\bibfnamefont {D.}~\bibnamefont {Shlivko}}\ and\ \bibinfo {author} {\bibfnamefont {P.~J.}\ \bibnamefont {Steinhardt}},\ }\href {\doibase 10.1016/j.physletb.2024.138826} {\bibfield  {journal} {\bibinfo  {journal} {Phys. Lett. B}\ }\textbf {\bibinfo {volume} {855}},\ \bibinfo {pages} {138826} (\bibinfo {year} {2024})},\ \Eprint {http://arxiv.org/abs/2405.03933} {arXiv:2405.03933 [astro-ph.CO]} \BibitemShut {NoStop}%
\bibitem [{\citenamefont {Ye}\ \emph {et~al.}(2024)\citenamefont {Ye}, \citenamefont {Martinelli}, \citenamefont {Hu},\ and\ \citenamefont {Silvestri}}]{Ye:2024ywg}%
  \BibitemOpen
  \bibfield  {author} {\bibinfo {author} {\bibfnamefont {G.}~\bibnamefont {Ye}}, \bibinfo {author} {\bibfnamefont {M.}~\bibnamefont {Martinelli}}, \bibinfo {author} {\bibfnamefont {B.}~\bibnamefont {Hu}}, \ and\ \bibinfo {author} {\bibfnamefont {A.}~\bibnamefont {Silvestri}},\ }\href@noop {} {\  (\bibinfo {year} {2024})},\ \Eprint {http://arxiv.org/abs/2407.15832} {arXiv:2407.15832 [astro-ph.CO]} \BibitemShut {NoStop}%
\bibitem [{\citenamefont {Li}\ \emph {et~al.}(2024)\citenamefont {Li}, \citenamefont {Wu}, \citenamefont {Du}, \citenamefont {Jin}, \citenamefont {Li}, \citenamefont {Zhang},\ and\ \citenamefont {Zhang}}]{Li:2024qso}%
  \BibitemOpen
  \bibfield  {author} {\bibinfo {author} {\bibfnamefont {T.-N.}\ \bibnamefont {Li}}, \bibinfo {author} {\bibfnamefont {P.-J.}\ \bibnamefont {Wu}}, \bibinfo {author} {\bibfnamefont {G.-H.}\ \bibnamefont {Du}}, \bibinfo {author} {\bibfnamefont {S.-J.}\ \bibnamefont {Jin}}, \bibinfo {author} {\bibfnamefont {H.-L.}\ \bibnamefont {Li}}, \bibinfo {author} {\bibfnamefont {J.-F.}\ \bibnamefont {Zhang}}, \ and\ \bibinfo {author} {\bibfnamefont {X.}~\bibnamefont {Zhang}},\ }\href@noop {} {\  (\bibinfo {year} {2024})},\ \Eprint {http://arxiv.org/abs/2407.14934} {arXiv:2407.14934 [astro-ph.CO]} \BibitemShut {NoStop}%
\bibitem [{\citenamefont {Yang}\ \emph {et~al.}(2024)\citenamefont {Yang}, \citenamefont {Ren}, \citenamefont {Wang}, \citenamefont {Lu}, \citenamefont {Zhang}, \citenamefont {Cai},\ and\ \citenamefont {Saridakis}}]{Yang:2024kdo}%
  \BibitemOpen
  \bibfield  {author} {\bibinfo {author} {\bibfnamefont {Y.}~\bibnamefont {Yang}}, \bibinfo {author} {\bibfnamefont {X.}~\bibnamefont {Ren}}, \bibinfo {author} {\bibfnamefont {Q.}~\bibnamefont {Wang}}, \bibinfo {author} {\bibfnamefont {Z.}~\bibnamefont {Lu}}, \bibinfo {author} {\bibfnamefont {D.}~\bibnamefont {Zhang}}, \bibinfo {author} {\bibfnamefont {Y.-F.}\ \bibnamefont {Cai}}, \ and\ \bibinfo {author} {\bibfnamefont {E.~N.}\ \bibnamefont {Saridakis}},\ }\href {\doibase 10.1016/j.scib.2024.07.029} {\bibfield  {journal} {\bibinfo  {journal} {Sci. Bull.}\ }\textbf {\bibinfo {volume} {69}},\ \bibinfo {pages} {2698} (\bibinfo {year} {2024})},\ \Eprint {http://arxiv.org/abs/2404.19437} {arXiv:2404.19437 [astro-ph.CO]} \BibitemShut {NoStop}%
\bibitem [{\citenamefont {Calderon}\ \emph {et~al.}(2024)\citenamefont {Calderon} \emph {et~al.}}]{DESI:2024aqx}%
  \BibitemOpen
  \bibfield  {author} {\bibinfo {author} {\bibfnamefont {R.}~\bibnamefont {Calderon}} \emph {et~al.} (\bibinfo {collaboration} {DESI}),\ }\href {\doibase 10.1088/1475-7516/2024/10/048} {\bibfield  {journal} {\bibinfo  {journal} {JCAP}\ }\textbf {\bibinfo {volume} {10}},\ \bibinfo {pages} {048} (\bibinfo {year} {2024})},\ \Eprint {http://arxiv.org/abs/2405.04216} {arXiv:2405.04216 [astro-ph.CO]} \BibitemShut {NoStop}%
\bibitem [{\citenamefont {Park}\ \emph {et~al.}(2024{\natexlab{b}})\citenamefont {Park}, \citenamefont {de~Cruz~Perez},\ and\ \citenamefont {Ratra}}]{Chan-GyungPark:2024brx}%
  \BibitemOpen
  \bibfield  {author} {\bibinfo {author} {\bibfnamefont {C.-G.}\ \bibnamefont {Park}}, \bibinfo {author} {\bibfnamefont {J.}~\bibnamefont {de~Cruz~Perez}}, \ and\ \bibinfo {author} {\bibfnamefont {B.}~\bibnamefont {Ratra}},\ }\href@noop {} {\  (\bibinfo {year} {2024}{\natexlab{b}})},\ \Eprint {http://arxiv.org/abs/2410.13627} {arXiv:2410.13627 [astro-ph.CO]} \BibitemShut {NoStop}%
\bibitem [{\citenamefont {Wolf}\ \emph {et~al.}(2024)\citenamefont {Wolf}, \citenamefont {Ferreira},\ and\ \citenamefont {Garc\'\i{}a-Garc\'\i{}a}}]{Wolf:2024stt}%
  \BibitemOpen
  \bibfield  {author} {\bibinfo {author} {\bibfnamefont {W.~J.}\ \bibnamefont {Wolf}}, \bibinfo {author} {\bibfnamefont {P.~G.}\ \bibnamefont {Ferreira}}, \ and\ \bibinfo {author} {\bibfnamefont {C.}~\bibnamefont {Garc\'\i{}a-Garc\'\i{}a}},\ }\href@noop {} {\  (\bibinfo {year} {2024})},\ \Eprint {http://arxiv.org/abs/2409.17019} {arXiv:2409.17019 [astro-ph.CO]} \BibitemShut {NoStop}%
\bibitem [{\citenamefont {Pourojaghi}\ \emph {et~al.}(2024)\citenamefont {Pourojaghi}, \citenamefont {Malekjani},\ and\ \citenamefont {Davari}}]{Pourojaghi:2024bxa}%
  \BibitemOpen
  \bibfield  {author} {\bibinfo {author} {\bibfnamefont {S.}~\bibnamefont {Pourojaghi}}, \bibinfo {author} {\bibfnamefont {M.}~\bibnamefont {Malekjani}}, \ and\ \bibinfo {author} {\bibfnamefont {Z.}~\bibnamefont {Davari}},\ }\href@noop {} {\  (\bibinfo {year} {2024})},\ \Eprint {http://arxiv.org/abs/2408.10704} {arXiv:2408.10704 [astro-ph.CO]} \BibitemShut {NoStop}%
\bibitem [{\citenamefont {Sabogal}\ \emph {et~al.}(2024)\citenamefont {Sabogal}, \citenamefont {Silva}, \citenamefont {Nunes}, \citenamefont {Kumar}, \citenamefont {Di~Valentino},\ and\ \citenamefont {Giar\`e}}]{Sabogal:2024yha}%
  \BibitemOpen
  \bibfield  {author} {\bibinfo {author} {\bibfnamefont {M.~A.}\ \bibnamefont {Sabogal}}, \bibinfo {author} {\bibfnamefont {E.}~\bibnamefont {Silva}}, \bibinfo {author} {\bibfnamefont {R.~C.}\ \bibnamefont {Nunes}}, \bibinfo {author} {\bibfnamefont {S.}~\bibnamefont {Kumar}}, \bibinfo {author} {\bibfnamefont {E.}~\bibnamefont {Di~Valentino}}, \ and\ \bibinfo {author} {\bibfnamefont {W.}~\bibnamefont {Giar\`e}},\ }\href@noop {} {\  (\bibinfo {year} {2024})},\ \Eprint {http://arxiv.org/abs/2408.12403} {arXiv:2408.12403 [astro-ph.CO]} \BibitemShut {NoStop}%
\bibitem [{\citenamefont {Jiang}\ \emph {et~al.}(2024)\citenamefont {Jiang}, \citenamefont {Pedrotti}, \citenamefont {da~Costa},\ and\ \citenamefont {Vagnozzi}}]{Jiang:2024xnu}%
  \BibitemOpen
  \bibfield  {author} {\bibinfo {author} {\bibfnamefont {J.-Q.}\ \bibnamefont {Jiang}}, \bibinfo {author} {\bibfnamefont {D.}~\bibnamefont {Pedrotti}}, \bibinfo {author} {\bibfnamefont {S.~S.}\ \bibnamefont {da~Costa}}, \ and\ \bibinfo {author} {\bibfnamefont {S.}~\bibnamefont {Vagnozzi}},\ }\href@noop {} {\  (\bibinfo {year} {2024})},\ \Eprint {http://arxiv.org/abs/2408.02365} {arXiv:2408.02365 [astro-ph.CO]} \BibitemShut {NoStop}%
\bibitem [{\citenamefont {Dinda}\ and\ \citenamefont {Maartens}(2024)}]{Dinda:2024ktd}%
  \BibitemOpen
  \bibfield  {author} {\bibinfo {author} {\bibfnamefont {B.~R.}\ \bibnamefont {Dinda}}\ and\ \bibinfo {author} {\bibfnamefont {R.}~\bibnamefont {Maartens}},\ }\href@noop {} {\  (\bibinfo {year} {2024})},\ \Eprint {http://arxiv.org/abs/2407.17252} {arXiv:2407.17252 [astro-ph.CO]} \BibitemShut {NoStop}%
\bibitem [{\citenamefont {Hern\'andez-Almada}\ \emph {et~al.}(2024)\citenamefont {Hern\'andez-Almada}, \citenamefont {Mendoza-Mart\'\i{}nez}, \citenamefont {Garc\'\i{}a-Aspeitia},\ and\ \citenamefont {Motta}}]{Hernandez-Almada:2024ost}%
  \BibitemOpen
  \bibfield  {author} {\bibinfo {author} {\bibfnamefont {A.}~\bibnamefont {Hern\'andez-Almada}}, \bibinfo {author} {\bibfnamefont {M.~L.}\ \bibnamefont {Mendoza-Mart\'\i{}nez}}, \bibinfo {author} {\bibfnamefont {M.~A.}\ \bibnamefont {Garc\'\i{}a-Aspeitia}}, \ and\ \bibinfo {author} {\bibfnamefont {V.}~\bibnamefont {Motta}},\ }\href {\doibase 10.1016/j.dark.2024.101668} {\bibfield  {journal} {\bibinfo  {journal} {Phys. Dark Univ.}\ }\textbf {\bibinfo {volume} {46}},\ \bibinfo {pages} {101668} (\bibinfo {year} {2024})},\ \Eprint {http://arxiv.org/abs/2407.09430} {arXiv:2407.09430 [astro-ph.CO]} \BibitemShut {NoStop}%
\bibitem [{\citenamefont {Menci}\ \emph {et~al.}(2024)\citenamefont {Menci}, \citenamefont {Sen},\ and\ \citenamefont {Castellano}}]{Menci:2024hop}%
  \BibitemOpen
  \bibfield  {author} {\bibinfo {author} {\bibfnamefont {N.}~\bibnamefont {Menci}}, \bibinfo {author} {\bibfnamefont {A.~A.}\ \bibnamefont {Sen}}, \ and\ \bibinfo {author} {\bibfnamefont {M.}~\bibnamefont {Castellano}},\ }\href@noop {} {\  (\bibinfo {year} {2024})},\ \Eprint {http://arxiv.org/abs/2410.22940} {arXiv:2410.22940 [astro-ph.CO]} \BibitemShut {NoStop}%
\bibitem [{\citenamefont {Ramadan}\ \emph {et~al.}(2024{\natexlab{b}})\citenamefont {Ramadan}, \citenamefont {Sakstein},\ and\ \citenamefont {Rubin}}]{Ramadan:2024kmn}%
  \BibitemOpen
  \bibfield  {author} {\bibinfo {author} {\bibfnamefont {O.~F.}\ \bibnamefont {Ramadan}}, \bibinfo {author} {\bibfnamefont {J.}~\bibnamefont {Sakstein}}, \ and\ \bibinfo {author} {\bibfnamefont {D.}~\bibnamefont {Rubin}},\ }\href {\doibase 10.1103/PhysRevD.110.L041303} {\bibfield  {journal} {\bibinfo  {journal} {Phys. Rev. D}\ }\textbf {\bibinfo {volume} {110}},\ \bibinfo {pages} {L041303} (\bibinfo {year} {2024}{\natexlab{b}})},\ \Eprint {http://arxiv.org/abs/2405.18747} {arXiv:2405.18747 [astro-ph.CO]} \BibitemShut {NoStop}%
\bibitem [{\citenamefont {Berghaus}\ \emph {et~al.}(2024)\citenamefont {Berghaus}, \citenamefont {Kable},\ and\ \citenamefont {Miranda}}]{Berghaus:2024kra}%
  \BibitemOpen
  \bibfield  {author} {\bibinfo {author} {\bibfnamefont {K.~V.}\ \bibnamefont {Berghaus}}, \bibinfo {author} {\bibfnamefont {J.~A.}\ \bibnamefont {Kable}}, \ and\ \bibinfo {author} {\bibfnamefont {V.}~\bibnamefont {Miranda}},\ }\href@noop {} {\  (\bibinfo {year} {2024})},\ \Eprint {http://arxiv.org/abs/2404.14341} {arXiv:2404.14341 [astro-ph.CO]} \BibitemShut {NoStop}%
\bibitem [{\citenamefont {Qu}\ \emph {et~al.}(2024{\natexlab{b}})\citenamefont {Qu}, \citenamefont {Surrao}, \citenamefont {Bolliet}, \citenamefont {Hill}, \citenamefont {Sherwin},\ and\ \citenamefont {Jense}}]{Qu:2024lpx}%
  \BibitemOpen
  \bibfield  {author} {\bibinfo {author} {\bibfnamefont {F.~J.}\ \bibnamefont {Qu}}, \bibinfo {author} {\bibfnamefont {K.~M.}\ \bibnamefont {Surrao}}, \bibinfo {author} {\bibfnamefont {B.}~\bibnamefont {Bolliet}}, \bibinfo {author} {\bibfnamefont {J.~C.}\ \bibnamefont {Hill}}, \bibinfo {author} {\bibfnamefont {B.~D.}\ \bibnamefont {Sherwin}}, \ and\ \bibinfo {author} {\bibfnamefont {H.~T.}\ \bibnamefont {Jense}},\ }\href@noop {} {\  (\bibinfo {year} {2024}{\natexlab{b}})},\ \Eprint {http://arxiv.org/abs/2404.16805} {arXiv:2404.16805 [astro-ph.CO]} \BibitemShut {NoStop}%
\bibitem [{\citenamefont {Adolf}\ \emph {et~al.}(2024)\citenamefont {Adolf}, \citenamefont {Hirsch}, \citenamefont {Krieg}, \citenamefont {P\"as},\ and\ \citenamefont {Tabet}}]{Adolf:2024twn}%
  \BibitemOpen
  \bibfield  {author} {\bibinfo {author} {\bibfnamefont {P.}~\bibnamefont {Adolf}}, \bibinfo {author} {\bibfnamefont {M.}~\bibnamefont {Hirsch}}, \bibinfo {author} {\bibfnamefont {S.}~\bibnamefont {Krieg}}, \bibinfo {author} {\bibfnamefont {H.}~\bibnamefont {P\"as}}, \ and\ \bibinfo {author} {\bibfnamefont {M.}~\bibnamefont {Tabet}},\ }\href {\doibase 10.1088/1475-7516/2024/08/048} {\bibfield  {journal} {\bibinfo  {journal} {JCAP}\ }\textbf {\bibinfo {volume} {08}},\ \bibinfo {pages} {048} (\bibinfo {year} {2024})},\ \Eprint {http://arxiv.org/abs/2406.09964} {arXiv:2406.09964 [astro-ph.CO]} \BibitemShut {NoStop}%
\bibitem [{\citenamefont {Bhattacharya}\ \emph {et~al.}(2024{\natexlab{a}})\citenamefont {Bhattacharya}, \citenamefont {Borghetto}, \citenamefont {Malhotra}, \citenamefont {Parameswaran}, \citenamefont {Tasinato},\ and\ \citenamefont {Zavala}}]{Bhattacharya:2024hep}%
  \BibitemOpen
  \bibfield  {author} {\bibinfo {author} {\bibfnamefont {S.}~\bibnamefont {Bhattacharya}}, \bibinfo {author} {\bibfnamefont {G.}~\bibnamefont {Borghetto}}, \bibinfo {author} {\bibfnamefont {A.}~\bibnamefont {Malhotra}}, \bibinfo {author} {\bibfnamefont {S.}~\bibnamefont {Parameswaran}}, \bibinfo {author} {\bibfnamefont {G.}~\bibnamefont {Tasinato}}, \ and\ \bibinfo {author} {\bibfnamefont {I.}~\bibnamefont {Zavala}},\ }\href {\doibase 10.1088/1475-7516/2024/09/073} {\bibfield  {journal} {\bibinfo  {journal} {JCAP}\ }\textbf {\bibinfo {volume} {09}},\ \bibinfo {pages} {073} (\bibinfo {year} {2024}{\natexlab{a}})},\ \Eprint {http://arxiv.org/abs/2405.17396} {arXiv:2405.17396 [astro-ph.CO]} \BibitemShut {NoStop}%
\bibitem [{\citenamefont {Bhattacharya}\ \emph {et~al.}(2024{\natexlab{b}})\citenamefont {Bhattacharya}, \citenamefont {Borghetto}, \citenamefont {Malhotra}, \citenamefont {Parameswaran}, \citenamefont {Tasinato},\ and\ \citenamefont {Zavala}}]{Bhattacharya:2024kxp}%
  \BibitemOpen
  \bibfield  {author} {\bibinfo {author} {\bibfnamefont {S.}~\bibnamefont {Bhattacharya}}, \bibinfo {author} {\bibfnamefont {G.}~\bibnamefont {Borghetto}}, \bibinfo {author} {\bibfnamefont {A.}~\bibnamefont {Malhotra}}, \bibinfo {author} {\bibfnamefont {S.}~\bibnamefont {Parameswaran}}, \bibinfo {author} {\bibfnamefont {G.}~\bibnamefont {Tasinato}}, \ and\ \bibinfo {author} {\bibfnamefont {I.}~\bibnamefont {Zavala}},\ }\href@noop {} {\  (\bibinfo {year} {2024}{\natexlab{b}})},\ \Eprint {http://arxiv.org/abs/2410.21243} {arXiv:2410.21243 [astro-ph.CO]} \BibitemShut {NoStop}%
\bibitem [{\citenamefont {Guth}(1981)}]{Guth:1980zm}%
  \BibitemOpen
  \bibfield  {author} {\bibinfo {author} {\bibfnamefont {A.~H.}\ \bibnamefont {Guth}},\ }\href {\doibase 10.1103/PhysRevD.23.347} {\bibfield  {journal} {\bibinfo  {journal} {Phys. Rev. D}\ }\textbf {\bibinfo {volume} {23}},\ \bibinfo {pages} {347} (\bibinfo {year} {1981})}\BibitemShut {NoStop}%
\bibitem [{\citenamefont {Linde}(1982)}]{Linde:1981mu}%
  \BibitemOpen
  \bibfield  {author} {\bibinfo {author} {\bibfnamefont {A.~D.}\ \bibnamefont {Linde}},\ }\href {\doibase 10.1016/0370-2693(82)91219-9} {\bibfield  {journal} {\bibinfo  {journal} {Phys. Lett. B}\ }\textbf {\bibinfo {volume} {108}},\ \bibinfo {pages} {389} (\bibinfo {year} {1982})}\BibitemShut {NoStop}%
\bibitem [{\citenamefont {Albrecht}\ and\ \citenamefont {Steinhardt}(1982)}]{Albrecht:1982wi}%
  \BibitemOpen
  \bibfield  {author} {\bibinfo {author} {\bibfnamefont {A.}~\bibnamefont {Albrecht}}\ and\ \bibinfo {author} {\bibfnamefont {P.~J.}\ \bibnamefont {Steinhardt}},\ }\href {\doibase 10.1103/PhysRevLett.48.1220} {\bibfield  {journal} {\bibinfo  {journal} {Phys. Rev. Lett.}\ }\textbf {\bibinfo {volume} {48}},\ \bibinfo {pages} {1220} (\bibinfo {year} {1982})}\BibitemShut {NoStop}%
\bibitem [{\citenamefont {Vilenkin}(1983)}]{Vilenkin:1983xq}%
  \BibitemOpen
  \bibfield  {author} {\bibinfo {author} {\bibfnamefont {A.}~\bibnamefont {Vilenkin}},\ }\href {\doibase 10.1103/PhysRevD.27.2848} {\bibfield  {journal} {\bibinfo  {journal} {Phys. Rev. D}\ }\textbf {\bibinfo {volume} {27}},\ \bibinfo {pages} {2848} (\bibinfo {year} {1983})}\BibitemShut {NoStop}%
\bibitem [{\citenamefont {Starobinsky}(1980)}]{Starobinsky:1980te}%
  \BibitemOpen
  \bibfield  {author} {\bibinfo {author} {\bibfnamefont {A.~A.}\ \bibnamefont {Starobinsky}},\ }\href {\doibase 10.1016/0370-2693(80)90670-X} {\bibfield  {journal} {\bibinfo  {journal} {Phys. Lett. B}\ }\textbf {\bibinfo {volume} {91}},\ \bibinfo {pages} {99} (\bibinfo {year} {1980})}\BibitemShut {NoStop}%
\bibitem [{\citenamefont {Qu}\ \emph {et~al.}(2024{\natexlab{c}})\citenamefont {Qu} \emph {et~al.}}]{ACT:2023dou}%
  \BibitemOpen
  \bibfield  {author} {\bibinfo {author} {\bibfnamefont {F.~J.}\ \bibnamefont {Qu}} \emph {et~al.} (\bibinfo {collaboration} {ACT}),\ }\href {\doibase 10.3847/1538-4357/acfe06} {\bibfield  {journal} {\bibinfo  {journal} {Astrophys. J.}\ }\textbf {\bibinfo {volume} {962}},\ \bibinfo {pages} {112} (\bibinfo {year} {2024}{\natexlab{c}})},\ \Eprint {http://arxiv.org/abs/2304.05202} {arXiv:2304.05202 [astro-ph.CO]} \BibitemShut {NoStop}%
\bibitem [{\citenamefont {Balkenhol}\ \emph {et~al.}(2023)\citenamefont {Balkenhol} \emph {et~al.}}]{SPT-3G:2022hvq}%
  \BibitemOpen
  \bibfield  {author} {\bibinfo {author} {\bibfnamefont {L.}~\bibnamefont {Balkenhol}} \emph {et~al.} (\bibinfo {collaboration} {SPT-3G}),\ }\href {\doibase 10.1103/PhysRevD.108.023510} {\bibfield  {journal} {\bibinfo  {journal} {Phys. Rev. D}\ }\textbf {\bibinfo {volume} {108}},\ \bibinfo {pages} {023510} (\bibinfo {year} {2023})},\ \Eprint {http://arxiv.org/abs/2212.05642} {arXiv:2212.05642 [astro-ph.CO]} \BibitemShut {NoStop}%
\bibitem [{\citenamefont {Lewis}\ and\ \citenamefont {Challinor}(2006)}]{Lewis:2006fu}%
  \BibitemOpen
  \bibfield  {author} {\bibinfo {author} {\bibfnamefont {A.}~\bibnamefont {Lewis}}\ and\ \bibinfo {author} {\bibfnamefont {A.}~\bibnamefont {Challinor}},\ }\href {\doibase 10.1016/j.physrep.2006.03.002} {\bibfield  {journal} {\bibinfo  {journal} {Phys. Rept.}\ }\textbf {\bibinfo {volume} {429}},\ \bibinfo {pages} {1} (\bibinfo {year} {2006})},\ \Eprint {http://arxiv.org/abs/astro-ph/0601594} {arXiv:astro-ph/0601594} \BibitemShut {NoStop}%
\bibitem [{\citenamefont {Giar\`e}\ \emph {et~al.}(2021)\citenamefont {Giar\`e}, \citenamefont {Di~Valentino}, \citenamefont {Melchiorri},\ and\ \citenamefont {Mena}}]{Giare:2020vzo}%
  \BibitemOpen
  \bibfield  {author} {\bibinfo {author} {\bibfnamefont {W.}~\bibnamefont {Giar\`e}}, \bibinfo {author} {\bibfnamefont {E.}~\bibnamefont {Di~Valentino}}, \bibinfo {author} {\bibfnamefont {A.}~\bibnamefont {Melchiorri}}, \ and\ \bibinfo {author} {\bibfnamefont {O.}~\bibnamefont {Mena}},\ }\href {\doibase 10.1093/mnras/stab1442} {\bibfield  {journal} {\bibinfo  {journal} {Mon. Not. Roy. Astron. Soc.}\ }\textbf {\bibinfo {volume} {505}},\ \bibinfo {pages} {2703} (\bibinfo {year} {2021})},\ \Eprint {http://arxiv.org/abs/2011.14704} {arXiv:2011.14704 [astro-ph.CO]} \BibitemShut {NoStop}%
\bibitem [{\citenamefont {Di~Valentino}\ and\ \citenamefont {Melchiorri}(2022)}]{DiValentino:2021imh}%
  \BibitemOpen
  \bibfield  {author} {\bibinfo {author} {\bibfnamefont {E.}~\bibnamefont {Di~Valentino}}\ and\ \bibinfo {author} {\bibfnamefont {A.}~\bibnamefont {Melchiorri}},\ }\href {\doibase 10.3847/2041-8213/ac6ef5} {\bibfield  {journal} {\bibinfo  {journal} {Astrophys. J. Lett.}\ }\textbf {\bibinfo {volume} {931}},\ \bibinfo {pages} {L18} (\bibinfo {year} {2022})},\ \Eprint {http://arxiv.org/abs/2112.02993} {arXiv:2112.02993 [astro-ph.CO]} \BibitemShut {NoStop}%
\bibitem [{\citenamefont {D'Eramo}\ \emph {et~al.}(2022)\citenamefont {D'Eramo}, \citenamefont {Di~Valentino}, \citenamefont {Giar\`e}, \citenamefont {Hajkarim}, \citenamefont {Melchiorri}, \citenamefont {Mena}, \citenamefont {Renzi},\ and\ \citenamefont {Yun}}]{DEramo:2022nvb}%
  \BibitemOpen
  \bibfield  {author} {\bibinfo {author} {\bibfnamefont {F.}~\bibnamefont {D'Eramo}}, \bibinfo {author} {\bibfnamefont {E.}~\bibnamefont {Di~Valentino}}, \bibinfo {author} {\bibfnamefont {W.}~\bibnamefont {Giar\`e}}, \bibinfo {author} {\bibfnamefont {F.}~\bibnamefont {Hajkarim}}, \bibinfo {author} {\bibfnamefont {A.}~\bibnamefont {Melchiorri}}, \bibinfo {author} {\bibfnamefont {O.}~\bibnamefont {Mena}}, \bibinfo {author} {\bibfnamefont {F.}~\bibnamefont {Renzi}}, \ and\ \bibinfo {author} {\bibfnamefont {S.}~\bibnamefont {Yun}},\ }\href {\doibase 10.1088/1475-7516/2022/09/022} {\bibfield  {journal} {\bibinfo  {journal} {JCAP}\ }\textbf {\bibinfo {volume} {09}},\ \bibinfo {pages} {022} (\bibinfo {year} {2022})},\ \Eprint {http://arxiv.org/abs/2205.07849} {arXiv:2205.07849 [astro-ph.CO]} \BibitemShut {NoStop}%
\bibitem [{\citenamefont {Di~Valentino}\ \emph {et~al.}(2023)\citenamefont {Di~Valentino}, \citenamefont {Gariazzo}, \citenamefont {Giar\`e}, \citenamefont {Melchiorri}, \citenamefont {Mena},\ and\ \citenamefont {Renzi}}]{DiValentino:2022edq}%
  \BibitemOpen
  \bibfield  {author} {\bibinfo {author} {\bibfnamefont {E.}~\bibnamefont {Di~Valentino}}, \bibinfo {author} {\bibfnamefont {S.}~\bibnamefont {Gariazzo}}, \bibinfo {author} {\bibfnamefont {W.}~\bibnamefont {Giar\`e}}, \bibinfo {author} {\bibfnamefont {A.}~\bibnamefont {Melchiorri}}, \bibinfo {author} {\bibfnamefont {O.}~\bibnamefont {Mena}}, \ and\ \bibinfo {author} {\bibfnamefont {F.}~\bibnamefont {Renzi}},\ }\href {\doibase 10.1103/PhysRevD.107.103528} {\bibfield  {journal} {\bibinfo  {journal} {Phys. Rev. D}\ }\textbf {\bibinfo {volume} {107}},\ \bibinfo {pages} {103528} (\bibinfo {year} {2023})},\ \Eprint {http://arxiv.org/abs/2212.11926} {arXiv:2212.11926 [astro-ph.CO]} \BibitemShut {NoStop}%
\bibitem [{\citenamefont {Giar\`e}\ \emph {et~al.}(2023)\citenamefont {Giar\`e}, \citenamefont {Mena},\ and\ \citenamefont {Di~Valentino}}]{Giare:2023aix}%
  \BibitemOpen
  \bibfield  {author} {\bibinfo {author} {\bibfnamefont {W.}~\bibnamefont {Giar\`e}}, \bibinfo {author} {\bibfnamefont {O.}~\bibnamefont {Mena}}, \ and\ \bibinfo {author} {\bibfnamefont {E.}~\bibnamefont {Di~Valentino}},\ }\href {\doibase 10.1103/PhysRevD.108.103539} {\bibfield  {journal} {\bibinfo  {journal} {Phys. Rev. D}\ }\textbf {\bibinfo {volume} {108}},\ \bibinfo {pages} {103539} (\bibinfo {year} {2023})},\ \Eprint {http://arxiv.org/abs/2307.14204} {arXiv:2307.14204 [astro-ph.CO]} \BibitemShut {NoStop}%
\bibitem [{\citenamefont {Aghanim}\ \emph {et~al.}(2020{\natexlab{c}})\citenamefont {Aghanim} \emph {et~al.}}]{Planck:2018lbu}%
  \BibitemOpen
  \bibfield  {author} {\bibinfo {author} {\bibfnamefont {N.}~\bibnamefont {Aghanim}} \emph {et~al.} (\bibinfo {collaboration} {Planck}),\ }\href {\doibase 10.1051/0004-6361/201833886} {\bibfield  {journal} {\bibinfo  {journal} {Astron. Astrophys.}\ }\textbf {\bibinfo {volume} {641}},\ \bibinfo {pages} {A8} (\bibinfo {year} {2020}{\natexlab{c}})},\ \Eprint {http://arxiv.org/abs/1807.06210} {arXiv:1807.06210 [astro-ph.CO]} \BibitemShut {NoStop}%
\bibitem [{\citenamefont {Ye}\ \emph {et~al.}(2023)\citenamefont {Ye}, \citenamefont {Jiang},\ and\ \citenamefont {Piao}}]{Ye:2023zel}%
  \BibitemOpen
  \bibfield  {author} {\bibinfo {author} {\bibfnamefont {G.}~\bibnamefont {Ye}}, \bibinfo {author} {\bibfnamefont {J.-Q.}\ \bibnamefont {Jiang}}, \ and\ \bibinfo {author} {\bibfnamefont {Y.-S.}\ \bibnamefont {Piao}},\ }\href {\doibase 10.1103/PhysRevD.108.063512} {\bibfield  {journal} {\bibinfo  {journal} {Phys. Rev. D}\ }\textbf {\bibinfo {volume} {108}},\ \bibinfo {pages} {063512} (\bibinfo {year} {2023})},\ \Eprint {http://arxiv.org/abs/2305.18873} {arXiv:2305.18873 [astro-ph.CO]} \BibitemShut {NoStop}%
\bibitem [{\citenamefont {Shaikh}\ \emph {et~al.}(2024)\citenamefont {Shaikh} \emph {et~al.}}]{ACT:2023skz}%
  \BibitemOpen
  \bibfield  {author} {\bibinfo {author} {\bibfnamefont {S.}~\bibnamefont {Shaikh}} \emph {et~al.} (\bibinfo {collaboration} {ACT, DES}),\ }\href {\doibase 10.1093/mnras/stad3987} {\bibfield  {journal} {\bibinfo  {journal} {Mon. Not. Roy. Astron. Soc.}\ }\textbf {\bibinfo {volume} {528}},\ \bibinfo {pages} {2112} (\bibinfo {year} {2024})},\ \Eprint {http://arxiv.org/abs/2309.04412} {arXiv:2309.04412 [astro-ph.CO]} \BibitemShut {NoStop}%
\bibitem [{\citenamefont {Marques}\ \emph {et~al.}(2024)\citenamefont {Marques} \emph {et~al.}}]{ACT:2023ipp}%
  \BibitemOpen
  \bibfield  {author} {\bibinfo {author} {\bibfnamefont {G.~A.}\ \bibnamefont {Marques}} \emph {et~al.} (\bibinfo {collaboration} {ACT, DES}),\ }\href {\doibase 10.1088/1475-7516/2024/01/033} {\bibfield  {journal} {\bibinfo  {journal} {JCAP}\ }\textbf {\bibinfo {volume} {01}},\ \bibinfo {pages} {033} (\bibinfo {year} {2024})},\ \Eprint {http://arxiv.org/abs/2306.17268} {arXiv:2306.17268 [astro-ph.CO]} \BibitemShut {NoStop}%
\bibitem [{\citenamefont {Kim}\ \emph {et~al.}(2024)\citenamefont {Kim} \emph {et~al.}}]{ACT:2024okh}%
  \BibitemOpen
  \bibfield  {author} {\bibinfo {author} {\bibfnamefont {J.}~\bibnamefont {Kim}} \emph {et~al.} (\bibinfo {collaboration} {ACT, DESI}),\ }\href@noop {} {\  (\bibinfo {year} {2024})},\ \Eprint {http://arxiv.org/abs/2407.04606} {arXiv:2407.04606 [astro-ph.CO]} \BibitemShut {NoStop}%
\bibitem [{\citenamefont {Sailer}\ \emph {et~al.}(2024)\citenamefont {Sailer} \emph {et~al.}}]{Sailer:2024coh}%
  \BibitemOpen
  \bibfield  {author} {\bibinfo {author} {\bibfnamefont {N.}~\bibnamefont {Sailer}} \emph {et~al.},\ }\href@noop {} {\  (\bibinfo {year} {2024})},\ \Eprint {http://arxiv.org/abs/2407.04607} {arXiv:2407.04607 [astro-ph.CO]} \BibitemShut {NoStop}%
\bibitem [{\citenamefont {Wenzl}\ \emph {et~al.}(2024)\citenamefont {Wenzl} \emph {et~al.}}]{ACT:2024npz}%
  \BibitemOpen
  \bibfield  {author} {\bibinfo {author} {\bibfnamefont {L.}~\bibnamefont {Wenzl}} \emph {et~al.} (\bibinfo {collaboration} {ACT}),\ }\href@noop {} {\  (\bibinfo {year} {2024})},\ \Eprint {http://arxiv.org/abs/2405.12795} {arXiv:2405.12795 [astro-ph.CO]} \BibitemShut {NoStop}%
\bibitem [{\citenamefont {Calabrese}\ \emph {et~al.}(2008)\citenamefont {Calabrese}, \citenamefont {Slosar}, \citenamefont {Melchiorri}, \citenamefont {Smoot},\ and\ \citenamefont {Zahn}}]{Calabrese:2008rt}%
  \BibitemOpen
  \bibfield  {author} {\bibinfo {author} {\bibfnamefont {E.}~\bibnamefont {Calabrese}}, \bibinfo {author} {\bibfnamefont {A.}~\bibnamefont {Slosar}}, \bibinfo {author} {\bibfnamefont {A.}~\bibnamefont {Melchiorri}}, \bibinfo {author} {\bibfnamefont {G.~F.}\ \bibnamefont {Smoot}}, \ and\ \bibinfo {author} {\bibfnamefont {O.}~\bibnamefont {Zahn}},\ }\href {\doibase 10.1103/PhysRevD.77.123531} {\bibfield  {journal} {\bibinfo  {journal} {Phys. Rev. D}\ }\textbf {\bibinfo {volume} {77}},\ \bibinfo {pages} {123531} (\bibinfo {year} {2008})},\ \Eprint {http://arxiv.org/abs/0803.2309} {arXiv:0803.2309 [astro-ph]} \BibitemShut {NoStop}%
\bibitem [{\citenamefont {Ade}\ \emph {et~al.}(2014)\citenamefont {Ade} \emph {et~al.}}]{Planck:2013pxb}%
  \BibitemOpen
  \bibfield  {author} {\bibinfo {author} {\bibfnamefont {P.~A.~R.}\ \bibnamefont {Ade}} \emph {et~al.} (\bibinfo {collaboration} {Planck}),\ }\href {\doibase 10.1051/0004-6361/201321591} {\bibfield  {journal} {\bibinfo  {journal} {Astron. Astrophys.}\ }\textbf {\bibinfo {volume} {571}},\ \bibinfo {pages} {A16} (\bibinfo {year} {2014})},\ \Eprint {http://arxiv.org/abs/1303.5076} {arXiv:1303.5076 [astro-ph.CO]} \BibitemShut {NoStop}%
\bibitem [{\citenamefont {Di~Valentino}\ \emph {et~al.}(2013)\citenamefont {Di~Valentino}, \citenamefont {Galli}, \citenamefont {Lattanzi}, \citenamefont {Melchiorri}, \citenamefont {Natoli}, \citenamefont {Pagano},\ and\ \citenamefont {Said}}]{DiValentino:2013mt}%
  \BibitemOpen
  \bibfield  {author} {\bibinfo {author} {\bibfnamefont {E.}~\bibnamefont {Di~Valentino}}, \bibinfo {author} {\bibfnamefont {S.}~\bibnamefont {Galli}}, \bibinfo {author} {\bibfnamefont {M.}~\bibnamefont {Lattanzi}}, \bibinfo {author} {\bibfnamefont {A.}~\bibnamefont {Melchiorri}}, \bibinfo {author} {\bibfnamefont {P.}~\bibnamefont {Natoli}}, \bibinfo {author} {\bibfnamefont {L.}~\bibnamefont {Pagano}}, \ and\ \bibinfo {author} {\bibfnamefont {N.}~\bibnamefont {Said}},\ }\href {\doibase 10.1103/PhysRevD.88.023501} {\bibfield  {journal} {\bibinfo  {journal} {Phys. Rev. D}\ }\textbf {\bibinfo {volume} {88}},\ \bibinfo {pages} {023501} (\bibinfo {year} {2013})},\ \Eprint {http://arxiv.org/abs/1301.7343} {arXiv:1301.7343 [astro-ph.CO]} \BibitemShut {NoStop}%
\bibitem [{\citenamefont {Ade}\ \emph {et~al.}(2016{\natexlab{a}})\citenamefont {Ade} \emph {et~al.}}]{Planck:2015fie}%
  \BibitemOpen
  \bibfield  {author} {\bibinfo {author} {\bibfnamefont {P.~A.~R.}\ \bibnamefont {Ade}} \emph {et~al.} (\bibinfo {collaboration} {Planck}),\ }\href {\doibase 10.1051/0004-6361/201525830} {\bibfield  {journal} {\bibinfo  {journal} {Astron. Astrophys.}\ }\textbf {\bibinfo {volume} {594}},\ \bibinfo {pages} {A13} (\bibinfo {year} {2016}{\natexlab{a}})},\ \Eprint {http://arxiv.org/abs/1502.01589} {arXiv:1502.01589 [astro-ph.CO]} \BibitemShut {NoStop}%
\bibitem [{\citenamefont {Addison}\ \emph {et~al.}(2016)\citenamefont {Addison}, \citenamefont {Huang}, \citenamefont {Watts}, \citenamefont {Bennett}, \citenamefont {Halpern}, \citenamefont {Hinshaw},\ and\ \citenamefont {Weiland}}]{Addison:2015wyg}%
  \BibitemOpen
  \bibfield  {author} {\bibinfo {author} {\bibfnamefont {G.~E.}\ \bibnamefont {Addison}}, \bibinfo {author} {\bibfnamefont {Y.}~\bibnamefont {Huang}}, \bibinfo {author} {\bibfnamefont {D.~J.}\ \bibnamefont {Watts}}, \bibinfo {author} {\bibfnamefont {C.~L.}\ \bibnamefont {Bennett}}, \bibinfo {author} {\bibfnamefont {M.}~\bibnamefont {Halpern}}, \bibinfo {author} {\bibfnamefont {G.}~\bibnamefont {Hinshaw}}, \ and\ \bibinfo {author} {\bibfnamefont {J.~L.}\ \bibnamefont {Weiland}},\ }\href {\doibase 10.3847/0004-637X/818/2/132} {\bibfield  {journal} {\bibinfo  {journal} {Astrophys. J.}\ }\textbf {\bibinfo {volume} {818}},\ \bibinfo {pages} {132} (\bibinfo {year} {2016})},\ \Eprint {http://arxiv.org/abs/1511.00055} {arXiv:1511.00055 [astro-ph.CO]} \BibitemShut {NoStop}%
\bibitem [{\citenamefont {Aghanim}\ \emph {et~al.}(2017)\citenamefont {Aghanim} \emph {et~al.}}]{Planck:2016tof}%
  \BibitemOpen
  \bibfield  {author} {\bibinfo {author} {\bibfnamefont {N.}~\bibnamefont {Aghanim}} \emph {et~al.} (\bibinfo {collaboration} {Planck}),\ }\href {\doibase 10.1051/0004-6361/201629504} {\bibfield  {journal} {\bibinfo  {journal} {Astron. Astrophys.}\ }\textbf {\bibinfo {volume} {607}},\ \bibinfo {pages} {A95} (\bibinfo {year} {2017})},\ \Eprint {http://arxiv.org/abs/1608.02487} {arXiv:1608.02487 [astro-ph.CO]} \BibitemShut {NoStop}%
\bibitem [{\citenamefont {Renzi}\ \emph {et~al.}(2018)\citenamefont {Renzi}, \citenamefont {Di~Valentino},\ and\ \citenamefont {Melchiorri}}]{Renzi:2017cbg}%
  \BibitemOpen
  \bibfield  {author} {\bibinfo {author} {\bibfnamefont {F.}~\bibnamefont {Renzi}}, \bibinfo {author} {\bibfnamefont {E.}~\bibnamefont {Di~Valentino}}, \ and\ \bibinfo {author} {\bibfnamefont {A.}~\bibnamefont {Melchiorri}},\ }\href {\doibase 10.1103/PhysRevD.97.123534} {\bibfield  {journal} {\bibinfo  {journal} {Phys. Rev. D}\ }\textbf {\bibinfo {volume} {97}},\ \bibinfo {pages} {123534} (\bibinfo {year} {2018})},\ \Eprint {http://arxiv.org/abs/1712.08758} {arXiv:1712.08758 [astro-ph.CO]} \BibitemShut {NoStop}%
\bibitem [{\citenamefont {Di~Valentino}\ \emph {et~al.}(2018)\citenamefont {Di~Valentino}, \citenamefont {Linder},\ and\ \citenamefont {Melchiorri}}]{DiValentino:2017rcr}%
  \BibitemOpen
  \bibfield  {author} {\bibinfo {author} {\bibfnamefont {E.}~\bibnamefont {Di~Valentino}}, \bibinfo {author} {\bibfnamefont {E.~V.}\ \bibnamefont {Linder}}, \ and\ \bibinfo {author} {\bibfnamefont {A.}~\bibnamefont {Melchiorri}},\ }\href {\doibase 10.1103/PhysRevD.97.043528} {\bibfield  {journal} {\bibinfo  {journal} {Phys. Rev. D}\ }\textbf {\bibinfo {volume} {97}},\ \bibinfo {pages} {043528} (\bibinfo {year} {2018})},\ \Eprint {http://arxiv.org/abs/1710.02153} {arXiv:1710.02153 [astro-ph.CO]} \BibitemShut {NoStop}%
\bibitem [{\citenamefont {Di~Valentino}\ \emph {et~al.}(2019)\citenamefont {Di~Valentino}, \citenamefont {Melchiorri},\ and\ \citenamefont {Silk}}]{DiValentino:2019qzk}%
  \BibitemOpen
  \bibfield  {author} {\bibinfo {author} {\bibfnamefont {E.}~\bibnamefont {Di~Valentino}}, \bibinfo {author} {\bibfnamefont {A.}~\bibnamefont {Melchiorri}}, \ and\ \bibinfo {author} {\bibfnamefont {J.}~\bibnamefont {Silk}},\ }\href {\doibase 10.1038/s41550-019-0906-9} {\bibfield  {journal} {\bibinfo  {journal} {Nature Astron.}\ }\textbf {\bibinfo {volume} {4}},\ \bibinfo {pages} {196} (\bibinfo {year} {2019})},\ \Eprint {http://arxiv.org/abs/1911.02087} {arXiv:1911.02087 [astro-ph.CO]} \BibitemShut {NoStop}%
\bibitem [{\citenamefont {Addison}\ \emph {et~al.}(2024)\citenamefont {Addison}, \citenamefont {Bennett}, \citenamefont {Halpern}, \citenamefont {Hinshaw},\ and\ \citenamefont {Weiland}}]{Addison:2023fqc}%
  \BibitemOpen
  \bibfield  {author} {\bibinfo {author} {\bibfnamefont {G.~E.}\ \bibnamefont {Addison}}, \bibinfo {author} {\bibfnamefont {C.~L.}\ \bibnamefont {Bennett}}, \bibinfo {author} {\bibfnamefont {M.}~\bibnamefont {Halpern}}, \bibinfo {author} {\bibfnamefont {G.}~\bibnamefont {Hinshaw}}, \ and\ \bibinfo {author} {\bibfnamefont {J.~L.}\ \bibnamefont {Weiland}},\ }\href {\doibase 10.3847/1538-4357/ad6d61} {\bibfield  {journal} {\bibinfo  {journal} {Astrophys. J.}\ }\textbf {\bibinfo {volume} {974}},\ \bibinfo {pages} {187} (\bibinfo {year} {2024})},\ \Eprint {http://arxiv.org/abs/2310.03127} {arXiv:2310.03127 [astro-ph.CO]} \BibitemShut {NoStop}%
\bibitem [{\citenamefont {Park}\ and\ \citenamefont {Ratra}(2019)}]{Park:2017xbl}%
  \BibitemOpen
  \bibfield  {author} {\bibinfo {author} {\bibfnamefont {C.-G.}\ \bibnamefont {Park}}\ and\ \bibinfo {author} {\bibfnamefont {B.}~\bibnamefont {Ratra}},\ }\href {\doibase 10.3847/1538-4357/ab3641} {\bibfield  {journal} {\bibinfo  {journal} {Astrophys. J.}\ }\textbf {\bibinfo {volume} {882}},\ \bibinfo {pages} {158} (\bibinfo {year} {2019})},\ \Eprint {http://arxiv.org/abs/1801.00213} {arXiv:1801.00213 [astro-ph.CO]} \BibitemShut {NoStop}%
\bibitem [{\citenamefont {Handley}(2021)}]{Handley:2019tkm}%
  \BibitemOpen
  \bibfield  {author} {\bibinfo {author} {\bibfnamefont {W.}~\bibnamefont {Handley}},\ }\href {\doibase 10.1103/PhysRevD.103.L041301} {\bibfield  {journal} {\bibinfo  {journal} {Phys. Rev. D}\ }\textbf {\bibinfo {volume} {103}},\ \bibinfo {pages} {L041301} (\bibinfo {year} {2021})},\ \Eprint {http://arxiv.org/abs/1908.09139} {arXiv:1908.09139 [astro-ph.CO]} \BibitemShut {NoStop}%
\bibitem [{\citenamefont {Efstathiou}\ and\ \citenamefont {Gratton}(2020)}]{Efstathiou:2020wem}%
  \BibitemOpen
  \bibfield  {author} {\bibinfo {author} {\bibfnamefont {G.}~\bibnamefont {Efstathiou}}\ and\ \bibinfo {author} {\bibfnamefont {S.}~\bibnamefont {Gratton}},\ }\href {\doibase 10.1093/mnrasl/slaa093} {\bibfield  {journal} {\bibinfo  {journal} {Mon. Not. Roy. Astron. Soc.}\ }\textbf {\bibinfo {volume} {496}},\ \bibinfo {pages} {L91} (\bibinfo {year} {2020})},\ \Eprint {http://arxiv.org/abs/2002.06892} {arXiv:2002.06892 [astro-ph.CO]} \BibitemShut {NoStop}%
\bibitem [{\citenamefont {Di~Valentino}\ \emph {et~al.}(2021{\natexlab{d}})\citenamefont {Di~Valentino}, \citenamefont {Melchiorri},\ and\ \citenamefont {Silk}}]{DiValentino:2020hov}%
  \BibitemOpen
  \bibfield  {author} {\bibinfo {author} {\bibfnamefont {E.}~\bibnamefont {Di~Valentino}}, \bibinfo {author} {\bibfnamefont {A.}~\bibnamefont {Melchiorri}}, \ and\ \bibinfo {author} {\bibfnamefont {J.}~\bibnamefont {Silk}},\ }\href {\doibase 10.3847/2041-8213/abe1c4} {\bibfield  {journal} {\bibinfo  {journal} {Astrophys. J. Lett.}\ }\textbf {\bibinfo {volume} {908}},\ \bibinfo {pages} {L9} (\bibinfo {year} {2021}{\natexlab{d}})},\ \Eprint {http://arxiv.org/abs/2003.04935} {arXiv:2003.04935 [astro-ph.CO]} \BibitemShut {NoStop}%
\bibitem [{\citenamefont {Benisty}\ and\ \citenamefont {Staicova}(2021)}]{Benisty:2020otr}%
  \BibitemOpen
  \bibfield  {author} {\bibinfo {author} {\bibfnamefont {D.}~\bibnamefont {Benisty}}\ and\ \bibinfo {author} {\bibfnamefont {D.}~\bibnamefont {Staicova}},\ }\href {\doibase 10.1051/0004-6361/202039502} {\bibfield  {journal} {\bibinfo  {journal} {Astron. Astrophys.}\ }\textbf {\bibinfo {volume} {647}},\ \bibinfo {pages} {A38} (\bibinfo {year} {2021})},\ \Eprint {http://arxiv.org/abs/2009.10701} {arXiv:2009.10701 [astro-ph.CO]} \BibitemShut {NoStop}%
\bibitem [{\citenamefont {Vagnozzi}\ \emph {et~al.}(2021{\natexlab{a}})\citenamefont {Vagnozzi}, \citenamefont {Di~Valentino}, \citenamefont {Gariazzo}, \citenamefont {Melchiorri}, \citenamefont {Mena},\ and\ \citenamefont {Silk}}]{Vagnozzi:2020rcz}%
  \BibitemOpen
  \bibfield  {author} {\bibinfo {author} {\bibfnamefont {S.}~\bibnamefont {Vagnozzi}}, \bibinfo {author} {\bibfnamefont {E.}~\bibnamefont {Di~Valentino}}, \bibinfo {author} {\bibfnamefont {S.}~\bibnamefont {Gariazzo}}, \bibinfo {author} {\bibfnamefont {A.}~\bibnamefont {Melchiorri}}, \bibinfo {author} {\bibfnamefont {O.}~\bibnamefont {Mena}}, \ and\ \bibinfo {author} {\bibfnamefont {J.}~\bibnamefont {Silk}},\ }\href {\doibase 10.1016/j.dark.2021.100851} {\bibfield  {journal} {\bibinfo  {journal} {Phys. Dark Univ.}\ }\textbf {\bibinfo {volume} {33}},\ \bibinfo {pages} {100851} (\bibinfo {year} {2021}{\natexlab{a}})},\ \Eprint {http://arxiv.org/abs/2010.02230} {arXiv:2010.02230 [astro-ph.CO]} \BibitemShut {NoStop}%
\bibitem [{\citenamefont {Vagnozzi}\ \emph {et~al.}(2021{\natexlab{b}})\citenamefont {Vagnozzi}, \citenamefont {Loeb},\ and\ \citenamefont {Moresco}}]{Vagnozzi:2020dfn}%
  \BibitemOpen
  \bibfield  {author} {\bibinfo {author} {\bibfnamefont {S.}~\bibnamefont {Vagnozzi}}, \bibinfo {author} {\bibfnamefont {A.}~\bibnamefont {Loeb}}, \ and\ \bibinfo {author} {\bibfnamefont {M.}~\bibnamefont {Moresco}},\ }\href {\doibase 10.3847/1538-4357/abd4df} {\bibfield  {journal} {\bibinfo  {journal} {Astrophys. J.}\ }\textbf {\bibinfo {volume} {908}},\ \bibinfo {pages} {84} (\bibinfo {year} {2021}{\natexlab{b}})},\ \Eprint {http://arxiv.org/abs/2011.11645} {arXiv:2011.11645 [astro-ph.CO]} \BibitemShut {NoStop}%
\bibitem [{\citenamefont {Di~Valentino}\ \emph {et~al.}(2021{\natexlab{e}})\citenamefont {Di~Valentino}, \citenamefont {Melchiorri}, \citenamefont {Mena}, \citenamefont {Pan},\ and\ \citenamefont {Yang}}]{DiValentino:2020kpf}%
  \BibitemOpen
  \bibfield  {author} {\bibinfo {author} {\bibfnamefont {E.}~\bibnamefont {Di~Valentino}}, \bibinfo {author} {\bibfnamefont {A.}~\bibnamefont {Melchiorri}}, \bibinfo {author} {\bibfnamefont {O.}~\bibnamefont {Mena}}, \bibinfo {author} {\bibfnamefont {S.}~\bibnamefont {Pan}}, \ and\ \bibinfo {author} {\bibfnamefont {W.}~\bibnamefont {Yang}},\ }\href {\doibase 10.1093/mnrasl/slaa207} {\bibfield  {journal} {\bibinfo  {journal} {Mon. Not. Roy. Astron. Soc.}\ }\textbf {\bibinfo {volume} {502}},\ \bibinfo {pages} {L23} (\bibinfo {year} {2021}{\natexlab{e}})},\ \Eprint {http://arxiv.org/abs/2011.00283} {arXiv:2011.00283 [astro-ph.CO]} \BibitemShut {NoStop}%
\bibitem [{\citenamefont {Yang}\ \emph {et~al.}(2021)\citenamefont {Yang}, \citenamefont {Pan}, \citenamefont {Di~Valentino}, \citenamefont {Mena},\ and\ \citenamefont {Melchiorri}}]{Yang:2021hxg}%
  \BibitemOpen
  \bibfield  {author} {\bibinfo {author} {\bibfnamefont {W.}~\bibnamefont {Yang}}, \bibinfo {author} {\bibfnamefont {S.}~\bibnamefont {Pan}}, \bibinfo {author} {\bibfnamefont {E.}~\bibnamefont {Di~Valentino}}, \bibinfo {author} {\bibfnamefont {O.}~\bibnamefont {Mena}}, \ and\ \bibinfo {author} {\bibfnamefont {A.}~\bibnamefont {Melchiorri}},\ }\href {\doibase 10.1088/1475-7516/2021/10/008} {\bibfield  {journal} {\bibinfo  {journal} {JCAP}\ }\textbf {\bibinfo {volume} {10}},\ \bibinfo {pages} {008} (\bibinfo {year} {2021})},\ \Eprint {http://arxiv.org/abs/2101.03129} {arXiv:2101.03129 [astro-ph.CO]} \BibitemShut {NoStop}%
\bibitem [{\citenamefont {Cao}\ \emph {et~al.}(2021)\citenamefont {Cao}, \citenamefont {Ryan},\ and\ \citenamefont {Ratra}}]{Cao:2021ldv}%
  \BibitemOpen
  \bibfield  {author} {\bibinfo {author} {\bibfnamefont {S.}~\bibnamefont {Cao}}, \bibinfo {author} {\bibfnamefont {J.}~\bibnamefont {Ryan}}, \ and\ \bibinfo {author} {\bibfnamefont {B.}~\bibnamefont {Ratra}},\ }\href {\doibase 10.1093/mnras/stab942} {\bibfield  {journal} {\bibinfo  {journal} {Mon. Not. Roy. Astron. Soc.}\ }\textbf {\bibinfo {volume} {504}},\ \bibinfo {pages} {300} (\bibinfo {year} {2021})},\ \Eprint {http://arxiv.org/abs/2101.08817} {arXiv:2101.08817 [astro-ph.CO]} \BibitemShut {NoStop}%
\bibitem [{\citenamefont {Dhawan}\ \emph {et~al.}(2021)\citenamefont {Dhawan}, \citenamefont {Alsing},\ and\ \citenamefont {Vagnozzi}}]{Dhawan:2021mel}%
  \BibitemOpen
  \bibfield  {author} {\bibinfo {author} {\bibfnamefont {S.}~\bibnamefont {Dhawan}}, \bibinfo {author} {\bibfnamefont {J.}~\bibnamefont {Alsing}}, \ and\ \bibinfo {author} {\bibfnamefont {S.}~\bibnamefont {Vagnozzi}},\ }\href {\doibase 10.1093/mnrasl/slab058} {\bibfield  {journal} {\bibinfo  {journal} {Mon. Not. Roy. Astron. Soc.}\ }\textbf {\bibinfo {volume} {506}},\ \bibinfo {pages} {L1} (\bibinfo {year} {2021})},\ \Eprint {http://arxiv.org/abs/2104.02485} {arXiv:2104.02485 [astro-ph.CO]} \BibitemShut {NoStop}%
\bibitem [{\citenamefont {Dinda}(2022)}]{Dinda:2021ffa}%
  \BibitemOpen
  \bibfield  {author} {\bibinfo {author} {\bibfnamefont {B.~R.}\ \bibnamefont {Dinda}},\ }\href {\doibase 10.1103/PhysRevD.105.063524} {\bibfield  {journal} {\bibinfo  {journal} {Phys. Rev. D}\ }\textbf {\bibinfo {volume} {105}},\ \bibinfo {pages} {063524} (\bibinfo {year} {2022})},\ \Eprint {http://arxiv.org/abs/2106.02963} {arXiv:2106.02963 [astro-ph.CO]} \BibitemShut {NoStop}%
\bibitem [{\citenamefont {Gonzalez}\ \emph {et~al.}(2021)\citenamefont {Gonzalez}, \citenamefont {Benetti}, \citenamefont {von Marttens},\ and\ \citenamefont {Alcaniz}}]{Gonzalez:2021ojp}%
  \BibitemOpen
  \bibfield  {author} {\bibinfo {author} {\bibfnamefont {J.~E.}\ \bibnamefont {Gonzalez}}, \bibinfo {author} {\bibfnamefont {M.}~\bibnamefont {Benetti}}, \bibinfo {author} {\bibfnamefont {R.}~\bibnamefont {von Marttens}}, \ and\ \bibinfo {author} {\bibfnamefont {J.}~\bibnamefont {Alcaniz}},\ }\href {\doibase 10.1088/1475-7516/2021/11/060} {\bibfield  {journal} {\bibinfo  {journal} {JCAP}\ }\textbf {\bibinfo {volume} {11}},\ \bibinfo {pages} {060} (\bibinfo {year} {2021})},\ \Eprint {http://arxiv.org/abs/2104.13455} {arXiv:2104.13455 [astro-ph.CO]} \BibitemShut {NoStop}%
\bibitem [{\citenamefont {Akarsu}\ \emph {et~al.}(2023{\natexlab{c}})\citenamefont {Akarsu}, \citenamefont {Di~Valentino}, \citenamefont {Kumar}, \citenamefont {Ozyigit},\ and\ \citenamefont {Sharma}}]{Akarsu:2021max}%
  \BibitemOpen
  \bibfield  {author} {\bibinfo {author} {\bibfnamefont {O.}~\bibnamefont {Akarsu}}, \bibinfo {author} {\bibfnamefont {E.}~\bibnamefont {Di~Valentino}}, \bibinfo {author} {\bibfnamefont {S.}~\bibnamefont {Kumar}}, \bibinfo {author} {\bibfnamefont {M.}~\bibnamefont {Ozyigit}}, \ and\ \bibinfo {author} {\bibfnamefont {S.}~\bibnamefont {Sharma}},\ }\href {\doibase 10.1016/j.dark.2022.101162} {\bibfield  {journal} {\bibinfo  {journal} {Phys. Dark Univ.}\ }\textbf {\bibinfo {volume} {39}},\ \bibinfo {pages} {101162} (\bibinfo {year} {2023}{\natexlab{c}})},\ \Eprint {http://arxiv.org/abs/2112.07807} {arXiv:2112.07807 [astro-ph.CO]} \BibitemShut {NoStop}%
\bibitem [{\citenamefont {Cao}\ and\ \citenamefont {Ratra}(2022)}]{Cao:2022ugh}%
  \BibitemOpen
  \bibfield  {author} {\bibinfo {author} {\bibfnamefont {S.}~\bibnamefont {Cao}}\ and\ \bibinfo {author} {\bibfnamefont {B.}~\bibnamefont {Ratra}},\ }\href {\doibase 10.1093/mnras/stac1184} {\bibfield  {journal} {\bibinfo  {journal} {Mon. Not. Roy. Astron. Soc.}\ }\textbf {\bibinfo {volume} {513}},\ \bibinfo {pages} {5686} (\bibinfo {year} {2022})},\ \Eprint {http://arxiv.org/abs/2203.10825} {arXiv:2203.10825 [astro-ph.CO]} \BibitemShut {NoStop}%
\bibitem [{\citenamefont {Glanville}\ \emph {et~al.}(2022)\citenamefont {Glanville}, \citenamefont {Howlett},\ and\ \citenamefont {Davis}}]{Glanville:2022xes}%
  \BibitemOpen
  \bibfield  {author} {\bibinfo {author} {\bibfnamefont {A.}~\bibnamefont {Glanville}}, \bibinfo {author} {\bibfnamefont {C.}~\bibnamefont {Howlett}}, \ and\ \bibinfo {author} {\bibfnamefont {T.~M.}\ \bibnamefont {Davis}},\ }\href {\doibase 10.1093/mnras/stac2891} {\bibfield  {journal} {\bibinfo  {journal} {Mon. Not. Roy. Astron. Soc.}\ }\textbf {\bibinfo {volume} {517}},\ \bibinfo {pages} {3087} (\bibinfo {year} {2022})},\ \Eprint {http://arxiv.org/abs/2205.05892} {arXiv:2205.05892 [astro-ph.CO]} \BibitemShut {NoStop}%
\bibitem [{\citenamefont {Bel}\ \emph {et~al.}(2022)\citenamefont {Bel}, \citenamefont {Larena}, \citenamefont {Maartens}, \citenamefont {Marinoni},\ and\ \citenamefont {Perenon}}]{Bel:2022iuf}%
  \BibitemOpen
  \bibfield  {author} {\bibinfo {author} {\bibfnamefont {J.}~\bibnamefont {Bel}}, \bibinfo {author} {\bibfnamefont {J.}~\bibnamefont {Larena}}, \bibinfo {author} {\bibfnamefont {R.}~\bibnamefont {Maartens}}, \bibinfo {author} {\bibfnamefont {C.}~\bibnamefont {Marinoni}}, \ and\ \bibinfo {author} {\bibfnamefont {L.}~\bibnamefont {Perenon}},\ }\href {\doibase 10.1088/1475-7516/2022/09/076} {\bibfield  {journal} {\bibinfo  {journal} {JCAP}\ }\textbf {\bibinfo {volume} {09}},\ \bibinfo {pages} {076} (\bibinfo {year} {2022})},\ \Eprint {http://arxiv.org/abs/2206.03059} {arXiv:2206.03059 [astro-ph.CO]} \BibitemShut {NoStop}%
\bibitem [{\citenamefont {Yang}\ \emph {et~al.}(2023)\citenamefont {Yang}, \citenamefont {Giar\`e}, \citenamefont {Pan}, \citenamefont {Di~Valentino}, \citenamefont {Melchiorri},\ and\ \citenamefont {Silk}}]{Yang:2022kho}%
  \BibitemOpen
  \bibfield  {author} {\bibinfo {author} {\bibfnamefont {W.}~\bibnamefont {Yang}}, \bibinfo {author} {\bibfnamefont {W.}~\bibnamefont {Giar\`e}}, \bibinfo {author} {\bibfnamefont {S.}~\bibnamefont {Pan}}, \bibinfo {author} {\bibfnamefont {E.}~\bibnamefont {Di~Valentino}}, \bibinfo {author} {\bibfnamefont {A.}~\bibnamefont {Melchiorri}}, \ and\ \bibinfo {author} {\bibfnamefont {J.}~\bibnamefont {Silk}},\ }\href {\doibase 10.1103/PhysRevD.107.063509} {\bibfield  {journal} {\bibinfo  {journal} {Phys. Rev. D}\ }\textbf {\bibinfo {volume} {107}},\ \bibinfo {pages} {063509} (\bibinfo {year} {2023})},\ \Eprint {http://arxiv.org/abs/2210.09865} {arXiv:2210.09865 [astro-ph.CO]} \BibitemShut {NoStop}%
\bibitem [{\citenamefont {Stevens}\ \emph {et~al.}(2023)\citenamefont {Stevens}, \citenamefont {Khoraminezhad},\ and\ \citenamefont {Saito}}]{Stevens:2022evv}%
  \BibitemOpen
  \bibfield  {author} {\bibinfo {author} {\bibfnamefont {J.}~\bibnamefont {Stevens}}, \bibinfo {author} {\bibfnamefont {H.}~\bibnamefont {Khoraminezhad}}, \ and\ \bibinfo {author} {\bibfnamefont {S.}~\bibnamefont {Saito}},\ }\href {\doibase 10.1088/1475-7516/2023/07/046} {\bibfield  {journal} {\bibinfo  {journal} {JCAP}\ }\textbf {\bibinfo {volume} {07}},\ \bibinfo {pages} {046} (\bibinfo {year} {2023})},\ \Eprint {http://arxiv.org/abs/2212.09804} {arXiv:2212.09804 [astro-ph.CO]} \BibitemShut {NoStop}%
\bibitem [{\citenamefont {Favale}\ \emph {et~al.}(2023)\citenamefont {Favale}, \citenamefont {G\'omez-Valent},\ and\ \citenamefont {Migliaccio}}]{Favale:2023lnp}%
  \BibitemOpen
  \bibfield  {author} {\bibinfo {author} {\bibfnamefont {A.}~\bibnamefont {Favale}}, \bibinfo {author} {\bibfnamefont {A.}~\bibnamefont {G\'omez-Valent}}, \ and\ \bibinfo {author} {\bibfnamefont {M.}~\bibnamefont {Migliaccio}},\ }\href {\doibase 10.1093/mnras/stad1621} {\bibfield  {journal} {\bibinfo  {journal} {Mon. Not. Roy. Astron. Soc.}\ }\textbf {\bibinfo {volume} {523}},\ \bibinfo {pages} {3406} (\bibinfo {year} {2023})},\ \Eprint {http://arxiv.org/abs/2301.09591} {arXiv:2301.09591 [astro-ph.CO]} \BibitemShut {NoStop}%
\bibitem [{\citenamefont {Giar\`e}\ \emph {et~al.}(2024{\natexlab{e}})\citenamefont {Giar\`e}, \citenamefont {Di~Valentino},\ and\ \citenamefont {Melchiorri}}]{Giare:2023ejv}%
  \BibitemOpen
  \bibfield  {author} {\bibinfo {author} {\bibfnamefont {W.}~\bibnamefont {Giar\`e}}, \bibinfo {author} {\bibfnamefont {E.}~\bibnamefont {Di~Valentino}}, \ and\ \bibinfo {author} {\bibfnamefont {A.}~\bibnamefont {Melchiorri}},\ }\href {\doibase 10.1103/PhysRevD.109.103519} {\bibfield  {journal} {\bibinfo  {journal} {Phys. Rev. D}\ }\textbf {\bibinfo {volume} {109}},\ \bibinfo {pages} {103519} (\bibinfo {year} {2024}{\natexlab{e}})},\ \Eprint {http://arxiv.org/abs/2312.06482} {arXiv:2312.06482 [astro-ph.CO]} \BibitemShut {NoStop}%
\bibitem [{\citenamefont {Ishak}(2019)}]{Ishak:2018his}%
  \BibitemOpen
  \bibfield  {author} {\bibinfo {author} {\bibfnamefont {M.}~\bibnamefont {Ishak}},\ }\href {\doibase 10.1007/s41114-018-0017-4} {\bibfield  {journal} {\bibinfo  {journal} {Living Rev. Rel.}\ }\textbf {\bibinfo {volume} {22}},\ \bibinfo {pages} {1} (\bibinfo {year} {2019})},\ \Eprint {http://arxiv.org/abs/1806.10122} {arXiv:1806.10122 [astro-ph.CO]} \BibitemShut {NoStop}%
\bibitem [{\citenamefont {Heisenberg}(2019)}]{Heisenberg:2018vsk}%
  \BibitemOpen
  \bibfield  {author} {\bibinfo {author} {\bibfnamefont {L.}~\bibnamefont {Heisenberg}},\ }\href {\doibase 10.1016/j.physrep.2018.11.006} {\bibfield  {journal} {\bibinfo  {journal} {Phys. Rept.}\ }\textbf {\bibinfo {volume} {796}},\ \bibinfo {pages} {1} (\bibinfo {year} {2019})},\ \Eprint {http://arxiv.org/abs/1807.01725} {arXiv:1807.01725 [gr-qc]} \BibitemShut {NoStop}%
\bibitem [{\citenamefont {Akrami}\ \emph {et~al.}(2021)\citenamefont {Akrami} \emph {et~al.}}]{CANTATA:2021asi}%
  \BibitemOpen
  \bibfield  {author} {\bibinfo {author} {\bibfnamefont {Y.}~\bibnamefont {Akrami}} \emph {et~al.} (\bibinfo {collaboration} {CANTATA}),\ }\href {\doibase 10.1007/978-3-030-83715-0} {\emph {\bibinfo {title} {{Modified Gravity and Cosmology. An Update by the CANTATA Network}}}},\ edited by\ \bibinfo {editor} {\bibfnamefont {E.~N.}\ \bibnamefont {Saridakis}}, \bibinfo {editor} {\bibfnamefont {R.}~\bibnamefont {Lazkoz}}, \bibinfo {editor} {\bibfnamefont {V.}~\bibnamefont {Salzano}}, \bibinfo {editor} {\bibfnamefont {P.}~\bibnamefont {Vargas~Moniz}}, \bibinfo {editor} {\bibfnamefont {S.}~\bibnamefont {Capozziello}}, \bibinfo {editor} {\bibfnamefont {J.}~\bibnamefont {Beltr\'an~Jim\'enez}}, \bibinfo {editor} {\bibfnamefont {M.}~\bibnamefont {De~Laurentis}}, \ and\ \bibinfo {editor} {\bibfnamefont {G.~J.}\ \bibnamefont {Olmo}}\ (\bibinfo  {publisher} {Springer},\ \bibinfo {year} {2021})\ \Eprint {http://arxiv.org/abs/2105.12582} {arXiv:2105.12582 [gr-qc]} \BibitemShut {NoStop}%
\bibitem [{\citenamefont {Nojiri}\ \emph {et~al.}(2017)\citenamefont {Nojiri}, \citenamefont {Odintsov},\ and\ \citenamefont {Oikonomou}}]{Nojiri:2017ncd}%
  \BibitemOpen
  \bibfield  {author} {\bibinfo {author} {\bibfnamefont {S.}~\bibnamefont {Nojiri}}, \bibinfo {author} {\bibfnamefont {S.~D.}\ \bibnamefont {Odintsov}}, \ and\ \bibinfo {author} {\bibfnamefont {V.~K.}\ \bibnamefont {Oikonomou}},\ }\href {\doibase 10.1016/j.physrep.2017.06.001} {\bibfield  {journal} {\bibinfo  {journal} {Phys. Rept.}\ }\textbf {\bibinfo {volume} {692}},\ \bibinfo {pages} {1} (\bibinfo {year} {2017})},\ \Eprint {http://arxiv.org/abs/1705.11098} {arXiv:1705.11098 [gr-qc]} \BibitemShut {NoStop}%
\bibitem [{\citenamefont {Zhang}\ \emph {et~al.}(2007)\citenamefont {Zhang}, \citenamefont {Liguori}, \citenamefont {Bean},\ and\ \citenamefont {Dodelson}}]{Zhang:2007nk}%
  \BibitemOpen
  \bibfield  {author} {\bibinfo {author} {\bibfnamefont {P.}~\bibnamefont {Zhang}}, \bibinfo {author} {\bibfnamefont {M.}~\bibnamefont {Liguori}}, \bibinfo {author} {\bibfnamefont {R.}~\bibnamefont {Bean}}, \ and\ \bibinfo {author} {\bibfnamefont {S.}~\bibnamefont {Dodelson}},\ }\href {\doibase 10.1103/PhysRevLett.99.141302} {\bibfield  {journal} {\bibinfo  {journal} {Phys. Rev. Lett.}\ }\textbf {\bibinfo {volume} {99}},\ \bibinfo {pages} {141302} (\bibinfo {year} {2007})},\ \Eprint {http://arxiv.org/abs/0704.1932} {arXiv:0704.1932 [astro-ph]} \BibitemShut {NoStop}%
\bibitem [{\citenamefont {Amendola}\ \emph {et~al.}(2008)\citenamefont {Amendola}, \citenamefont {Kunz},\ and\ \citenamefont {Sapone}}]{Amendola:2007rr}%
  \BibitemOpen
  \bibfield  {author} {\bibinfo {author} {\bibfnamefont {L.}~\bibnamefont {Amendola}}, \bibinfo {author} {\bibfnamefont {M.}~\bibnamefont {Kunz}}, \ and\ \bibinfo {author} {\bibfnamefont {D.}~\bibnamefont {Sapone}},\ }\href {\doibase 10.1088/1475-7516/2008/04/013} {\bibfield  {journal} {\bibinfo  {journal} {JCAP}\ }\textbf {\bibinfo {volume} {04}},\ \bibinfo {pages} {013} (\bibinfo {year} {2008})},\ \Eprint {http://arxiv.org/abs/0704.2421} {arXiv:0704.2421 [astro-ph]} \BibitemShut {NoStop}%
\bibitem [{\citenamefont {Zhao}\ \emph {et~al.}(2009)\citenamefont {Zhao}, \citenamefont {Pogosian}, \citenamefont {Silvestri},\ and\ \citenamefont {Zylberberg}}]{Zhao:2008bn}%
  \BibitemOpen
  \bibfield  {author} {\bibinfo {author} {\bibfnamefont {G.-B.}\ \bibnamefont {Zhao}}, \bibinfo {author} {\bibfnamefont {L.}~\bibnamefont {Pogosian}}, \bibinfo {author} {\bibfnamefont {A.}~\bibnamefont {Silvestri}}, \ and\ \bibinfo {author} {\bibfnamefont {J.}~\bibnamefont {Zylberberg}},\ }\href {\doibase 10.1103/PhysRevD.79.083513} {\bibfield  {journal} {\bibinfo  {journal} {Phys. Rev. D}\ }\textbf {\bibinfo {volume} {79}},\ \bibinfo {pages} {083513} (\bibinfo {year} {2009})},\ \Eprint {http://arxiv.org/abs/0809.3791} {arXiv:0809.3791 [astro-ph]} \BibitemShut {NoStop}%
\bibitem [{\citenamefont {Linder}(2005)}]{Linder:2005in}%
  \BibitemOpen
  \bibfield  {author} {\bibinfo {author} {\bibfnamefont {E.~V.}\ \bibnamefont {Linder}},\ }\href {\doibase 10.1103/PhysRevD.72.043529} {\bibfield  {journal} {\bibinfo  {journal} {Phys. Rev. D}\ }\textbf {\bibinfo {volume} {72}},\ \bibinfo {pages} {043529} (\bibinfo {year} {2005})},\ \Eprint {http://arxiv.org/abs/astro-ph/0507263} {arXiv:astro-ph/0507263} \BibitemShut {NoStop}%
\bibitem [{\citenamefont {Linder}\ and\ \citenamefont {Cahn}(2007)}]{Linder:2007hg}%
  \BibitemOpen
  \bibfield  {author} {\bibinfo {author} {\bibfnamefont {E.~V.}\ \bibnamefont {Linder}}\ and\ \bibinfo {author} {\bibfnamefont {R.~N.}\ \bibnamefont {Cahn}},\ }\href {\doibase 10.1016/j.astropartphys.2007.09.003} {\bibfield  {journal} {\bibinfo  {journal} {Astropart. Phys.}\ }\textbf {\bibinfo {volume} {28}},\ \bibinfo {pages} {481} (\bibinfo {year} {2007})},\ \Eprint {http://arxiv.org/abs/astro-ph/0701317} {arXiv:astro-ph/0701317} \BibitemShut {NoStop}%
\bibitem [{\citenamefont {Nguyen}\ \emph {et~al.}(2023)\citenamefont {Nguyen}, \citenamefont {Huterer},\ and\ \citenamefont {Wen}}]{Nguyen:2023fip}%
  \BibitemOpen
  \bibfield  {author} {\bibinfo {author} {\bibfnamefont {N.-M.}\ \bibnamefont {Nguyen}}, \bibinfo {author} {\bibfnamefont {D.}~\bibnamefont {Huterer}}, \ and\ \bibinfo {author} {\bibfnamefont {Y.}~\bibnamefont {Wen}},\ }\href {\doibase 10.1103/PhysRevLett.131.111001} {\bibfield  {journal} {\bibinfo  {journal} {Phys. Rev. Lett.}\ }\textbf {\bibinfo {volume} {131}},\ \bibinfo {pages} {111001} (\bibinfo {year} {2023})},\ \Eprint {http://arxiv.org/abs/2302.01331} {arXiv:2302.01331 [astro-ph.CO]} \BibitemShut {NoStop}%
\bibitem [{\citenamefont {Specogna}\ \emph {et~al.}(2024)\citenamefont {Specogna}, \citenamefont {Di~Valentino}, \citenamefont {Levi~Said},\ and\ \citenamefont {Nguyen}}]{Specogna:2023nkq}%
  \BibitemOpen
  \bibfield  {author} {\bibinfo {author} {\bibfnamefont {E.}~\bibnamefont {Specogna}}, \bibinfo {author} {\bibfnamefont {E.}~\bibnamefont {Di~Valentino}}, \bibinfo {author} {\bibfnamefont {J.}~\bibnamefont {Levi~Said}}, \ and\ \bibinfo {author} {\bibfnamefont {N.-M.}\ \bibnamefont {Nguyen}},\ }\href {\doibase 10.1103/PhysRevD.109.043528} {\bibfield  {journal} {\bibinfo  {journal} {Phys. Rev. D}\ }\textbf {\bibinfo {volume} {109}},\ \bibinfo {pages} {043528} (\bibinfo {year} {2024})},\ \Eprint {http://arxiv.org/abs/2305.16865} {arXiv:2305.16865 [astro-ph.CO]} \BibitemShut {NoStop}%
\bibitem [{\citenamefont {Di~Valentino}\ \emph {et~al.}(2016)\citenamefont {Di~Valentino}, \citenamefont {Melchiorri},\ and\ \citenamefont {Silk}}]{DiValentino:2015bja}%
  \BibitemOpen
  \bibfield  {author} {\bibinfo {author} {\bibfnamefont {E.}~\bibnamefont {Di~Valentino}}, \bibinfo {author} {\bibfnamefont {A.}~\bibnamefont {Melchiorri}}, \ and\ \bibinfo {author} {\bibfnamefont {J.}~\bibnamefont {Silk}},\ }\href {\doibase 10.1103/PhysRevD.93.023513} {\bibfield  {journal} {\bibinfo  {journal} {Phys. Rev. D}\ }\textbf {\bibinfo {volume} {93}},\ \bibinfo {pages} {023513} (\bibinfo {year} {2016})},\ \Eprint {http://arxiv.org/abs/1509.07501} {arXiv:1509.07501 [astro-ph.CO]} \BibitemShut {NoStop}%
\bibitem [{\citenamefont {Ade}\ \emph {et~al.}(2016{\natexlab{b}})\citenamefont {Ade} \emph {et~al.}}]{Planck:2015bue}%
  \BibitemOpen
  \bibfield  {author} {\bibinfo {author} {\bibfnamefont {P.~A.~R.}\ \bibnamefont {Ade}} \emph {et~al.} (\bibinfo {collaboration} {Planck}),\ }\href {\doibase 10.1051/0004-6361/201525814} {\bibfield  {journal} {\bibinfo  {journal} {Astron. Astrophys.}\ }\textbf {\bibinfo {volume} {594}},\ \bibinfo {pages} {A14} (\bibinfo {year} {2016}{\natexlab{b}})},\ \Eprint {http://arxiv.org/abs/1502.01590} {arXiv:1502.01590 [astro-ph.CO]} \BibitemShut {NoStop}%
\bibitem [{\citenamefont {Andrade}\ \emph {et~al.}(2024)\citenamefont {Andrade}, \citenamefont {Capistrano}, \citenamefont {Di~Valentino},\ and\ \citenamefont {Nunes}}]{Andrade:2023pws}%
  \BibitemOpen
  \bibfield  {author} {\bibinfo {author} {\bibfnamefont {U.}~\bibnamefont {Andrade}}, \bibinfo {author} {\bibfnamefont {A.~a. J.~S.}\ \bibnamefont {Capistrano}}, \bibinfo {author} {\bibfnamefont {E.}~\bibnamefont {Di~Valentino}}, \ and\ \bibinfo {author} {\bibfnamefont {R.~C.}\ \bibnamefont {Nunes}},\ }\href {\doibase 10.1093/mnras/stae402} {\bibfield  {journal} {\bibinfo  {journal} {Mon. Not. Roy. Astron. Soc.}\ }\textbf {\bibinfo {volume} {529}},\ \bibinfo {pages} {831} (\bibinfo {year} {2024})},\ \Eprint {http://arxiv.org/abs/2309.15781} {arXiv:2309.15781 [astro-ph.CO]} \BibitemShut {NoStop}%
\bibitem [{\citenamefont {Akrami}\ \emph {et~al.}(2020)\citenamefont {Akrami} \emph {et~al.}}]{Planck:2020olo}%
  \BibitemOpen
  \bibfield  {author} {\bibinfo {author} {\bibfnamefont {Y.}~\bibnamefont {Akrami}} \emph {et~al.} (\bibinfo {collaboration} {Planck}),\ }\href {\doibase 10.1051/0004-6361/202038073} {\bibfield  {journal} {\bibinfo  {journal} {Astron. Astrophys.}\ }\textbf {\bibinfo {volume} {643}},\ \bibinfo {pages} {A42} (\bibinfo {year} {2020})},\ \Eprint {http://arxiv.org/abs/2007.04997} {arXiv:2007.04997 [astro-ph.CO]} \BibitemShut {NoStop}%
\bibitem [{\citenamefont {Carron}\ \emph {et~al.}(2022)\citenamefont {Carron}, \citenamefont {Mirmelstein},\ and\ \citenamefont {Lewis}}]{Carron:2022eyg}%
  \BibitemOpen
  \bibfield  {author} {\bibinfo {author} {\bibfnamefont {J.}~\bibnamefont {Carron}}, \bibinfo {author} {\bibfnamefont {M.}~\bibnamefont {Mirmelstein}}, \ and\ \bibinfo {author} {\bibfnamefont {A.}~\bibnamefont {Lewis}},\ }\href {\doibase 10.1088/1475-7516/2022/09/039} {\bibfield  {journal} {\bibinfo  {journal} {JCAP}\ }\textbf {\bibinfo {volume} {09}},\ \bibinfo {pages} {039} (\bibinfo {year} {2022})},\ \Eprint {http://arxiv.org/abs/2206.07773} {arXiv:2206.07773 [astro-ph.CO]} \BibitemShut {NoStop}%
\bibitem [{\citenamefont {Rosenberg}\ \emph {et~al.}(2022)\citenamefont {Rosenberg}, \citenamefont {Gratton},\ and\ \citenamefont {Efstathiou}}]{Rosenberg:2022sdy}%
  \BibitemOpen
  \bibfield  {author} {\bibinfo {author} {\bibfnamefont {E.}~\bibnamefont {Rosenberg}}, \bibinfo {author} {\bibfnamefont {S.}~\bibnamefont {Gratton}}, \ and\ \bibinfo {author} {\bibfnamefont {G.}~\bibnamefont {Efstathiou}},\ }\href {\doibase 10.1093/mnras/stac2744} {\bibfield  {journal} {\bibinfo  {journal} {Mon. Not. Roy. Astron. Soc.}\ }\textbf {\bibinfo {volume} {517}},\ \bibinfo {pages} {4620} (\bibinfo {year} {2022})},\ \Eprint {http://arxiv.org/abs/2205.10869} {arXiv:2205.10869 [astro-ph.CO]} \BibitemShut {NoStop}%
\bibitem [{\citenamefont {Tristram}\ \emph {et~al.}(2024)\citenamefont {Tristram} \emph {et~al.}}]{Tristram:2023haj}%
  \BibitemOpen
  \bibfield  {author} {\bibinfo {author} {\bibfnamefont {M.}~\bibnamefont {Tristram}} \emph {et~al.},\ }\href {\doibase 10.1051/0004-6361/202348015} {\bibfield  {journal} {\bibinfo  {journal} {Astron. Astrophys.}\ }\textbf {\bibinfo {volume} {682}},\ \bibinfo {pages} {A37} (\bibinfo {year} {2024})},\ \Eprint {http://arxiv.org/abs/2309.10034} {arXiv:2309.10034 [astro-ph.CO]} \BibitemShut {NoStop}%
\bibitem [{\citenamefont {Chatrchyan}\ \emph {et~al.}(2024)\citenamefont {Chatrchyan}, \citenamefont {Niedermann}, \citenamefont {Poulin},\ and\ \citenamefont {Sloth}}]{Chatrchyan:2024xjj}%
  \BibitemOpen
  \bibfield  {author} {\bibinfo {author} {\bibfnamefont {A.}~\bibnamefont {Chatrchyan}}, \bibinfo {author} {\bibfnamefont {F.}~\bibnamefont {Niedermann}}, \bibinfo {author} {\bibfnamefont {V.}~\bibnamefont {Poulin}}, \ and\ \bibinfo {author} {\bibfnamefont {M.~S.}\ \bibnamefont {Sloth}},\ }\href@noop {} {\  (\bibinfo {year} {2024})},\ \Eprint {http://arxiv.org/abs/2408.14537} {arXiv:2408.14537 [astro-ph.CO]} \BibitemShut {NoStop}%
\bibitem [{\citenamefont {Allali}\ and\ \citenamefont {Notari}(2024)}]{Allali:2024aiv}%
  \BibitemOpen
  \bibfield  {author} {\bibinfo {author} {\bibfnamefont {I.~J.}\ \bibnamefont {Allali}}\ and\ \bibinfo {author} {\bibfnamefont {A.}~\bibnamefont {Notari}},\ }\href@noop {} {\  (\bibinfo {year} {2024})},\ \Eprint {http://arxiv.org/abs/2406.14554} {arXiv:2406.14554 [astro-ph.CO]} \BibitemShut {NoStop}%
\bibitem [{\citenamefont {Roy~Choudhury}\ and\ \citenamefont {Okumura}(2024)}]{RoyChoudhury:2024wri}%
  \BibitemOpen
  \bibfield  {author} {\bibinfo {author} {\bibfnamefont {S.}~\bibnamefont {Roy~Choudhury}}\ and\ \bibinfo {author} {\bibfnamefont {T.}~\bibnamefont {Okumura}},\ }\href@noop {} {\  (\bibinfo {year} {2024})},\ \Eprint {http://arxiv.org/abs/2409.13022} {arXiv:2409.13022 [astro-ph.CO]} \BibitemShut {NoStop}%
\bibitem [{\citenamefont {Ishak}\ \emph {et~al.}(2024)\citenamefont {Ishak} \emph {et~al.}}]{DESI:2024yrg}%
  \BibitemOpen
  \bibfield  {author} {\bibinfo {author} {\bibfnamefont {M.}~\bibnamefont {Ishak}} \emph {et~al.} (\bibinfo {collaboration} {DESI}),\ }\href@noop {} {\  (\bibinfo {year} {2024})},\ \Eprint {http://arxiv.org/abs/2411.12026} {arXiv:2411.12026 [astro-ph.CO]} \BibitemShut {NoStop}%
\bibitem [{\citenamefont {Adame}\ \emph {et~al.}(2024{\natexlab{c}})\citenamefont {Adame} \emph {et~al.}}]{DESI:2024hhd}%
  \BibitemOpen
  \bibfield  {author} {\bibinfo {author} {\bibfnamefont {A.~G.}\ \bibnamefont {Adame}} \emph {et~al.} (\bibinfo {collaboration} {DESI}),\ }\href@noop {} {\  (\bibinfo {year} {2024}{\natexlab{c}})},\ \Eprint {http://arxiv.org/abs/2411.12022} {arXiv:2411.12022 [astro-ph.CO]} \BibitemShut {NoStop}%
\bibitem [{\citenamefont {Ma}\ and\ \citenamefont {Bertschinger}(1995)}]{Ma:1995ey}%
  \BibitemOpen
  \bibfield  {author} {\bibinfo {author} {\bibfnamefont {C.-P.}\ \bibnamefont {Ma}}\ and\ \bibinfo {author} {\bibfnamefont {E.}~\bibnamefont {Bertschinger}},\ }\href {\doibase 10.1086/176550} {\bibfield  {journal} {\bibinfo  {journal} {Astrophys. J.}\ }\textbf {\bibinfo {volume} {455}},\ \bibinfo {pages} {7} (\bibinfo {year} {1995})},\ \Eprint {http://arxiv.org/abs/astro-ph/9506072} {arXiv:astro-ph/9506072} \BibitemShut {NoStop}%
\bibitem [{\citenamefont {Hojjati}\ \emph {et~al.}(2011)\citenamefont {Hojjati}, \citenamefont {Pogosian},\ and\ \citenamefont {Zhao}}]{Hojjati:2011ix}%
  \BibitemOpen
  \bibfield  {author} {\bibinfo {author} {\bibfnamefont {A.}~\bibnamefont {Hojjati}}, \bibinfo {author} {\bibfnamefont {L.}~\bibnamefont {Pogosian}}, \ and\ \bibinfo {author} {\bibfnamefont {G.-B.}\ \bibnamefont {Zhao}},\ }\href {\doibase 10.1088/1475-7516/2011/08/005} {\bibfield  {journal} {\bibinfo  {journal} {JCAP}\ }\textbf {\bibinfo {volume} {08}},\ \bibinfo {pages} {005} (\bibinfo {year} {2011})},\ \Eprint {http://arxiv.org/abs/1106.4543} {arXiv:1106.4543 [astro-ph.CO]} \BibitemShut {NoStop}%
\bibitem [{\citenamefont {Caldwell}\ \emph {et~al.}(2007)\citenamefont {Caldwell}, \citenamefont {Cooray},\ and\ \citenamefont {Melchiorri}}]{Caldwell:2007cw}%
  \BibitemOpen
  \bibfield  {author} {\bibinfo {author} {\bibfnamefont {R.}~\bibnamefont {Caldwell}}, \bibinfo {author} {\bibfnamefont {A.}~\bibnamefont {Cooray}}, \ and\ \bibinfo {author} {\bibfnamefont {A.}~\bibnamefont {Melchiorri}},\ }\href {\doibase 10.1103/PhysRevD.76.023507} {\bibfield  {journal} {\bibinfo  {journal} {Phys. Rev. D}\ }\textbf {\bibinfo {volume} {76}},\ \bibinfo {pages} {023507} (\bibinfo {year} {2007})},\ \Eprint {http://arxiv.org/abs/astro-ph/0703375} {arXiv:astro-ph/0703375} \BibitemShut {NoStop}%
\bibitem [{\citenamefont {Hu}\ and\ \citenamefont {Sawicki}(2007)}]{Hu:2007pj}%
  \BibitemOpen
  \bibfield  {author} {\bibinfo {author} {\bibfnamefont {W.}~\bibnamefont {Hu}}\ and\ \bibinfo {author} {\bibfnamefont {I.}~\bibnamefont {Sawicki}},\ }\href {\doibase 10.1103/PhysRevD.76.104043} {\bibfield  {journal} {\bibinfo  {journal} {Phys. Rev. D}\ }\textbf {\bibinfo {volume} {76}},\ \bibinfo {pages} {104043} (\bibinfo {year} {2007})},\ \Eprint {http://arxiv.org/abs/0708.1190} {arXiv:0708.1190 [astro-ph]} \BibitemShut {NoStop}%
\bibitem [{\citenamefont {Koyama}\ and\ \citenamefont {Maartens}(2006)}]{Koyama:2005kd}%
  \BibitemOpen
  \bibfield  {author} {\bibinfo {author} {\bibfnamefont {K.}~\bibnamefont {Koyama}}\ and\ \bibinfo {author} {\bibfnamefont {R.}~\bibnamefont {Maartens}},\ }\href {\doibase 10.1088/1475-7516/2006/01/016} {\bibfield  {journal} {\bibinfo  {journal} {JCAP}\ }\textbf {\bibinfo {volume} {01}},\ \bibinfo {pages} {016} (\bibinfo {year} {2006})},\ \Eprint {http://arxiv.org/abs/astro-ph/0511634} {arXiv:astro-ph/0511634} \BibitemShut {NoStop}%
\bibitem [{\citenamefont {Tsujikawa}(2007)}]{Tsujikawa:2007gd}%
  \BibitemOpen
  \bibfield  {author} {\bibinfo {author} {\bibfnamefont {S.}~\bibnamefont {Tsujikawa}},\ }\href {\doibase 10.1103/PhysRevD.76.023514} {\bibfield  {journal} {\bibinfo  {journal} {Phys. Rev. D}\ }\textbf {\bibinfo {volume} {76}},\ \bibinfo {pages} {023514} (\bibinfo {year} {2007})},\ \Eprint {http://arxiv.org/abs/0705.1032} {arXiv:0705.1032 [astro-ph]} \BibitemShut {NoStop}%
\bibitem [{\citenamefont {Bonvin}\ and\ \citenamefont {Pogosian}(2023)}]{Bonvin:2022tii}%
  \BibitemOpen
  \bibfield  {author} {\bibinfo {author} {\bibfnamefont {C.}~\bibnamefont {Bonvin}}\ and\ \bibinfo {author} {\bibfnamefont {L.}~\bibnamefont {Pogosian}},\ }\href {\doibase 10.1038/s41550-023-02003-y} {\bibfield  {journal} {\bibinfo  {journal} {Nature Astron.}\ }\textbf {\bibinfo {volume} {7}},\ \bibinfo {pages} {1127} (\bibinfo {year} {2023})},\ \Eprint {http://arxiv.org/abs/2209.03614} {arXiv:2209.03614 [astro-ph.CO]} \BibitemShut {NoStop}%
\bibitem [{\citenamefont {Wang}\ and\ \citenamefont {Steinhardt}(1998)}]{Wang:1998gt}%
  \BibitemOpen
  \bibfield  {author} {\bibinfo {author} {\bibfnamefont {L.-M.}\ \bibnamefont {Wang}}\ and\ \bibinfo {author} {\bibfnamefont {P.~J.}\ \bibnamefont {Steinhardt}},\ }\href {\doibase 10.1086/306436} {\bibfield  {journal} {\bibinfo  {journal} {Astrophys. J.}\ }\textbf {\bibinfo {volume} {508}},\ \bibinfo {pages} {483} (\bibinfo {year} {1998})},\ \Eprint {http://arxiv.org/abs/astro-ph/9804015} {arXiv:astro-ph/9804015} \BibitemShut {NoStop}%
\bibitem [{\citenamefont {Tsujikawa}\ \emph {et~al.}(2009)\citenamefont {Tsujikawa}, \citenamefont {Gannouji}, \citenamefont {Moraes},\ and\ \citenamefont {Polarski}}]{Tsujikawa:2009ku}%
  \BibitemOpen
  \bibfield  {author} {\bibinfo {author} {\bibfnamefont {S.}~\bibnamefont {Tsujikawa}}, \bibinfo {author} {\bibfnamefont {R.}~\bibnamefont {Gannouji}}, \bibinfo {author} {\bibfnamefont {B.}~\bibnamefont {Moraes}}, \ and\ \bibinfo {author} {\bibfnamefont {D.}~\bibnamefont {Polarski}},\ }\href {\doibase 10.1103/PhysRevD.80.084044} {\bibfield  {journal} {\bibinfo  {journal} {Phys. Rev. D}\ }\textbf {\bibinfo {volume} {80}},\ \bibinfo {pages} {084044} (\bibinfo {year} {2009})},\ \Eprint {http://arxiv.org/abs/0908.2669} {arXiv:0908.2669 [astro-ph.CO]} \BibitemShut {NoStop}%
\bibitem [{\citenamefont {Gannouji}\ \emph {et~al.}(2009)\citenamefont {Gannouji}, \citenamefont {Moraes},\ and\ \citenamefont {Polarski}}]{Gannouji:2008wt}%
  \BibitemOpen
  \bibfield  {author} {\bibinfo {author} {\bibfnamefont {R.}~\bibnamefont {Gannouji}}, \bibinfo {author} {\bibfnamefont {B.}~\bibnamefont {Moraes}}, \ and\ \bibinfo {author} {\bibfnamefont {D.}~\bibnamefont {Polarski}},\ }\href {\doibase 10.1088/1475-7516/2009/02/034} {\bibfield  {journal} {\bibinfo  {journal} {JCAP}\ }\textbf {\bibinfo {volume} {02}},\ \bibinfo {pages} {034} (\bibinfo {year} {2009})},\ \Eprint {http://arxiv.org/abs/0809.3374} {arXiv:0809.3374 [astro-ph]} \BibitemShut {NoStop}%
\bibitem [{\citenamefont {Torrado}\ and\ \citenamefont {Lewis}(2021)}]{Torrado:2020dgo}%
  \BibitemOpen
  \bibfield  {author} {\bibinfo {author} {\bibfnamefont {J.}~\bibnamefont {Torrado}}\ and\ \bibinfo {author} {\bibfnamefont {A.}~\bibnamefont {Lewis}},\ }\href {\doibase 10.1088/1475-7516/2021/05/057} {\bibfield  {journal} {\bibinfo  {journal} {JCAP}\ }\textbf {\bibinfo {volume} {05}},\ \bibinfo {pages} {057} (\bibinfo {year} {2021})},\ \Eprint {http://arxiv.org/abs/2005.05290} {arXiv:2005.05290 [astro-ph.IM]} \BibitemShut {NoStop}%
\bibitem [{\citenamefont {Zucca}\ \emph {et~al.}(2019)\citenamefont {Zucca}, \citenamefont {Pogosian}, \citenamefont {Silvestri},\ and\ \citenamefont {Zhao}}]{Zucca:2019xhg}%
  \BibitemOpen
  \bibfield  {author} {\bibinfo {author} {\bibfnamefont {A.}~\bibnamefont {Zucca}}, \bibinfo {author} {\bibfnamefont {L.}~\bibnamefont {Pogosian}}, \bibinfo {author} {\bibfnamefont {A.}~\bibnamefont {Silvestri}}, \ and\ \bibinfo {author} {\bibfnamefont {G.-B.}\ \bibnamefont {Zhao}},\ }\href {\doibase 10.1088/1475-7516/2019/05/001} {\bibfield  {journal} {\bibinfo  {journal} {JCAP}\ }\textbf {\bibinfo {volume} {05}},\ \bibinfo {pages} {001} (\bibinfo {year} {2019})},\ \Eprint {http://arxiv.org/abs/1901.05956} {arXiv:1901.05956 [astro-ph.CO]} \BibitemShut {NoStop}%
\bibitem [{\citenamefont {Wang}\ \emph {et~al.}(2023)\citenamefont {Wang}, \citenamefont {Mirpoorian}, \citenamefont {Pogosian}, \citenamefont {Silvestri},\ and\ \citenamefont {Zhao}}]{Wang:2023tjj}%
  \BibitemOpen
  \bibfield  {author} {\bibinfo {author} {\bibfnamefont {Z.}~\bibnamefont {Wang}}, \bibinfo {author} {\bibfnamefont {S.~H.}\ \bibnamefont {Mirpoorian}}, \bibinfo {author} {\bibfnamefont {L.}~\bibnamefont {Pogosian}}, \bibinfo {author} {\bibfnamefont {A.}~\bibnamefont {Silvestri}}, \ and\ \bibinfo {author} {\bibfnamefont {G.-B.}\ \bibnamefont {Zhao}},\ }\href {\doibase 10.1088/1475-7516/2023/08/038} {\bibfield  {journal} {\bibinfo  {journal} {JCAP}\ }\textbf {\bibinfo {volume} {08}},\ \bibinfo {pages} {038} (\bibinfo {year} {2023})},\ \Eprint {http://arxiv.org/abs/2305.05667} {arXiv:2305.05667 [astro-ph.CO]} \BibitemShut {NoStop}%
\bibitem [{\citenamefont {Gelman}\ and\ \citenamefont {Rubin}(1992)}]{Gelman:1992zz}%
  \BibitemOpen
  \bibfield  {author} {\bibinfo {author} {\bibfnamefont {A.}~\bibnamefont {Gelman}}\ and\ \bibinfo {author} {\bibfnamefont {D.~B.}\ \bibnamefont {Rubin}},\ }\href {\doibase 10.1214/ss/1177011136} {\bibfield  {journal} {\bibinfo  {journal} {Statist. Sci.}\ }\textbf {\bibinfo {volume} {7}},\ \bibinfo {pages} {457} (\bibinfo {year} {1992})}\BibitemShut {NoStop}%
\bibitem [{\citenamefont {Lewis}(2019)}]{Lewis:2019xzd}%
  \BibitemOpen
  \bibfield  {author} {\bibinfo {author} {\bibfnamefont {A.}~\bibnamefont {Lewis}},\ }\href@noop {} {\  (\bibinfo {year} {2019})},\ \Eprint {http://arxiv.org/abs/1910.13970} {arXiv:1910.13970 [astro-ph.IM]} \BibitemShut {NoStop}%
\bibitem [{\citenamefont {Pogosian}\ \emph {et~al.}(2022)\citenamefont {Pogosian}, \citenamefont {Raveri}, \citenamefont {Koyama}, \citenamefont {Martinelli}, \citenamefont {Silvestri}, \citenamefont {Zhao}, \citenamefont {Li}, \citenamefont {Peirone},\ and\ \citenamefont {Zucca}}]{Pogosian:2021mcs}%
  \BibitemOpen
  \bibfield  {author} {\bibinfo {author} {\bibfnamefont {L.}~\bibnamefont {Pogosian}}, \bibinfo {author} {\bibfnamefont {M.}~\bibnamefont {Raveri}}, \bibinfo {author} {\bibfnamefont {K.}~\bibnamefont {Koyama}}, \bibinfo {author} {\bibfnamefont {M.}~\bibnamefont {Martinelli}}, \bibinfo {author} {\bibfnamefont {A.}~\bibnamefont {Silvestri}}, \bibinfo {author} {\bibfnamefont {G.-B.}\ \bibnamefont {Zhao}}, \bibinfo {author} {\bibfnamefont {J.}~\bibnamefont {Li}}, \bibinfo {author} {\bibfnamefont {S.}~\bibnamefont {Peirone}}, \ and\ \bibinfo {author} {\bibfnamefont {A.}~\bibnamefont {Zucca}},\ }\href {\doibase 10.1038/s41550-022-01808-7} {\bibfield  {journal} {\bibinfo  {journal} {Nature Astron.}\ }\textbf {\bibinfo {volume} {6}},\ \bibinfo {pages} {1484} (\bibinfo {year} {2022})},\ \Eprint {http://arxiv.org/abs/2107.12992} {arXiv:2107.12992 [astro-ph.CO]} \BibitemShut {NoStop}%
\end{thebibliography}%

 \end{document}